\documentstyle[12pt,a4,psfig,here]{book}
\newcommand{\be}{\begin{equation}}
\newcommand{\ee}{\end{equation}}
\newcommand{\bea}{\begin{eqnarray}}
\newcommand{\eea}{\end{eqnarray}}
\newcommand{\bean}{\begin{eqnarray*}}
\newcommand{\eean}{\end{eqnarray*}}
\newcommand{\ba}{\begin{array}}
\newcommand{\ea}{\end{array}}

\newcommand{\slashs}[1]{\not{\!#1}}
\newcommand{\bra}[1]{\left< #1 \right|}
\newcommand{\ket}[1]{\left| #1 \right>}
\newcommand{\mbk}[1]{\left< #1 \right>}
\newcommand{\pbk}[1]{\left[ #1 \right]}
\def\Li{{\rm Li}_2}
\def\Tr{{\rm Tr}}
\def\J#1{J_{#1}}
\def\N#1{N_{#1}}
\def\L#1{L_{#1}}
\def\K#1{K_{#1}}
\def\uint{\int dy\,dz\:\:}
\def\pint{\int{dy\,dz\over\sqrt{(1-y)^2-a}}\:\:}
\def\tint{\int{dy\,dz \over (1-y)^2-a}\:\:}
\def\kint{\int{dy\,dz\over\{(1-y)^2-a\}^{3/2}}\:\:}
\begin{document}
%-------------------- title page -----------------------------
\begin{titlepage}

\vspace*{4cm}

\begin{center}
{\huge \bf Spin Correlations}\\ 
{\huge \bf in Top Quark Production}\\
{\huge \bf at \boldmath{$e^+ e^-$} Linear Colliders}
\end{center}

\vspace{3cm}

\begin{center}
{\large \bf \sl Takashi Nasuno}
\end{center}
\vspace{1cm}

\begin{center}
{\large \bf \it Dept. of Physics, Hiroshima University}\\
{\large \bf \it Higashi-Hiroshima 739-8526, Japan} 
\end{center}

\vspace{2cm}

\begin{center}
%{\large \it \today}
{\large \it March, $1999$, Thesis}
\end{center}
\end{titlepage}
\clearpage
%--------------------------------------------------------------
\baselineskip 21pt
\pagenumbering{roman}
\setcounter{page}{2}
%------------------------------------------------------------------
%\input ack.tex
\begin{center}
{\LARGE \bf ACKNOWLEDGMENT}
\end{center}

\vspace{2cm}

It is a great pleasure to thank the many people
who have contributed to the completion of this thesis, through their
knowledge, guidance, support and friendship.
First of all, I would like to thank my thesis supervisor, Jiro Kodaira,
for many discussions, his assistances and enlightening comments on my
works over years.
My understanding of physics have been improved by his enthusiasm and practical
viewpoints.
Moreover he gave me many chances to meet impressive physicists and
opportunities to discuss with them. 
First, thanks to his supports, I could take part in
the 24th SLAC Summer Institute (SSI) on Particle Physics, August 19-30,
1996, where I studied new aspects of physics and felt the passion of many
physicists.
After SSI, I had a nice meet with Stephen Parke and got a chance to
collaborate with him. 
Second, the International Symposium on
``QCD Corrections and New Physics'', he held at Hiroshima in 1997,
led me to deep understanding of physics which became a base of this
thesis.
Again I would like to express my heartfelt thanks to Jiro Kodaira.
 
Stephen Parke constantly stimulated my interest, not only in top quark physics,
but in many other branches of physics, 
and kindly provided me with modern calculations of the polarized $t
\bar{t}$ cross-sections.
His supports are greatly acknowledged from my heart.
%Moreover I would like to thank him.
%Since I am not good at speaking English, I guess my English made him
%trouble.
%Nevertheless, he always listened to me without bothering my talk.
%I will never forget his kindness.

I wish to express my gratitude to Haruo Ui, Taizo Muta and Michihiro
Hirata for many valuable instructions.
Their supports encouraged me and made me possible to complete this thesis.
I also benefited from discussions with Tetsuya Onogi, Takuya Morozumi,
Yuichiro Kiyo and Michihiro Hori.
Tetsuya Onogi and Takuya Morozumi helped me understand the implications of 
top quark physics.
And I would like to thank Yuichiro Kiyo and Michihiro Hori for their
friendship and the countless discussions till far into the night.
They were good collaborators for me to get a new understanding of physics.
I also thank Hideko Masuda and Mami Kohno for their assistances.
Without their help, I probably would not spend a good time in Hiroshima
University.
I would like to express my deepest gratitude to Tsuneo Uematsu in Kyoto
University and Kazuhiro Tanaka in Juntendo University.  
I was always encouraged by their physics and the discussion with them.

Finally in my personal side, I would like to thank my parents and my
sister for their continuous supports and encouragements which made me possible
to continue my researches.
\newpage
%----------------------- abstract & words ---------------------
%\input abst.tex
\begin{center}
{\LARGE \bf ABSTRACT}
\end{center}

\vspace{2cm}
The object of this thesis is to study the spin-spin correlations for top
quark productions and their decay at polarized $e^+e^-$ linear colliders. 
For the analyses of the spin-spin correlations in top quark production,
we introduce a generic spin basis and show which spin basis is the
most optimal.
We discuss the spin-angular correlations for the top quark decay, and 
demonstrate that the directions of electro-weak decay products of
polarized top quarks are strongly correlated to the top quark spin.
We also explain how to measure the top quark spin using decay products. 
The spin correlations using the generic spin basis are extensively
studied at the next to leading order of QCD for top quark productions at
$e^+e^-$ colliders.  
The radiative corrections induce an anomalous $\gamma/Z$ magnetic moment
for the top quarks and allow for real gluon emissions.
Therefore they might change the leading order analyses.
We show that higher order effects on the top quark spin orientation are
very small. 
The final results are that the top (or anti-top) quarks are produced
in an essentially unique spin configuration for a particular spin basis
even after including the ${\cal O}( \alpha_{s} )$ QCD corrections.
Our results imply that measuring the spin correlation for top quarks is
very promising to reveal new aspects of the Standard Model, as well as
new physics beyond the Standard Model.
\newpage
%----------------------------------------------------------------
%\input words.tex
\vspace*{18cm}

{\it 
\hspace*{4cm} ``Truth emerges more readily from error than confusion.'' \\
\hspace*{10cm}                                    -Francis Bacon-
} 

\newpage
%----------------------- contents -----------------------------
\tableofcontents
\listoffigures
\listoftables
%----------------------- text ---------------------------------
\newpage
\pagenumbering{arabic}
\setcounter{page}{1}
\baselineskip 20pt
%----------------------- introduction ---------------------------------
%\input intro.tex
\chapter{Introduction}

High-energy $e^{+} e^{-}$ colliders have been essential instruments to
search for the fundamental constituents of matter and their interactions. 
Merged with the experimental observations at hadron colliders, the
structures and properties of the ``fundamental''particles are adequately
described by the Standard Model.
The matter particles, which are leptons and quarks, can be classified
into three generations with identical symmetries.
It has been known that the electro-weak and strong forces are described
by gauge field theories.
The strong force is described by the ${\rm SU}(3)_{c}$ gauge theory,
which is called Quantum Chromo Dynamics (QCD)~\cite{QCD}.
The vector bosons of this model are grouped into eight colored massless
gluons which mediate the strong interactions.
The QCD has only one arbitrary or free parameter, $\Lambda_{QCD}$.
Nevertheless, it explains, all hadronic dynamics including confinement,
asymptotic freedom, nucleon structure, and baryon and meson
spectroscopy. 
On the other hand, the electro-weak theory has many arbitrary parameters and
is described with the gauge group of ${\rm SU}(2)_{L} \otimes {\rm
U}(1)_{Y}$~\cite{EW}, where ${\rm SU}(2)_{L}$ is the weak isospin group, acting
on left-handed fermions, and ${\rm U}(1)_{Y}$ is the hypercharge group.
At ``low'' energy (at the scale $\sim~250$ GeV), the ${\rm
SU}(2)_{L} \otimes {\rm U}(1)_{Y}$ symmetry is spontaneously broken.
This mechanism is referred to as Higgs mechanism~\cite{HIGGS}, which
induces mass of gauge bosons, $W^{\pm},Z$, and other fermions. 
The heavy gauge bosons, $W^{\pm},Z$, were discovered at CERN in
the middle of 1980's.
Thanks to these impressive successes, the gauge theory with the group
${\rm SU}(3)_{c} \otimes {\rm SU}(2)_{L} \otimes {\rm U}(1)_{Y}$
has been called the ``Standard Model''. 

The Standard Model has been tested and checked
through precision experiments with remarkable accuracy during the past
few decades.
With those numerous data, the Standard Model has been successful in
describing the properties of particles and the structure of their basic
interactions.  
Despite this success, however, the Standard Model has not been
considered as the final theory of nature.
As a matter of fact, the Higgs mechanism is still hypothetical and has
not been established experimentally.  
The properties of the latest member of the matter, top quark, have not
yet been fully revealed. 
To make a further progress and to step on to a new stage, we must investigate
both the ``strengths'' and the ``weaknesses'' of the Standard Model and
must understand it in detail.
There are two complementary and sometimes mutually related strategies to
explore the new aspects of the Standard Model and to reveal signals
of new physics beyond it. 
The first is to look for ``new'' phenomena at high energy collider
facilities, the $e^+ e^-$ colliders LEP II, the $e p$ 
colliders HERA, $p \bar{p}$ colliders TEVATRON, the $p p$ colliders LHC
and $e^+ e^-$ linear colliders JLC.
The high energy colliders bring us a good chance to have a clue to new
physics.
The second is to utilize good probes, which might be Higgs particle and top
quarks.
Precision studies of these heavy particles also may reveal new aspects
of the Standard Model and clues to the physics beyond the Standard
Model, since these particles are considered to have important
information on the mechanism of breaking of the electro-weak symmetry.
In this thesis, we will focus on the top quark as a fascinating probe.
A high energy $e^+e^-$ linear collider provides a very impressive tool
to carry out detailed studies of top quarks.

The first stage of top quark physics was opened with its discovery
at FNAL~\cite{CDF,D0}.
The top quark with a very large mass $173.8 \pm 5.2$ GeV~\cite{PDG}
makes us expect that the top quark plays a special role in particle
physics.
It might contain a hint of new physics beyond the Standard Model.
The presence of Higgs boson and its Yukawa coupling to fermions play a
fundamental role in the Standard Model,
and the Yukawa coupling is proportion to the fermion mass. 
Therefore the top quark will likely offer a possibility for its direct
measurement, which has not yet been established experimentally. 
Among many other possibilities, for instance, T or CP violation in the
top quark sector and the possibility of the presence of anomalous
couplings have been discussed in many papers~\cite{SEE}.
The very heavy top quark will provide us with a unique opportunity to
understand nature deeply.

Now, the following fact is typical to heavy top quarks that the top
quark rapidly decays through electro-weak interactions before
hadronization effects~\cite{k,bdkkz} come in.
Therefore there are significant angular correlation between the decay
products of the top quark and the spin of the top quark~\cite{Decays}. 
This implies that if the produced top and anti-top quarks correlate
their spins, there will be a sizable angular correlations between all
the particles, both incoming and outgoing, in these events.
There are many papers on the angular correlations for top
quark events~\cite{bop,kly} produced at both $e^+ e^-$
colliders~\cite{eecollider} and hadron colliders~\cite{hcollider}.
In most of these works, the top quark spin is decomposed in the
helicity basis. 
For ultra-relativistic particles, this approach is appropriate.
However, in general, 
the helicity basis is not a unique choice for massive particles.
It may depend on the energy (speed) of the produced quark which
spin basis is the most optimal to study the spin correlations.
Mahlon and Parke~\cite{mp} have considered various decompositions of
the top quark spin which result in a large asymmetry at hadron
colliders. 
Parke and Shadmi~\cite{ps} extended this study to the $e^+ e^-$ annihilation
process at the leading order in the perturbation theory and found that
the \lq\lq off-diagonal\rq\rq\ basis is the most efficient decomposition
of the top (anti-top) quark spin. 
In this spin basis, the top quarks are produced in an essentially unique
spin configuration.
This result is of great interest and importance from both theoretical
and phenomenological viewpoints.
However, it is of crucial importance to
estimate the radiative corrections to this process which are dominated
by QCD.
The radiative corrections, in general, add two effects to the leading
order analysis: the first is that a new vertex structure (anomalous
$\gamma/Z$ magnetic moment) is induced by the loop corrections to the
tree level vertex, the second is that a (hard) real gluon emission from
the final quarks can flip the spin and change the momentum of parent
quarks.
Therefore, compared to the radiative corrections to physical quantities
which are spin independent,
it is possible that spin-dependent quantities may be particularly
sensitive to the effects of QCD radiative corrections.
The analytical study of QCD radiative corrections to heavy
quark production was pioneered in Ref.~\cite{gnt}
(see {\it e.g.} Ref.~\cite{gkl} for a recent article).
Polarized heavy quark productions, in helicity basis, have been calculated
by many authors\cite{s,tung}.
Recently, the polarized cross-sections for not only longitudinally but also
transversely polarized heavy quarks have been investigated~\cite{TRS} at
the next to leading order of QCD corrections.
Meanwhile, polarization phenomena in top quark
productions near threshold have also been analyzed~\cite{THRES}.

The main part of this thesis is a discussion on the polarized top quark
production at the QCD one-loop level.
We focus on the issue of what is the optimal decomposition of the
top quark spin for  $e^+ e^-$ colliders.
(The physics of top quark production at muon colliders and $e^+e^-$
colliders is identical provided the energy is not tuned to the Higgs boson
resonance.)
We calculate the cross-section in a \lq\lq generic\rq\rq\ spin
basis which includes the helicity basis as a special case.

The thesis is organized as follows. 
In Section 2, we review some techniques for calculating helicity
amplitudes, since these techniques are useful and effective to
investigate, in particular, the spin dependent amplitudes. 
One can see that these calculational methods are indispensable techniques
to analyze the spin correlations.
In Section 3, we discuss the spin-spin correlations in top quark pair
productions and the spin-angular correlations in the decay process of top 
quarks at the leading order in perturbation theory.
The measurement of top quark spin is also discussed.
We demonstrate that top quark pairs are produced in an essentially
unique spin configuration in polarized $e^+ e^-$ linear colliders,
and that the angular distribution of top quark decay products is
strongly correlated to the direction of top quark spin axis.
 In section 4, we examine the QCD corrections to the polarized top
(anti-top) quark production, 
and present our analytic calculations of one-loop
corrections to the polarized top quark production in a generic spin basis. 
We also give the numerical results both in the helicity, beam-line and the
off-diagonal bases.  
We compare the full one-loop results with those in the leading order and
soft gluon approximations. 
The final chapter contains a summary.
In Appendices, the notations, conventions and some useful formulae for
the helicity amplitude we use in this thesis are included.
The phase space integrals which are needed in Section 4 are summarized.
The unpolarized total cross-section for top pair production, using our
results, is also given as a cross check. 

\newpage
%----------------------- spinor helicity ------------------------------
%\input hel-amp.tex
\chapter{Spinor Helicity Method}
%
%%%%%%%%%%%%%%%%%%%%%%%%%%%%%%%%%%%%%%%%%%%%%%%%%%%%%%%%
%    History & usefulness
%%%%%%%%%%%%%%%%%%%%%%%%%%%%%%%%%%%%%%%%%%%%%%%%%%%%%%%%
\indent
With the progress of precision in experimental data,
we must calculate higher order Feynman diagrams in the
perturbation theory in order to refine the theoretical predictions. 
For scattering process with more than two particles in the final state,
the spin projection operator methods quickly become unwieldy.
This difficulty is overcome by employing the spinor helicity method, 
which is a powerful technique for computing helicity amplitudes for
multi particle process involving spin $1/2$ and spin $1$ particles.

The technique of helicity amplitudes in the high-energy limit (which
corresponds to massless limit) was pioneered in the paper by Bjorken and
Chen~\cite{BJ}, and later was extended to the massless gauge theories by
the CALKUL collaboration~\cite{CC}.
Xu, Zhang and Chang~\cite{XZC} had proposed a new definition of the
polarization vector of the gauge boson based on the helicity spinor basis and
the decomposition of amplitudes into gauge-invariant subsets.
They generalized the helicity amplitude techniques developed by the
CALKUL collaboration to the non-abelian case. 
Meanwhile, a technique~\cite{SUSY}, which utilizes the super symmetry
connecting vector particle scattering with scalar particle scattering,
was suggested, and the combined use of these techniques was shown to be
more advantageous by Parke and Taylor.
The helicity amplitude method, which is useful in the calculation of
cross-sections for processes with many final state particles, is
reviewed in the article~\cite{PARKE1} by Parke and Mangano.
After these developments, these techniques for massless particles
were extended to the case of massive particles by Kleiss and Stirling
and other authors~\cite{MASSIVE}. 
This useful extension makes us possible to analyze and calculate the
processes involving massive particles more easily.

These techniques enable us to obtain easily analytically compact
expressions for various amplitudes, so they are extremely useful for the
analysis of the process we are interested in.
We use these techniques for top quark pair productions and their
decay.     
In this chapter, we present a brief introduction to the spinor
helicity method following Ref.~\cite{PARKE1}.
\section{Massless Spinor States \label{sec:massless-sp}}

In this section, we derive and summarize expressions for the spinor
products~\cite{XZC,PARKE1,HABER} for massless fermions.
The spinor helicity method utilizing the spinor products is extremely
efficient for QCD with only massless quarks, 
and the massless case is a basis for the extension to the massive 
spinor case.
We demonstrate an example using this technique in the next
section.
%In this case, these techniques is a ideal for QCD where light quarks 
%almost can be neglected, and is a basis for the extension to the massive 
%spinor.
The extension of this technique to massive fermions is discussed in
Sec.~{\ref{sec:MASSIVE}}.

We will begin with massless helicity spinors and denote their 
chiral projection by,
\be
\begin{array}{rcl}
\ket{p \pm} & \equiv & u_{\pm}(p) 
                    ~ = ~ v_{\mp}(p)
                    ~ = ~ \frac{1 \pm \gamma_{5}}{2} u(p), \\
\bra{p \pm} & \equiv & \bar{u}_{\pm}(p)
                    ~ = ~ \bar{v}_{\mp}(p). 
\end{array}
\ee
\noindent
These massless spinors have the following properties.
\begin{enumerate}
\item
Dirac Equation:
\be
\hat{p} \ket{p \pm} ~=~ \bra{p \pm} \hat{p} ~=~ 0,
(\hat{p} \equiv \gamma^{\mu} p_{\mu}).
\ee
\item
Chirality Conditions:
\be
\begin{array}{lcr}
(1 \pm \gamma_{5}) \ket{p \mp} & = & 0,\\
\bra{p \pm} (1 \pm \gamma_{5}) & = & 0.
\end{array}
\ee
\item
Normalization and Completeness:
\bea
\bra{p \pm} \gamma^{\mu} \ket{p \pm} &=& 2 p^{\mu}, \\
\ket{p +}\bra{p +} + \ket{p -}\bra{p -} &=& \hat{p}. 
\eea
\item
Massless Spinor Relations:
\bea
\ket{p \pm} \bra{p \pm} &=& \frac{1 \pm \gamma_{5}}{2} \hat{p}, 
\label{eqn:rel1}\\
\mbk{p + |q +} &=& \mbk{p -|q -} = 0, 
\label{eqn:rel2}\\
\mbk{p - |q +} &=& - \mbk{q -|p +}, 
\label{eqn:rel3}\\
\mbk{p - |p +} &=&  \mbk{p +|p -} = 0.
\label{eqn:rel4} 
\eea
\item
Chisholm Identity:
\bea
\bra{p \pm} \gamma^{\mu} \ket{q \pm} 
\left[\gamma_{\mu} \right] =   
2[ \ket{q \pm} \bra{p \pm} +  \ket{p \mp} \bra{q \mp}].
%, \\
%\mbk{A + |\gamma^{\mu}|B +}
%\mbk{C - |\gamma_{\mu}|D -} &=&
%\mbk{A + |D -} \mbk{C - |B +}.
\eea
\end{enumerate}
Equations (\ref{eqn:rel2})$\sim$(\ref{eqn:rel4}) show that most
of spinor products vanish, and only a few spinor products 
become non-zero.
Therefore, it is convenient to introduce the following notation for spinor
products:
\bea
\begin{array}{lcccr}
\mbk{p q} & \equiv & \mbk{p -|q +} & = & - \mbk{q p}, \\
\pbk{p q} & \equiv & \mbk{p +|q -} & = & - \pbk{q p}. 
\end{array}
\eea
These spinor products satisfy
\bea
\mbk{p q}^* = - \pbk{p q}, 
\eea
where all spinors are assumed to have positive energy.
The squared spinor products become 
\bea
\left| \mbk{p q} \right|^2 ~=~ \left| \pbk{p q}\right|^2 
                           ~=~ 2 p \cdot q. 
\eea
The spinor product $\mbk{p~q}$ plays a fundamental role in this paper. 
Other properties of the spinor products are summarized in the Appendix
D.
\section{Circular Polarizations of Massless Bosons 
\label{sec:massless-gb}}

We shall explain how to construct the polarization vector of massless
gauge bosons by employing the spinor techniques~\cite{CC,PARKE1,HABER}.
The polarization vectors, $\varepsilon_{\pm}^{\mu}(k)$,
corresponding to states of definite helicity, satisfy,
\bea
\varepsilon_{\pm}(k) \cdot k 
                        & = & 0  
\label{eqn:pv1}, \\
\varepsilon_{\mp}^{\mu}(k) 
                        & = & [\varepsilon_{\pm}^{\mu}(k) ]^*, 
\label{eqn:pv2} \\
\varepsilon_{\lambda}(k) \cdot \varepsilon_{\lambda'}^{*}(k)  
                        & = & - \delta_{\lambda \lambda'} 
\label{eqn:pv3},
\eea
where $k^{\mu}$ is the momentum of massless gauge boson, and
the subscripts $\lambda$, $\lambda'$ denote the helicity states
($\pm$).

In the four dimensions, the physical Hilbert space of massless vector is
isomorphic to the physical Hilbert space of massless spinor, since they
both lie in the one-dimensional representations of SO(2).
This isomorphism is realized through a linear transformation which
relates the vector states to like-helicity fermion states:
\bea
\begin{array}{lcl}
\varepsilon_{-}^{\mu}(k) & = & N \bra{k +} \gamma^{\mu} \ket{p +}, \\
\varepsilon_{+}^{\mu}(k) & = & N^* \bra{k -} \gamma^{\mu} \ket{p -}, 
\end{array}
\eea
where $\varepsilon_{\mu}^{\pm}(k)$ is the polarization vector of an
incoming massless gauge boson, the momentum $p$ is a light-like arbitrary 
reference momentum and assumed to be not parallel to the momentum $k$.
The factor $N$ is a normalization constant. 
The factor $N$ is determined from Eqn.(\ref{eqn:pv3}):
\bea
\varepsilon^{-}(k) \cdot \varepsilon^{+}(k) 
   ~=~  - 2 |N|^2 p \cdot k = -1. 
\eea
Therefore, the polarization vectors of massless gauge bosons are given by
\bea
\varepsilon^{\mu}_{-}(k) &=& \frac{e^{i \phi(k,p)}}{\sqrt{2}} 
      \frac{ \bra{k +} \gamma^{\mu} \ket{p +}}
           { \mbk{p k}}, \\
\varepsilon^{\mu}_{+}(k) &=& - \frac{e^{- i \phi(k,p)}}{\sqrt{2}} 
      \frac{ \bra{k -} \gamma^{\mu} \ket{p -}}
           { \pbk{p k}},
\eea
where $e^{i \phi(k,p)}$ is a phase factor which depends on the
momentum $k$, and the reference momentum $p$.
In this thesis, we set this phase factor unity and we obtain the
polarization vector:
\bea
\varepsilon^{\mu}_{\mp}(k) ~=~ \mp
      \frac{1}{\sqrt{2}} 
      \frac{ \bra{k \mp} \gamma^{\mu} \ket{p \mp}}
           { \mbk{p \pm| k \mp}}.
\label{eqn:mlvb}
\eea  
In Eqn. (\ref{eqn:mlvb}), the reference momentum $p$ can be chosen
freely in a convenient way because of gauge invariance of the theory.
The proof proceeds in the following way.
By changing the reference momentum $p$ in Eqn.(\ref{eqn:mlvb}) to
another reference momentum $q$, we have
\bea
\varepsilon_{\pm}^{\mu}(k,p) 
   &=& \varepsilon_{\pm}^{\mu,q} + \beta_{\pm}(p,q,k) k^{\mu} ,
\label{eqn:trans} \\
\beta_{-}(p,q,k) &=& \frac{\sqrt{2} \mbk{q p}}
                   {\mbk{q k} \mbk{p k}}, \\
\beta_{+}(p,q,k) &=& (\beta_{+})^*.
\eea
When the polarization vector $\varepsilon_{\pm}^{\mu}(k)$ is replaced by
its momentum $k$, the scattering amplitude ${\cal M}$ should vanish:
\bea
{\cal M} ~=~ {\cal M}_{\mu} \varepsilon_{\pm}^{\mu} ~\rightarrow~
      {\cal M}_{\mu} k^{\mu} ~=~ 0.
\eea
Therefore the second term in the Eqn.(\ref{eqn:trans}) does not
contribute to the helicity amplitude.
This implies that we can choose a different reference momentum
$p$ for each polarization vector in the process,
and furthermore different reference momentum for each gauge invariant subset of
the full amplitude, without worrying about the relative phase.

\begin{figure}[H]
\begin{center}
\begin{tabular}{cc}
\leavevmode\psfig{file=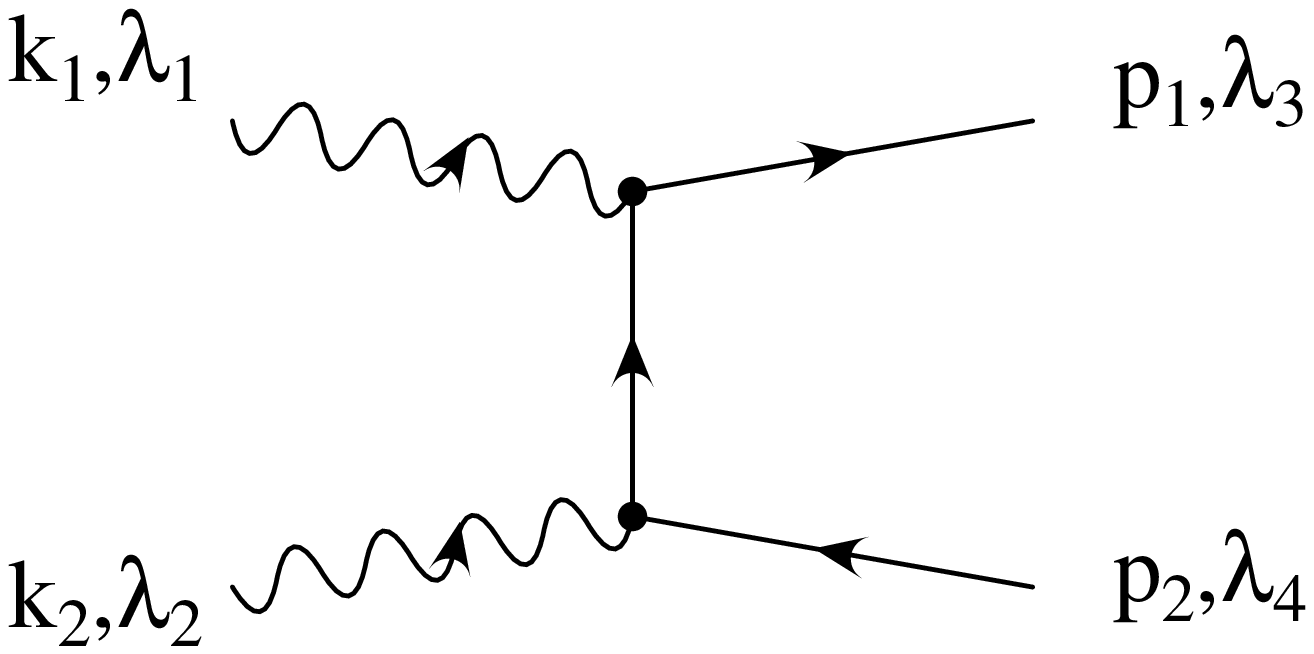,width=6.5cm} &
\leavevmode\psfig{file=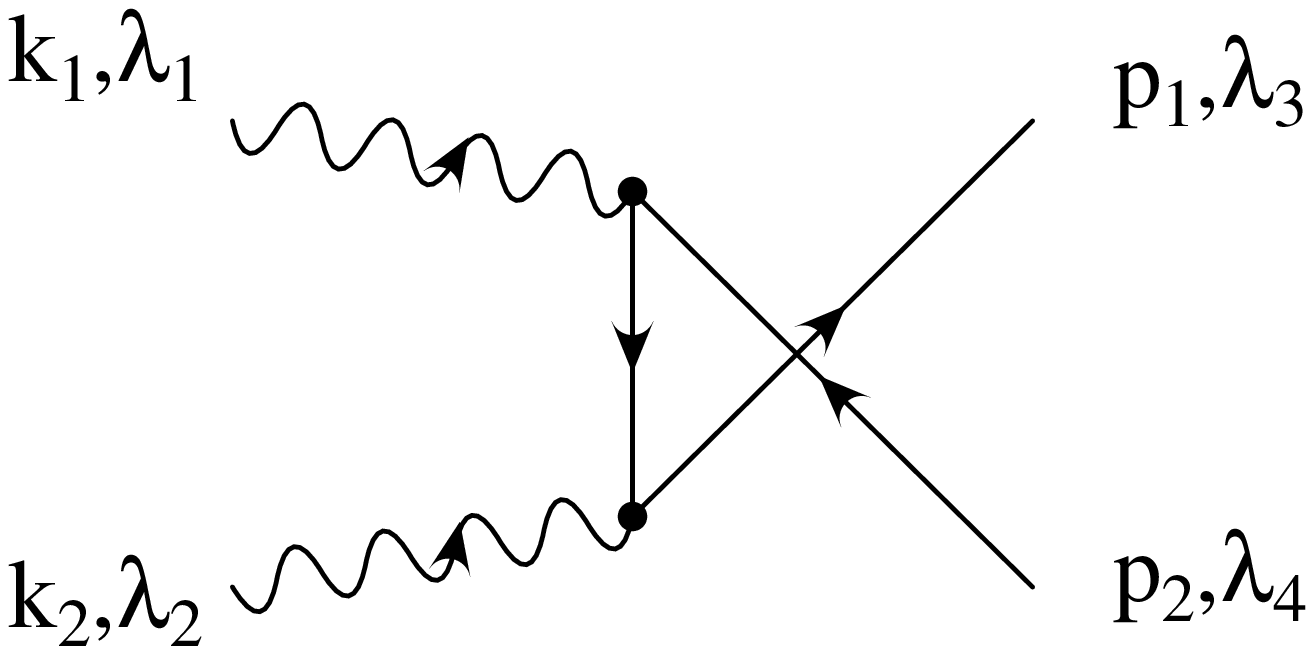,width=6.5cm} \\
(a) & (b)
\end{tabular}
\caption[The tree level contributions to the $\gamma \gamma \to q \bar{q}$
 process]{The tree level contributions to the $\gamma \gamma \to q \bar{q}$
 process}
\label{fig:ex-process}
\end{center}
\end{figure}

Now we show an application of these techniques to a real
process~\cite{PARKE1,HABER}. 
The process we consider as an example is the photon-photon scattering process:
\bea
\gamma (k_1,\lambda_1) + \gamma (k_2,\lambda_2) 
~\rightarrow~ q (p_1,\lambda_3) + \bar{q}(p_2,\lambda_4),
\nonumber 
\eea
where $k_1,k_2$ are the momenta of real photons, $p_1,p_2$ are the
momenta of final massless quarks, and $\lambda_{i}, (i=1,2,3,4)$ denotes 
the helicity state ($\pm$).
Two diagrams contribute to this process at the leading order,
$t$ channel (Fig.\ref{fig:ex-process} (a)) and $u$ channel fermion
(Fig.\ref{fig:ex-process} (b)) exchanges.
% Fig
The corresponding amplitudes are 
\bea
{\cal M}(\lambda_3,\lambda_4;\lambda_1,\lambda_2) 
&=& 
{\cal M}_t(\lambda_3,\lambda_4;\lambda_1,\lambda_2) +
{\cal M}_u(\lambda_3,\lambda_4;\lambda_1,\lambda_2)~,\\ 
{\cal M}_t(\lambda_3,\lambda_4;\lambda_1,\lambda_2) 
&=& 
- e^2 Q_q^2 \bar{u}(p_1,\lambda_3) 
\Bigl[ 
\hat{\varepsilon}_{1}(k_1,\lambda_1)
\frac{\hat{p}_1 - \hat{k}_1}{(p_1 - k_1)^2}
\hat{\varepsilon}_{2}(k_2,\lambda_2)
                       \Bigr]
v(p_2,\lambda_4),
\nonumber \\
{\cal M}_u(\lambda_3,\lambda_4;\lambda_1,\lambda_2)
&=& 
- e^2 Q_q^2 \bar{u}(p_1,\lambda_3) 
\Bigl[ 
\hat{\varepsilon}_{2}(k_2,\lambda_2)
\frac{\hat{p}_1 - \hat{k}_2}{(p_1 - k_2)^2}
\hat{\varepsilon}_{1}(k_1,\lambda_1) 
                       \Bigr]
v(p_2,\lambda_4),\nonumber
\eea
where $Q_q$ is the quark charge in the unit of electron charge $e$.
It is easy to show that the helicity amplitudes with
$\lambda_1=\lambda_2$, and with $\lambda_3=\lambda_4$ vanish.
Therefore, only four helicity amplitudes corresponding to,
\bea
{\cal M}(\lambda_3,\lambda_4;\lambda_1,\lambda_2) ~=~
{\cal M}(\pm,\mp ; +, -),{\cal M}(\pm ,\mp ; -, +),
\eea 
should be considered.
Here we consider the amplitude ${\cal M}( +, -; -, +)$,
and use the photon polarization vectors,
\bea
\hat{\varepsilon}(k_1,-1) 
&=& \frac{1}{\sqrt{2} \mbk{p_2~k_1}}
\mbk{k_1 +|\gamma^{\mu}|p_2 +} \gamma_{\mu} \nonumber \\
&=& 
\frac{\sqrt{2}}{\mbk{p_2~k_1}} 
(\ket{k_1 -} \bra{p_2 -} + \ket{p_2 +} \bra{k_1 +})~,\\
%------------------
\hat{\varepsilon}(k_2,+1) &=& - \frac{1}{\sqrt{2} \pbk{p_1~k_2}}
\mbk{k_2 -|\gamma^{\mu}|p_1 -} \gamma_{\mu} \nonumber \\
&=&
- \frac{\sqrt{2}}{\pbk{p_1~k_2}} 
(\ket{k_2 +} \bra{p_1 +} + \ket{p_1 -} \bra{k_2 -})~,
\eea
where the reference momentum of the polarization vector for the photon
with momentum $k_{1}/k_{2}$ has been chosen to be the momentum
$p_{2}/p_{1}$.
We obtain,
\bea
{\cal M}_t(+, -; -, +) 
&=& 
e^2 Q_q^2 
\frac{
\mbk{p_1~k_1} 
\mbk{p_2 ~k_1} \pbk{k_1~p_1}
\mbk{k_2~p_2} 
     }{ (p_{1} \cdot k_1) \mbk{p_2~k_1} \pbk{p_1 k_2}} 
\nonumber \\
&=&
- 2 e^2 Q_q^2 
\frac{\pbk{k_1~p_1}  \mbk{k_2~p_2}}
{ \mbk{k_1~p_1}^{*} \mbk{p_1~k_2} }
\nonumber \\
&=& -2 e^2 Q_q^2 e^{i \phi(k_1,p_1)}
\frac{\mbk{k_2~p_2}}{ \mbk{p_1~k_2} }, \nonumber \\
{\cal M}_u(+, -; -, +) 
&=& 
e^2 Q_q^2 
\frac{
\mbk{p_1~p_1} 
\mbk{p_2 ~k_1} 
\bra{k_2 -} (\hat{p}_1 - \hat{k}_2) \ket{p_1 -} \mbk{p_2 ~p_2}
     }{ (p_{1} \cdot k_2) \mbk{p_2~k_1} \pbk{p_1 k_2}} 
\nonumber \\
&=& 0, \nonumber
\eea
where $e^{i \phi(k_1,p_1)}$ is the phase factor and given by
$e^{i \phi(k_1,p_1)}~=~\pbk{k_1~p_1}/ \mbk{k_1~p_1}^{*}$.
The amplitude ${\cal M}_u$ gives zero, because $\mbk{p_2 p_2} =
\mbk{p_1  p_1}=0$.
Thus we immediately arrive at the final result,
\bea
|{\cal M}(+, -; -, +) |^2
&=& 4 e^4 Q_q^4 \frac{k_2 \cdot p_2}{k_1 \cdot p_2}.
\eea
By using the invariance under the CP transformation,
the remaining non-zero helicity amplitudes are,
\bea
|{\cal M}(+, -; -, +) |^2
~=~
|{\cal M}(-, +; +, -) |^2
&=& 4 e^4 Q_q^4 \frac{k_2 \cdot p_2}{k_2 \cdot p_1},
\\
|{\cal M}(+, -; +, -) |^2
~=~
|{\cal M}(-, +; -, +) |^2
&=& 4 e^4 Q_q^4 \frac{k_1 \cdot p_2}{k_1 \cdot p_1}.
\eea
This example shows the efficiency of the spinor helicity method to
calculate the cross-section.
The squared amplitude for $\gamma \gamma \to q \bar{q}$ process,
averaged over initial spins and summed over final and colors ($N_C=3$)
reads,
\bea
|{\cal M}|^2 ~=~ 2 N_{C} e^4 Q_{q}^4 (\frac{t^2 + u^2}{u t}).
\eea
Here we use the Mandelstam valuables, $t=(k_1 - p_1)^2$ and $u=(k_1 - p_2)^2$.
\clearpage
%%%%%%%%%%%%%%%%%%%%%%%%%%%%%%%%%%%%%%%%%%%%%%%%%%%%%%%%
\section{Massive Spinor States}
\label{sec:MASSIVE}
%%%%%%%%%%%%%%%%%%%%%%%%%%%%%%%%%%%%%%%%%%%%%%%%%%%%%%%%

We consider, in this section, massive spinor states and show that the
massive spinor states also can be described in terms of massless spinor
states \cite{BJ,PARKE1}. 
At first, we summarize the properties of the massive spinors
$u(p,s),~v(p,s)$ with momentum $p$ and polarization $s$.
\begin{enumerate}
\item
Dirac Equation:
\bea
( \hat{p} - m ) u(p,s) &=& 0, \quad
\bar{u}(p,s) ( \hat{p} - m ) ~=~ 0, \\
( \hat{p} + m ) v(p,s) &=& 0, \quad
\bar{v}(p,s) ( \hat{p} + m ) ~=~ 0. 
\eea
\item
Normalization, Orthogonality and Completeness:
\bea
     \begin{array}{lcr}
\bar{u}(p,s1)~u(p,s2) 
           &=& 2 m \delta_{s1, s2},  
           \label{eqn:const1}\\
\bar{v}(p,s1)~v(p,s2) 
           &=& - 2 m \delta_{s1,s2}, 
           \label{eqn:const2}
     \end{array}
     \label{eqn:const3}\\
     \begin{array}{lcr}
\bar{u}(p,s1)~v(p,s2) &=& 0, \\
\bar{v}(p,s1)~u(p,s2) &=& 0, 
     \end{array}
     \label{eqn:const4}\\
     \begin{array}{lcr}
\sum_{s} u(p,s)~\bar{u}(p,s) 
           &=& \hat{p} + m, \\
\sum_{s} v(p,s)~\bar{v}(p,s) 
           &=& \hat{p} - m. 
     \end{array}
     \label{eqn:const5}
\eea
\end{enumerate}
The spin vector $s^{\mu}$ of massive fermion has the form in the
fermion's rest frame: 
\bea
s^{\mu} ~=~ (0,~\mbox{{\boldmath $s$}}), 
\eea
with $\mbox{{\boldmath $s$}} \cdot \mbox{{\boldmath $s$}} ~=~ 1$.
The spin vector, when the fermion is moving with momentum $p$, become
\bea
s^{\mu} ~=~ \left(
\frac{\mbox{{\boldmath $p$}} \cdot \mbox{{\boldmath $s$}}}{m},~
            \mbox{{\boldmath $s$}} + 
\frac{\mbox{{\boldmath $p$}}~ 
     (\mbox{{\boldmath $p$}} \cdot \mbox{{\boldmath $s$}})}
                 {m (E + m)} \right),
\eea
where $p=(E,~${\boldmath $p$}$)$ and spin vector
satisfies $p \cdot s ~=~0, s \cdot s ~=~-1$. 
The projection operators~\cite{BJ&Drell} are given by
\bea
u(p,s) \bar{u}(p,s) &=& (\hat{p} + m)\frac{1 + \gamma_{5} \hat{s}}{2} ~u(p),\\
v(p,s) \bar{v}(p,s) &=& (\hat{p} - m)\frac{1 + \gamma_{5} \hat{s}}{2} ~v(p).
\eea
\noindent
Therefore, the polarized fermion states are projected as, 
\bea
u(p,\uparrow) &=& \frac{1 + \gamma_{5} \hat{s}}{2} ~u(p),  \label{eqn:u-up}\\
u(p,\downarrow) &=& \frac{1 - \gamma_{5} \hat{s}}{2} ~u(p),\label{eqn:u-dn}\\
v(p,\uparrow) &=& \frac{1 + \gamma_{5} \hat{s}}{2} ~v(p),  \label{eqn:v-up}\\ 
v(p,\downarrow) &=& \frac{1 - \gamma_{5} \hat{s}}{2} ~v(p) \label{eqn:v-dn},
\eea
where spin states, $\uparrow$/$\downarrow$ refers to fermions with spin 
in $+${\boldmath $s$}$/-${\boldmath $s$} direction in their rest frame.
Now, we rewrite Eqns.(\ref{eqn:u-up})$\sim$(\ref{eqn:v-dn}) in terms of 
massless helicity spinor states and show how to describe the massive
spinor states in terms of massless helicity states.
Let us consider, for example, the state $u(p,\uparrow)$ and rewrite it using
properties satisfied by gamma matrices,
\bea
u(p, \uparrow) &=& \frac{1 + \gamma_5}{2}
                   \frac{1 + \gamma_5\hat{s}}{2}~u(p)
                 + \frac{1 - \gamma_5}{2} 
                   \frac{1 + \gamma_5\hat{s}}{2}~u(p), \nonumber \\
               &=& \frac{1 + \gamma_5}{2} 
                   \frac{1 + \hat{s}}{2}~u(p)
                 + \frac{1 - \gamma_5}{2} 
                   \frac{1 - \hat{s}}{2}~u(p). \label{eqn:trns1}
\eea
In Eqn.(\ref{eqn:trns1}), the quantities $[(1 \pm \hat{s})/2]~u(p)$ can be
identified with massless spinors having momentum $(p \pm m s)/2$
due to the following reasons~\cite{KOD-PARKE}
\begin{enumerate}
\item{Dirac Equation:}
At first, note that the momentum $p_i~(i=1,2)$ is light-like,
\bea
p_{1} &\equiv& \frac{p + m s}{2},~~(p_{1}^2 ~=~0), \\
p_{2} &\equiv& \frac{p - m s}{2},~~(p_{2}^2 ~=~0), 
\eea
and simple relations between $p_i,~p$ and $s$,
\bea
      p  &=& p_1 + p_2,\\
     m s &=& p_1 - p_2.
\eea
Next it is easy to show,
\[
 \hat{p}_1 \frac{1 + \hat{s}}{2} u(p)
~=~
 \hat{p}_2 \frac{1 - \hat{s}}{2} u(p)
~=~0.
\] 
Therefore, the quantities $[(1 \pm \hat{s})/2]~u(p)$ are proportional to
     the massless spinors $Q(p_{1})$, $Q(p_{2})$ respectively.
\bea
Q(p_{1}) &=& A \frac{(1 + \hat{s})}{2} u(p) ~,\\
Q(p_{2}) &=& B \frac{(1 - \hat{s})}{2} u(p) ~,\\
\eea
where factors $A,~B$ are normalization constant.
\item{Projection Property:}
The normalization factors $A$ and $B$ will be fixed by constructing the
     projection operators, which read
\bea
    Q(p_1)\bar{Q}(p_1) ~=~ |A|^2 \hat{p}_1, \nonumber \\
    Q(p_2)\bar{Q}(p_2) ~=~ |B|^2 \hat{p}_2, \nonumber
\eea
\end{enumerate}
Therefore, we conclude that $[(1 \pm \hat{s})/2]~u(p)$ are the massless
spinors $Q(p_i)$ except for phase factors.
Above consideration tells us
\bea
u(p, \uparrow) &=& 
\frac{1 + \gamma_5}{2}\, Q(p_1 ) 
 + e^{i\phi(p_{1},p_{2})}\,\frac{1 - \gamma_5}{2}\, Q(p_2 ), \nonumber \\
              &\equiv& 
\ket{p_1 +} + e^{i\phi(p_{1},p_{2})} \ket{p_2 -}.
\eea
Here, we have chosen the overall phase factor such that there is no phase 
in front of $\ket{p_1 +}$.
Therefore, $\phi(p_{1},p_{2})$ represents the relative phase between the
states $\ket{p_1 +}$ and $\ket{p_2 -}$,
This phase factor can be determined by the normalization condition for
$u(p,\uparrow)$ in Eqn.(\ref{eqn:const1}).
It is not difficult to obtain, 
\bea
 e^{i \phi} = \frac{\mbk{p_2~ p_1} }{m}, 
\eea
where we have used the relation:
\bea
  | \mbk{ p_2~ p_1 } |^2 = 2 \,p_1 \cdot p_2 = m^2. 
\eea
In this way, all massive spinor states can be written in terms of
massless chiral spinors as (for the anti-fermions, we use the
superscript ``bar''for their momenta), 

\bea
u(p,\uparrow)   &=& \ket{p_1 +} + \ket{p_2 - }
             \frac{ \mbk{p_2~ p_1} }{m} , \\    
u(p,\downarrow) &=& \ket{p_1 -} + \ket{p_2 + }
             \frac{ \pbk{p_2~ p_1} }{m}, \\
\bar{u}(p,\uparrow)   &=& \bra{p_1 +} + 
             \frac{ \pbk{p_1~ p_2} }{m} \bra{p_2 -},\\ 
\bar{u}(p,\downarrow) &=& \bra{p_1 - } + 
             \frac{ \mbk{p_1~ p_2} }{m} \bra{p_2 +},\\
v(\bar{p},\uparrow)   &=& \ket{ \bar{p}_1 - } - \ket{ \bar{p}_2 +}
             \frac{ \pbk{ \bar{p}_2~ \bar{p}_1 } }{m}, \\
v(\bar{p},\downarrow) &=& \ket{ \bar{p}_1 + } - \ket{ \bar{p}_2 -}
             \frac{ \mbk{ \bar{p}_2~ \bar{p}_1 } }{m}, \\
\bar{v}(\bar{p},\uparrow)   &=& \bra{ \bar{p}_1 - } - 
             \frac{ \mbk{ \bar{p}_1~ \bar{p}_2 } }{m} 
             \bra{ \bar{p}_2 +}, \\ 
\bar{v}(\bar{p},\downarrow) &=& \bra{ \bar{p}_1 + } - 
             \frac{ \pbk{ \bar{p}_1~ \bar{p}_2 } }{m} 
             \bra{\bar{p}_2 - }.
\eea 
The light like momenta $p_{1}$, $p_{2}$, $\bar{p}_{1}$ and
$\bar{p}_{2}$ are defined by,
\bea
p_{1} &\equiv& \frac{p + m s_{p}}{2},~~
p_{2} ~\equiv~ \frac{p - m s_{p}}{2},~~
 \\
\bar{p}_{1} &\equiv& \frac{\bar{p} + m s_{\bar{p}}}{2},~~
\bar{p}_{2} ~\equiv~ \frac{\bar{p} - m s_{\bar{p}}}{2}.
\eea 

For the convenience of later calculations,
let us write down $p_i$ explicitly in the rest and moving frame of the
massive fermions.
In the rest frame, the four momentum of the massive fermion and the spin vector
take
\bea
p    &=& (m,~\mbox{{\boldmath $0$}}),~~\\
\label{eqn:rest-mom}
s_{p}&=& (0,~\mbox{{\boldmath $\omega$}}),
~~(|\mbox{{\boldmath $\omega$}}|^2~=~1).
\label{eqn:rest-spin}
\eea
So, the light like momenta $p_{1},~p_{2}$ are
\bea
p_{1}~=~\frac{m}{2}(1,~\mbox{{\boldmath $\omega$}}),~~
p_{2}~=~\frac{m}{2}(1,~- \mbox{{\boldmath $\omega$}}).
\label{eqn:rest-mom-12}
\eea
These equations show that the direction of the particle's spin is in the
same direction of the spatial part of $p_{1}$ or 
in the opposite direction of the spatial part of $p_{2}$.
For the moving fermion, $p=(E,~{\cal P}\mbox{{\boldmath $n$}})$,
($|\mbox{{\boldmath $n$}}|^2=1$), $p_i$ are obtained by boosting
Eqn.(\ref{eqn:rest-mom})$\sim$(\ref{eqn:rest-mom-12}).
The spin vector and decomposed light-like momenta $p_{1},~p_{2}$ become
\bea
s_{p}&=& 
\biggl(
{\cal P} 
\frac{(\mbox{{\boldmath $n$}} \cdot \mbox{{\boldmath $\omega$}})}{m}, 
\mbox{{\boldmath $\omega$}} + 
\frac{ {\cal P}^2 
(\mbox{{\boldmath $n$}} \cdot \mbox{{\boldmath $\omega$}}) 
 ~\mbox{{\boldmath $n$}}}{m (m + E)} 
\biggr), \\
p_{1} &=&
\frac{1}{2} 
\biggl(
E + {\cal P}~ 
(\mbox{{\boldmath $n$}} \cdot \mbox{{\boldmath $\omega$}}), 
{\cal P}~ \mbox{{\boldmath $n$}} + 
      m ~ \mbox{{\boldmath $\omega$}} + 
\frac{ {\cal P}^2 
(\mbox{{\boldmath $n$}} \cdot \mbox{{\boldmath $\omega$}}) 
~ \mbox{{\boldmath $n$}}}{(m + E)} 
\biggr), \\
p_{2} &=&
\frac{1}{2} 
\biggl(
E - {\cal P}~ 
(\mbox{{\boldmath $n$}} \cdot \mbox{{\boldmath $\omega$}}), 
{\cal P}~\mbox{{\boldmath $n$}} - 
      m ~\mbox{{\boldmath $\omega$}} - 
\frac{ {\cal P}^2 
~(\mbox{{\boldmath $n$}} \cdot \mbox{{\boldmath $\omega$}}) 
~ \mbox{{\boldmath $n$}}}{(m + E)} 
\biggr).
\eea
Here we consider the familiar helicity state, that is, the spin of
massive particle is projected along its direction of motion, 
\bea
\mbox{{\boldmath $n$}} &=& \mbox{{\boldmath $\omega$}}, \\
p_{1} &=& \frac{1}{2} 
\biggl(
E + {\cal P},
\left[ {\cal P} + m + \frac{{\cal P}^2}{(m + E)} 
\right] \mbox{{\boldmath $n$}}
\biggr), \\
p_{2} &=&
\frac{1}{2} \biggl(
E - {\cal P},
\left[ {\cal P} - m - \frac{{\cal P}^2}{(m + E)} 
\right] \mbox{{\boldmath $n$}}
\biggr). 
\eea
In the large momentum limit (which is equivalent to the massless limit), 
$p_{1}~=~p$, $p_{2}~=~0$. 
Then, the spinors become the chiral eigenstates:
\bea
u(p,R) &\to& \ket{p_{1} + },~
u(p,L) ~\to~ \ket{p_{1} - }~, \\
v(\bar{p},R) &\to& \ket{\bar{p}_{1} - },~
v(\bar{p},L) ~\to~ \ket{\bar{p}_{1} + }~, \\
\bar{u}(p,R) &\to& \bra{p_{1} - },~
\bar{u}(p,L) ~\to~ \bra{p_{1} + }~, \\
\bar{v}(\bar{p},R) &\to& \bra{\bar{p}_{1} + },~
\bar{v}(\bar{p},L) ~\to~ \bra{\bar{p}_{1} - }~,
\eea
where we have labeled helicity states by R (L) instead of $\uparrow$
($\downarrow$).
This demonstrates the relation between the right-handed (left-handed)
helicity state, $u(p,R)$ ($u(p,L)$), in the massless limit and 
the right-handed (left-handed) chiral state.
\clearpage
\section{Circular Polarizations of Massive Bosons}
The polarization vectors for massive gauge
boson~\cite{MASSIVE,PARKE1,PARKE2} can also be described in terms of
massless spinors. 
Transversely polarized states satisfy
Eqns.(\ref{eqn:pv1}) $\sim$ (\ref{eqn:pv3}) in Sec.\ref{sec:massless-gb}.
For massive gauge boson, we must also take into account
the longitudinally polarized state.
Let  $\varepsilon^{\mu}_{\pm}$ denote transversely polarized states
and $\varepsilon_{0}^{\mu}$ is the longitudinally one.
The properties of polarization vectors for a massive gauge boson with
momentum $k$ and mass $m$ are
\bea
\varepsilon_{\lambda}(k) \cdot k  &=& 0, \label{eqn:mpv1} \\
\varepsilon_{\lambda}(k) \cdot \varepsilon^*_{\lambda'}(k) &=&
- \delta_{\lambda \lambda'}  \label{eqn:mpv2},\\
\sum_{\lambda} 
\varepsilon_{\lambda}^{\mu}(k) \varepsilon_{\lambda}^{\nu *}(k) &=& 
- g^{\mu \nu} + \frac{k^{\mu} k^{\nu}}{m^2} \label{eqn:mpv3}, 
\eea
where $\lambda$ and $\lambda'$ denote $+,-$ and $0$.

Let $s$ be the spin vector associated with the massive vector boson, which
satisfies,
\bea
k \cdot s ~=~0,~~(k^2 ~=~ m^2).
\eea
Here three polarization vectors $\varepsilon_{\lambda}$ are defined such
that in the vector boson rest frame, the boson has the spin projection 
$\lambda=(+,0,-)$ with respect to the spatial part of the spin vector $s$.
Three polarization vectors $\varepsilon^{\mu}_{\lambda}(k)$
are given in terms of two light-like vectors,
\bea
k_{1}^{\mu} &=& \frac{k^{\mu} - m s^{\mu}}{2}, \\
k_{2}^{\mu} &=& \frac{k^{\mu} + m s^{\mu}}{2}.
\eea
Then, we get polarization vectors for the incoming gauge boson:
\bea
\varepsilon_{\pm}^{\mu} &=& 
\frac{\bra{k_1 \pm} \gamma^{\mu} \ket{k_2 \pm} }{\sqrt{2} m}, \\
\varepsilon_{0}^{\mu} &=& 
\frac{
        \bra{k_1 +} \gamma^{\mu} \ket{k_1 +} -
        \bra{k_2 +} \gamma^{\mu} \ket{k_2 +} }{2 m} ~=~
\frac{k_1^{\mu} - k_2^{\mu}}{m}. 
\eea
In this expressions, the helicity basis is given by choosing the
spatial part of the spin vector $s$ to be in the same direction as the
spatial part of the momentum vector $k$.

\newpage
%----------------------- Leading Order ------------------------------
%\input lo.tex
%----------------------- Leading Order---------------------------------
\chapter{Spin Correlations at Leading Order}
In 1994, the top quark was discovered by the CDF and D0
collaborations \cite{CDF,D0}.
The measured top quark mass is approximately $175$ GeV,
which is nearly twice the mass scale of electro-weak symmetry
breaking.
Thus the top quark with large mass brings a good opportunity to
understand electro-weak symmetry breaking and to reveal physics beyond
the Standard Model.
The top quark decays electroweakly before hadronizing because
its width is much greater than the hadronization timescale set
by $\Lambda_{QCD}$ \cite{bdkkz}.
Therefore, there are significant angular correlations between the
decay products of the top quark and the spin of the top quark.
These angular correlations depend sensitively on the top quark
couplings to the $Z^0$ boson and photon, and to the $W$ boson and $b$
quark. 
These angular distributions are considered as good information to
constrain the couplings of top quarks in the Standard Model.
Most works~\cite{bop,kly} are performed by using top quark spin
decomposed in the helicity basis.
In the case of ultra-relativistic particles, this decomposition is appropriate.
However this basis is not necessarily optimal to investigate the spin
correlations since the produced top quarks are non-relativistic at
realistic linear colliders: {\it e.g.} $\beta ~=~ v/c \sim 0.5$
for a $400$ GeV collider.
As a matter of fact, Mahlon and Parke~\cite{mp} have shown that this
decomposition in helicity basis is far from the optimal decomposition
for top quark productions at hadron colliders.
Parke and Shadmi~\cite{ps} have extended these analyses to lepton
colliders.

In this chapter, we discuss the spin-spin correlations for the top quark
pair production at polarized $e^+ e^-$ linear colliders at the
leading order in perturbation theory.
We introduce the ``generic spin basis''
and investigate which decomposition of top quark spin is optimal
to study spin correlations for moderate energies. 
The spin-angular correlations for the top (anti-top) quark decay 
are also discussed.

\section{Spin-Spin Correlations for Top Pair production}

In this section, we derive the leading order spin dependent
differential cross-sections for top quark pair productions in a generic
spin basis.
We find the most optimal spin decomposition for top
and anti-top quark to study spin correlations.
The one-loop analyses will be given in the next chapter. 
% Fig.1
\begin{figure}[H]
\begin{center}
\leavevmode\psfig{file=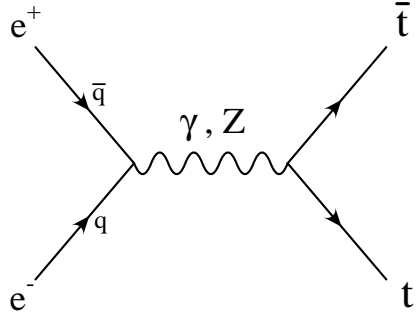,width=6cm}
\caption[The tree level contributions to the $e^- e^+ \to t \bar{t}$
 process]{The tree level contributions to the $e^- e^+ \to t \bar{t}$
 process}
\label{fig:tree-process}
\end{center}
\end{figure}
The Feynman graph for the $e^{+}e^{-} \rightarrow t \bar{t}$ process at the
leading order in perturbation theory is given in
Fig.\ref{fig:tree-process}.
We calculate the amplitude using the spinor helicity method.
The momenta of electron, positron, top quark and anti-top quark
are labeled by $q$, $\bar{q}$, $t$ and $\bar{t}$ respectively,
and $s_{t}$ ($s_{\bar{t}}$) represents the spin vector of top
(anti-top) quark.
As explained in the previous chapter, we decompose the top (anti-top)
quark momentum $t$ $(\bar{t})$ into a sum of two massless momenta
$t_{1},t_{2}$ ($\bar{t}_{1},\bar{t}_{2}$),
\bea
t     &=& t_{1}+t_{2},~(\bar{t} ~=~ \bar{t}_{1}+ \bar{t}_{2}), \\
m s_t &=& t_{1}-t_{2},~(m s_{\bar{t}} ~=~ \bar{t}_{1}-\bar{t}_{2}).  
\eea 
The amplitudes for $e^-_{L} e^+_{R}$ ($e^-_{R}e^+_{L}$)
scattering processes are obtained to be,
\bea
\lefteqn{
{\cal M}
       (e^-_L e^+_R \rightarrow t_{s_{t}} \bar{t}_{s_{\bar{t}}})
        } \nonumber \\
&=&
\frac{4 \pi \alpha}{s} 
\bra{\bar{q}-} \gamma^{\mu}_{L} \ket{q-}
\bar{u}(t,s_t)
\gamma_{\mu} \left[ f_{LL} \gamma_{L} + f_{LR} \gamma_{R} \right] 
v(\bar{t},s_{\bar{t}})~, \label{eqn:ampli1} \\
\lefteqn{
{\cal M}
       (e^-_R e^+_L \rightarrow t_{s_{t}} \bar{t}_{s_{\bar{t}}})
        } \nonumber \\
&=&
\frac{4 \pi \alpha}{s} 
\bra{\bar{q}+} \gamma^{\mu}_{R} \ket{q+}
\bar{u}(t,s_t)
\gamma_{\mu} \left[ f_{RL} \gamma_{L} + f_{RR} \gamma_{R} \right] 
v(\bar{t},s_{\bar{t}})~, \label{eqn:ampli2}
\eea 
where $e^-_{L}/e^+_L$ ($e^-_R/e^+_R$) means the electron/positron with
left-handed (right-handed) helicity state. 
We have neglected the electron (positron) mass. 
The quantity $\alpha$ is the fine structure constant $\alpha=e^{2}/(4 \pi)$,
and $s$ is the total energy,
$\gamma^{\mu}_{R/L} \equiv  
\gamma^{\mu} \gamma_{R/L}$, $\gamma_{R/L} \equiv (1 \pm \gamma_5)/2$.
The quantities $f_{IJ}$'s are the sum of photon and $Z^0$ boson couplings
to fermions (electron and top quark) including the propagator and are
given by  
\[ f_{IJ} = - Q_t + Q_e^I Q_t^J \, \frac{1}{\sin^2 \theta_W}\,
              \frac{s}{s - M_Z^2 + i M_Z \Gamma_{Z}} \ .\]
Here $M_Z$ is the $Z$ boson mass, $\Gamma_{Z}$ is the width of $Z$
boson, $\theta_W$ is the Weinberg angle and $I,J \in (L,R)$.  
The parameter $Q_t = 2/3$ is the electric charge of the top
quark in the unit of electron charge $e$.
The electron couplings to the $Z$ boson are given by
\be
   Q_e^L = \frac{2 \sin^2 \theta_W - 1}{2 \cos \theta_W} \quad ,\quad 
   Q_e^R = \frac{ \sin^2 \theta_W}{ \cos \theta_W} \ .
\ee
The top quark couplings to the $Z$ boson are given by
\be
     Q_t^L = \frac{3 - 4 \sin^2 \theta_W}{6 \cos \theta_W} \quad ,\quad 
     Q_t^R = - \frac{2 \sin^2 \theta_W}{3 \cos \theta_W} \ .\label{zcoupling}
\ee
The squared amplitude for $e^-_Le^+_R$ scattering reads
\bea
\lefteqn{ 
|{\cal M}(e^-_L e^+_R \rightarrow t_{\uparrow} \bar{t}_{\uparrow})|^2 }
\nonumber \\
&=& 4 \left(\frac{4 \pi \alpha}{s} \right)^2
\biggl[
|f_{LL}|^2 (2 q \cdot t_{1})(2 \bar{q} \cdot \bar{t}_{2}) +
|f_{LR}|^2 (2 \bar{q} \cdot t_{2})(2 q \cdot \bar{t}_{1}) 
\nonumber \\
&& \mbox{\hspace*{3cm}}~+~
\frac{1}{m^2}
\bigl\{ 
f_{LL}f^*_{LR} 
        \Tr[\gamma_{L} q t_{1} t_{2} \bar{q} \bar{t}_{2} \bar{t}_{1}]
       + c.c
\bigr\}
\biggr]~, \label{eqn:samp1} \\
%----------------------
\lefteqn{
|{\cal M}(e^-_L e^+_R \rightarrow t_{\uparrow} \bar{t}_{\downarrow})|^2 }
\nonumber \\
&=& 4 \left(\frac{4 \pi \alpha}{s} \right)^2
\biggl[
|f_{LL}|^2 (2 q \cdot t_{1})(2 \bar{q} \cdot \bar{t}_{1}) +
|f_{LR}|^2 (2 \bar{q} \cdot t_{2})(2 q \cdot \bar{t}_{2}) 
\nonumber \\
&& \mbox{\hspace*{3cm}}~+~
\frac{1}{m^2}
\bigl\{ 
f_{LL}f^*_{LR} 
        \Tr[\gamma_{L} q t_{1} t_{2} \bar{q} \bar{t}_{1} \bar{t}_{2}]
       + c.c
\bigr\}
\biggr]~,\label{eqn:samp2} \\
%----------------------
\lefteqn{ 
|{\cal M}(e^-_L e^+_R \rightarrow t_{\downarrow} \bar{t}_{\uparrow})|^2 }
\nonumber \\
&=& 4 \left(\frac{4 \pi \alpha}{s} \right)^2
\biggl[
|f_{LL}|^2 (2 q \cdot t_{2})(2 \bar{q} \cdot \bar{t}_{2}) +
|f_{LR}|^2 (2 \bar{q} \cdot t_{1})(2 q \cdot \bar{t}_{1}) 
\nonumber \\
&& \mbox{\hspace*{3cm}}~+~
\frac{1}{m^2}
\bigl\{ 
f_{LL}f^*_{LR} 
        \Tr[\gamma_{L} q t_{2} t_{1} \bar{q} \bar{t}_{2} \bar{t}_{1}]
       + c.c
\bigr\}
\biggr]~, \label{eqn:samp3} \\
%----------------------
\lefteqn{
|{\cal M}(e^-_L e^+_R \rightarrow t_{\downarrow} \bar{t}_{\downarrow})|^2 }
\nonumber \\
&=& 4 \left(\frac{4 \pi \alpha}{s} \right)^2
\biggl[
|f_{LL}|^2 (2 q \cdot t_{2})(2 \bar{q} \cdot \bar{t}_{1}) +
|f_{LR}|^2 (2 \bar{q} \cdot t_{1})(2 q \cdot \bar{t}_{2}) 
\nonumber \\
&& \mbox{\hspace*{3cm}}~+~
\frac{1}{m^2}
\bigl\{ 
f_{LL}f^*_{LR} 
        \Tr[\gamma_{L} q t_{2} t_{1} \bar{q} \bar{t}_{1} \bar{t}_{2}]
       + c.c
\bigr\}
\biggr]~, \label{eqn:samp4}
\eea
where all momenta, $p$, under the ``Tr'' operator are understood to be
$\hat{p}$.
The squared amplitude for $e^-_{R}e^+_L$ scattering are given by
interchanging $L,~R$ as well as $\uparrow,~\downarrow$ in
Eqn.(\ref{eqn:samp1})$\sim$(\ref{eqn:samp4}).   

\begin{figure}[H]
\begin{center}
\begin{tabular}{cc}
\leavevmode\psfig{file=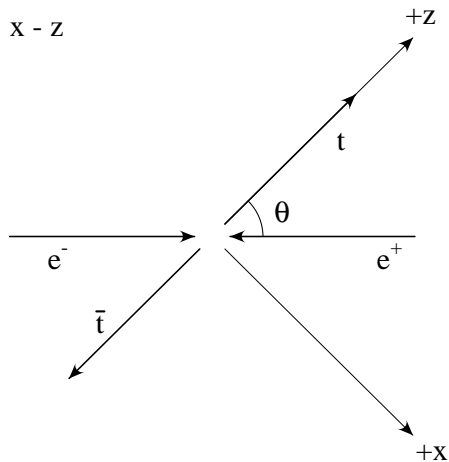,width=6cm} 
\end{tabular}
\caption{$e^- e^+ \to t \bar{t}$ process in the center of mass frame.}
\label{fig:ZMF}
\end{center}
\end{figure}

\begin{figure}[H]
\begin{center}
\begin{tabular}{cc}
\leavevmode\psfig{file=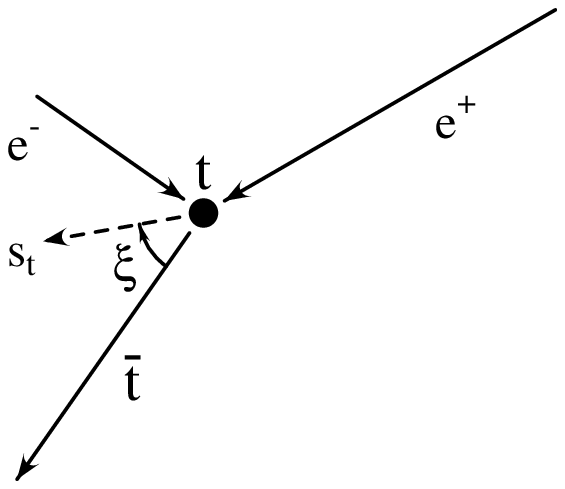,width=5.5cm} &
\leavevmode\psfig{file=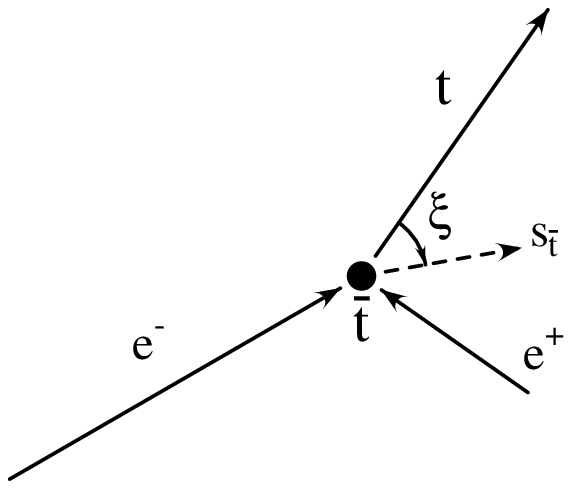,width=5.5cm} 
\end{tabular}
\caption[The generic spin basis for the top (anti-top)
quark in its rest frame.]{The generic spin basis for the top (anti-top)
quark in its rest frame.  ${\bf s}_t$ (${\bf s}_{\bar{t}}$) is the top
 (anti-top) spin axis.} 
\label{fig:spin-def}
\end{center}
\end{figure}

Before discussing the spin basis, let us fix the coordinate system.
In the center of mass (CM) frame, we take the production plane to be
$x-z$ plane and set the direction of the motion of the top quark to be
$+z$ direction (see Fig.\ref{fig:ZMF}).
%%%%%%%%%%%%%%%%%%%%%%%
% Fig.3
%%%%%%%%%%%%%%%%%%%%%%%
%%%%%%%%%%%%%%%%%%%%%%%
\noindent
%We investigate the differential cross-section for a generic spin basis.
%We use a generic spin basis~\cite{ps} with the spin in the production
%plane and define the spins of the top and anti-top quarks by the
%parameter $\xi$ as shown in Fig.\ref{fig:spin-def}.
Now, the generic spin basis we consider here is based on the following facts:
First, CP is conserved in the Standard Model.
Hence the spins of top and ant-top quark can be defined to be back-to-back.
Second, there is no transverse polarization of the top
(anti-top) quark at the tree level in the Standard Model.
%therefore we consider the spins of top and anti-top quarks in the
%production plain.
Here the ``transverse'' means the direction  normal to the production
plane\cite{TP}.
% and the component is caused by the imaginally part of squared
%amplitude~
Hence, the spin vectors are on the production plane and they are
parametrized by one parameter.
%the $Z$ boson width is negligible in the region of center of mass (CM)
%energy $\sqrt{s}$ above the production threshold for top quarks.  
%except for width.
We define the generic basis in the following way;
For top quark, we got to its rest frame and its spin is decomposed along
the direction ${\bf s}_t$ in the rest frame of the top quark which makes
an angle $\xi$ with the anti-top quark momentum in the clockwise
direction.   
Similarly, the anti-top quark spin states are defined in the anti-top
rest frame along the direction ${\bf s}_{\bar{t}}$ having the same
angle $\xi$ from the direction of the top quark momentum (see
Fig.\ref{fig:spin-def}). 
\clearpage
%%%%%%%%%%%%%%%%%%%%%%%
% Fig.1
%%%%%%%%%%%%%%%%%%%%%%%
\noindent
%We use the following notation in this thesis: the state
%$t_{\uparrow}\,\bar{t}_{\uparrow}\,(t_{\downarrow}\,\bar{t}_{\downarrow})$
%refers to a top with spin in the $+ {\bf s}_t \,(- {\bf s}_t )$
%direction in the top rest frame and an anti-top
%with spin $+ {\bf s}_{\bar{t}} \,(- {\bf s}_{\bar{t}} )$ in the anti-top
%rest frame. 
%           
In the top (anti-top) quark rest frame, the top (anti-top) quark momentum
$t_{RF}$, $(\bar{t}_{RF})$ and the spin vector become $s^{RF}_{t}$
($s^{RF}_{\bar{t}}$) as,
\bea
\begin{array}{cc}
t^{\mu}_{RF} ~=~ m (1,0,0,0), & 
\bar{t}^{\mu}_{RF} ~=~m (1,0,0,0),\\
(s^{RF}_{t})^{\mu}   ~=~ (0,- \sin \xi, 0, - \cos \xi), &
(s^{RF}_{\bar{t}})^{\mu} ~=~ (0, \sin \xi, 0, \cos \xi).
\end{array}
\eea
Now let us decompose the top (anti-top) quark momentum $t_{RF}$
($\bar{t}_{RF}$) in terms of 
massless momenta $t^{RF}_{1}$ and $t^{RF}_{2}$ ($\bar{t}^{RF}_{1}$
and $\bar{t}^{RF}_{2}$), 
\bea
t_{RF} &=& t^{RF}_{1} + t^{RF}_{2},
~~m s^{RF}_{t} ~=~t^{RF}_{1} - t^{RF}_{2},\\
\bar{t}_{RF} &=& \bar{t}^{RF}_{1} + \bar{t}^{RF}_{2},
~~m s_{\bar{t}}^{RF} ~=~\bar{t}^{RF}_{1} - \bar{t}^{RF}_{2}.
\eea 
From these relations, the four massless momenta are expressed as
\bea
(t^{RF}_1)^{\mu}&=&\frac{m}{2}(1,- \sin \xi, 0, - \cos \xi),\\
(t^{RF}_2)^{\mu}&=&\frac{m}{2}(1,  \sin \xi, 0,  \cos \xi), \\
(\bar{t}^{RF}_1)^{\mu}&=&\frac{m}{2}(1, \sin \xi, 0, \cos \xi),\\
(\bar{t}^{RF}_2)^{\mu}&=&\frac{m}{2}(1, - \sin \xi, 0,  - \cos \xi).
\eea
%In Fig.\ref{fig:ZMF}, we consider the top pair production in center of mass
%(CM) frame and set the direction of the motion of the
%top quark to $+z$ axis, production plain to $x-z$ plain.
%The momenta $t_{1}^{RF},~t_{2}^{RF}$ are boosted 
To get back to the CM frame, we boost back all quantities in the
$+z~(-z)$ direction with the top (anti-top) quark speed $\beta ~=~ \sqrt{1 - 4
m^2/s}$. 
%The kinematical variables for the anti-top quark can be
%obtained similarly.
The momenta of particles in the CM frame are given by,
\bea
t^{\mu} &=& m \gamma ( 1, 0, 0, \beta),~~
\bar{t}^{\mu} ~=~ m \gamma ( 1, 0, 0, - \beta),\\ 
q^{\mu} &=& m \gamma ( 1, \sin \theta, 0, \cos \theta)~,~~
\bar{q}^{\mu} ~=~ m \gamma ( 1, -\sin \theta, 0, -\cos \theta)~,\\
t_1^{\mu} &=& \frac{m}{2} 
( \gamma (1 - \beta \cos \xi), - \sin \xi, 0, \gamma(\beta - \cos
\xi))~,\\
t_2^{\mu} &=& \frac{m}{2} 
( \gamma (1 + \beta \cos \xi), - \sin \xi, 0, \gamma(\beta + \cos
\xi))~,\\
\bar{t}_1^{\mu} &=& \frac{m}{2} 
( \gamma (1 - \beta \cos \xi), - \sin \xi, 0, -\gamma(\beta - \cos
\xi))~,\\
\bar{t}_2^{\mu} &=& \frac{m}{2} 
( \gamma (1 + \beta \cos \xi), - \sin \xi, 0, -\gamma(\beta + \cos \xi))~,
\eea
\noindent
where $\gamma = 1/\sqrt{1 - \beta^2}$ and 
$\theta$ is the scattering angle of the top quark with respect to the
electron in the CM frame [see Fig.{\ref{fig:ZMF}}].

Let us write down the squared amplitudes using kinematical variables in
the CM frame.
Since each term in Eqn.(\ref{eqn:samp1}) reads as
\bea
&&
(2 q \cdot t_1)(2 \bar{q} \cdot \bar{t}_2) \nonumber \\
&& =
m^4  \gamma^4 \bigl[
(1 - \beta \cos \theta)^2 - 
\{ (\cos \theta - \beta) \cos \xi - 
   \sqrt{1 - \beta^2}\sin \theta \sin \xi \}^2
              \bigr]~, \qquad \\
&&
(2 q \cdot \bar{t}_1)(2 \bar{q} \cdot t_2) \nonumber \\
&& =
m^4 \gamma^4 \bigl[ 
(1 + \beta \cos \theta)^2 - 
\{ (\beta + \cos \theta) \cos \xi +
   \sqrt{1 - \beta^2} \sin \theta \sin \xi \}^2
             \bigr]~, \qquad \\
&&
\Tr[\gamma_L q t_1 t_2 \bar{q} \bar{t}_2 \bar{t}_2] ~=~
\Tr[\gamma_R q t_1 t_2 \bar{q} \bar{t}_2 \bar{t}_2] \nonumber \\
&&  =
m^6 \gamma^4 
\bigl[ 1 - \beta^2 - 
      \{ \sqrt{1 - \beta^2} \cos \theta \cos \xi + 
                       \sin \theta \sin \xi \}^2 
\bigr]~, \qquad
\eea
the squared amplitude becomes
\bea
    |{\cal M}
    (e^-_L e^+_R \rightarrow t_{\uparrow} \bar{t}_{\uparrow})|^2 
&=& 16 \pi^2 \alpha^2 | A_{LR} \cos \xi - B_{LR} \sin \xi|^2, 
    \label{eqn:samp-CMuu}
\eea
where the quantities $A_{LR}$ and $B_{LR}$ are defined by
\bea
    A_{LR} &=& \Bigl[ ( f_{LL} + f_{LR} ) \sqrt{1 - \beta^2}\,
            \sin \theta \Bigr] /2 \ ,\\
    B_{LR} &=& \Bigl[ f_{LL} ( \cos \theta + \beta )
             + f_{LR} ( \cos \theta - \beta ) \Bigr] /2  
          .\label{eqn:largecoupling}
\eea
Similarly, other squared amplitudes are
\bea
%&&
    |{\cal M} 
    (e^-_L e^+_R \rightarrow t_{\downarrow} \bar{t}_{\downarrow})|^2 
%    \nonumber \\
%&& 
%\qquad =
&=&
16 \pi^2 \alpha^2 | A_{LR} \cos \xi - B_{LR} \sin \xi|^2~, 
    \label{eqn:asmp-CMdd}\\
%
%&&
    |{\cal M}
    (e^-_L e^+_R \rightarrow  t_{\uparrow} \bar{t}_{\downarrow}
                   {\rm~or~}  t_{\downarrow} \bar{t}_{\uparrow}  )|^2 
%\nonumber \\
%&& 
%\qquad =
&=&
16 \pi^2 \alpha^2
    | A_{LR} \sin \xi + B_{LR} \cos \xi \pm  D_{LR}|^2~,
\qquad    \label{eqn:asmp-CMud}
\eea
with
\bea
    D_{LR} &=& \Bigl[ f_{LL} ( 1 + \beta \cos\theta )
             + f_{LR} ( 1 - \beta \cos\theta ) \Bigr] /2~.
\eea
The differential cross-section is given by
\[
\frac{d \sigma}{d \cos \theta}
~=~\frac{1}{32 \pi}\frac{\beta}{s} \sum_{N_C}|{\cal M}|^2 ~, 
\]
where $N_c$ is the color degree of freedom, $N_c=3$ and $s$ is the total energy.
Thus, we arrive at the final expression of the polarized
cross-sections for the top quark production,
\bea
    \frac{d \sigma}{d \cos \theta}
    (e^-_L e^+_R \rightarrow t_{\uparrow} \bar{t}_{\uparrow}) 
&=& \frac{d \sigma}{d \cos \theta}
    (e^-_L e^+_R \rightarrow t_{\downarrow} \bar{t}_{\downarrow}) 
    \nonumber \\
&=& \frac{3 \pi \alpha^2}{2 s} \beta | A_{LR} \cos \xi - B_{LR} \sin \xi|^2, 
    \label{eqn:GSB}\\
    \frac{d \sigma}{d \cos \theta}
    (e^-_L e^+_R \rightarrow  t_{\uparrow} \bar{t}_{\downarrow}
                   {\rm~or~}  t_{\downarrow} \bar{t}_{\uparrow}  ) 
&=&
    \frac{3 \pi \alpha^2}{2 s} \beta 
    | A_{LR} \sin \xi + B_{LR} \cos \xi \pm  D_{LR}|^2, 
    \nonumber 
\eea

\begin{figure}[H]
\begin{center}
\begin{tabular}{cc}
\leavevmode\psfig{file=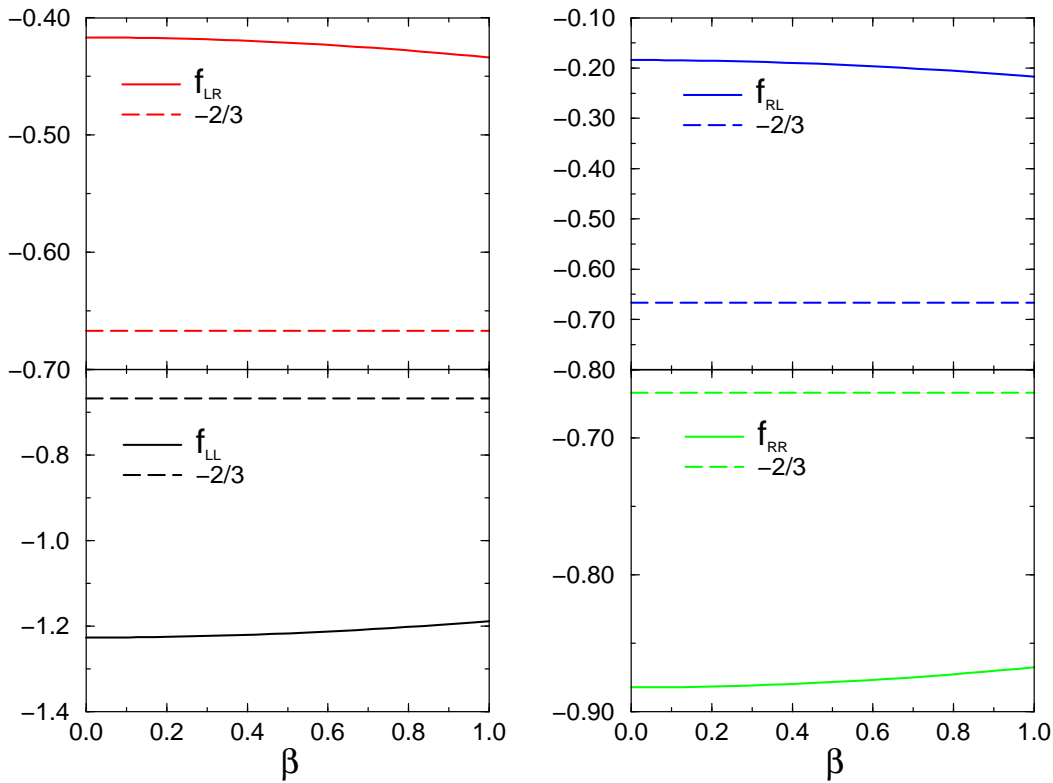,width=14cm,angle=-90} 
\end{tabular}
\caption{Top quark speed, $\beta$, dependence of $f_{LL},f_{LR},f_{RL}$
 and $f_{RR}$.}
\label{fig:para-f}
\end{center}
\end{figure}

Now, consider numerical behavior of the parameter $f_{IJ}$.
The production threshold for top quarks is far above the $Z$ boson mass.
The contribution of the $Z$ boson width to $f_{IJ}$ is suppressed by a factor 
\[
 \frac{M_Z \Gamma_Z}{(s - M_Z^2)^2 + (M_Z \Gamma_Z)^2} ~\sim~ (0.0015)~,     
\]
when $\sqrt{s}=400~{\rm GeV},~M_Z = 91 ~{\rm GeV},~\Gamma_Z=2.49~{\rm
GeV}$.
Therefore, in our discussions bellow, we neglect the $Z$ boson width.
% is negligible and we take only the real part of $f_{IJ}$. 
We use the values for the parameters of the standard model in Table
\ref{tbl:param}. 
\begin{center}
\begin{table}[H]
\begin{tabular}{|c|c|c|c|}
\hline 
$    m    $ &    $M_{Z}$   & $\alpha$  & $\sin^2 \theta_W$ 
\\ \hline \hline 
$ 175 $ [GeV] & $91.187$ [GeV] & $ 1/128 $ & $0.2315$ \\ \hline
\end{tabular}
\caption{Input parameters of the Standard Model.}
\label{tbl:param}
\end{table}
\end{center}
We show the $\beta$ dependence of $f_{IJ}$'s in Fig.\ref{fig:para-f}.
\clearpage
%
%%%%%%%%%%%%%%%%%%%%%%%
% Fig.4
%%%%%%%%%%%%%%%%%%%%%%%
%%%%%%%%%%%%%%%%%%%%%%%
\noindent
For the top quark production, $f_{IJ}$ have a weak $\beta$ dependence and are
given by
\bea
 -1.23 \leq f_{LL} \leq - 1.19 &,& -0.882 \leq f_{LR}\leq -0.867~, 
\nonumber \\
 -0.433 \leq f_{RR} \leq  -0.417  &,& - 0.217 \leq f_{RL} \leq -0.184~.
\nonumber
\eea
When fermions do not couple to $Z$ boson and couple only to the photon, 
all parameters $f_{IJ}$'s go to $-2/3$.
The difference, $f_{IJ} - (- 2/3)$, shows the contribution from the $Z$ boson,
and we can investigate their effects in all energy region.

Since we are interested in maximizing the spin correlations
of top quark pairs, we vary the spin angle, $\xi$, 
to find an appropriate spin basis.
Following the articles~\cite{mp,ps}, we introduce ``helicity'',
``beamline'' and ``off-diagonal'' bases and derive the polarized
cross-sections in these bases from Eqn.(\ref{eqn:GSB}).  
%
%\clearpage

\begin{figure}[H]
\begin{center}
        \leavevmode\psfig{file=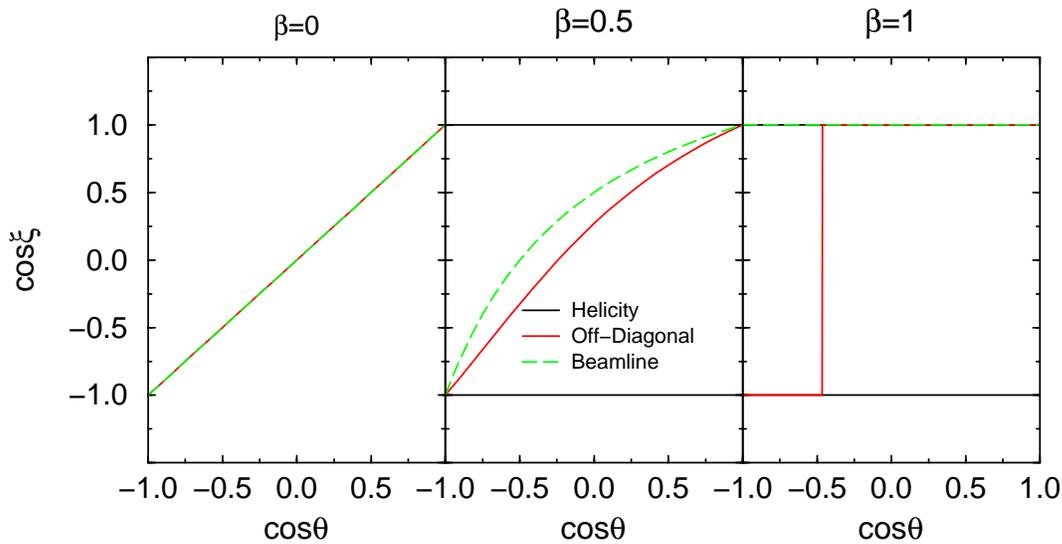,angle=-90,width=14cm}
\caption[The dependence of the spin angle, $\xi$ on the scattering angle 
 $\theta$.]{The dependence
 of the spin angle, $\xi$ on the scattering angle 
 $\theta$ in helicity, beamline and off-diagonal bases at $\beta=0,0.5$
 and $1$.
It is noted that the off-diagonal basis at $\beta=1$
changes from $\cos \xi=-1$ to 
$\cos \xi=+1$ at $\cos \theta = -(f_{LL} - f_{LR})/(f_{LL} + f_{LR})$.}    
\label{fig:basis}
\end{center} 
\end{figure}

\begin{enumerate}
\item{Helicity basis}

The helicity basis, the most familiar spin basis, is given by 
\bea
 \cos \xi ~=~\pm 1,
\label{eqn:HEL}
\eea
for which the top quark spin is defined along its direction of motion.
It is noted that the helicity basis can not be used in the threshold energy
region, because the top (anti-top) quark momentum is undefined in this region.
Substituting Eqn.(\ref{eqn:HEL}) to the general polarized cross-section
in Eqn.(\ref{eqn:GSB}), we can obtain the well known polarized
cross-section:
\bea
    \frac{d \sigma}{d \cos \theta}
    (e^-_L e^+_R \rightarrow t_L \bar{t}_L) 
&=& \frac{d \sigma}{d \cos \theta}
    (e^-_L e^+_R \rightarrow t_R \bar{t}_R) 
    \nonumber \\
&=& \frac{3 \pi \alpha^2}{8 s} \beta 
    | f_{LL} + f_{LR}|^2 (1 - \beta^2) \sin^2 \theta, 
    \label{eqn:HEL-CS}\\
    \frac{d \sigma}{d \cos \theta}
    (e^-_L e^+_R \rightarrow t_R \bar{t}_L
                   {\rm ~or~}  t_L \bar{t}_R ) 
&=&
    \frac{3 \pi \alpha^2}{8 s} \beta
    | f_{LL}(1 \mp \beta) + f_{LR}(1 \pm \beta) |^2
    (1 \mp \cos \theta)^2. 
    \nonumber 
\eea
Here $t_L$ ($t_R$) state corresponds to $t_{\uparrow}$ ($t_{\downarrow}$)
state with $\cos \xi = +1$.
%Similarly, with $\cos \xi = -1$, $t_L$ ($t_R$) state corresponds to
%$t_{\downarrow}$ ($t_{\uparrow}$) state.

\item{Beamline basis}

In the rest frame of top quark, there are three natural choices for 
the direction of the top quark spin; direction of the electron, the
positron or the anti-top quark momentum.
In the threshold region, the anti-top quark momentum direction,
which is undefined, is excluded leaving only the electron or
positron momentum directions.
The situation is the same for the anti-top quark spin. 
The natural choice (see the discussion at the end of this section)is
that the top (anti-top) quark spin is in the direction 
of the positron (electron) momentum in its rest frame.
This spin basis is called \lq\lq beamline basis \rq\rq \cite{mp,ps,pk1},
The spin angle $\xi$ is determined by 
\bea
\cos \xi ~=~ \frac{\cos \theta + \beta}
                  {1 + \beta \cos \theta}.
\label{eqn:beam-def}
\eea
The polarized cross-sections in the beamline basis are
\bea
    \frac{d \sigma}{d \cos \theta}
    (e^-_L e^+_R \rightarrow t_{\uparrow} \bar{t}_{\uparrow}) 
&=& \frac{d \sigma}{d \cos \theta}
    (e^-_L e^+_R \rightarrow t_{\downarrow} \bar{t}_{\downarrow}) 
    \nonumber \\
&=& \left( \frac{3 \pi \alpha^2}{2 s} \beta \right)
    | f_{LR} |^2 
    \frac{\beta^2 (1 - \beta^2) \sin^2 \theta}
         {(1 + \beta \cos \theta)^2} , 
    \nonumber \\
    \frac{d \sigma}{d \cos \theta}
    (e^-_L e^+_R \rightarrow t_{\downarrow} \bar{t}_{\uparrow}) 
&=&
    \left( \frac{3 \pi \alpha^2}{2 s} \beta \right)
    |f_{LR}|^2 
      \frac{\beta^4 \sin^4 \theta} 
           {(1 + \beta \cos \theta)^2}, 
    \label{eqn:BEAM-CS}\\
    \frac{d \sigma}{d \cos \theta}
    (e^-_L e^+_R \rightarrow t_{\uparrow} \bar{t}_{\downarrow}) 
&=&
    \left( \frac{3 \pi \alpha^2}{2 s} \beta \right)
    \Biggl| 
      f_{LL}(1 + \beta \cos \theta ) + 
      f_{LR}\frac{(1 - \beta)^2}{(1 + \beta \cos \theta)} 
    \Biggr|^2.
    \nonumber 
\eea

\item{Off-diagonal basis}

There exists the third possibility called \lq\lq off-diagonal\rq\rq\
basis which makes the contributions from the like-spin configuration
vanish~\cite{ps}. 
The spin angle, $\xi$, is determined by
\be
 \tan \xi = \frac{A_{LR}}{B_{LR}}
   = \frac{( f_{LL} + f_{LR} ) \sqrt{1 - \beta^2}\, \sin \theta}
          {f_{LL} ( \cos \theta + \beta )
             + f_{LR} ( \cos \theta - \beta )} \ .
\label{eqn:off-def}
\ee
In this basis, we obtain the polarized cross-sections,
\bea
  \frac{d \sigma}{d \cos \theta}
  ( e^-_L e^+_R \to t_{\uparrow} \bar{t}_{\uparrow} )
  ~=~ 
  \frac{d \sigma}{d \cos \theta}
  ( e^-_L e^+_R \to t_{\downarrow} \bar{t}_{\downarrow} )
  ~\equiv~ 0 \qquad, \qquad \quad \nonumber \\
  \frac{d \sigma}{d \cos \theta}
  ( e^-_L e^+_R \to t_{\uparrow} \bar{t}_{\downarrow} 
           \  {\rm or} 
                      \  t_{\downarrow} \bar{t}_{\uparrow})
    \hspace*{6.3cm} 
    \nonumber \\  
   = \left( \frac{3 \pi \alpha^2}{2 s} \beta \right)
    \Bigl( \sqrt{A_{LR}^2 + B_{LR}^2} \mp D_{LR} \Bigr)^2
    \hspace*{4.6cm} 
    \label{eqn:OFF-CS} \\
   = \left( \frac{3 \pi \alpha^2}{8 s} \beta \right)
      \biggl[f_{LL} (1 + \beta \cos \theta) + 
             f_{LR} (1 - \beta \cos  \theta) 
  \qquad \qquad \qquad
  \nonumber \\
  \mp \sqrt{ \left[ f_{LL} (1 + \beta \cos \theta)+ 
                    f_{LR} (1 - \beta \cos \theta) \right]^2 -
                    4 f_{LL}f_{LR} \beta^2  \sin ^2 \theta }
      \biggr]^2
  \quad
\nonumber.
\eea
It is to be noted that we can consider another \lq\lq off-diagonal
\rq\rq\ basis, 
\bea
 \tan \xi = \frac{A_{RL}}{B_{RL}}
   = \frac{( f_{RR} + f_{RL} ) \sqrt{1 - \beta^2}\, \sin \theta}
          {f_{RR} ( \cos \theta + \beta )
             + f_{RL} ( \cos \theta - \beta )} \,, \label{eqn:offdiaxi2}
\eea  
which makes the contributions from the like-spin 
configuration vanish in the $e^{-}_{R}e^{+}_{L}$ scattering
process.
So, we can consider two \lq\lq off-diagonal \rq\rq\ bases for top quark
pair production because $f_{LR}/f_{LL}~\neq~f_{RL}/f_{RR}$.
However, the difference between 
$f_{LR}/f_{LL}$ and $f_{RL}/f_{RR}$ is not large numerically (see
     Fig.\ref{fig:para-f}). 
Therefore we call the basis defined by 
Eqn.(\ref{eqn:off-def}) as ``off-diagonal'' basis in the following
discussions.
\end{enumerate}
Let us consider the difference between above three bases.
In Fig.\ref{fig:basis}, we plot the dependence of the spin angle $\xi$ on
the scattering angle $\theta$ for the helicity, beamline and off-diagonal
bases at $\beta=0$ (threshold), $\beta=0.5$ ($\sqrt{s}\sim400$ GeV)
and $\beta=1$ (ultra high energy).
\clearpage
%%%%%%%%%%%%%%%%%%%%%%%
% Fig.5
%%%%%%%%%%%%%%%%%%%%%%%%%%
\noindent
When $\beta$ is zero, the helicity basis can not be defined.
However, the beamline and off-diagonal bases are defined also at
threshold and they are given by the same equation, $\xi = \theta$.
This fact says that it is correct that the top (anti-top) quark
spin decomposes into the direction of positron (electron) momentum in the
threshold region. 
When $\beta = 0.5$, we can see apparently the difference among these
three bases.
In particular, for the central region in which $\cos \theta \simeq 0$,
there are remarkable differences between the three bases.
However, the difference between the beamline basis and off-diagonal
basis are small for all scattering angle compared to the helicity basis.
The beamline and off-diagonal bases become close to the helicity basis
with $\cos \xi = \pm 1$ near $\cos \theta = \pm 1$.
When $\beta=1$, the beamline basis coincides with the helicity basis
with $\cos \xi = + 1$.
While, the off-diagonal basis is equal to the helicity basis with $\cos
\xi = -1$ from $\cos \theta =-1$ to $\cos \theta = -(f_{LL} -
f_{LR})/(f_{LL} + f_{LR})$, and equal to the helicity basis with
$\cos \xi = +1$ form $\cos \theta=-(f_{LL} - f_{LR})/(f_{LL} + f_{LR})$
to $ \cos \theta =+ 1$.
The off-diagonal basis corresponds to the different helicity bases at
$\cos \theta \simeq 1$ and $\cos \theta \simeq - 1$.

%\clearpage

\begin{figure}[H]
\begin{center}
\leavevmode\psfig{file=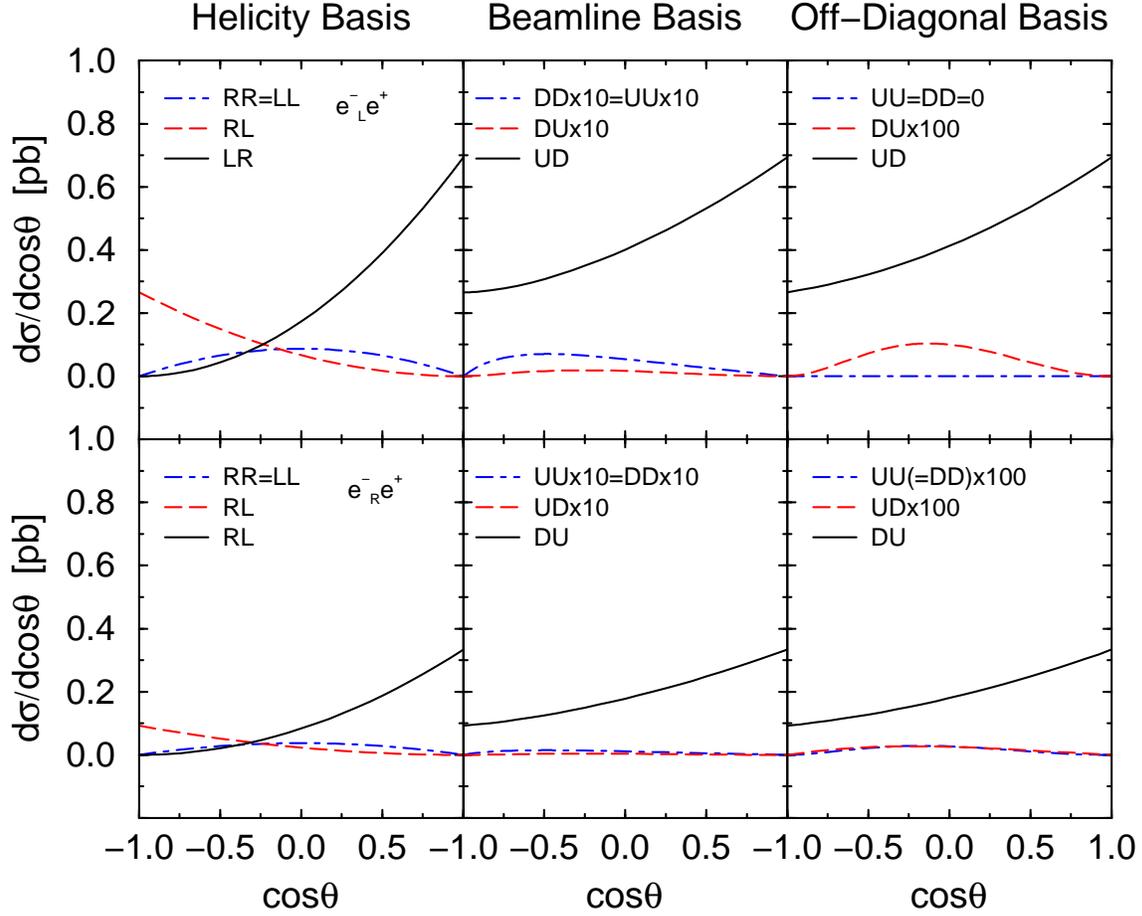,width=15cm,angle=-90}  
\caption[The polarized differential cross-sections.]{The polarized
 differential cross-sections in the helicity, 
 beamline and  off-diagonal bases at $\sqrt{s}~=~400~{\rm GeV}$ for the 
  $e^{-}_{L} e^+ (e^{-}_{R} e^+) \rightarrow t \bar{t}$ process.
We plot the 
 $t_{L} \bar{t}_{L}$ (LL), $t_{R} \bar{t}_{R}$ (RR),
 $t_{R} \bar{t}_{L}$ (RL), $t_{R} \bar{t}_{L}$ (RL) productions 
for the helicity basis and  
 $t_{\uparrow} \bar{t}_{\downarrow}$ (UD),
 $t_{\downarrow} \bar{t}_{\uparrow}$ (DU),
 $t_{\uparrow} \bar{t}_{\uparrow}$ (UU) and 
 $t_{\downarrow} \bar{t}_{\downarrow}$ (DD) productions for the beamline
 and the off-diagonal bases.
Note that the sub-dominant and sub-sub-dominant configurations for both $e^-_L e^+$ and $e^-_R e^+$ process are amplified by the factor of 10 in the beamline basis and by the factor of 100 in the off-diagonal basis. }
\label{fig:diff}
\end{center}
\end{figure}

Now we will show the polarized cross-sections for the helicity, the beamline 
and the off-diagonal bases in Fig.\ref{fig:diff} at $\sqrt{s}=400$ GeV.
\clearpage
%%%%%%%%%%%%%%%%%%%%%%%
% Fig.6
%%%%%%%%%%%%%%%%%%%%%%%
\noindent
%%%%%%%%%%%%%%%%%%%%%%%
\begin{center}
\begin{table}[H]
\begin{tabular}{|c||c|c|c|}
\hline 
$e^{-}_L e^+$ 
                     & Helicity & Beamline & Off-Diagonal \\ \hline
\hline
Dominant Frac.       & $0.5305$ (LR)& $0.97846$ ~(UD)~  & $0.99876$ (UD)\\ \hline
Sub-Dominant Frac.   & $0.2033$ (RL)& $0.00964$ (UU,DD) & $0.00124$ (DU)\\ \hline
\hline 
$e^{-}_R e^+$ 
                     & Helicity & Beamline & Off-Diagonal \\ \hline
\hline
Dominant Frac.       & $0.5804$ (RL)& $0.98979$ ~(DU)~  & $0.99744$ ~(DU)~ \\ \hline
Sub-Dominant Frac.   & $0.1608$ (LR)& $0.00457$ (UU,DD) & $0.00089$ (UU,DD)\\ \hline
\end{tabular}
\caption[The fraction of $e^{-}_{L/R} e^+$ cross-sections for the
 dominant and the sub-dominant spin configurations.]{The fraction of
 $e^{-}_{L/R} e^+$ cross-sections for the 
 dominant and the sub-dominant spin configurations in the helicity,
 beamline and off-diagonal bases at $\sqrt{s}=400$ GeV.}
\label{tbl:frac}
\end{table}
\end{center}
The dominant component for the helicity basis is $t_{L} \bar{t}_{R}$
(LR) for $e^{-}_{L} e^+$ scattering and  
$t_{R} \bar{t}_{L}$ (RL) for $e^{-}_{R} e^+$ scattering at $\sqrt{s}=400$ GeV.
This dominant component occupies $53 \%$ of the total cross-section for
$e^{-}_{L} e^+$ scattering and $58 \%$ of the total cross-section for
$e^{-}_{R} e^+$ as shown in Table \ref{tbl:frac}.
Other components make up more than $40 \%$ of the total cross-section.
Thus, the helicity basis does not show strong spin correlations for the
polarized cross-sections at this energy.
In contrast, in the beamline and the off-diagonal bases, only one
component dominates the total cross-section.
The dominant component for the beamline and the off-diagonal bases is $t_{\uparrow}
\bar{t}_{\downarrow}$ (UD) and makes up more than $97 \%$ of the total
cross-section for $e^{-}_{L} e^+$ scattering.
For $e^{-}_{R} e^+$ scattering, the dominant contribution in the
beamline and off-diagonal bases is $t_{\downarrow}
\bar{t}_{\uparrow}$ (DU) component and make up more than $98 \%$ of the
total cross-section.
Other components give almost zero contribution.
Apparently the top (anti-top) quark spin is strongly correlated with
positron (electron) spin in the beamline and off-diagonal bases.

Here we focus on the beamline and off-diagonal bases and try to explain 
(1){\sl why the top (anti-top) quark spin is associated with the positron
(electron) spin } and  
(2){\sl why the polarized cross-section in the
beamline and off-diagonal bases show similar behaviors}. 
To answer these questions, recall the expression Eqn.(\ref{eqn:ampli1}) for
$e^-_L e^+ \rightarrow t_{s_{t}} \bar{t}_{s_{\bar{t}}} $.
\bea
\lefteqn{
{\cal M}
       (e^-_L e^+_R \rightarrow t_{s_{t}} \bar{t}_{s_{\bar{t}}})
        } \nonumber \\
&=&
\frac{4 \pi \alpha}{s} \bra{\bar{q}-} \gamma^{\mu}_{L} \ket{q-}
\bar{u}(t,s_t)
\gamma_{\mu} \left[ f_{LL} \gamma_{L} + f_{LR} \gamma_{R} \right] 
v(\bar{t},s_{\bar{t}}). 
\label{eqn:ampli-next}
\eea 
Using Fierz transformations in Eqn.(\ref{eqn:ampli-next}), we obtain,
\bea
{\cal M}
       (e^-_L e^+_R \rightarrow t_{s_{t}} \bar{t}_{s_{\bar{t}}})
&=& \frac{8 \pi \alpha}{s} f_{LL}
\biggl[ \,\bar{u}(t,s_t) \ket{\bar{q}+} \bra{q+} v(\bar{t},s_{\bar{t}}) 
    \nonumber \\ 
&& \hspace*{.6cm}
+ \frac{f_{LR}}{f_{LL}} 
\,\bar{u}(t,s_t) \ket{q-} \bra{\bar{q}+} v(\bar{t},s_{\bar{t}}) \,
\biggr].
\label{eqn:fierz}
\eea
\clearpage
%%%%%%%%%%%%%%%%%%%%%%%
% Fig.7
%%%%%%%%%%%%%%%%%%%%%%%
\begin{figure}[H]
\begin{center}
\begin{tabular}{cc}
\leavevmode\psfig{file=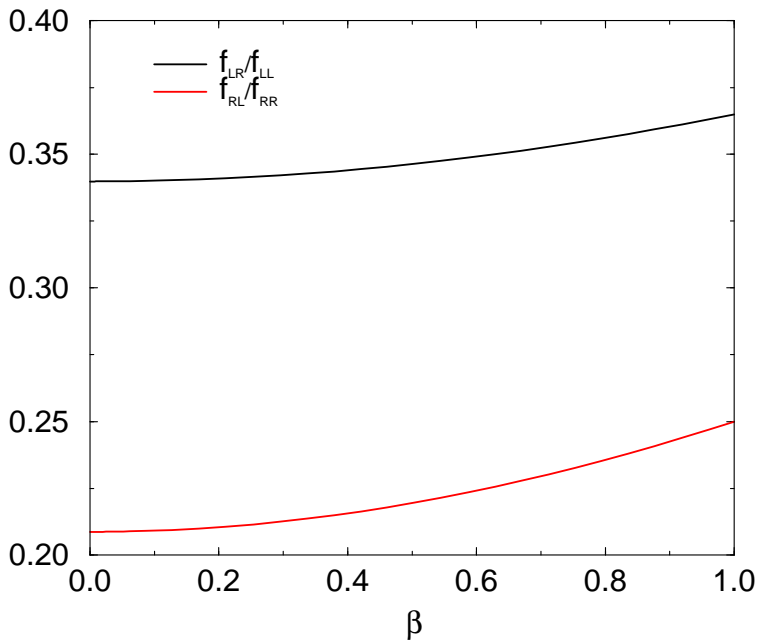,width=10cm,angle=-90} 
\end{tabular}
\caption{Top quark speed, $\beta$, dependence of ratios, $f_{LR}/f_{LL}$ and
 $f_{RL}/f_{RR}$.}
\label{fig:para-r}
\end{center}
\end{figure}
\noindent
Figure (\ref{fig:para-r}) shows the top quark speed $\beta$ dependence of
the ratio $f_{IJ}$'s.
For the top quark production, the $f_{IJ}$'s have a weak energy
dependence;
Numerically the parameters $f_{IJ}$ fall in the range,
\bea
0.34 \leq \frac{f_{LR}}{f_{LL}} \leq 0.37 , \\
0.21 \leq \frac{f_{RL}}{f_{RR}} \leq 0.25 . 
\eea
Because the ratio $|f_{LR}/f_{LL}|$ is less than 
1, the first term of  Eqn.(\ref{eqn:fierz}) is dominant.
This implies that the top (ant-top) quark spin is associated with
positron (electron) spin.  

We now come to the question (2).
Now remember the definition of the spin angle $\xi$ in the beamline and the
off-diagonal bases in Eqns.(\ref{eqn:beam-def}) and (\ref{eqn:off-def}).
The spin angle $\xi$ for the beamline basis is obtained from
Eqn.(\ref{eqn:off-def}) with $f_{LR} ~=~ 0$.
Then we focus on the factor $f_{LR}$ to answer the question (2).
Since the factor $|f_{LR}|$ is smaller than $|f_{LL}|$,
let us take an approximation of $f_{LR}=0$ in
Eqns.(\ref{eqn:BEAM-CS}) and (\ref{eqn:OFF-CS}). 
The polarized cross-sections in the beamline and off-diagonal bases become
the same expressions as 
\bea
    \frac{d \sigma}{d \cos \theta}
    (e^-_L e^+_R \rightarrow t_{\uparrow} \bar{t}_{\uparrow}) 
&=& \frac{d \sigma}{d \cos \theta}
    (e^-_L e^+_R \rightarrow t_{\downarrow} \bar{t}_{\downarrow}) 
~=~ 0,
    \nonumber \\
    \frac{d \sigma}{d \cos \theta}
    (e^-_L e^+_R \rightarrow t_{\downarrow} \bar{t}_{\uparrow}) 
&=& 0,
    \label{eqn:BEAM-OFF}\\
    \frac{d \sigma}{d \cos \theta}
    (e^-_L e^+_R \rightarrow t_{\uparrow} \bar{t}_{\downarrow}) 
&=&
    \left( \frac{3 \pi \alpha^2}{2 s} \beta \right)
    |f_{LL}|^2 (1 + \beta \cos \theta )^2.
    \nonumber 
\eea
Thus the difference between the polarized cross-sections in the beamline
    and the off-diagonal bases is ${\cal
    O}(|f_{LR}/f_{LL}|)$.
The subdominant contributions, (UU, DD, DU), come from the ${\cal
    O}(|f_{LR}/f_{LL}|)$ corrections.  
This is the reason why polarized cross-sections in the beamline basis
behaves like that in the off-diagonal basis.

\begin{figure}[H]
\begin{center}
        \leavevmode\psfig{file=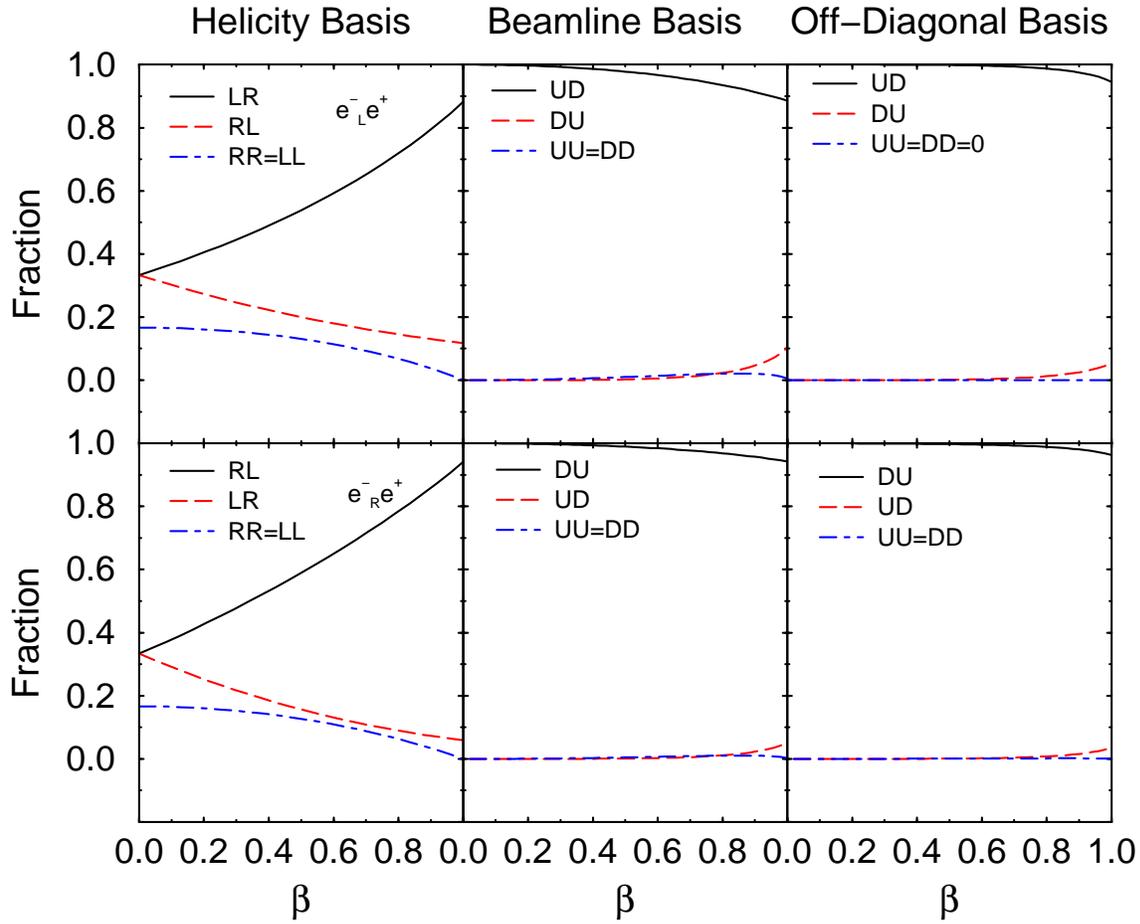,angle=-90,width=15cm}
\caption[The fraction of top quark pairs as a function of the top quark speed, $\beta$.]{The fraction of top quark pairs in the helicity, beamline and
 off-diagonal bases, as a function of the top quark speed, $\beta$.}
\label{fig:fraction}
\end{center} 
\end{figure}

Finally we discuss the top quark speed $\beta$ dependence of the
polarized cross-sections in the helicity, beamline and off-diagonal
bases.
We define the fraction $\sigma(e^-_L e^+ \rightarrow
t_{s_{t}}t_{s_{\bar{t}}})/\sigma_T$ of top pair productions for each spin
configurations, where $\sigma_T$ is the total cross-section for
the process, $e^-_Le^+ \rightarrow t \bar{t}$ and the suffix 
$s_{t}$ ($s_{\bar{t}}$) represents the spin of top (anti-top) quark.
We present the fraction in the helicity, beamline and off-diagonal bases in
Fig.\ref{fig:fraction}.
\clearpage
%%%%%%%%%%%%%%%%%%%%%%%
% Fig.8
%%%%%%%%%%%%%%%%%%%%%%%%%%
\noindent
In the helicity basis, as the CM energy get larger, the spin correlation becomes
stronger for $e^-_Le^+$ and $e^-_Re^+$ scattering.
%In Fig.\ref{fig:fraction}, the helicity basis is a useful
%basis in the ultra high energy limit. 
The polarized cross-sections in the beamline basis coincide with those in the
helicity basis in the limit $\beta$ to $1$.
This is in agreement with the fact that 
the helicity and beamline bases become identical in the limit of $\beta$ to
$1$ (see the discussion on Fig.\ref{fig:basis}). 
In the beamline and off-diagonal bases, the UD ($t_{\uparrow}
\bar{t}_{\downarrow}$) channel dominates for $e^-_L e^+$ process
for all values $\beta$ as shown in Fig.\ref{fig:fraction}.
(The DU ($t_{\downarrow} \bar{t}_{\uparrow}$) channel for $e^-_R e^+$
gives the dominant fraction in the beamline and off-diagonal bases.)
However, there is a slight difference between in the beamline basis and in the
off-diagonal basis near $\beta=1$.
This is due to the fact that the off-diagonal basis 
corresponds to the helicity basis with either $\cos \xi = +1$ or with $\cos
\xi = -1$, depending on the scattering angle,
while the beamline basis coincides the helicity 
basis with $\cos \xi =1$ in the limit $\beta$ to $1$. 
\clearpage

\section{Top Quark Decays at Leading Order}
In the previous section, we have shown that the top quark pairs are
produced in an unique spin configurations at polarized $e^+e^-$ linear
colliders in a appropriate basis. 
We discuss in this section the spin-angular correlations for the top
quark decay and measurement of the top quark spin.
Because of the heavy mass of top quark, the width of top quark amounts
to nearly $1.57$ GeV~\cite{Decays}.
In consequence, the top quark decays rapidly through the electro-weak
interaction before the hadronization (which is governed by the scale
$\Lambda_{QCD}$) take place. 
Therefore, we can obtain information of the top quark spin by measuring 
its decay products without being suffered from the complicated
hadronization effects \cite{BIGI}. 
We show the electro-weak decay products of polarized top quark
are strongly correlated to its spin axis at the leading order in
perturbation theory of the Standard Model.  
%%%%%%%%%%%%%%%%%%%%%%%
% Fig.8
\begin{figure}[H]
\begin{center}
\begin{tabular}{cc}
        \leavevmode\psfig{file=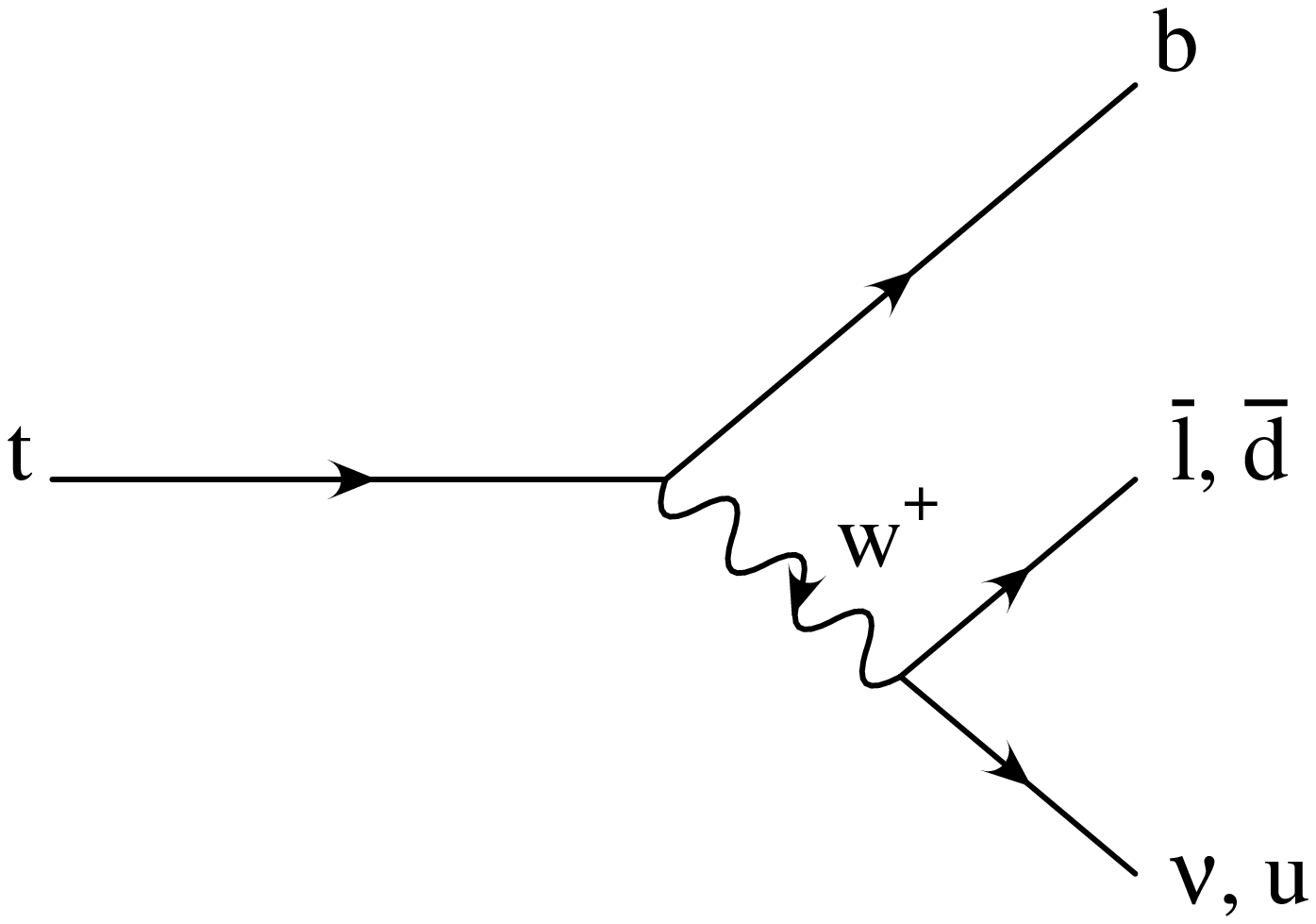,width=7cm} &
        \leavevmode\psfig{file=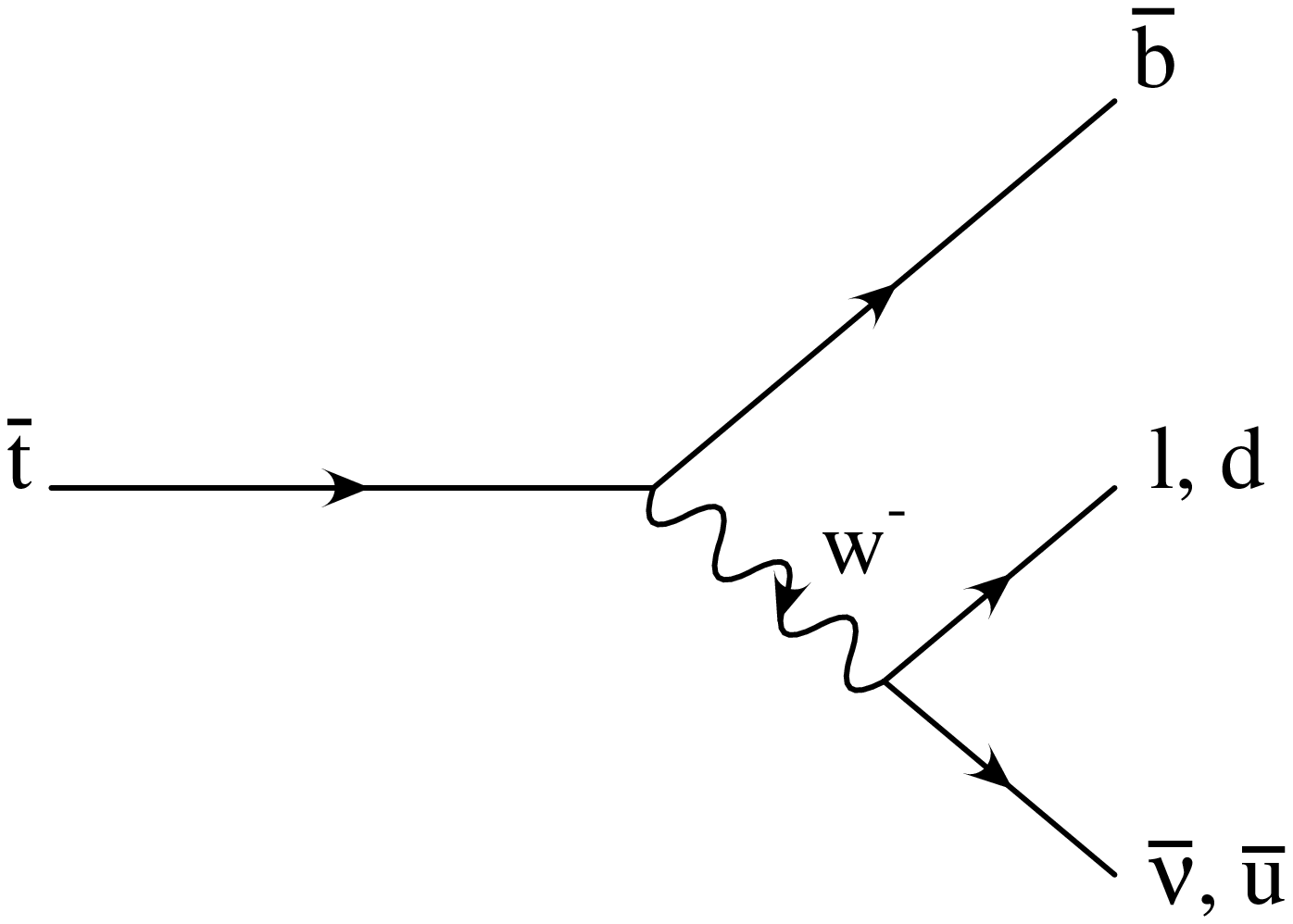,width=7cm} 
\end{tabular}
\caption[Top and Anti-Top Quark Decay Process]{The hadronic and
 semi-leptonic decay process of top (anti-top) quark}
\label{fig:feynd-d}
\end{center}
\end{figure}
%%%%%%%%%%%%%%%%%%%%%%%%%%
The top (anti-top) quarks decay dominantly,
\begin{eqnarray}
 t       ~ \rightarrow ~ W^{+} b,~ 
 \bar{t} ~ \rightarrow ~ W^{-} \bar{b}. 
\end{eqnarray}
For the $W^{+}$ ($W^{-}$) decay, there are hadronic decay
and leptonic decay modes (see Fig. \ref{fig:feynd-d});
\begin{eqnarray}
\begin{array}{cccccc}
 W^{+}       & \rightarrow & u \bar{d} ~{\rm or }~ c \bar{s}, &
 W^{-}       & \rightarrow & d \bar{u} ~{\rm or }~ s \bar{c}, \\ 
 W^{+}       & \rightarrow & \bar{l}~ \nu_{l}, &
 W^{-}       & \rightarrow & l ~ \bar{\nu}_{l}. 
\end{array}
\label{eqn:had-lep}
\end{eqnarray}
Now let us calculate the decay distribution for the top quark decay.
(The anti-top quark decay can be calculated similarly.)
%Note that for the anti-top quark decay, one can derive it using same
%manner as top quark decay.
Momenta of the particles are assigned by its symbol, and
we decompose the top momentum $t$ into a sum of two massless momenta
$t_{1}$ and $t_{2}$,  
\[
t ~=~ t_{1} + t_{2}.
\]
The spin vector of top quark is related to the momenta $t_1,~t_2$
by the following relation,
\[
s_{t} ~=~ \frac{1}{2 m}(t_{1} - t_{2}).
\]
The amplitudes for $t_{\uparrow}/t_{\downarrow} \to b \bar{l} \nu_{l}$
are given by,
\bea
{\cal M}(t_{\uparrow} \to b \bar{l} \nu_{l})
& = & - g^2 
\frac{V_{tb}\mbk{b~\nu} [\bar{l}~t_{2}]}
{(t - b)^2 - M_{W}^2 + i M_{W} \Gamma_{W}}
\frac{\mbk{t_2~t_1}}{m}~,
\label{eqn:t-d-amp1} \\
{\cal M}(t_{\uparrow} \to b \bar{l} \nu_{l})
& = & - g^2 
\frac{V_{tb}\mbk{b~\nu} [\bar{l}~t_{1}]}
{(t - b)^2 - M_{W}^2 + i M_{W} \Gamma_{W}}
~.
\label{eqn:t-d-amp2}
\eea
The mass and the width of $W$ boson are $M_{W}$ and $\Gamma_{W}$,
and ${\rm SU(2)}$ coupling $g$ is related to the electro-magnetic coupling 
through $g=e/\sin \theta_{W}$.
$V_{tb}$ is the Cabibbo-Kobayashi-Maskawa (CKM) matrix element.
For the hadronic decay of the top quark, one should replace
$\bar{l}$ with $\bar{d}$ and $\nu$ with $u$ in
Eqns.(\ref{eqn:t-d-amp1}) and (\ref{eqn:t-d-amp2}).
%The rate of the hadronic decay mode is approximately 9 times larger than 
%the rate of each leptonic decay mode.
%However, we must take into account the efficiency to identify the  
%d-type quark from u-type (or vice versa) in the process to utilize the
%hadronic decay products as a spin analyzer. 
%Therefore, we consider the the semi-leptonic decay mode of the top
%(anti-top) quark to simplify our discussions.
%
%********************************************************************
%%%%%%%%%%%%%%%%%%%%%%%%
%% Fig.8
%\begin{figure}[H]
%\begin{center}
%\begin{tabular}{cc}
%        \leavevmode\psfig{file=top-d.ps,width=7cm} &
%        \leavevmode\psfig{file=antitop-d.ps,width=7cm} 
%\end{tabular}
%\caption[Top and Anti-Top Quarks Decay Process]{The hadronic and
% semi-leptonic decay process of top (anti-top) quark}
%\label{fig:feynd-d}
%\end{center}
%\end{figure}
%%%%%%%%%%%%%%%%%%%%%%%%%%%
The squared matrix elements for the top (anti-top) quark decay, $t
\rightarrow b \bar{l} \nu_{l}$, are given by 
\bea
|{\cal M}(t_{\uparrow} \rightarrow b \bar{l} \nu_{l})|^2
 &=& 
\frac{4 g^4 |V_{tb}|^2}
     {[(\bar{l} + \nu)^2 - M_{W}^2]^2 + M_{W}^2 \Gamma^2_{W}}
    (b \cdot \nu)(\bar{l} \cdot t_{2})~, 
\label{eqn:decay-t2}\\
|{\cal M}(t_{\downarrow}\rightarrow b \bar{l} \nu_{l})|^2
 &=& 
\frac{4 g^4 |V_{tb}|^2}
     {[(\bar{l} + \nu)^2 - M_{W}^2]^2 + M_{W}^2 \Gamma^2_{W}}
    (b \cdot \nu)(\bar{l} \cdot t_{1}) ~.
\label{eqn:decay-t1}
%\\
%|{\cal M}_{\uparrow}( \bar{t} \rightarrow \bar{b} l \bar{\nu}_{l})|^2
% &=& 
%\frac{4 g^4 |V_{tb}|^2}
%     {[(l + \bar{\nu})^2 - M_{W}^2]^2 + M_{W}^2 \Gamma^2_{W}}
%    (\bar{b} \cdot \bar{\nu})(l \cdot \bar{t}_{1})~, 
%\label{eqn:decay-tbar1}\\
%|{\cal M}_{\downarrow}( \bar{t} \rightarrow \bar{b} l \bar{\nu}_{l})|^2
% &=& 
%\frac{4 g^4 |V_{tb}|^2}
%     {[(l + \bar{\nu})^2 - M_{W}^2]^2 + M_{W}^2 \Gamma^2_{W}}
%    (\bar{b} \cdot \bar{\nu})(l \cdot \bar{t}_{2})~. 
%\label{eqn:decay-tbar2}
\eea
To calculate the decay distributions, we take the top (anti-top) quark rest
frame in which the direction of top quark spin axis coincides with the
$z-$axis.
%We consider the decay process of the top quark, since the decay
%process of anti-top quark is calculated in the same way for the top
%quark case.
The kinematical variables for top quark decay process are,  
\bea
\begin{array}{rcl}
s_{t}^{\mu} &=& (0,{\bf s}_{t})~=~(0,0,0,1)~,  \\
t^{\mu}  &=& m (1,0,0,0)~,\\
t_{1} &=& (m/2)(1,0,0,1)~,\\
t_{2} &=& (m/2)(1,0,0,-1)~,\\
\bar{l}^{\mu}  &=& E_{\bar{l}} (1,\sin \theta_{\bar{l}} \cos \phi,
                      \sin \theta_{\bar{l}} \sin \phi,
                      \cos \theta_{\bar{l}})~. \\
\end{array}
\eea
Here the lepton mass is neglected, and the energy of lepton
is denoted by $E_{\bar{l}}$.
The variables $\theta_{\bar{l}},~\phi$ are angles to specify the
orientation of top quark spin as in Fig.\ref{fig:d-config}.
\clearpage
%%%%%%%%%%%%%%%%%%%%%%%%%%
% Fig.8
\begin{figure}[H]
\begin{center}
        \leavevmode\psfig{file=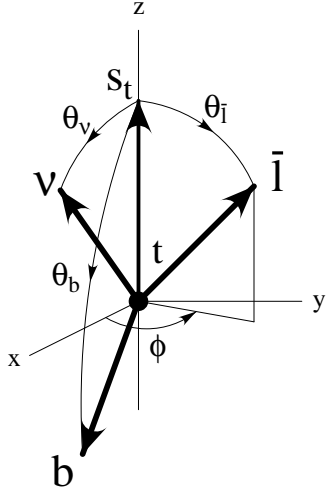,width=4.3cm}
\caption[The Momentum Configurations in Top Quark Decay]{The
 momentum configurations for decay products, lepton and neutrino, in the
 top quark rest frame. 
The direction of top quark spin is defined by the direction of $+z$ axis.}
\label{fig:d-config}
\end{center} 
\end{figure}
%%%%%%%%%%%%%%%%%%%%%%%%%%

It is convenient to introduce the scaled variables, $w, y, \bar{y}, x_{\bar{l}}$
and $x_{\nu}$.
\bea
w^2 &=& (\bar{l} + \nu)^2,~
y ~=~ \frac{w^2}{m^2},~
\bar{y} ~=~ \frac{M_{W}^2}{m^2}, \nonumber \\
\gamma &=& \frac{\Gamma_{W}}{M_{W}},~
x_{\bar{l}} ~=~ \frac{2 E_{\bar{l}}}{m},~
x_{\nu} ~=~ \frac{2 E_{\nu}}{m}. \nonumber 
\eea
The three particle phase space integral can be parametrized in terms of scaled
variables $x_{\bar{l}}$, $y$, and the angles $\theta_{\bar{l}}$, $\phi$.
\bea
\int (PS)_{3} 
&=& \int \frac{d^3 b}{(2 \pi)^3 (2 b^0)}
         \frac{d^3 \bar{l}}{(2 \pi)^3 (2 \bar{l}^0)}
         \frac{d^3 \nu}{(2 \pi)^3 (2 \nu^0)}
         (2 \pi)^4
         \delta^{4}(t - b - \bar{l} - \nu), \nonumber \\
&=& \frac{m^2}{2 (4 \pi)^4}
    \int_{0}^{1} d x_{\bar{l}} 
    \int_{0}^{x_{\bar{l}}} d y
    \int d \cos \theta_{\bar{l}} d \phi.
\eea
The differential distribution for the lepton, $\bar{l}$, in the decay of the
polarized top quark can be written as 
\bea
\frac{d \Gamma (s_{t})}{d x_{\bar{l}} d y d \cos \theta_{\bar{l}}}
    (t \rightarrow b \bar{l}\nu_{l})
~ = ~
\frac{m g^4}{16 (4 \pi)^3} \left(\frac{|V_{tb}|^2}{\bar{y}^2}\right)
\frac{F_{0}(s_{t})}{(1 - y/\bar{y})^2 + \gamma^2},
\label{eqn:tri-d}
\eea
where $s_{t}$ denotes top quark spin states ($\uparrow$ or $\downarrow$),
and the function $F_{0}(s_{t})$ is given by
\bea
\begin{array}{rcl}
F_{0}(\uparrow)   &=& x_{\bar{l}} (1 - x_{\bar{l}})(1 + \cos \theta_{\bar{l}}), \\
F_{0}(\downarrow) &=& x_{\bar{l}} (1 - x_{\bar{l}})(1 - \cos \theta_{\bar{l}}). \\
\end{array}
\eea
After integrating over variable $y$ in
Eqn.(\ref{eqn:tri-d}), we obtain the double differential
($x_{\bar{l}}-\theta_{\bar{l}}$) distribution.
\bea
\frac{d \Gamma (s_{t})}{d x_{\bar{l}} d \cos \theta_{\bar{l}}}
    (t \rightarrow b \bar{l} \nu_{l})
~ = ~
\frac{m g^4}{16 (4 \pi)^3} \left(\frac{1}{\bar{y} \gamma}\right)
F_{0}(s_{t})
\arctan \left[ \frac{\gamma x_{\bar{l}}}
                    {(1 + \gamma^2) \bar{y} - x_{\bar{l}}} \right].
\label{eqn:dbl-d}
\eea
This result shows that the double differential distribution in written
in a factrized form, the energy distribution and the angular distribution.

The above formula Eqn.\ref{eqn:dbl-d} will be reduced to a simple form
if we take into account that 
the $W$ boson propagator in Eqn.(\ref{eqn:tri-d}) can be approximated as 
for the process $t \to b \bar{l} \nu$.
%$\Gamma_{W} \simeq 2.06~{\rm GeV}$, $M_{W} \simeq 80.41~{\rm GeV}$,
\[
 \frac{1}{(1 - y/\bar{y})^2 + \gamma^2}
~=~ \frac{\pi}{\gamma} \delta \left( 1 - \frac{y}{\bar{y}}\right)
~+~ {\cal O}( \gamma )~.    
\]
In fact, the value of $\gamma$ is small, $\sim 0.02$ and  we can neglect the
 contribution from terms of order ${\cal O}(\gamma)$ (the narrow width
 approximation). 
%Therefore, we can take the internal particle, $W$ boson, on-sell in this
% approximation, and deal with this process simple.
In the narrow width approximation for $W$ boson, $\gamma \rightarrow 0$,
Eqn.(\ref{eqn:dbl-d}) is reduced to,
\noindent
\bea
\frac{1}{\Gamma_{T}}
\frac{d \Gamma}{d x_{\bar{l}}~d \cos \theta_{\bar{l}}} &=&
\frac{1}{2} {\cal F}(x_{\bar{l}}) \left[ 1 +
\alpha_{\bar{l}} \cos \theta_{\bar{l}} \right], 
\label{eqn:factri-d}\\
{\cal F}(x_{\bar{l}}) &=&
\frac{3 x_{\bar{l}} (1 - x_{\bar{l}})}{(1 - \bar{y})^2(1 + 2 \bar{y})}~,
\eea
with the region of kinematical valuable $x_{\bar{l}}$,
\bea
\bar{y} ~ \leq ~ x_{\bar{l}} ~ \leq ~1.  
\eea
Here $\Gamma_{T}$ is the total decay width,
and the parameter $\alpha_{\bar{l}}$ ($\alpha_{\bar{l}}=1$ for the
$\uparrow$ quark spin state and $\alpha_{\bar{l}}=-1$ for the
$\downarrow$ state.) denotes the strength of
correlation between the momentum direction of decay particle
$\bar{l}$ and the direction of top quark spin axis.
%And the correlation coefficient $\alpha_{\bar{l}}$ is $+1$ for the decay
%process, $t_{\uparrow} \rightarrow b \bar{l} \nu_{l}$, and $-1$ for the
%decay process, 
%$t_{\downarrow} \rightarrow b \bar{l} \nu_{l}$.
%
%As we shown the double differential decay distribution of top quark,
%which is evaluated by tagging the energy and the angle of the lepton,
%the differential distribution is written by simple form
%Eqn.(\ref{eqn:factri-d}).
%In general, however,

In the same way, we can calculate the differential distribution for any
decay product $i$ ($i~=~b,~\bar{l},~\nu,~u,~\bar{d},~W$).
The differential angular distribution, in general, is written as after
integrating over $x_i$ ~\cite{JEZA,KUHN,GREG},
\bea
\frac{1}{\Gamma_{T}}
\frac{d\Gamma}{d \cos\theta_{i}}
& = &
\frac{1 + \alpha_{i} \cos\theta_{i}
     }{2},
\label{eqn:DECAY-DIS}
\eea
where $\theta_{i}$ denotes the angle between the top quark spin axis
${\bf s}_{t}$ and the direction of motion of decay product $i$ ($i=b,
~\bar{l},~\nu,~u,~\bar{d}$,~W).
The correlation coefficient $\alpha_{i}$ may be computed from the
squared matrix elements (\ref{eqn:decay-t2})$\sim$(\ref{eqn:decay-t1}).
(see Refs.\cite{JEZA,KUHN} for the details.)
\clearpage
\begin{table}[H]
\begin{center}
%\begin{tabular}{|c||c|r|}
%\hline
%$i$                    & $\alpha_{i}$ & $\alpha_{i}$
%\\ \hline\hline
%$\bar{l}$ or $\bar{d}$ & $1$ & $1.00$ \\ \hline
%$\nu$ or $u$           & 
%$\frac{(\bar{y} - 1)(2 \bar{y}^2 + 11\bar{y} -1) - 12 \bar{y}^2 \ln \bar{y}}
%      {(2\bar{y} + 1)(\bar{y} - 1)^2}$  
%                       & $-0.318$ \\ \hline
%$W^+$                  & $- \frac{\bar{y} - 2}{\bar{y} + 2}$ 
%                       & $0.406$ \\ \hline
%$b$                    & $\frac{\bar{y} - 2}{\bar{y} + 2}$ 
%                       & $-0.406$ \\ \hline
%\end{tabular}
\bea
\begin{array}{|c||c|r|}
\hline
i & \alpha_{i} & \alpha_{i} 
\\ \hline\hline
\bar{l}{\rm ~or~}\bar{d} & 1 & 1.00 \\ \hline
\nu {\rm ~or~} u           & 
\frac{
  (\bar{y} - 1)(2 \bar{y}^2 + 11\bar{y} -1) 
- 12 \bar{y}^2 \ln \bar{y}}
{(2\bar{y} + 1)(\bar{y} - 1)^2}  
                       & -0.318 \\ \hline
W^+                  & - \frac{\bar{y} - 2}{\bar{y} + 2} 
                       & 0.406 \\ \hline
b                    & \frac{\bar{y} - 2}{\bar{y} + 2} 
                       & -0.406 \\ \hline
\end{array}
\nonumber
\eea
\caption[Correlation coefficients $\alpha_{i}$ for the polarized top
 quark decay in the narrow width approximation for $W$
 boson.]{Correlation coefficients $\alpha_{i}$ for the polarized top  
 quark decay in the narrow width approximation for $W$ boson.
The $b$ quark mass is neglected, and the values $m=175.0$ 
 GeV and $M_{W}=80.41$ GeV are used.} 
\label{tbl:alpha}
\end{center}
\end{table}
\noindent
For the spin-up top quark, the correlation coefficients $\alpha_{i}$ 
are given in Table.\ref{tbl:alpha}.
For the spin-down top quark, the correlation coefficients have the
opposite sign.  
%
%\newpage

\begin{figure}[H]
\begin{center}
        \leavevmode\psfig{file=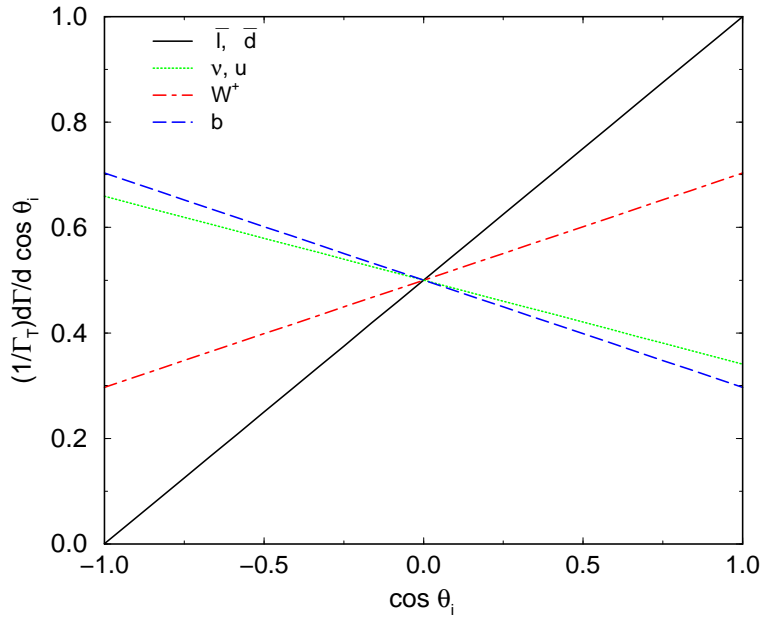,width=10cm,angle=-90}
\caption[The Angular Distributions of Top Decay Products]{The angular distributions of decay products $i$, ($\bar{l}, \bar{d}, u, \nu_{l},b$ and $W^{+}$), in the top quark decay process.
The $\theta_{i}$ is the angle between the direction of top quark spin and 
 the moving direction of decay product $i$.}
\label{fig:angle-d}
\end{center} 
\end{figure}

Using the values in Table.\ref{tbl:alpha}, we plot the single particle spin-angular
correlation Eqn.(\ref{eqn:DECAY-DIS}) in Fig.\ref{fig:angle-d}.

\noindent
%In Fig.\ref{fig:angle-d}, 
The lepton and d-type quark are maximally
correlated with the top quark spin axis compared with other particles,
$u, \nu_{l},b$ and $W^{+}$.
It is surprising that the charged lepton or $d$-type quark come from the 
decay of the $W$ boson which is less correlated with the top quark spin.
%This comes from the fact that the cancellation of the charged lepton or
%$d$-type quark requires a subtle cancellation between the amplitudes
%involving transverse and longitudinal $W$ boson for the top quark decay
%in the Standard Model~\cite{ps}.

Here it might be interesting and instructive to investigate spin
component of $W$ boson contributes to the decay process, ($t \to b \bar{l}
\nu$).
To answer this question, we consider the top quark cascade decay $t \to
b W^+ \to b \bar{l} \nu$ ~\cite{ps,pk1,pk2}. 
The differential decay distribution of a polarized top quark depends on three
angles.
First is the angle, $\chi_{W}^{t}$, which specifies the moving
direction of the $W$ boson with respect to the top quark spin in the top 
quark rest frame.
Second is the angle, $\pi - \chi_l^{W}$, between the direction of motion
of $b$ quark and lepton in the $W$ boson rest frame.
Finally the azimuthal angle, $\Phi$, between the lepton and top quark spin
around the direction of $W$ boson.
The differential decay distribution is 
\bea
\lefteqn{
\frac{1}{\Gamma_T}
\frac{d^3 \Gamma}{ d \cos \chi_W^{t} d \cos \chi_l^{W} d \Phi}
        } \nonumber \\
&=& \frac{3}{16 \pi (m^2 + 2 M_W^2)}
\Bigl[ m^2(1 + \cos \chi_W^t) \sin^2 \chi_l^{W} 
+ M_W^2 (1 - \cos \chi_W^{t})(1 - \cos \chi_l^{W})^2 
\nonumber \\ 
&& \qquad \qquad \qquad \qquad
+~2 m M_W(1 - \cos \chi_l^{W}) \sin \chi_W^t \sin \chi_l^W \cos \Phi 
\Bigr]~.
\label{eqn:dbl-td}
\eea
%where $\Gamma_T$ stands for the total decay width.
In Eqn.(\ref{eqn:dbl-td}), the first and the second terms are
contributions from the longitudinal and transverse $W$ boson, and
the third term represents the interference between the
longitudinal and transverse $W$ boson.
The contribution of the transverse $W$ boson is suppressed by the
factor $M_W^2/m^2$, and the interference contribution, by the
factor $2 M_W/m$ compared to the contribution of the longitudinal $W$.
Note that the interference term does not contribute to the total decay width
because of the factor $\cos \Phi$.
\clearpage
\begin{figure}[H]
\begin{center}
\begin{tabular}{cc}
        \leavevmode\psfig{file=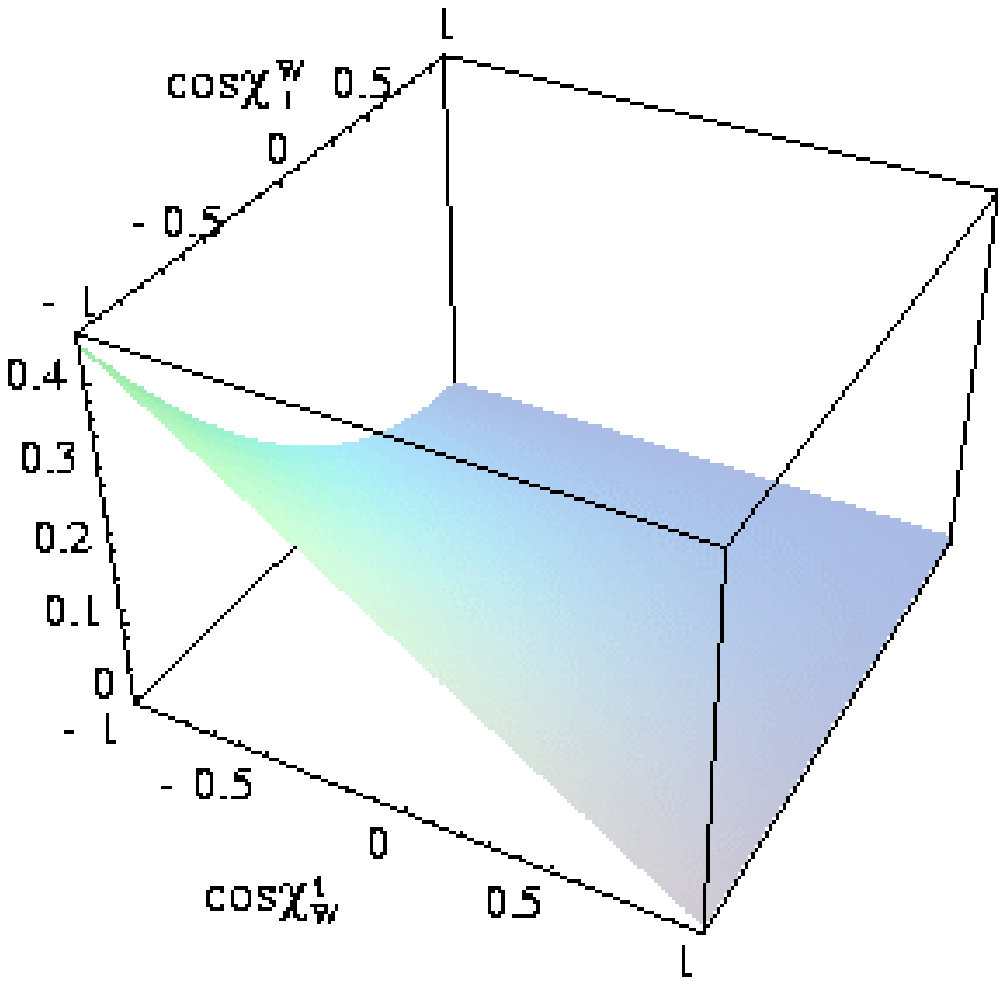,width=7cm} &
        \leavevmode\psfig{file=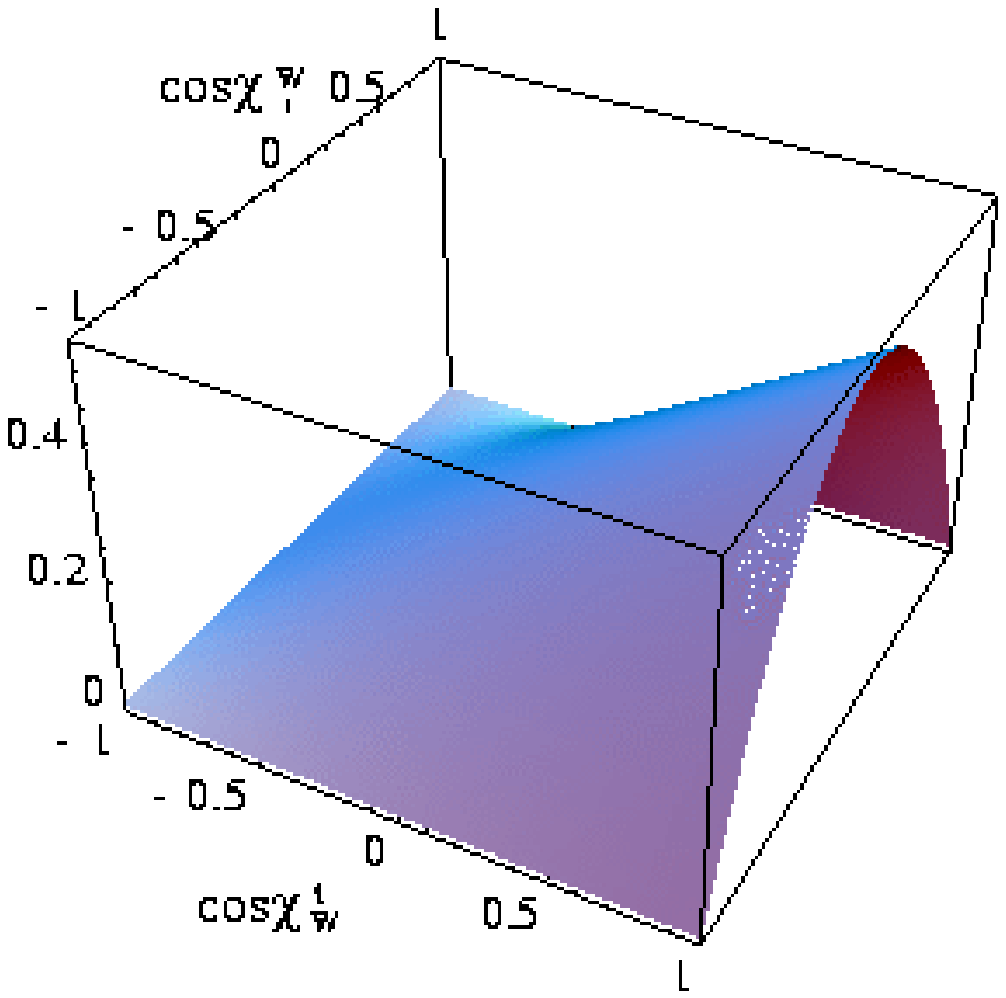,width=7cm} \\
(a) & (b)
\end{tabular}
\caption[The 3-dimensional plot of the top quark differential angular 
 distribution, $\frac{1}{\Gamma_T}$ $\frac{d^2 \Gamma}{d \cos \chi_l^W d \cos \chi_W^t}$.]{The 3-dimensional plot of the top quark differential angular 
 distribution, 
$\frac{1}{\Gamma_T} \frac{d^2 \Gamma}{d \cos \chi_l^W d \cos \chi_W^t}$.
(a) is the contribution from the longitudinal $W$ boson
and (b) from the transverse $W$ boson (b).}
\label{fig:angle-3d-w}
\end{center} 
\end{figure}
\begin{figure}[H]
\begin{center}
\begin{tabular}{ccl}
        \leavevmode\psfig{file=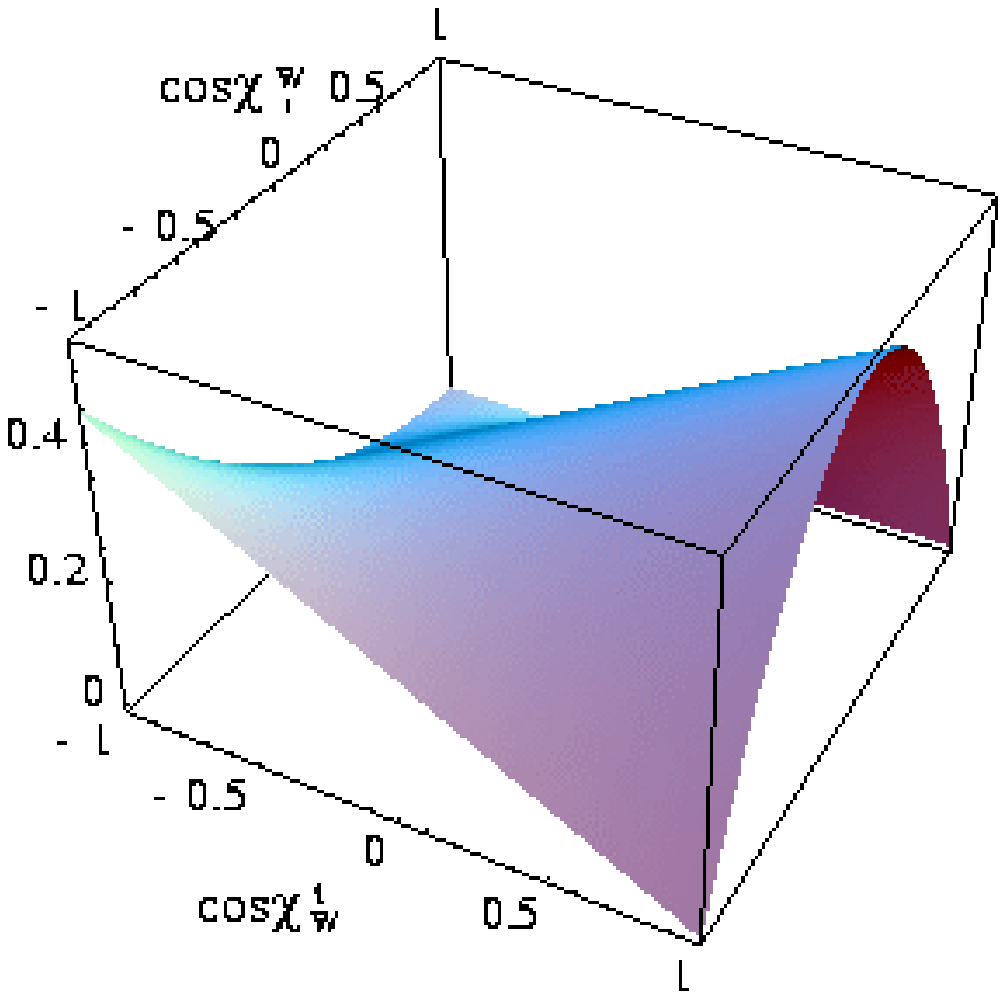,width=7cm}  &
        \leavevmode\psfig{file=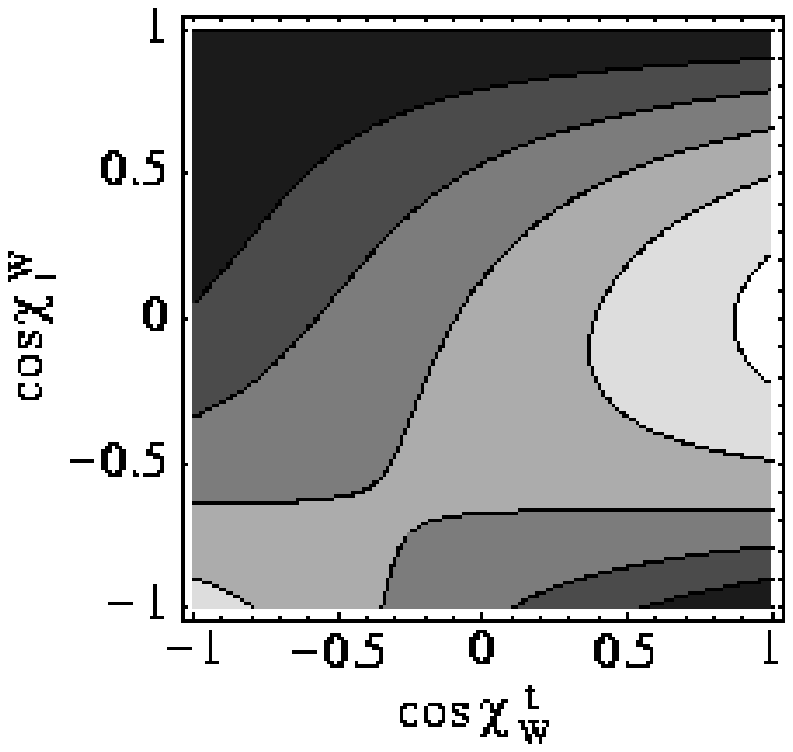,width=7.5cm} &
        \leavevmode\psfig{file=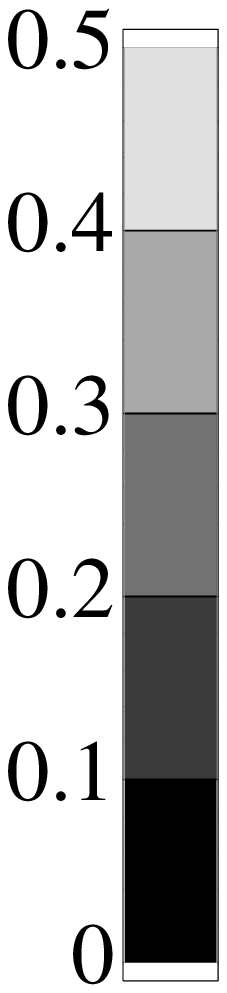,width=1.cm}
\end{tabular}
\caption[The 3-dimensional and contour plot of the top quark differential angular 
 distribution, 
$\frac{1}{\Gamma_T} \frac{d^2 \Gamma}{d \cos \chi_l^W d \cos \chi_W^t}$.]{The
 3-dimensional and contour plot of the top quark differential angular  
 distribution, 
$\frac{1}{\Gamma_T} \frac{d^2 \Gamma}{d \cos \chi_l^W d \cos \chi_W^t}$.
Each contour-line shows the angular distribution with
the values $0.1,0.2,0.3,0.4,0.5$.} 
\label{fig:angle-3d}
\end{center} 
\end{figure}
\noindent

We plot the $3$-dimensional differential angular distribution in 
$\cos \chi_l^W$ verses $\cos \chi_W^t$ in Fig.\ref{fig:angle-3d-w}.
Fig.\ref{fig:angle-3d-w}(a) presents the contribution of longitudinal $W$ 
boson, Fig.\ref{fig:angle-3d-w}(b) shows the transverse $W$ boson.
Fig.\ref{fig:angle-3d} shows $3$-dimensional and contour plots of the
differential angular distribution in $\cos \chi_l^W$ verses $\cos
\chi_W^t$.
When the $W$ boson momentum is parallel to the
top quark spin $\cos \chi_{W}^t=1$, only the longitudinal $W$ boson
contributes to the angular distribution.
When the $W$ boson momentum is anti-parallel to the
top quark spin $\cos \chi_{W}^t=-1$, the transverse $W$ boson purely
contributes. 
In the region, $-1 \leq \cos \chi_W^t \leq 1$, both the longitudinal and 
the transverse $W$ boson contribute and significantly.
Although the contribution of transverse $W$ boson
to the total decay width is small, $\sim 30 \%$, the transverse $W$
boson has a significant impact on the angular distribution of the top
quark decay products. 

%To see how the information of the top quark spin are transmitted to the 
%decay products in the processes $t \rightarrow b \bar{l} \nu$, 
%we consider the top quark cascade decay $t \to b W^+ \to b \bar{l} \nu$
%~\cite{ps,pk1}.
%remember that the longitudinal $W_{L}$ dominates in the decay process $t
%\rightarrow b W$.   
%(When the longitudinal polarization vector $\epsilon^{\mu}_{L}(k)$ 
%becomes parallel to its momentum $k^{\mu}$, the contribution of 
%$W_{L}$ has a large factor $m/M_{W}$, $M_{W}$ is the W boson mass.)
%The direction of the $W_{L}$ boson momentum is that of the 
%top quark spin due to the angular momentum conservation.   
%In the subsequent decay, the lepton $\bar{l}$ emitted from $W_{L}$ 
%are highly boosted. Hence the leptons $\bar{l}$ can go away into 
%the direction of the top quark spin.
%Similar analysis for the anti-top quark tells us that the favorable 
%direction of the momentum $l$ is the opposite to the direction of 
%the anti-top quark spin.
%

Now we move the to the discussion on the measurement~\cite{PARKE1,HKN} of
the top quark spin. 
As mentioned before, there are two top quark decay modes, hadronic decay 
and semi-leptonic decay.
The rate of the hadronic decay mode is approximately 9 times larger than 
the rate of each leptonic decay mode.
However, we must take into account the efficiency to identify the  
d-type quark from u-type (or vice versa) in order to utilize the
hadronic decay products. 
Therefore, we consider the the semi-leptonic decay mode of the top
(anti-top) quark to simplify our discussions.
%The most efficient spin analyzer is the charged lepton, for which 
%$\alpha_{\bar{l}}=1$. 
From Eqn.(\ref{eqn:DECAY-DIS}), we can obtain the
probability $P\left(\uparrow|~\mbox{in the cone}\right)$ 
that the top quark has spin up when we pick up events which satisfy the
condition $ \cos\theta_{i}> y $,
\bea
P
\left(\uparrow|~\mbox{in the cone}
\right)  
& = &
\frac{
\left(1 + \left < s_{t} \right > \right) \left( 2+\alpha_{i}(1+y)
\right)
      }
{4+ 2 \left< s_{t} \right> \alpha_{i} (1+y)},
\label{eqn:PROB-UP}
\eea
where the probability function $P(A|B)$ denotes the conditional
probability of the event A, and 
$\left< s_{t} \right>$ denotes the average value of the top quark
spin defined by 
\bea
\left<s_{t}\right>
= 
\sum_{s_{t}=-1,1} s_{t} \times P(s_{t}).
\eea
$P(s_{t})$ is the probability that the top quarks with spin $s_{t}$
are produced.

It is straightforward to extend this analysis to the top pair
production process. The double decay distribution for the decay
product ``$i$'' from the top quark and the decay product``$\bar{i}$'' 
from the anti-top quark is given by 
\bea
\frac{d^{2}\Gamma }{d\cos\theta_{i} d \cos\theta_{\bar{i}} }
&\simeq &
\frac{1}{4}
\left \{ 
1 + \alpha_{i} \left< s_{t} \right> \cos\theta_{i}
+ \alpha_{\bar{i}} \left< s_{\bar{t}} \right> \cos\theta_{\bar{i}}
\right.
\nonumber \\
&& \quad \left.
+ \alpha_{i}\alpha_{\bar{i}}
\left< s_{t}s_{\bar{t}} \right> \cos\theta_{i}\cos\theta_{\bar{i}} 
\right \}, 
\label{eqn:DOUB-DECAY-DIS}
\eea
where $\theta_{i}$ $(\theta_{\bar{i}})$ is the angle between 
the direction of the top (anti-top) quark spin and the momentum of
decay product $i$ ($\bar{i}$) in the top (anti-top) quark
rest frame.
As can be seen in the above expression, it is obvious how the spin 
correlation between the top and anti-top quarks is related 
to the distribution of their decay products. 

When we collect events for which the charged lepton $i$ from
the top quark lies in the cone defined by $\cos\theta_{i}>y$, 
the effective $\alpha$ in the expression Eqn.(\ref{eqn:DECAY-DIS}) 
for anti-top quark becomes,
\bea
\alpha^{eff}_{\bar{i}}
& = &
\frac{ 
\{ 
2 \left < s_{\bar{t}} \right > + \alpha_{i} \left < s_{t}s_{ \bar{t}}
\right > (1+y) 
\}
      }{ 2+\alpha_{i} \left <s_{t}\right > (1+y) }
\times \alpha_{\bar{i}},
\label{eqn:EFF-ALPHA}
\eea
which determine the $\bar{i}$ distribution under the condition 
mentioned above.
Then the distribution of the decay product $\bar{i}$ from the 
anti-top quark is given by
\bea
\frac{1}{\Gamma_{T}}
\frac{d \Gamma
      }{ d \cos\theta_{\bar{i} }}
& = &
\frac{
1 + \alpha^{eff}_{\bar{i} } 
    \cos \theta_{\bar{i} }
     }{2}.
\label{enq:DECAY-DIS-C}
\eea
By observing this distribution, we can obtain the averaged value of the
spin $ \left <s_{t}\right >$ and the spin correlation 
$\left<s_{t}s_{\bar{t}}\right>$ between the top and anti-top quarks. 
\newpage
%----------------------- QCD Corr. ------------------------------------
%\input qcd.tex
%----------------------- QCD Corr. ------------------------------------
\chapter{QCD Corrections to Spin Correlations}
A basic consequence of the large top quark mass is its short lifetime.
The top quark lifetime is $\tau~\sim~10^{-24}$ second and top quark
decays very rapidly to the $b$ quark and the $W$ boson before QCD long
distance effects come in.
Therefore, top quarks are produced essentially as unbounded fermion.
However they still feel the strong interactions and will radiate gluons.
Especially, there are two ``new'' effects, which are caused by the QCD
radiative correction.
First, the QCD corrections induce the new structure to $\gamma/Z - t
-\bar{t}$ vertex.
Second, the emission of a real gluon with spin $1$ leads to the
spin-flip effects.
Thus it is important and urgent to study how the tree level picture will
be affected by QCD correction.  
The QCD ${\cal O}(\alpha_{s})$ corrections to the top quark productions
have been studied in several papers~\cite{s,tung}, including
analyses of the effects on the production angle distributions and their
polarization. 
However, in these works, the top quark spin is decomposed in the
helicity basis.
In previous chapter, we studied that the top quark spin decomposition in 
the helicity basis is not appropriate at moderate energies, and
the ``off-diagonal'' basis is optimal spin decomposition in this energy
region at the leading order.

In this chapter, we present the differential cross-sections for
top quark productions in the ``generic'' spin basis at QCD one-loop 
level, and investigate which is the optimal decomposition of
top quark spin at QCD one-loop level at the energy far above the
threshold for top quarks. 
We quantitatively analyze spin-spin correlations for top quarks, 
and examine the spin flip effects caused by the QCD radiative corrections.

%This chapter is motivated the fact
%(1) QCD correction to the vertex $t-\bar{t}-\gamma/Z$ induce the new
%vertex structure.
%(2) As the hard gluon emission leads to the spin flip effects, 
%the hard gluon emission may not change the result at the leading order.
%Therefore we qualitatively analyze the spin-spin correlation in a
%generic spin basis, answer to the above two questions.
%\clearpage
 
\section{Vertex Corrections and Soft Gluon Case}

In this section, we derive the QCD correction at ${\cal
O}(\alpha_s)$ to the spin dependent differential
cross-section for top quark pair production in the soft gluon
approximation (SGA)~\cite{KNP1,KNP2}.
In this approximation, only the QCD vertex corrections
modify the spin correlations of the top quarks.
Therefore, in order to investigate a new structure induced by the QCD
corrections, it is instructive to consider the soft gluon approximation.
We give analytic expressions for the polarized cross-sections at QCD
one-loop level in the soft-gluon approximation using the generic spin basis. 
\subsection{QCD Corrections to Vertex $\gamma/Z - t - \bar{t}$}
\begin{figure}[H]
\begin{center}
\begin{tabular}{cc}
\leavevmode\psfig{file=ttbar21.ps,width=5cm} &
\leavevmode\psfig{file=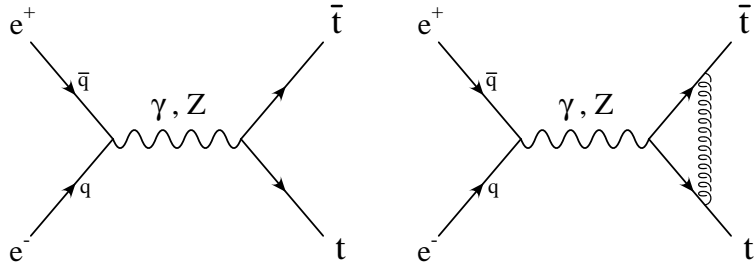,width=5cm} 
\end{tabular}
\caption{The tree level and the QCD one-loop contributions
to the $e^- e^+ \to t \bar{t}$ process.}
\label{fig:process1}
\end{center}
\end{figure}

%
%The ultraviolet singularities are removed by appropriate counterterms
%fixed by one shell renormalization condition for the top quark,
%The full one-loop analysis will be given in the next section.
%We use the same generic spin basis as in Ref.~\cite{ps}.
%Therefore we use a generic spin basis with the spin of the 
%top quark and anti-top quark in the production plane.
%We define the spins of the top and anti-top quarks by
%the parameter $\xi$ as given in Fig. \ref{fig:spin-def}.
%%%%%%%%%%%%%%%%%%%%%%%%
%\begin{figure}[H]
%\begin{center}
%\begin{tabular}{cc}
%\leavevmode\psfig{file=ttbar11.ps,width=5cm} &
%\leavevmode\psfig{file=ttbar12.ps,width=5cm} 
%\end{tabular}
%\caption{The generic spin basis for the top (anti-top)
%quark in its rest frame.  ${\bf s}_t$ (${\bf s}_{\bar{t}}$) is the top
%(anti-top) spin axis.}
%\end{center}
%\end{figure}
%%%%%%%%%%%%%%%%%%%%%%%%
%\noindent
%The top quark spin is decomposed along the direction ${\bf s}_t$
%in the rest frame of the top quark which makes an angle
%$\xi$ with the anti-top quark momentum in the clockwise direction.   
%Similarly, the anti-top quark spin states are defined in the anti-top
%rest frame along the direction ${\bf s}_{\bar{t}}$ having the same
%angle $\xi$ from the direction of the top quark momentum.
%We use the following notation in this thesis: the state
%$t_{\uparrow}\,\bar{t}_{\uparrow}\,(t_{\downarrow}\,\bar{t}_{\downarrow})$
%refers to a top with spin in the $+ {\bf s}_t \,(- {\bf s}_t )$
%direction in the top rest frame and an anti-top
%with spin $+ {\bf s}_{\bar{t}} \,(- {\bf s}_{\bar{t}} )$
%in the anti-top rest frame.
%\clearpage
The QCD corrections to the cross-sections
are given by the interference between the tree and the one-loop
vertex diagram in Fig.\ref{fig:process1}.
%%%%%%%%%%%%%%%%%%%%%%%
%%%%%%%%%%%%%%%%%%%%%%%
\noindent
There are both infrared and ultra-violet singularities in the one-loop
integrals of the virtual corrections.
We treat the ultra-violet singularities within the frame work of dimensional
reguralization in $D = 4 - 2 \varepsilon$, and 
introduce an infinitesimal mass for the gluon to avoid infrared
singularities.

At the one-loop level, the $\gamma -t-\bar{t}$ and $Z-t-\bar{t}$ vertex
functions can be written in terms of three form factors
$A_0,B_0,C_0$ as follows:
\bea
   \Gamma^{\gamma}_{\mu} &=& e Q_t
        \left[ ( 1 + A_0 ) \, \gamma_{\mu} + B_0 \, \frac{t_{\mu} -
              \bar{t}_{\mu}}{2 m} \right] \ ,\label{gammavertex}\\
   \Gamma^Z_{\mu} &=& \frac{e}{\sin \theta_W}
        \biggl[ \left\{ Q_t^L \, ( 1 + A_0 ) + ( Q_t^L - Q_t^R ) \, B_0 \right\}
                (\gamma_L )_{\mu} \nonumber\\
         & & \qquad\quad
          + \left\{ Q_t^R \, ( 1 + A_0 ) -  ( Q_t^L - Q_t^R ) \, B_0 \right\}
                (\gamma_R )_{\mu} \nonumber\\
          & & \qquad\qquad\qquad + \frac{Q_t^L + Q_t^R}{2} \, B_0\, 
           \frac{t_{\mu} - \bar{t}_{\mu}}{2 m} + \frac{Q_t^L - Q_t^R}{2} 
           \, C_0 \, \frac{t_{\mu} + \bar{t}_{\mu}}{2 m} \gamma_5 \biggr]
              \ ,\label{zvertex}
\eea
where $Q_t$, $\theta_W$, $m$ and $t_{\mu}\, (\bar{t}_{\mu})$ 
have been defined in the previous chapter.
The form factor $A_0$ represents the enhancement effect to the tree vertex
structure. 
So, this factor does not change the tree-level results except
for the multiplicative enhancement.
Meanwhile, the factors $B_0,~C_0$ are the coefficients of new vertex
structures induced by the QCD corrections.
In particular, the factor $C_0$ represents the contribution from only the QCD
corrected $Z - t -\bar{t}$ vertex.
\bea
A_0 &=& 
\hat{\alpha}_{s} \Biggl[
C_{UV} 
- \ln \frac{m^2}{\mu^2} +
 3 \beta \ln \left( - \frac{1 + \beta}{1 - \beta} \right) +
 \frac{1 + \beta^2}{\beta}
\Bigl\{ 
    \ln \left( - \frac{1 + \beta}{1 - \beta} \right) 
    \ln \frac{\lambda^2}{m^2} \nonumber \\ 
&+& \frac{1}{2} \ln \left( - \frac{1 - \beta^2}{4 \beta^2} \right)
                \ln \left( - \frac{1 + \beta}{1 - \beta} \right) +
    \Li \left(\frac{1 + \beta}{2 \beta}\right) - 
    \Li \left(- \frac{1 - \beta}{2 \beta}\right) 
\Bigr\} \Biggl] ,\qquad \\
B_0 &=&  \hat{\alpha}_s
          \frac{1 - \beta^2}{\beta} \ln \left( - \frac{1 + \beta}{1 -
                      \beta} \right) , \\
C_0 &=& \hat{\alpha}_s
          \left[ ( 2 + \beta^2 ) \frac{1 - \beta^2}{\beta}
          \ln \left( - \frac{1 + \beta}{1 - \beta} \right) 
        - 2 (1 - \beta^2 ) \right]  ,
\eea
with 
\bea
C_{UV} &=& \frac{1}{\varepsilon} + \ln 4 \pi - \gamma_{E}~,\\
\beta  &=& \sqrt{1 - \frac{4 m^2}{s + i \epsilon} }~.
\eea
where $\varepsilon=2 - D/2$, $D$ is the space-time dimension, 
$\gamma_{E}\simeq0.57721$ is the Euler constant, 
$\beta$ is the speed of the produced top (anti-top)
quark and the strong coupling constant is $\hat{\alpha}_s \equiv
\frac{C_2 (R)}{4 \pi} \alpha_s = \frac{C_2 (R)}{(4 \pi )^2} g^2$
with $C_2 (R) = 4/3$ for SU(3) color.

We have introduced an infinitesimal mass $\lambda$ for the gluon to avoid
infrared singularities.
The wave function renormalization factor $z_{2}$ in the one-shell
renormalization scheme is calculated to be,
\bea
z_{2} &=& 1 + \hat{\alpha}_{s}
\left( - C_{UV} + \ln \frac{m^2}{\mu^2} - 4 
       - 2 \ln \frac{\lambda^2}{m^2} \right).
\eea
After multiplying the wave function renormalization factor, the \lq\lq
renormalized\rq\rq\ form factors $A,B$ and $C$ read
\bea
    A &=& A_R + A_I \nonumber \\
      &=&  
     \hat{\alpha}_s     
     \left[ \left( \frac{1 + \beta^2}{\beta} \ln
                  \frac{1 + \beta}{1 - \beta} - 2 \right)
              \ln \frac{\lambda^2}{m^2} - 4 + 3 \beta
              \ln \frac{1 + \beta}{1 - \beta} \right.
       + \frac{1 + \beta^2}{\beta}  \nonumber \\
      &&\times  \left. 
            \left\{ \frac{1}{2} \ln^2 \frac{1 + \beta}{1 - \beta}
           + \ln^2 \frac{1 + \beta}{2 \beta} -
                       \ln^2 \frac{1 - \beta}{2 \beta}
          + 2 {\rm Li}_2 \left( \frac{1 - \beta}{1 + \beta} \right)
          + \frac{2}{3} \pi^2 \right\} \right] \nonumber \\
      && + i \pi \hat{\alpha}_s \left[ 
           -3 \beta - \frac{1 + \beta^2}{\beta} 
           -  \Bigl( \ln \frac{1 - \beta}{2 \beta}+
                     \ln \frac{1 + \beta}{2 \beta}
              \Bigr)                        
           - \frac{1 + \beta^2}{\beta} 
             \ln \Bigl( \frac{\lambda^2}{m^2} \Bigr)
                  \right]~
,\label{afactor}
\qquad \\
    B     &=& B_R + B_I \nonumber \\
          &=& \hat{\alpha}_s
          \frac{1 - \beta^2}{\beta} \ln \frac{1 + \beta}{1 -
                      \beta} 
          - 
           i \pi \hat{\alpha}_s \frac{1 - \beta^2}{\beta}
\ ,\label{bfactor}\\
    C     &=& C_R + C_I \nonumber \\
          &=& \hat{\alpha}_s
          \left[ ( 2 + \beta^2 ) \frac{1 - \beta^2}{\beta}
          \ln \frac{1 + \beta}{1 - \beta} -2 (1 - \beta^2 ) \right]
          - i \pi \hat{\alpha}_{s} \frac{2 + \beta^2}{\beta}        
\ , \label{cfactor}
\eea
where suffices $R,~I$ stand for the real part and the imaginally part of
the form factors $A,~B,~C$ respectively.
In the above expressions, we take only the real part of the
form factors.
The reasons are
(1) The contribution of $Z$ boson width is negligible in the region
of CM energy $\sqrt{s}$ far above the production threshold for top quarks.
(2) We do not consider the transverse top quark polarization normal to the
scattering plane, since we
are interested in how QCD corrections modify the tree  
level spin correlations and which spin basis is the most effective for
spin correlation studies 
(Although it is known~\cite{kly,TRS} that the transverse top quark polarization
becomes non-zero when the higher order QCD corrections are included and
this transverse polarization is very important and
related to the phenomena of CP violation,
we do not consider transverse polarization of the top quarks). 

We get the amplitudes for the $e^-_{L/R} e^+_{R/L} \to t_{s_{t}}
\bar{t}_{s_{\bar{t}}}$ process using above renormalized vertex: 
\bea
\lefteqn{
{\cal M}(e^-_L e^+_R \to t_{s_{t}} \bar{t}_{s_{\bar{t}}})}
\nonumber \\
&=& \left( \frac{4 \pi \alpha}{s} \right)
\bra{\bar{q} -} (\gamma_{L})_{\mu} \ket{q -}
%\times 
\bar{u}(t,s_{t})
\Biggl[ 
  \Bigl\{ f_{LL}(1 + A_{R} + B_{R}) - f_{LR} B_{R} \Bigr\}
  \gamma_{L}^{\mu} \nonumber \\
&& 
\mbox{\hspace{2cm}} + 
\Bigl\{ f_{LR}(1 + A_{R} + B_{R}) - f_{LL} B_{R} \Bigr\}
\gamma_{R}^{\mu}
\nonumber \\
&&
\mbox{\hspace{2cm}} + 
\frac{B_{R}}{2}(f_{LL} + f_{LR}) \frac{t^{\mu} - \bar{t}^{\mu}}{2 m} 
\nonumber \\
&&
\mbox{\hspace{2cm}} + 
\frac{C_{R}}{2}(f_{LL} - f_{LR}) \frac{t^{\mu} + \bar{t}^{\mu}}{2 m} 
\gamma_{5}           \Biggr] v(\bar{t},s_{\bar{t}}), \\
%
%-------------------------------------------------------------
%
\lefteqn{
{\cal M}(e^-_R e^+_L \to t_{s_{t}} \bar{t}_{s_{\bar{t}}})}
\nonumber \\
&=& \left( \frac{4 \pi \alpha}{s} \right)
\bra{\bar{q} +} (\gamma_{R})_{\mu} \ket{q +}
%\times 
\bar{u}(t,s_{t})
\Biggl[ 
  \Bigl\{ f_{RR}(1 + A_{R} + B_{R}) - f_{RL} B_{R} \Bigr\}
  \gamma_{R}^{\mu} \nonumber \\
&& 
\mbox{\hspace{2cm}} + 
\Bigl\{ f_{RL}(1 + A_{R} + B_{R}) - f_{RR} B_{R} \Bigr\}
\gamma_{L}^{\mu}
\nonumber \\
&&
\mbox{\hspace{2cm}} + 
\frac{B_{R}}{2}(f_{RR} + f_{RL}) \frac{t^{\mu} - \bar{t}^{\mu}}{2 m} 
\nonumber \\
&&
\mbox{\hspace{2cm}} + 
\frac{C_{R}}{2}(f_{RR} - f_{RL}) \frac{t^{\mu} + \bar{t}^{\mu}}{2 m} 
\gamma_{5} 
           \Biggr] v(\bar{t},s_{\bar{t}}),    
\eea
where the electron (positron) mass is neglected.
Note that the contribution from $C_{R}$ vanish because it is proportion
to the electron mass in the amplitudes.

Since we do not consider the transverse polarization of
top quark,
we can define the spin of top and anti-top quark in the production
plane. 
We use the generic spin basis defined in Chapter 3 and
again, we decompose the top (anti-top) quark momentum $t$ ($\bar{t}$)
into the two massless momenta $t_1,~t_2$ ($\bar{t}_1,~\bar{t}_2$).

Using spinor helicity method, we can easily obtain the squared
amplitudes at one-loop level, 
\bea
\lefteqn{
|{\cal M}(e^-_L e^+ \to t_{\uparrow} \bar{t}_{\uparrow})|^2 }
\nonumber \\
&=& \left( \frac{4 \pi \alpha}{s} \right)^2
\Biggl[
4 \Bigl\{ 
   |f_{LL}|^2 (1 + 2 A_{R} + 2 B_{R}) - 2 f_{LL} f_{LR} B_{R}
  \Bigr\}(2 q \cdot t_{1})(2 \bar{q} \cdot \bar{t}_{2}) \nonumber \\
&& + 
4 \Bigl\{ 
   |f_{LR}|^2 (1 + 2 A_{R} + 2 B_{R}) - 2 f_{LL} f_{LR} B_{R}
  \Bigr\}(2 q \cdot \bar{t}_{1})(2 \bar{q} \cdot {t}_{2}) \nonumber \\
&& + 
\frac{4}{m^2} \Bigl\{ 
   f_{LL} f_{LR} (1 + 2 A_{R} + 2 B_{R}) - 2 f_{LL} f_{LR} B_{R}
              \Bigr\}
  \Tr[ q t_{1} t_{2} \bar{q} \bar{t}_2 \bar{t}_1 ] \nonumber \\
&& + 
\frac{B_{R}}{2 m^2} (f_{LL} + f_{LR}) f_{LL}
  \Tr[ q (t - \bar{t}) \bar{q} \bar{t}_{2} 
         (t_{2} - \bar{t}_{1}) t_1 ] \nonumber \\
&& + 
\frac{B_{R}}{2 m^2} (f_{LL} + f_{LR}) f_{LR}
  \Tr[ q (t - \bar{t}) \bar{q} t_{2} 
         (t_{1} - \bar{t}_{2}) \bar{t}_1 ] \Biggr]~,
\label{eqn:SAMP1-QCD} \\
\lefteqn{
|{\cal M}(e^+ e^- \to t_{\uparrow} \bar{t}_{\downarrow})|^2 }
\nonumber \\
&=& \left( \frac{4 \pi \alpha}{s} \right)^2
\Biggl[
4 \Bigl\{ 
   |f_{LL}|^2 (1 + 2 A_{R} + 2 B_{R}) - 2 f_{LL} f_{LR} B_{R}
  \Bigr\}(2 q \cdot t_{1})(2 \bar{q} \cdot \bar{t}_{1}) \nonumber \\
&& + 
4 \Bigl\{ 
   |f_{LR}|^2 (1 + 2 A_{R} + 2 B_{R}) - 2 f_{LL} f_{LR} B_{R}
  \Bigr\}(2 q \cdot \bar{t}_{2})(2 \bar{q} \cdot {t}_{2}) \nonumber \\
&& + 
\frac{4}{m^2} \Bigl\{ 
   f_{LL} f_{LR} (1 + 2 A_{R} + 2 B_{R}) - 2 f_{LL} f_{LR} B_{R}
              \Bigr\}
  \Tr[ q t_{1} t_{2} \bar{q} \bar{t}_1 \bar{t}_2 ] \nonumber \\
&& + 
\frac{B_{R}}{2 m^2} (f_{LL} + f_{LR}) f_{LL}
  \Tr[ q (t - \bar{t}) \bar{q} \bar{t}_{1} 
         (t_{2} - \bar{t}_{2}) t_1 ] \nonumber \\
&& + 
\frac{B_{R}}{2 m^2} (f_{LL} + f_{LR}) f_{LR}
  \Tr[ q (t - \bar{t}) \bar{q} t_{2} 
         (t_{1} - \bar{t}_{1}) \bar{t}_2 ] \Biggr]~,
\label{eqn:SAMP2-QCD} \\
\lefteqn{
|{\cal M}(e^+ e^- \to t_{\downarrow} \bar{t}_{\uparrow})|^2 }
\nonumber \\
&=& \left( \frac{4 \pi \alpha}{s} \right)^2
\Biggl[
4 \Bigl\{ 
   |f_{LL}|^2 (1 + 2 A_{R} + 2 B_{R}) - 2 f_{LL} f_{LR} B_{R}
  \Bigr\}(2 q \cdot t_{2})(2 \bar{q} \cdot \bar{t}_{2}) \nonumber \\
&& + 
4 \Bigl\{ 
   |f_{LR}|^2 (1 + 2 A_{R} + 2 B_{R}) - 2 f_{LL} f_{LR} B_{R}
  \Bigr\}(2 q \cdot \bar{t}_{1})(2 \bar{q} \cdot {t}_{1}) \nonumber \\
&& + 
\frac{4}{m^2} \Bigl\{ 
   f_{LL} f_{LR} (1 + 2 A_{R} + 2 B_{R}) - 2 f_{LL} f_{LR} B_{R}
              \Bigr\}
  \Tr[ q t_{2} t_{1} \bar{q} \bar{t}_2 \bar{t}_1 ] \nonumber \\
&& + 
\frac{B_{R}}{2 m^2} (f_{LL} + f_{LR}) f_{LL}
  \Tr[ q (t - \bar{t}) \bar{q} \bar{t}_{2} 
         (t_{1} - \bar{t}_{1}) t_2 ] \nonumber \\
&& + 
\frac{B_{R}}{2 m^2} (f_{LL} + f_{LR}) f_{LR}
  \Tr[ q (t - \bar{t}) \bar{q} t_{1} 
         (t_{2} - \bar{t}_{2}) \bar{t}_1 ] \Biggr]~,
\label{eqn:SAMP3-QCD} \\
\lefteqn{
|{\cal M}(e^+ e^- \to t_{\downarrow} \bar{t}_{\downarrow})|^2 }
\nonumber \\
&=& \left( \frac{4 \pi \alpha}{s} \right)^2
\Biggl[
4 \Bigl\{ 
   |f_{LL}|^2 (1 + 2 A_{R} + 2 B_{R}) - 2 f_{LL} f_{LR} B_{R}
  \Bigr\}(2 q \cdot t_{2})(2 \bar{q} \cdot \bar{t}_{1}) \nonumber \\
&& + 
4 \Bigl\{ 
   |f_{LR}|^2 (1 + 2 A_{R} + 2 B_{R}) - 2 f_{LL} f_{LR} B_{R}
  \Bigr\}(2 q \cdot \bar{t}_{2})(2 \bar{q} \cdot {t}_{1}) \nonumber \\
&& + 
\frac{4}{m^2} \Bigl\{ 
   f_{LL} f_{LR} (1 + 2 A_{R} + 2 B_{R}) - 2 f_{LL} f_{LR} B_{R}
              \Bigr\}
  \Tr[ q t_{2} t_{1} \bar{q} \bar{t}_1 \bar{t}_2 ] \nonumber \\
&& + 
\frac{B_{R}}{2 m^2} (f_{LL} + f_{LR}) f_{LL}
  \Tr[ q (t - \bar{t}) \bar{q} \bar{t}_{1} 
         (t_{1} - \bar{t}_{2}) t_2 ] \nonumber \\
&& + 
\frac{B_{R}}{2 m^2} (f_{LL} + f_{LR}) f_{LR}
  \Tr[ q (t - \bar{t}) \bar{q} t_{1} 
         (t_{2} - \bar{t}_{1}) \bar{t}_2 ] \Biggr]~,
\label{eqn:SAMP4-QCD} 
\eea
The squared amplitudes for $e^-_R e^+_L$ scattering
process are obtained by interchanging suffices $L,~R$ as well as 
$\uparrow,~\downarrow$ in
Eqns.(\ref{eqn:SAMP1-QCD})$\sim$(\ref{eqn:SAMP4-QCD}).

The differential cross-section at the one-loop level 
in the CM frame as shown in Fig.{\ref{fig:ZMF}} is given by
\bea
   \lefteqn{\frac{d \sigma}{d \cos \theta}
        \left( e^-_L e^+_R \to t_{\uparrow} \bar{t}_{\uparrow} \right)
        = \frac{d \sigma}{d \cos \theta}
        \left( e^-_L e^+_R \to t_{\downarrow} \bar{t}_{\downarrow} \right)}
          \nonumber \\
    &=& \left( \frac{3 \pi \alpha^2}{2 s} \beta \right)
          ( A_{LR} \cos \xi - B_{LR} \sin \xi ) \label{upup1loop}\\
    & & \qquad \times \Bigl[ \, ( A_{LR} \cos \xi - B_{LR} \sin \xi )
                 ( 1 + 2 A_{R} + 2 B_R)  \nonumber\\
    & & \qquad\qquad\qquad\qquad - \, 2\, 
          (\gamma^2 A_{LR} \cos \xi - \bar{B}_{LR} \sin \xi )\, B_{R} \, \Bigr]
               \nonumber \ ,\\
   \lefteqn{\frac{d \sigma}{d \cos \theta}
        \left( e^-_L e^+_R \to t_{\uparrow} \bar{t}_{\downarrow} 
                 \  {\rm or} \  t_{\downarrow} \bar{t}_{\uparrow} \right)} 
             \nonumber\\
    &=& \left( \frac{3 \pi \alpha^2}{2 s} \beta \right)
          ( A_{LR} \sin \xi + B_{LR} \cos \xi 
               \pm D_{LR} ) \label{updown1loop}\\
    & & \qquad \times \Bigl[ \, ( A_{LR} \sin \xi + B_{LR} \cos \xi 
           \pm D_{LR} ) ( 1 + 2 A_{R} + 2 B_R)  \nonumber\\
    & & \qquad\qquad\qquad\qquad  - \, 2\, 
          (\gamma^2 A_{LR} \sin \xi + \bar{B}_{LR} \cos \xi
              \pm \bar{D}_{LR} )\, B_{R} \, \Bigr]
               \nonumber .
\eea
Here, the angle $\theta$ is the scattering angle of the top quark with 
respect to the incident electron, 
$\alpha$ is the QED fine structure constant and $\gamma =
1/\sqrt{1 - \beta^2}$. The quantities
$A_{LR},B_{LR},\bar{B}_{LR},D_{LR}$ and $\bar{D}_{LR}$ 
are defined by
\bea
    A_{LR} &=& \Bigl[ ( f_{LL} + f_{LR} ) \sqrt{1 - \beta^2}\,
            \sin \theta \Bigr] / \, 2 \ ,\nonumber\\
    B_{LR} &=& \Bigl[ f_{LL} ( \cos \theta + \beta )
             + f_{LR} ( \cos \theta - \beta ) \Bigr] / \, 2 = 
          \bar{B}_{LR} (- \beta ) \ ,\label{largecoupling}\\
    D_{LR} &=& \Bigl[ f_{LL} ( 1 + \beta \cos\theta )
             + f_{LR} ( 1 - \beta \cos\theta ) \Bigr] / \, 2 = 
          \bar{D}_{LR} (- \beta ) \ .\nonumber
\eea
The cross-sections Eqns.(\ref{upup1loop}) and (\ref{updown1loop}) 
contain an infrared singularities (in the form factor $A_R$) that
will be canceled by the contributions from the real gluon
emission. 
\clearpage
\subsection{Spin Correlations in the Soft Gluon Approximation}
Now we consider the cross-sections for the real gluon emission.
We isolate the soft gluon singularity by splitting the $t \bar{t} g$
phase space into a soft and a hard region.
Since the hard gluon with high momentum make the momenta of the top and
anti-top quarks change, 
the top and the anti-top quarks are not produced back to back.
Hence, in the hard gluon region, 
it is difficult to investigate the $\gamma/Z-t-\bar{t}$
vertex. 
Then, we concentrate on only the soft gluon region to study
the vertex structure induced by the QCD corrections~\cite{KNP1,KNP2}.

In the soft gluon approximation, it is very easy to
calculate the real gluon contribution. 
As is well known, the amplitude for the soft gluon emissions can be
written in the factorized form proportional to the tree amplitude. 
This means that the soft gluon emission does not change the spin
configurations or momentum of the produced heavy quark pairs from the
tree level values. 
Therefore, the QCD radiative corrections enter mainly through
the modifications of the vertex parts Eqns.(\ref{gammavertex}) and
(\ref{zvertex}).
The cross-section for the soft gluon emissions can be written as
\be
   \frac{d \sigma}{d \cos \theta} = J_{\rm IR}\ 
                     \frac{d \sigma_0}{d \cos \theta} \ ,\label{sgxsection}
\ee
where the subscript $0$ denotes the tree level
cross-section. The soft gluon contribution $J_{\rm IR}$ is defined by
\[   J_{\rm IR} \equiv  - 4 \pi C_2 (R) \alpha_s
           \int^{k^0 =\omega_{\rm max}} \frac{d^3 \vec{k}}{(2 \pi )^3 2 k^0}
           \left( \frac{t_{\mu}}{t \cdot k} - \frac{\bar{t}_{\mu}}
           {\bar{t} \cdot k} \right)^2 \ ,\]
where $\omega_{\rm max}$ is the cut-off of the soft
gluon energy, and
\[
 \omega_{\rm max} ~\leq~ \sqrt{s} - 2 m~,
\]
where the maximum value of $\omega_{\rm max}$ is constrained by the momentum
conservation.
At large $\omega_{\rm max}$, the soft-gluon approximation is 
spoiled, because it is the leading order of the amplitude in the
expansion of $\omega_{\rm max}$.
Therefore $\omega_{\rm max}$ should be
\bea
  \omega_{\rm max} &=& x_{\rm min}(\sqrt{s} - 2 m)~,
\eea
where $x_{\rm min}$ is a sufficiently small number.
The integral of $J_{\rm IR}$ can be easily performed and we obtain
\bean
     J_{\rm IR} &=& 2 \hat{\alpha}_s  
              \left[  \left( \frac{1 + \beta^2}{\beta} \ln 
     \frac{1 + \beta}{1 - \beta} -2 
                      \right) 
     \ln \frac{4 \omega_{\rm max}^2}
              {\lambda^2} \right.\\
       & & \qquad\qquad + \left. \frac{2}{\beta} \ln \frac{1 + \beta}{1 - \beta}
       - \frac{1 + \beta^2}{\beta} \left[ 2 {\rm Li}_2 
              \left( \frac{2 \beta}{1 + \beta} \right)
            + \frac{1}{2} \ln^2 \frac{1 + \beta}{1 - \beta} \right]
     \right]\ .
\eean

By adding the one-loop contributions
Eqns.(\ref{upup1loop}) and (\ref{updown1loop})
and the soft gluon ones Eqn.(\ref{sgxsection}), one can see that
the infrared singularities, $\ln \lambda$, are canceled out and
the finite results are obtained by replacing $2 A_R$ by
\bean
   \lefteqn{2 A_{R} + J_{\rm IR} 
}
%\equiv 2 \bar{A}}
\\       &=& 2 \hat{\alpha}_s     
     \left[ \left( \frac{1 + \beta^2}{\beta} \ln
                  \frac{1 + \beta}{1 - \beta} - 2 \right)
              \ln \frac{4 \omega_{\rm max}^2}{m^2} - 4 + 
           \frac{2 + 3 \beta^2}{\beta}
              \ln \frac{1 + \beta}{1 - \beta} \right.\\
       &+& \left. \frac{1 + \beta^2}{\beta}
            \left\{ \ln \frac{1 - \beta}{1 + \beta}
               \left( 3 \ln \frac{2 \beta}{1 +  \beta}
         +  \ln \frac{2 \beta}{1 - \beta} \right) 
         + 4 {\rm Li}_2 \left( \frac{1 - \beta}{1 + \beta} \right)
          + \frac{1}{3} \pi^2 \right\} \right]\ ,
\eean
in Eqns.(\ref{upup1loop}) and (\ref{updown1loop}).

Thus we obtain the ${\cal O}(\alpha_s)$ polarized cross-sections in the
soft-gluon approximation using the generic spin basis:
\bea
   \lefteqn{\frac{d \sigma}{d \cos \theta}
        \left( e^-_L e^+_R \to t_{\uparrow} \bar{t}_{\uparrow} \right)
        = \frac{d \sigma}{d \cos \theta}
        \left( e^-_L e^+_R \to t_{\downarrow} \bar{t}_{\downarrow} \right)}
          \nonumber \\
    &=& \left( \frac{3 \pi \alpha^2}{2 s} \beta \right)
          ( A_{LR} \cos \xi - B_{LR} \sin \xi ) \label{eqn:uu1loop}\\
    & & \qquad \times \Bigl[ \, ( A_{LR} \cos \xi - B_{LR} \sin \xi )
                 ( 1 + S_{I})  \nonumber\\
    & & \qquad\qquad\qquad\qquad - \, 2\, 
          (\gamma^2 A_{LR} \cos \xi - \bar{B}_{LR} \sin \xi )\, S_{II} \, \Bigr]
               \nonumber \ ,\\
   \lefteqn{\frac{d \sigma}{d \cos \theta}
        \left( e^-_L e^+_R \to t_{\uparrow} \bar{t}_{\downarrow} 
                 \  {\rm or} \  t_{\downarrow} \bar{t}_{\uparrow} \right)} 
             \nonumber\\
    &=& \left( \frac{3 \pi \alpha^2}{2 s} \beta \right)
          ( A_{LR} \sin \xi + B_{LR} \cos \xi 
               \pm D_{LR} ) \label{eqn:ud1loop}\\
    & & \qquad \times \Bigl[ \, ( A_{LR} \sin \xi + B_{LR} \cos \xi 
           \pm D_{LR} ) ( 1 + S_{I})  \nonumber\\
    & & \qquad\qquad\qquad\qquad  - \, 2\, 
          (\gamma^2 A_{LR} \sin \xi + \bar{B}_{LR} \cos \xi
              \pm \bar{D}_{LR} )\, S_{II} \, \Bigr]
               \nonumber ,
\eea
where
\bea
S_I %  \nonumber \\
&=&  2 \hat{\alpha}_s     
     \left[ \left( \frac{1 + \beta^2}{\beta} \ln
                  \frac{1 + \beta}{1 - \beta} - 2 \right)
              \ln \frac{4 \omega_{\rm max}^2}{m^2} - 4 + 
           \frac{3 + 2 \beta^2}{\beta}
              \ln \frac{1 + \beta}{1 - \beta} \right. \nonumber \\
    &+& \left. \frac{1 + \beta^2}{\beta}
            \left\{ \ln \frac{1 - \beta}{1 + \beta}
               \left( 3 \ln \frac{2 \beta}{1 +  \beta}
         +  \ln \frac{2 \beta}{1 - \beta} \right) 
         + 4 {\rm Li}_2 \left( \frac{1 - \beta}{1 + \beta} \right)
          + \frac{1}{3} \pi^2 \right\} \right]\ ,
\quad \qquad  \\
S_{II} &=& \hat{\alpha}_s
           \frac{1 - \beta^2}{\beta} \ln \frac{1 + \beta}{1 -
                      \beta}\ .
\eea
%
%The cross-sections for $e^-_R e^+_L$ can be obtained by interchanging
%$L,R$ as well as $\uparrow , \downarrow$ in the above formulae.

Since we are interested in maximizing the spin correlations
of the top quark pairs, we vary the spin angle, $\xi$, 
to find the appropriate spin basis.
As shown in the previous chapter, we analyze the polarized cross-section in 
``helicity'', ``beamline'' and ``off-diagonal'' bases at one-loop level
in the soft-gluon approximation.
\begin{enumerate}
\item Helicity basis:

The helicity basis is given by 
\[
 \cos \xi ~=~\pm 1. 
\]
Above condition leads to the following differential cross-sections;
\bea
&&
    \frac{d \sigma}{d \cos \theta}
    (e^-_L e^+_R \rightarrow t_L \bar{t}_L) 
~=~ \frac{d \sigma}{d \cos \theta}
    (e^-_L e^+_R \rightarrow t_R \bar{t}_R) 
    \nonumber \\
&&
= \frac{3 \pi \alpha^2}{8 s} 
    ( f_{LL} + f_{LR} )^2  \sin^2 \theta
\left[(1 - \beta^2)(1 + S_I) - 2 S_{II} \right] ~,\\
%%%%%%%%%%%%%%%
&&
    \frac{d \sigma}{d \cos \theta}
    (e^-_L e^+_R \rightarrow t_R \bar{t}_L
                   {\rm ~or~}  t_L \bar{t}_R ) 
\nonumber \\
&& =
    \frac{3 \pi \alpha^2}{8 s}  (1 \mp \cos \theta)
    \left[f_{LL}(1 \mp \beta) + f_{LR}(1 \pm \beta) \right]
\\
&& \times
    \left[
  \{ f_{LL}(1 \mp \beta) + f_{LR}(1 \pm \beta) \}(1 + S_{I})
- 2 
  \{ f_{LL}(1 \pm \beta) + f_{LR}(1 \mp \beta) \}S_{II}  
    \right]. 
    \nonumber 
\eea

\item Beamline basis:

In the beamline basis, the top (anti-top) quark spin is defined in the
positron (electron) direction in its rest frame.
The spin angle $\xi$ is obtained by 
\bea
\cos \xi ~=~ \frac{\cos \theta + \beta}
                  {1 + \beta \cos \theta}. 
\label{eqn:beam2-def} 
\eea
The differential polarized cross-sections in the beamline
basis are obtained by Eqns.(\ref{eqn:uu1loop}),
(\ref{eqn:ud1loop}) and (\ref{eqn:beam2-def}). 
\bea
&&
    \frac{d \sigma}{d \cos \theta}
    (e^-_L e^+_R \rightarrow t_{\uparrow} \bar{t}_{\uparrow}) 
~=~ \frac{d \sigma}{d \cos \theta}
    (e^-_L e^+_R \rightarrow t_{\downarrow} \bar{t}_{\downarrow}) 
    \nonumber \\
&&~~= \left( \frac{3 \pi \alpha^2}{2 s} \beta \right)
    \frac{\beta^2 \sin^2 \theta}{(1 + \beta \cos \theta)^2} 
\Biggl[  f_{LR}^2  (1 - \beta^2)(1 + S_{I}) 
         \nonumber \\
&& \qquad \qquad - 
     \Bigl[ 
             \{ 2 - \beta ( \beta - \cos \theta ) \}f_{LL}f_{LR}
           + \beta ( \beta + \cos \theta )f_{LR}^2
     \Bigr] S_{II}  
\Biggr ],  \\
&&
    \frac{d \sigma}{d \cos \theta}
    (e^-_L e^+_R \rightarrow t_{\downarrow} \bar{t}_{\uparrow}) 
\nonumber \\
&&~~=
    \left( \frac{3 \pi \alpha^2}{2 s} \beta \right)
    \frac{ \beta^4 \sin^4 \theta}{(1 + \beta \cos \theta)^2}
    \Biggl[
    f_{LR}^2 (1 + S_{I} + S_{II}) -
     f_{LL} f_{LR} S_{II}
    \Biggr]~, \\
%%%
&&
    \frac{d \sigma}{d \cos \theta}
    (e^-_L e^+_R \rightarrow t_{\uparrow} \bar{t}_{\downarrow})
\nonumber \\
&&~~=
    \left( \frac{3 \pi \alpha^2}{2 s} \beta \right)
    \Biggl[ 
      f_{LL}(1 + \beta \cos \theta ) + 
      f_{LR}\frac{(1 - \beta)^2}{(1 + \beta \cos \theta)} 
    \Biggr] \nonumber \\
&&~\times
    \Biggl[
    \Biggl\{ 
      f_{LL}(1 + \beta \cos \theta ) + 
      f_{LR}\frac{(1 - \beta)^2}{(1 + \beta \cos \theta)} 
    \Biggr\}(1 + S_{I}) \\
&&~~+ \left(\frac{1}{1 + \beta \cos \theta}\right) 
    \Bigl\{ 
      (f_{LL} - f_{LR})( \beta^2 (1 + \cos^2 \theta) - 2) - 
       4 f_{LR} (1 + \beta \cos \theta) 
    \Bigr\}S_{II} 
    \Biggr]~.
    \nonumber 
\eea

\item Off-diagonal basis:

At the tree level, we have shown that the \lq\lq off-diagonal\rq\rq\ basis 
makes the contributions from the like-spin 
configuration vanish~\cite{ps,pk1}. 
At order ${\cal O}(\alpha_s )$, we can find two off-diagonal bases for
the $e^-_Le^+_R$ scattering.
The first basis is given by
\bea
 A_{LR} \cos \xi - B_{LR} \sin \xi ~=~ 0~. \nonumber
\eea
This equation leads to the relation between the spin angle $\xi$ and
the scattering angle $\theta$,
\bea
 \tan \xi = \frac{A_{LR}}{B_{LR}}
   = \frac{( f_{LL} + f_{LR} ) \sqrt{1 - \beta^2}\, \sin \theta}
          {f_{LL} ( \cos \theta + \beta )
             + f_{LR} ( \cos \theta - \beta )} \ .
\label{offdiaxi}
\eea
This basis is equivalent to Eqn.(\ref{eqn:off-def}) in the
leading order analysis, and not modified by the
QCD corrections.
The first order QCD corrected cross-sections in this basis are
\bea
&& \frac{d \sigma}{d \cos \theta}
    \left( e^-_L e^+_R \to t_{\uparrow} \bar{t}_{\uparrow} 
           \ {\rm and} \ t_{\downarrow} \bar{t}_{\downarrow} \right)
    = 0 \ ,\label{upupsoft}\\
&& \frac{d \sigma}{d \cos \theta}
        \left( e^-_L e^+_R \to t_{\uparrow} \bar{t}_{\downarrow} 
                 \  {\rm or} \  t_{\downarrow} \bar{t}_{\uparrow} \right)
    = \left( \frac{3 \pi \alpha^2}{2 s} \beta \right) 
      \Bigl( \sqrt{A_{LR}^2 + B_{LR}^2} \mp D_{LR} \Bigr) 
 \label{updownsoft} \\ 
    &&\times \Biggl[ \Bigl( \sqrt{A_{LR}^2 + B_{LR}^2} \mp D_{LR} \Bigr)
        \Bigl( 1 + S_I \Bigr)
       - 2 \, \left( \frac{\gamma^2 A_{LR}^2 + B_{LR} \bar{B}_{LR}}
            {\sqrt{A_{LR}^2 + B_{LR}^2}} \mp \bar{D}_{LR} \right)
                S_{II} \Biggr]  \nonumber.  
\eea
The second basis is defined by,
\bea
( A_{LR} \cos \xi - B_{LR} \sin \xi ) ( 1 + S_{I}) 
- 2(\gamma^2 A_{LR} \cos \xi - \bar{B}_{LR} \sin \xi ) S_{II}
~=~0~. \qquad
\eea
Therefore the spin angle, $\xi$, satisfies the following relation,
\bea
\tan \xi ~=~ 
\frac{ (1 + S_I) - 2 \gamma^2 S_{II}}
     {B_{LR}(1 + S_{I}) - 2 \bar{B}_{LR} S_{II}} A_{LR}~. 
\label{offdiaxi2}
\eea
Since $S_{I}$ and $S_{II}$ are ${\cal O}(\alpha_s)$,
there is only difference of ${\cal O}(\alpha_s)$ between the
Eqn.(\ref{offdiaxi}) and (\ref{offdiaxi2}).
Hence, we use the off-diagonal basis defined by
Eqn.(\ref{offdiaxi}).
A similar result holds for $e^-_Re^+_L$ scattering. 
\end{enumerate}
%\clearpage

\begin{figure}[H]
\begin{center}
        \leavevmode\psfig{file=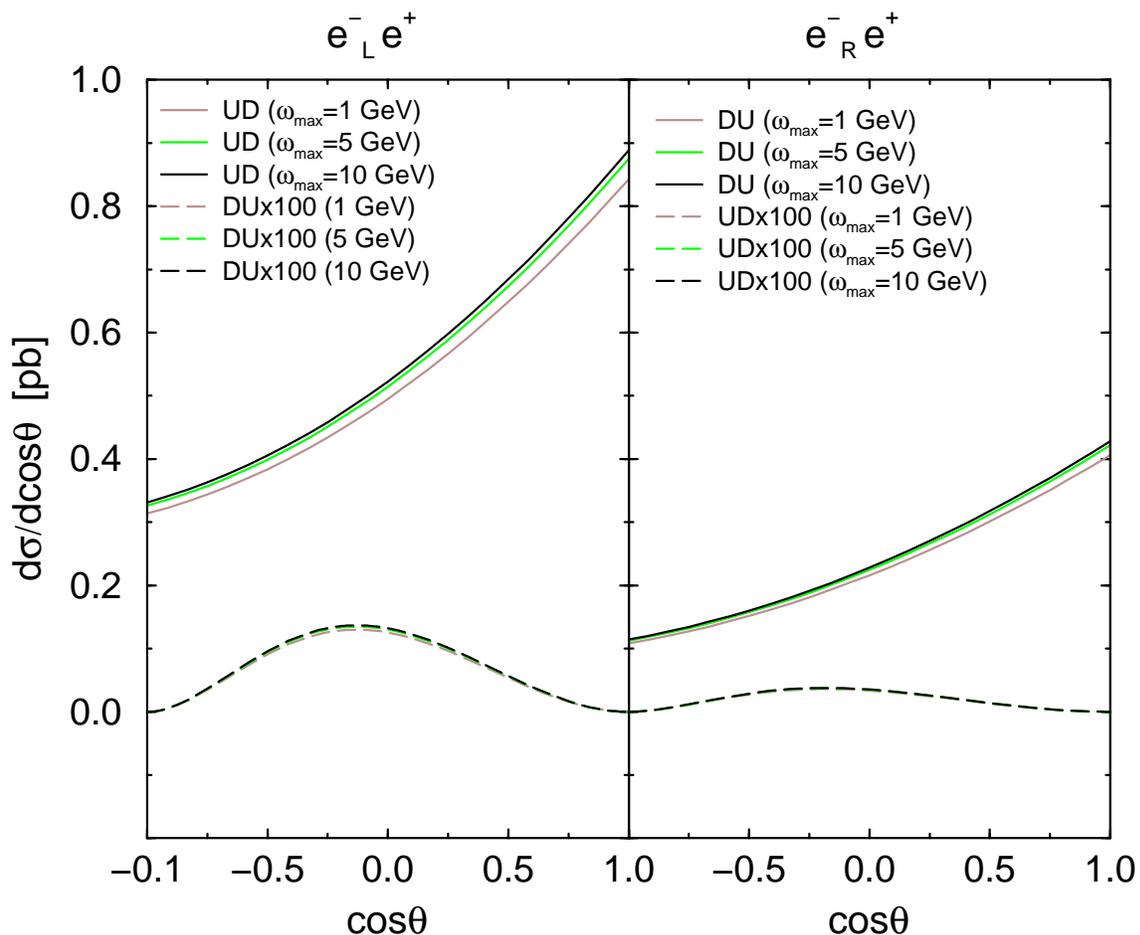,angle=-90,width=15cm} 
\caption[The $\omega_{\rm max}$ dependence of the cross-sections
in the off-diagonal basis at $\sqrt{s}=400 \, {\rm GeV}$.]{The
 $\omega_{\rm max}$ dependence of the cross-sections 
in the off-diagonal basis at $\sqrt{s}=400 \, {\rm GeV}$. The DU (UD)
 component for the $e^{-}_{L} e^{+}$  ($e^{-}_{R} e^{+}$) process at $\sqrt{s}=400 \, {\rm GeV}$ is multiplied by 100.} 
\label{fig:fig4-1}
\end{center} 
\end{figure}

In the soft-gluon approximation, the polarized cross-sections 
depend on the unknown value of $\omega_{\rm max}$ for the
soft gluon.
Before discussing the spin correlations in the soft-gluon approximation, 
we will examine the the $\omega_{\rm max}$ dependence of the cross
section.
Since we are interested in the cross-section mainly in the off-diagonal basis,
we present the $\omega_{\rm max}$ dependence of the differential
cross-section in the off-diagonal basis.
Fig.\ref{fig:fig4-1} shows the $\omega_{\rm max}$ dependence of the
differential cross-section, 
where the numerical values of parameters are taken from
Table.\ref{tbl:param} in Chapter 3. 
The running strong coupling constant is obtained by evolving
$\alpha_s(M_Z^2) = 0.118$ with $5$ flavor to $\alpha_s(s)$ with $6$
flavor. 
Here we use the renormalization equation for the running coupling constant
up to QCD two-loop level. 
We plot the cross-section at values for the $\omega_{\rm max}$; 
\[
 \omega_{\rm max} ~=~ x_{\rm min} (\sqrt{s} - 2 m)~,~
\left( 
       x_{\rm min} = \frac{1}{50},~\frac{1}{10},~\frac{1}{5} 
\right).
\]
The cross-sections behaves quite uniformly as the 
value of $\omega_{\rm max}$ is changed. 
%Thus above consideration 
This tells us that the behavior are qualitatively the same for any
reasonable value of $\omega_{\rm max}$. 
%\clearpage

%%%%%%%%%%%%%%%%%%%%%%%%%%%%

%%%%%%%%%%%%%%%%%%%%%%%%%%
\begin{figure}[H]
\begin{center}
        \leavevmode\psfig{file=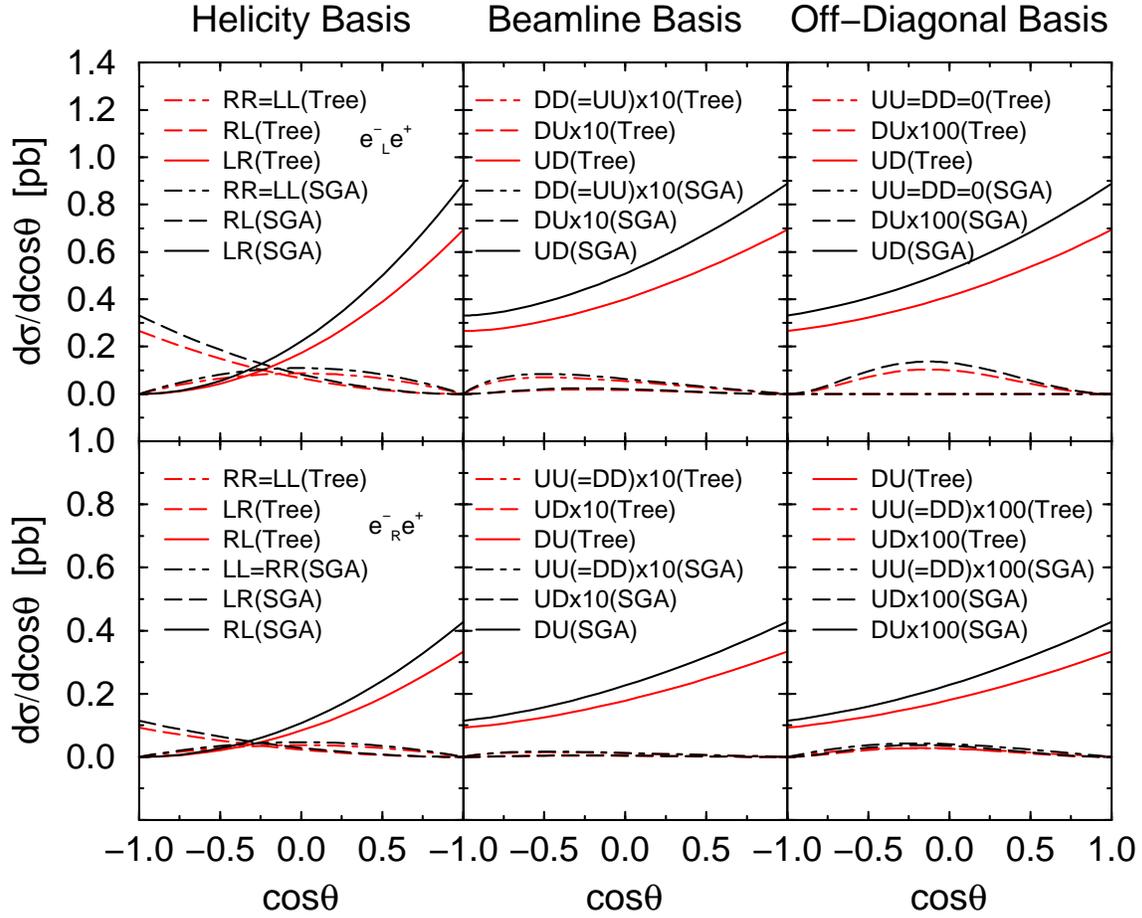,angle=-90,width=15cm}
\caption[The cross-sections in the helicity, beamline and off-diagonal
 bases at  $\sqrt{s}=400 \, {\rm GeV},\omega_{\rm max}=10 \,{\rm GeV}$]{The cross-sections in the helicity, beamline and off-diagonal bases at  $\sqrt{s}=400 \, {\rm GeV}, \omega_{\rm max}=10 \,{\rm GeV}$ for the 
$e^{-} e^+ \rightarrow t \bar{t}$ process:
 $t_L \bar{t}_R$ (LR),
 $t_R \bar{t}_L$ (RL), 
 $t_L \bar{t}_L$ (LL) and
 $t_R \bar{t}_R$ (RR) in the helicity,
 $t_{\uparrow} \bar{t}_{\downarrow}$ (UD),
 $t_{\downarrow} \bar{t}_{\uparrow}$ (DU), 
 $t_{\uparrow} \bar{t}_{\uparrow}$ (UU) and
 $t_{\downarrow} \bar{t}_{\downarrow}$ (DD) in the beamline and
 off-diagonal bases.
The suffices ``Tree'' and ``SGA'' mean the differential cross-section at the
tree level and at the one-loop level in the soft gluon
approximation. It should be noted that DU (UD,UU,DD) component for
the $e^{-}_{L} e^{+}$ ($e^{-}_{R} e^{+}$) process in off-diagonal basis
 is multiplied by 100,
and that DU, UU, DD (UD,UU,DD) components for
the $e^{-}_{L} e^{+}$ ($e^{-}_{R} e^{+}$) process in beamline basis
 is multiplied by 10}  
\label{fig:fig3}
\end{center} 
\end{figure}
%%%%%%%%%%%%%%%%%%%%%%%%%%
\noindent

\begin{center}
\begin{table}[H]
\begin{tabular}{|c||c|c|c|}
\hline 
$e^{-}_L e^+$ 
        & Helicity & Beamline & Off-Diagonal \\ \hline
\hline
Dominant Frac.       
        & $0.53610$ (LR)& $0.97964$ ~(UD)~  & $0.99871$ (UD)\\ \hline
Sub-Dominant Frac.   
        & $0.19992$ (RL)& $0.00905$ (UU,DD) & $0.00129$ (DU)\\ \hline
\hline 
$e^{-}_R e^+$ 
                     & Helicity & Beamline & Off-Diagonal \\ \hline
\hline
Dominant Frac.       
& $0.58709$ (RL)& $0.99084$ ~(DU)~  & $0.99699$ ~(DU)~ \\ \hline
Sub-Dominant Frac.   
& $0.15662$ (LR)& $0.00406$ (UU,DD) & $0.00110$ (UU,DD)\\ \hline
\end{tabular}
\caption[The fraction of $e^{-}_{L/R} e^+$ cross-sections for the
 dominant and the sub-dominant spin configurations at {\cal
 $O$}$(\alpha_s)$ in soft-gluon approximation.]{The fraction of
 $e^{-}_{L/R} e^+$ cross-sections for the 
 dominant and the sub-dominant spin configurations in the helicity,
 beamline and off-diagonal bases at {\cal $O$}$(\alpha_s)$ in soft-gluon
 approximation. Where $\sqrt{s}=400$ GeV and $\omega_{\rm max} = 10$ GeV.} 
\label{tbl:frac1}
\end{table}
\end{center}
\noindent

In Fig.\ref{fig:fig3}, we show the differential cross-sections in three
bases, at $\sqrt{s} = 400 \, {\rm GeV}$ and $\omega_{\rm max}=10~GeV$.
Table \ref{tbl:frac1} shows the fraction of the $e^-_{L/R} \to t \bar{t}$
cross-section for the dominant and the sub-dominant spin configurations.
%at $\sqrt{s}=400$ GeV, $\omega_{\rm max} = 10$ GeV.
The dominant components in the helicity basis 
are $t_{L} \bar{t}_{R}$ (LR) for $e^{-}_{L} e^+$ scattering and 
$t_{R} \bar{t}_{L}$ (RL) for $e^{-}_{R} e^+$ scattering at $\sqrt{s}=400$ GeV.
This component occupies $53 \%$ of total cross-section for $e^{-}_{L} e^+$
and $58 \%$ of total cross-section for $e^{-}_{R} e^+$ scattering. 
In the beamline basis, only UD $(t_{\uparrow} \bar{t}_{\downarrow})$
component is dominant for $e^-_L e^+_R$ cross-section, which makes up more 
than $97 \%$. 
On the other hand, DU $(t_{\downarrow} \bar{t}_{\uparrow})$
component is dominant for $e^-_R e^+_L$ cross-section, which makes up more 
than $99 \%$.
For $e^-_L e^+_R$ cross-section in the off-diagonal basis, the UU
$(t_{\uparrow} \bar{t}_{\uparrow})$  
and the DD $(t_{\downarrow} \bar{t}_{\downarrow})$ components are
identically zero.
The total cross-section is dominated by the UD
$(t_{\uparrow} \bar{t}_{\downarrow})$ component which amount to more
than $99 \%$ and DU $(t_{\downarrow} \bar{t}_{\uparrow})$ component
contributes less than $1~\%$.
For $e^-_Re^+_L$ scattering the UU $(t_{\uparrow} \bar{t}_{\uparrow})$ 
and the DD $(t_{\downarrow} \bar{t}_{\downarrow})$ components are 
non-zero because we have used the off-diagonal basis for $e^-_Le^+_R$
scattering. However the DU $(t_{\downarrow} \bar{t}_{\uparrow})$ component
is still more than $99\%$ of the total cross-section.
%Compare to the tree level cross-section, the enhancement of QCD
%corrections to the total cross-section is $\sim 30 \%$.
%Nevertheless, QCD corrections do not change the fraction of $e^{-}_{L/R}
%e^+$ cross-sections for the dominant and the sub-dominant spin
%configurations. 
Although there exist a magnetic 
moment modification to the $\gamma/Z-t-\bar{t}$ vertex from
QCD corrections, this does not change the behavior of the 
spin dependent cross-sections in the helicity, beamline and off-diagonal
bases. 
The QCD corrections, however, make the
differential cross-sections larger by $\sim 30 \%$ compared to the
tree level ones at this $\sqrt{s}$. 
Thus the off-diagonal and the beamline bases continue to display very
strong spin correlations for the top quark pairs even after taking 
the QCD corrections into account, at least in the soft gluon approximation.

%%%%%%%%%%%%%%%%%%%%%%%%%%%%%
%\begin{figure}[H]
%\begin{center}
%        \leavevmode\psfig{file=wdep1500.eps,angle=-90,width=14cm}
%\caption[The $\omega_{\rm max}$ dependence of the cross-sections
%in the off-diagonal basis at $1500 \, {\rm GeV}$.]{The $\omega_{\rm
%max}$ dependence of the cross-sections in the off-diagonal basis at
%$1500 \, {\rm GeV}$.} 
%\label{fig:fig4-2}
%\end{center} 
%\end{figure}
%%%%%%%%%%%%%%%%%%%%%%%%%%%%
\noindent

\clearpage

\section{Single Spin Correlations in $e^+ e^-$ Process}

The result in the soft gluon approximation imply that
the QCD corrections do not change the behavior of the 
spin dependent cross-sections in any basis. 
However, the following two points are shortcomings of this approximation.
First, soft gluon emission cannot change the spin 
of the heavy quarks.
Second, the heavy quark pairs are produced always back to back.
These points are invalid for the hard gluon,
and it is possible that the full 
${\cal O}(\alpha_s )$ QCD corrections might completely change the
conclusions of the previous section. 
In this section, we investigate the full ${\cal O}(\alpha_s )$
QCD corrections~\cite{pk1,KNP1}.
Since, in the presence of a hard gluon, 
the top and anti-top quarks are not generally produced back-to-back,
it is instructive to consider the single heavy quark spin correlations.
Namely, we investigate the inclusive cross-section for the production
of the top (or anti-top) quark in a particular spin configuration.
We have organized this section as follows.
After defining the kinematics and our conventions, we give the polarized
cross-section for the top quark using a generic spin basis closely
related to the spin basis of the previous section.
The numerical analysis will be relegated to the next section.

\subsection{Amplitudes and Kinematics}

The principle result of this section will be 
the inclusive cross-section for the polarized
top quark production, $e^+ e^- \to t_{\uparrow}{\rm ~or~}t_{\downarrow} ~+~ X$, 
where 
$X = \bar{t} {\rm ~or~} \bar{t} g$
(The cross-section for the anti-top quark inclusive production can be 
easily obtained from the results in this section.).

%%%%%%%%%%%%%%%%%%%%%%%%%%%%%%%%%
\begin{figure}[H]
\begin{center}
\begin{tabular}{cc}
\leavevmode\psfig{file=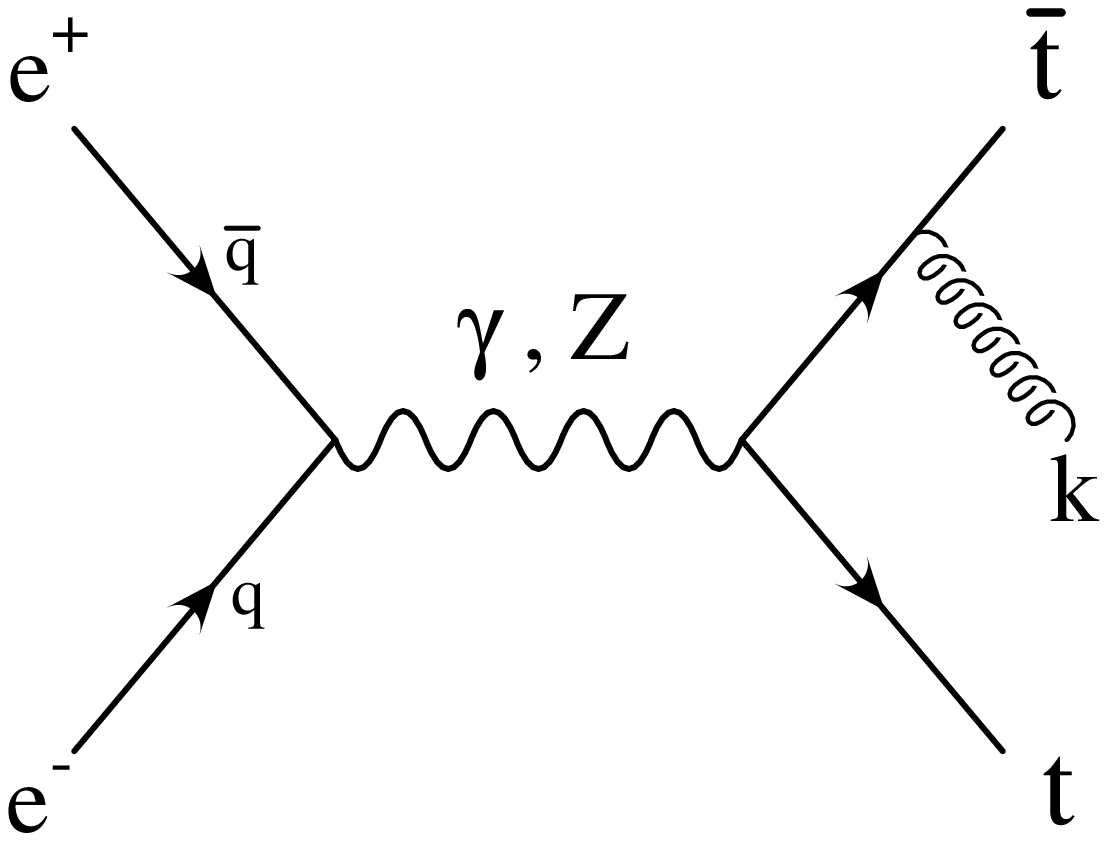,width=5cm} &
\leavevmode\psfig{file=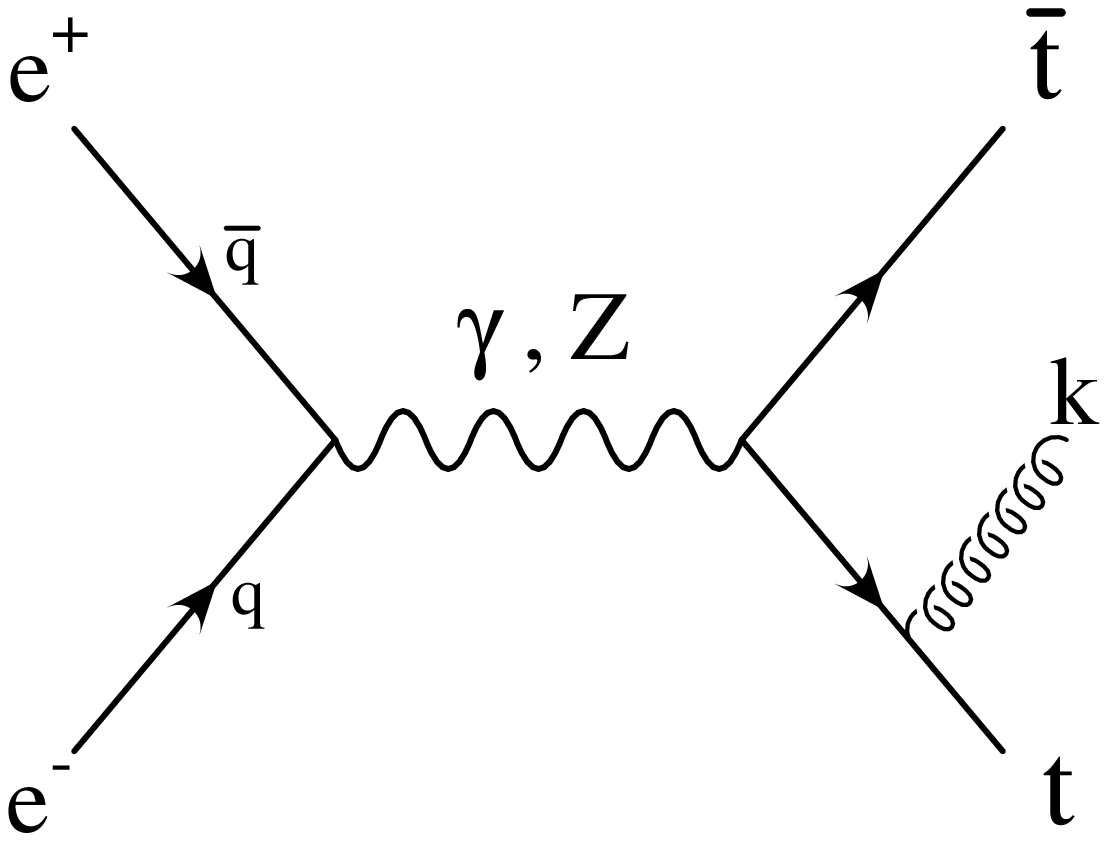,width=5cm} 
\end{tabular}
\caption{The real gluon emission contributions
to top quark pair production.}
\label{fig:fig5}
\end{center}
\end{figure}
%%%%%%%%%%%%%%%%%%%%%%%%%%%%%%%%%
\noindent
Since the vertex corrections are the same as in the previous
section, all that is required is the full real gluon emission contributions.
The real gluon emission diagrams to leading order 
in $\alpha_s$ are given in Fig.\ref{fig:fig5}.
Figure \ref{fig:fig5} also defines the momenta of particles.
We also use the spinor helicity method for massive fermions
to calculate the squares of these amplitudes for a polarized top 
quark (see Chapter 2).

The top quark momentum $t$ is decomposed into a sum of two massless
momenta $t_1 \,,\,t_2$ such that in the rest frame of the top
quark the spatial momentum of $t_1$ defines the spin axis for the top
quark.
\[    t = t_1 + t_2 \quad , \quad m s_t = t_1 - t_2 \ ,\]
where $s_t$ is the spin four vector of the top quark.
The amplitude for Fig.\ref{fig:fig5} is given by
\bea
 \lefteqn{M (e_L^- e_R^+ \to t_{s_t} \bar{t}_{s_{\bar{t}}} g)
   = 
\mbk{\bar{q} -| (\gamma_{L})_{\mu}|q -}}
%\bar{v}_{\uparrow} (\bar{q}) \gamma_L^{\mu} u_{\downarrow} (q)}
                \nonumber\\
   &\times& \bar{u}(t,s_{t}) \left[ \frac{1}{2 \bar{t} \cdot k}
             \left( \frac{a_{LL}}{2} \gamma_L^{\mu} +
                    \frac{a_{LR}}{2} \gamma_R^{\mu} \right)
               ( - \hat{\bar{t}} - \hat{k} + m ) \gamma_{\nu} \right.
                 \label{amplitude}\\
   & & \qquad\qquad + \left. \frac{1}{2 t \cdot k} \gamma_{\nu}
               ( \hat{t} + \hat{k} + m )
             \left( \frac{a_{LL}}{2} \gamma_L^{\mu} +
                    \frac{a_{LR}}{2} \gamma_R^{\mu} \right)
             \right] T^a v(\bar{t},s_{\bar{t}}) \varepsilon^{\nu}_a (k)
                \ ,\nonumber
\eea
where $\varepsilon$ is the
polarization vector of the gluon. $T^a$ is the color matrix, and
$\hat{p}= p^{\mu} \gamma_{\mu}$.
The coupling constants $a_{LI}$ are defined as follows:
\[  \frac{a_{LI}}{2} = \frac{e^2 g}{s} f_{LI} \ .\]

The expressions for the squares of the amplitudes given below
have been summed over the spins of the unobserved particles
(the anti-top quark and gluon) as well as the colors of the final
state particles. 
Let us write the square of the amplitude for the top quark
with spin \lq\lq up\rq\rq\ as
\bea
  \lefteqn{|M (e_L^- e_R^+ \to t_{\uparrow} \bar{t} g )|^2}\nonumber\\ 
    &=& N_c C_2 (R) \, \left[ \frac{1}{(\bar{t} \cdot k )^2}\, T_1 +
        \frac{1}{(\bar{t}\cdot k)(t\cdot k)}\, T_2 +
         \frac{1}{(t \cdot k )^2}\, T_3 \right] \ .\label{ampsquare}
\eea
After some calculation, we find
\bean
   T_1 &=& 4 \, |a_{LL}|^2 
        \left[ (\bar{t}\cdot k)(q\cdot k) - m^2 q\cdot (\bar{t} + k) 
          \right] \, (t_2 \cdot \bar{q}) \\
        &+& \, 4 \, |a_{LR}|^2
        \left[ (\bar{t}\cdot k)(\bar{q}\cdot k) - m^2 
            \bar{q} \cdot (\bar{t} + k) 
          \right] \, (t_1 \cdot q) \\
        &-&  \,  \left[ a_{LL} a_{LR}^* 
           ( (\bar{t}\cdot k) + m^2 ) {\rm Tr} (\gamma_R \, t_1 \, t_2 \,
            \bar{q} \, q ) + c.c. \right] \ ,\\
   T_2 &=& |a_{LL}|^2 
         \left[ 4 (t\cdot\bar{t})(\bar{t}\cdot q)(t_2 \cdot\bar{q})
         + (t_2 \cdot\bar{q}) {\rm Tr} (\gamma_R \, q \, k \, t \, \bar{t})
         + (\bar{t}\cdot q) {\rm Tr} (\gamma_L \, t_2 \, \bar{q} \, k \, 
       \bar{t}) \right] \\
        &+& \, |a_{LR}|^2 
         \left[ 4 \, (t\cdot\bar{t})(\bar{t}\cdot\bar{q})(t_1 \cdot q)
         + (t_1 \cdot q) {\rm Tr} (\gamma_L \, \bar{q} \, k \, t \, \bar{t})
         + (\bar{t}\cdot\bar{q}) {\rm Tr} (\gamma_R \, t_1 \, q \, k \, 
           \bar{t}) \right] \\
        &+& \, a_{LL} a_{LR}^* 
           \Bigl[ (t\cdot\bar{t})
                 {\rm Tr} (\gamma_R \, t_1 \, t_2 \, \bar{q} \, q)
          - (k\cdot q) {\rm Tr} (\gamma_R \, t_1 \, t_2 \, \bar{q} \, k)\\
        & & \qquad\qquad\qquad\quad
               + \frac{1}{2} {\rm Tr}
                 (\gamma_R \, t_2 \, t_1 \, q \, k \, t \, \bar{q}) 
               + \frac{1}{2} {\rm Tr}
                (\gamma_R \, t_1 \, t_2 \, \bar{q} \, q \, k \, \bar{t} ) 
              \Bigr] \\
        &+& \, a_{LL}^* a_{LR} 
           \Bigl[ (t\cdot\bar{t}) {\rm Tr}
                (\gamma_L \, t_2 \, t_1 \, q \, \bar{q})
          - (k\cdot\bar{q}) {\rm Tr} (\gamma_L \, t_2 \, t_1 \, q \, k)\\
        & & \qquad\qquad\qquad\quad
               + \frac{1}{2} {\rm Tr} 
           (\gamma_L \, t_1 \, t_2 \, \bar{q} \, k \, t \, q ) 
               + \frac{1}{2} {\rm Tr}
                (\gamma_L \, t_2 \, t_1 \, q \, \bar{q} \, k \, \bar{t} )
                \Bigr] \\
        &+& \, \, \, c.c. \ , \\
   T_3 &=& 2 \, |a_{LL}|^2
        \left[ - m^2 (k\cdot\bar{q}) - 2 (m^2 + (t\cdot k))
            (t_2 \cdot\bar{q}) + 2 ( (t + k)\cdot\bar{q})
                   (t_2 \cdot k) \right] \, (\bar{t}\cdot q )\\
       &+&  \, 2 \, |a_{LR}|^2
        \left[ - m^2 (k\cdot q) - 2 (m^2 + (t\cdot k))
            (t_1 \cdot q) + 2 ( (t + k)\cdot q) 
           (t_1 \cdot k )\right] \, (\bar{t}\cdot\bar{q}) \\
       &-&  \, \frac{m^2}{2}
           \left[ a_{LL} a_{LR}^* \{ 2 {\rm Tr}
                 (\gamma_R \, t_1 \, t_2 \, \bar{q} \, 
               q ) + {\rm Tr} (\gamma_R \, k \, t_2 \, \bar{q} \, q )  
                 + {\rm Tr} (\gamma_R \, t_1\, k \, \bar{q} \, q ) \} +
               c.c. \right] \ , 
\eean
where $\gamma_{R/L} \equiv \frac{1 \pm \gamma_5}{2}$ and
that all momentum, $p$, under the \lq\lq Tr\rq\rq\  operator are understood  
to be  $\hat{p}$. By interchanging the $t_1$ and
$t_2$ vectors in the above expressions,
we can get the amplitude square for the top quark
with spin \lq\lq down\rq\rq.
Since we neglect the $Z$ width, all the coupling constants $a_{LI}$ are real.

\begin{figure}[H]
\begin{center}
\leavevmode\psfig{file=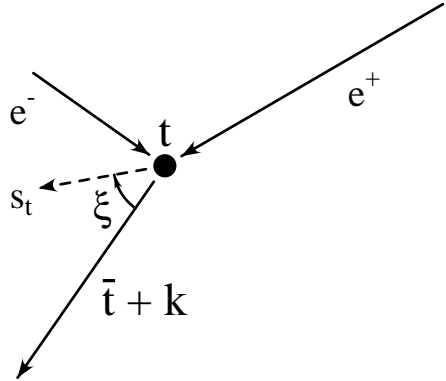,height=5cm}
\caption{The spin basis for the top
quark in the process  $e^- e^+ \to t \bar{t} g$.} 
\label{fig:fig6}
\end{center}
\end{figure}

\begin{figure}[H]
\begin{center}
\leavevmode\psfig{file=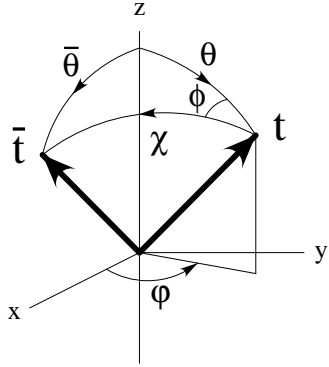,height=5cm}
\caption[The momentum (unit vectors) configuration of the top
and anti-top quarks in the CM frame.]{The momentum (unit vectors)
 configuration of the top and anti-top quarks in the CM frame. The
 momentum of $e^-$ ($e^+$) is in the $+z$ ($-z$) direction.} 
\label{fig:fig7}
\end{center}
\end{figure}

To define the spin basis for the top quark, we naturally extend the spin 
definition of the previous section to the present case. The top quark
spin is decomposed along the direction ${\bf s}_t$ in the rest frame of the
top quark which makes an angle $\xi$ with the sum of the anti-top quark
and the gluon momenta in the clockwise direction, see Fig.\ref{fig:fig6}.
%%%%%%%%%%%%%%%%%%%%%%%%%%%%%
%%%%%%%%%%%%%%%%%%%%%%%%%%%%%
\noindent
(We consider the case in which there is no transverse polarization.) 
To calculate the cross-section from Eqn.(\ref{ampsquare}), we take the
CM frame in which the $e^+ e^-$ beam line coincides with the $z$-axis,
\[  q = \frac{\sqrt{s}}{2} ( 1\,,\,0\,,\,0\,,\,1 )\quad , \quad
    \bar{q} = \frac{\sqrt{s}}{2} ( 1\,,\,0\,,\,0\,,\,-1 )\ .\]
We define the variables $x, y$ and $z$ which are related to CM 
energies of the gluon, top and anti-top quarks by  
\[  x \equiv 1 - \frac{2\, k\cdot (q + \bar{q})}{s}\ ,\ 
    y \equiv 1 - \frac{2\, t\cdot (q + \bar{q})}{s}\ ,\ 
    z \equiv 1 - \frac{2\, \bar{t}\cdot (q + \bar{q})}{s}\ .\] 
The momenta of the final state particles, in terms of these variables, are
\[   
k = \frac{\sqrt{s}}{2} 
       ( 1 - x \,,\, (1 - x) \, \mbox{{\boldmath $k$}} )\ , \,
t = \frac{\sqrt{s}}{2} 
       ( 1 - y \,,\, a(y) \, \mbox{{\boldmath $t$}} )\ ,\     
\bar{t} = \frac{\sqrt{s}}{2} 
       ( 1 - z \,,\, a(z) \, \mbox{\boldmath{$\bar{t}$}} \, )\ \ 
,\]
where $\mbox{{\boldmath $k$}}$, $\mbox{\boldmath $t$}$ and $\mbox{\boldmath
$\bar{t}$}$ mean the unit space vectors respectively and
\[  a(y) \equiv \sqrt{(1-y)^2 - a} \quad , \quad
        a(z) \equiv \sqrt{(1-z)^2 - a} \ ,\]
with $a \equiv 4 m^2 / s$. 
Fig.\ref{fig:fig7} defines the orientation of the top and anti-top momenta
and 
by energy-momentum conservation the momentum of the
gluon is also determined.
%%%%%%%%%%%%%%%%%%%%%%%%%%%%%%%%%
%%%%%%%%%%%%%%%%%%%%%%%%%%%%%%%%
\noindent
One can easily obtain the spin four vector
of the top quark in the CM frame by boosting the spin vector characterized
by $\xi$ in the top quark rest frame in the direction of top quark
momentum by $\beta (y)$ (the speed of the top quark in the CM frame).
The explicit form for $t_1 \, (t_2 = t - t_1 )$ is given by
\bea
      t_1^0 &=& \frac{m}{2} \left[ \, \gamma (y) \, (1 - \beta (y) \cos \xi )\,
               \right] \ ,\nonumber\\
      t_1^1 &=& \frac{m}{2} \left[ \,  \gamma (y) \, (\beta (y) - \cos \xi )
                  \sin \theta \cos \varphi + \sin \xi \cos \theta
                  \cos \varphi \, \right] \ ,\label{spinvecCM}\\
      t_1^2 &=& \frac{m}{2} \left[ \,  \gamma (y) \, (\beta (y) - \cos \xi )
                  \sin \theta \sin \varphi + \sin \xi \cos \theta
                  \sin \varphi \, \right] \ ,\nonumber\\
      t_1^3 &=& \frac{m}{2} \left[ \,  \gamma (y) \, (\beta (y) - \cos \xi )
                  \cos \theta - \sin \xi \sin \theta \, \right] \ , \nonumber
\eea
where
\[ \sqrt{a} \, \gamma (y) = 1 - y \quad , 
                    \quad \sqrt{a} \, \gamma (y) \, \beta (y) = a (y) \ .\]
If we eliminate the gluon momentum $k$ using the energy momentum
conservation and use the angular variables $\chi \,,\, \phi$ in
Fig.\ref{fig:fig7} to specify the orientation of the anti-top quark
(if one eliminates the anti-top
momentum, one can proceed in the similar way by introducing
other angular variables), the square of the amplitude
Eqn.(\ref{ampsquare}) can be written as
\be
  |M (e_L^- e_R^+ \to t_{\uparrow} \bar{t} g )|^2
     = \frac{s}{8}\, N_c C_2 (R) \left[ a_{LL}^2 M_1
           + a_{LR}^2 M_2 + a_{LL} a_{LR} M_3 \right] \ ,\label{ampsquareyz}
\ee 
where $M_i$ are the functions of $y\,,\,z$, angles defined above and
the spin orientation $\xi$.
\bean
   M_1 &=& \frac{8}{s} \Biggl[
        \frac{m^2}{(\bar{t} \cdot k)^2} 
        (2 t \cdot q - s)
        (t \cdot \bar{q} - \delta t \cdot \bar{t})
\nonumber \\
       &&~+
        \frac{1}{(t \cdot k)^2} 
         \Bigl[  
            m^2 (2 \bar{t} \cdot \bar{q} - s )(\bar{t} \cdot q) + 
            s (z + \frac{a}{2})(\bar{t} \cdot q)(\delta t \cdot \bar{q}) +
              (2 \bar{t} \cdot \bar{q} - s)(\bar{t} \cdot q)
              (\delta t \cdot k) 
         \Bigr]
\nonumber \\
       &&~+       
        \frac{1}{(t \cdot k)(\bar{t} \cdot k)}
        \Bigl[ 
           s \Bigl\{ \frac{s}{2} (1 - y \frac{a}{2}) -
                 (1 - \frac{a}{2}(t \cdot q)
             \Bigr\}(t \cdot \bar{q}) 
\nonumber \\
       && \hspace*{2.4cm}
        + ~s \Bigl\{ \frac{s}{2} (1 - z \frac{a}{2}) -
                (1 - \frac{a}{2}(\bar{t} \cdot \bar{q})
             \Bigr\}(\bar{t} \cdot q) 
\nonumber \\
       && \hspace*{2.4cm}
        - ~s \Bigl\{ \frac{s}{2} (1 - y - \frac{a}{2}) -
                   (1 - \frac{a}{2})(t \cdot q)
                      \Bigr\} (t \cdot \bar{q})
\nonumber \\
       && \hspace*{2.4cm}
        + ~s \Bigl \{ \frac{s}{2} (1 - z - \frac{a}{2}) -
                     2 (\bar{t} \cdot \bar{q})
             \Bigr \} (\bar{t} \cdot q) 
             (\delta t \cdot \bar{q}) \nonumber \\
       && \hspace*{2.4cm}
        + ~2 (\bar{t} \cdot \bar{q}) (\bar{t} \cdot q) (\delta t \cdot q)
        + (2 t \cdot \bar{q} - s) (\bar{t} \cdot q)
          (\delta t \cdot \bar{t})
         \Bigr]
         \Biggr]~, \nonumber \\
   M_2 &=&    M_1 ( q \leftrightarrow \bar{q} \, \,
                    \delta t \to - \delta t ) ~,  \nonumber \\
   M_3 &=& 
        \frac{8}{s} 
        \Biggl[~
        \frac{1}{(\bar{t} \cdot k)^2}
               \frac{s}{2}(y + \frac{a}{2})
            (
             - m^2 s - 2 (t \cdot \bar{q})(\delta t \cdot q)
             + 2 (t \cdot q)(\delta t \cdot \bar{q})
             )
\nonumber \\
       &&~ +
        \frac{m^2}{(t \cdot k)^2}
            \Bigl[ 
              - \frac{s^2}{2}(z + \frac{a}{2})
              -
            (s - 2 \bar{t} \cdot \bar{q}) (\delta t \cdot q)
          + (s - 2 \bar{t} \cdot      q ) (\delta t \cdot \bar{q})
             \Bigr]
\nonumber \\
       &&~ +   
 \frac{1}{(t \cdot k)(\bar{t} \cdot k)}
        \Bigl[ 
         m^2 \Bigl\{ 
          \frac{s}{2} (2 - a - y - z) - 4 (k \cdot q)(k \cdot \bar{q})
             \Bigr\}
\nonumber \\       
        &&~\hspace*{2.4cm} + 
     \left( - \frac{s^2}{2} z + 
              s (2 - a + z)(t \cdot \bar{q}) -
              s (z + \frac{a}{2})(k \cdot \bar{q})
     \right) (\delta t \cdot q) 
\nonumber \\       
        &&~\hspace*{2.4cm}- 
     \left( - \frac{s^2}{2} z + 
              s (2 - a + z)(t \cdot q) -
              s (z + \frac{a}{2})(k \cdot \bar{q})
      \right) (\delta t \cdot q)
\nonumber \\
        &&~\hspace*{2.4cm}+  
     \left( 
         s (1 - y) \Bigl\{s - (\bar{t} \cdot \bar{q})\Bigr\}
       - s (3 + z) (t \cdot \bar{q})
     \right)(\delta t \cdot k)
            \Bigr] 
            \Biggr],
\eean
where $\delta t$ is defined as 
\[  \delta t \equiv \frac{4}{s} ( t_1 - t_2 )\ .\]

We transform the scalar product of the momenta to the kinematical
variables in the CM frame.
The functions $M_i$'s are rewritten as
%-----------------------------------------
\bean
   M_1 &=& \frac{2}{yz} \left[ (1-y)^2 + a^2 (y) \cos^2 \theta + (1-z)^2
                   + a^2 (z) \cos^2 \bar{\theta} \right]\\
       &-& \, a \left( \frac{1}{yz} + \frac{1}{y^2} \right)
             (1 - y^2 + a^2 (y) \cos^2 \theta )
              - a \left( \frac{1}{yz} + \frac{1}{z^2} \right)
             (1 - z^2 + a^2 (z) \cos^2 \bar{\theta} )\\
       &+& \, 2 \left( \frac{2-2y-a}{yz} - \frac{a}{y^2} \right)
             a (y) \cos \theta -
              2 \left( \frac{2-2z-a}{yz} - \frac{a}{z^2} \right)
             a (z) \cos \bar{\theta}\\
       &+& \, \frac{1}{y z^2} (1 - z - a(z) \cos \bar{\theta})
               \left[ y (1 - a(z) \cos \bar{\theta}) - 
                  z (1 - a(y) \cos \theta ) \right] (\delta t \cdot \bar{t}) \\
       &-& \, \frac{1}{y z^2} (1 - z - a(z) \cos \bar{\theta})
               \left[ y - z + (y+z)( z - a(z) \cos \bar{\theta}) \right]
                  (\delta t \cdot ( \bar{q} + q  )) \\
       &+& \, \frac{a}{yz} \left( \frac{1}{y} + \frac{1}{z} \right) 
               \left[ y (1 - a(z) \cos \bar{\theta}) + 
                  z (1 + a(y) \cos \theta ) \right]
                  (\delta t \cdot \bar{q}) \\
       &-& \, \frac{2}{yz} \left[ (1 - y + a(y) \cos \theta ) + 
                   (1 - y - z) (1 - z - a(z) \cos \bar{\theta}) \right]
                   (\delta t \cdot \bar{q}) \ , \\
   M_2 &=&    M_1 ( \cos \theta \to - \cos \theta \,,\,
                    \cos \bar{\theta} \to - \cos \bar{\theta} \,,\,
                    \delta t \to - \delta t ) \ , \\
%-------------------------------------------------
   M_3 &=&  2a \left\{ \frac{4}{yz} - \frac{4}{y} - \frac{4}{z}
             - \frac{(y+z)^2}{yz} + \frac{(y+z)^2}{yz} \cos^2 \theta_k
             \right\} - 2 a^2 \left( \frac{1}{y} + \frac{1}{z} \right)^2\\
       &+& \, \frac{a}{y z} \left( \frac{1}{y} + \frac{1}{z} \right)
                 (y a(z) \cos \bar{\theta} - z a(y) \cos \theta)
                (\delta t \cdot ( \bar{q} + q )) \\ 
       &-& \, \frac{2}{y z}
              \left[ a(y) \cos \theta + (1 - y - z) a(z) \cos
               \bar{\theta} \right]
                (\delta t \cdot (\bar{q} + q )) \\ 
       &+& \, a \left( \frac{1}{y} + \frac{1}{z} \right)^2
                (\delta t \cdot (\bar{q} - q )) 
           +  \frac{2}{y z} \left[ z (y + z) - 2 (1 - y - z) \right]
                (\delta t \cdot (\bar{q} - q )) \\ 
       &+& \, \frac{2}{y z} \left[ (1 - y) a(z) \cos \bar{\theta} 
               + (3 + z) a(y) \cos \theta \right] (\delta t \cdot \bar{t}) \ .
\eean
In the above equations, $\theta_k$ is the angle between the $z$ axis
and the gluon momentum
\[  (y+z) \cos \theta_k = - a (y) \cos \theta - a (z) \cos \bar{\theta} \ .\]
Using Eqn.(\ref{spinvecCM})
we find that the products of $\delta t$ with 
momenta $q\,,\,\bar{q}$ and $\bar{t}$
are 
\bean
   \delta t \cdot q &=& \{ (1-y) \cos \theta - a (y) \} \cos \xi
             + \sqrt{a} \sin \theta \sin \xi \ ,\\
   \delta t \cdot \bar{q} &=& \{ - (1-y) \cos \theta - a (y) \} \cos \xi
             - \sqrt{a} \sin \theta \sin \xi \ ,\\
   \delta t \cdot \bar{t} &=& \{ - (1-z) a (y) + (1-y) a (z) \cos \chi \}
                    \cos \xi\\
      & & \qquad\qquad\qquad\qquad\qquad
                    +\, \sqrt{a} a (z) \sin \xi \sin \chi \cos \phi \ .
\eean
The unpolarized top quark production process is given by dropping the spin
dependent parts (terms proportional to $\delta t$)
in Eqn.(\ref{ampsquareyz}). 
As a check we have reproduced the results of Ref.~\cite{gnt} by putting 
$a_{LL} = a_{LR} = - \frac{2 e^2 g}{s} Q_q$.

The cross-section is given by
\be
    d \sigma (e_L^- e_R^+ \to t_{\uparrow} \bar{t} g ) = 
    \frac{1}{2s} |M (e_L^- e_R^+ \to t_{\uparrow} \bar{t} g )|^2
             (PS)_3 \ , \label{diffcrosssection}
\ee
where $(PS)_3$ is the three particle phase space.
\[  (PS)_3 = \frac{d^3 t}{(2\pi )^3 2 t^0}
             \frac{d^3 \bar{t}}{(2\pi )^3 2 \bar{t}^0}
             \frac{d^3 k}{(2\pi )^3 2 k^0} (2\pi )^4 \delta^4
             (t + \bar{t} + k - q - \bar{q} ) \ .\]
We introduce a small mass $\lambda$
for the gluon to regularize the infrared singularities.
It is easy to rewrite the above phase space integral as,
\bean
 \int (PS)_3 &=&  \frac{s}{(4 \pi)^5}
         \int^{y_+}_{y_-} dy \int^{z_+ (y)}_{z_- (y)} dz\\ 
      & & \qquad\qquad\qquad \times \int d\Omega d\cos \chi d\phi \, \delta
      \left( \cos \chi - \frac{y + z + yz + a - 1 - 2 a_{\lambda}}
                  {a(y) a(z)} \right) \ ,
\eean
where $d\Omega = d\cos \theta d\varphi$ is the solid angle for the top
quark and $a_{\lambda} \equiv \lambda^2 / s$.
The integration regions over $y$ and $z$ are determined by the
condition $|\cos \chi | \leq 1$,
\bean
     y_+ &=& 1 - \sqrt{a} \quad , \quad y_- = \sqrt{a a_{\lambda}}
              + a_{\lambda} \ ,\\
     z_{\pm}(y) &=& \frac{2}{4 y + a}
           \left[ y \left( 1 - y - \frac{a}{2} + a_{\lambda} \right)
          + a_{\lambda} \pm
           a (y) \sqrt{( y - a_{\lambda})^2 - a a_{\lambda}} \right] \ .
\eean
The integration over the angle $\phi$ is not difficult if one uses
the relation
\[ \cos \bar{\theta} = \cos \theta \cos \chi
          + \sin \theta \sin \chi \cos \phi \ .\]
The integrals we need are the followings:
\bean
  \int \cos\bar{\theta} \ d \phi &=& 2 \pi \cos\theta \cos\chi \ ,\\
  \int \cos^2\bar{\theta} \ d \phi &=& 2 \pi \left[ \cos^2\theta
      + \frac{1}{2} ( 1 - 3 \cos^2\theta ) \sin^2\chi \right] \ ,\\
  \int \cos^2\theta_k \ d \phi &=& 2 \pi \left[ \cos^2\theta
      + \frac{1}{2} ( 1 - 3 \cos^2\theta ) \ \frac{a^2 (z)}{(y + z)^2}
                       \  \sin^2\chi \right] \ ,\\
  \int \cos\bar{\theta} \cos\phi \ d \phi &=& \pi \sin\theta \sin\chi \ ,\\
  \int \cos^2\bar{\theta} \cos\phi \ d \phi &=&
             2 \pi \sin\theta \cos\theta \sin\chi \cos\chi \ .
\eean
Due to the $\delta$ function in the phase space integral,
the angle $\chi$ is a function of $y$ and $z$:
\bean
    \cos\chi &=& \frac{y + z + yz + a - 1}{a(y) a(z)}\ ,\\
    \sin^2\chi &=& \frac{4yz (1-z-y)- a (y+z)^2}{a^2 (y) a^2 (z)}\ ,
\eean
where we have put $a_{\lambda}$ to be zero since the $\lambda \to 0$
limit does not produce any singularities in the (squared) amplitude.
The remaining integrals to get the cross-section are over the
variables $y$ and $z$. According to the type of
integrand, we group the phase space integrals (after the integrations
over the angular variables) into
four distinct classes $\{J_i\}\,,\,\{N_i\}\,,\,\{L_i\}$ and $\{K_i\}$~\cite{tung}.
The individual integrals of these classes are summarized in Appendix.

\subsection{Cross-Section in the Generic Spin Basis}

We write the inclusive cross-section for the top quark
in the following form.
\be
 \frac{d \sigma}{d \cos \theta}
                  ( e_L^-  e_R^+ \to t_{\uparrow} X )
  = \frac{3 \pi \alpha^2}{4 s}
        \sum_{klmn} \left( D_{klmn} + \hat{\alpha}_s C_{klmn} \right)
              \cos^k \theta \sin^l \theta
             \cos^m \xi \sin^n \xi \ ,\label{txsection}
\ee
where $D_{klmn}$ are the contributions from the tree and the one-loop
diagrams and $C_{klmn}$ are from the real emission diagrams.
Let us first write down the $D_{klmn}$,
\bean
  D_{0000} &=& \beta [ f_{LL}^2 + f_{LR}^2 + 2 a f_{LL} f_{LR} ]
                    ( 1 + \hat{\alpha}_s V_{I} )\\
           & & \qquad\qquad\qquad - \beta [ 2 (f_{LL} + f_{LR})^2 - \beta^2
                (f_{LL} - f_{LR})^2 ] \hat{\alpha}_s V_{II} \ ,\\
  D_{2000} &=& \beta^3 (f_{LL}^2 + f_{LR}^2 )
                    ( 1 + \hat{\alpha}_s V_{I} ) + \beta^3
              (f_{LL} - f_{LR})^2 \hat{\alpha}_s V_{II} \ ,\\
  D_{1000} &=& 2 \beta^2 (f_{LL}^2 - f_{LR}^2 )
                    ( 1 + \hat{\alpha}_s V_{I} ) \ ,\\
  D_{0010} &=& \beta^2 (f_{LL}^2 - f_{LR}^2 )
                    ( 1 + \hat{\alpha}_s V_{I} ) \ ,\\
  D_{2010} &=& \beta^2 (f_{LL}^2 - f_{LR}^2 )
                    ( 1 + \hat{\alpha}_s V_{I} ) \ ,\\
  D_{1010} &=& \beta [ (f_{LL} + f_{LR} )^2 + 
              \beta^2 (f_{LL} - f_{LR} )^2  ]
                    ( 1 + \hat{\alpha}_s V_{I} )\\
           & & \qquad\qquad\qquad - 2 \beta [ (f_{LL} + f_{LR})^2 - \beta^2
                (f_{LL} - f_{LR})^2 ] \hat{\alpha}_s V_{II} \ ,\\
  D_{0101} &=& \frac{\beta}{\sqrt{a}} (f_{LL} + f_{LR} )^2 
                 [ a ( 1 + \hat{\alpha}_s V_{I} ) - ( 1 + a)
                       \hat{\alpha}_s V_{II} ] \ ,\\
  D_{1101} &=& \frac{\beta^2}{\sqrt{a}} (f_{LL}^2 - f_{LR}^2 ) 
                 [ a ( 1 + \hat{\alpha}_s V_{I} ) - ( 1 - a)
                       \hat{\alpha}_s V_{II}]\ ,
\eean
with
\[  \beta = \beta (0) = \sqrt{1-a} \quad , \quad
    \hat{\alpha}_s V_{I} = 2 A + 2 B_{R} \quad , \quad 
    \hat{\alpha}_s V_{II} = B_{R} \ .\]
$A_R\,,\,B_R$ are defined in Eqns.(\ref{afactor}) and (\ref{bfactor}).
 
For the $C_{klmn}$ we find,
\bean
  C_{0000} &=& 2\  (f_{LL}^2 + f_{LR}^2 )\  \left[ \ J_{\rm IR}^1 
                  - (4 - a) J_3 + (2 + a) J_2 + a J_1 
               + \frac{1}{4} R_1 \right]\\
           & & + \  4 a \,f_{LL} f_{LR} \ \left[ \ J_{\rm IR}^1 
               - 4 J_3 - J_2 - J_1 + \frac{1}{4} R_2 \right] \ ,\\
  C_{2000} &=& 2\  (f_{LL}^2 + f_{LR}^2 ) \ \left[ \ \ (1 - a) J_{\rm IR}^1
               - (4 - a) J_3 + (2 - a) J_2 - a J_1  - \frac{3}{4} R_1 \right] \\
           & & + \  4 a \,f_{LL} f_{LR} \ \left[ \  J_2 +  J_1
                   - \frac{3}{4} R_2 \right] \ ,\\
  C_{1000} &=& 2\ (f_{LL}^2 - f_{LR}^2 ) \ \Bigl[ \ (1 - a) J_{\rm IR}^2
             + \ a N_{10} - 2 (4 - 3a) N_9 \\
           & & \qquad\qquad\qquad\qquad\quad  - \, (4 - 5a) N_8
                     - 2 N_7  + 2 N_6 + 6 N_3 + 2 N_2 \Bigr] \ ,\\
  C_{0010} &=& \frac{1}{2} \ C_{1000} \\ 
           & & + \  (f_{LL}^2 - f_{LR}^2 ) \ \left[ \  - 4 N_6
               - a N_4 - a N_3 - a N_2 + (4 - a) N_1 + 
                  \frac{1}{2} R_3 \right] \ ,\\
  C_{2010} &=& C_{0010} - 2 (f_{LL}^2 - f_{LR}^2 ) R_3 \ ,\\
  C_{1010} &=& (f_{LL}^2 + f_{LR}^2 ) \ \Bigl[ \ 2 (2 - a) J_{\rm IR}^1
               + 2 a J_4 - 2 (8 - 5a) J_3\\
           & & \qquad\quad - \ 2 a (1 - a) L_8
                - 2a (1 - a) L_7 -  2 (4 + 3a) L_6 + 2 (4 - 3a) L_5\\ 
           & & \qquad\quad + \ 2a L_4  + a (10 - a) L_3
                 + (8 - 6a - a^2 ) L_2 - 2 (4 - 3a + a^2 ) L_1  \Bigr]\\
           & & + \  2 a \,f_{LL} f_{LR} \ \Bigl[ \ 2 J_{\rm IR}^1 - 8 J_3 
             + 4 L_6 + 4 L_5 + a L_3 +  a L_2 - 2 (2 - a) L_1 \Bigr] \ ,\\
  C_{0101} &=& \frac{\sqrt{a}}{2} \ (f_{LL}^2 + f_{LR}^2 ) \ \Bigl[ \
            4 J_{\rm IR}^1 - (8 + a) J_3  - (8 - a)(1 - a) L_7 - (12 + a) L_6 \\
           & & \qquad -\  2a L_5 + 2a L_4 + (8 + a) L_3 + (4 - 5a) L_2 
                  + 4 (2 - a) L_1  \Bigr]\\
           & & + \  \sqrt{a} \ f_{LL} f_{LR} \ \Bigl[ \ 4 J_{\rm IR}^1 
                     - (16 - a) J_3 - a (1 - a) L_7 + (4 + a) L_6\\
           & & \qquad\qquad\qquad + \ 2a L_5
                      + a L_3  - (4 - 3a) L_2 + 2a L_1 \Bigr]\ ,\\
  C_{1101} &=& \frac{\sqrt{a}}{2} \ C_{1000} + \sqrt{a} \ (f_{LL}^2 - f_{LR}^2 ) \\ 
           & & \times \ \Biggl[ \ 
              - a N_{10} - 9 a N_9 + 2 N_7 - 2 N_6 - \frac{1}{2} (12 + a) N_3
              + \frac{a}{2} N_2 - (12 + a) N_1\\ 
           & & \quad\quad - \ a (1 + a ) K_8 + a (1 - a) K_7
                   + 9 a (1 - a) K_6 - 3 a (5 + a) K_5\\
           & & \quad\quad - \  (12 + a) K_4 +
               (4 - 5a) K_3 - 3 (4 + 5a) K_2
                          + 3 (4 + a) (1 - a) K_1 \Biggr]\ ,
\eean
where we have defined
\bean
      J_{\rm IR}^1 &=& (2 - a) J_6 - a J_5 \ ,\\
      J_{\rm IR}^2 &=& 2 (2 - a) N_{13} - a N_{12} - a N_{11} \ ,\\
      R_1 &=&  - 4 a (1 - a) L_7 - 4 (2 - a) L_6 - 8 (1 - a) L_5\\
          & & + \ a^2 L_4 +  a (2 + 3a) L_3
        - a (2 - a) L_2 + (8 - 8a + 3a^2 ) L_1 \ ,\\
      R_2 &=& - 4 L_6 - 4 L_5 - a L_3  - a L_2 + 2(2 - a) L_1 \ ,\\ 
      R_3 &=&  a^2 N_{10} + 3a (2 + a) N_9 
                + 4 a N_3 - 2 a N_2 + 8 (1 + a ) N_1 \\
          & & + \ 2 a^2 K_8 - a^2 (1 - a) K_7 
                   - 3 a (1 - a)(2 + a) K_6 + 6 a (1 + 2a) K_5\\
          & & + \ 2 (4 + 3a) K_4 + (a^2 + 6a - 8) K_3
                   + 3 a (8 + a) K_2 - 12 a (1 - a) K_1 \ . 
\eean
Note that the integrals $J_{\rm IR}^1\,,\,J_{\rm IR}^2$ 
namely $J_5\,,\,J_6\,,\,N_{11}\,,\,N_{12}$ and $N_{13}$
contain the infrared singularity. This singularity is exactly canceled
out in the sum Eqn.(\ref{txsection}) by the contributions
from $D_{klmn}$.%, Eqn.(\ref{t1loopxsection}).
The numerical value of the coefficients $C_{klmn}$ and $D_{klmn}$ are
show in Appendix.
We are now ready to discuss our numerical results.

\newpage

\section{Numerical Results}

We are now in the position to give the cross-section for polarized top
quark production for any $e^+ e^-$ collider.
Since we did not specify the spin angle $\xi$ for the top quark, we can
predict the polarized cross-section for any top quark spin.
The spin configurations we will display are the helicity, the beamline
and the off-diagonal bases for center of mass energies
$\sqrt{s} = 400 \, {\rm GeV}$, $500 \, {\rm GeV}$, $800 \,{\rm GeV}$, 
$1000 \, {\rm GeV}$, $1500 \, {\rm GeV}$ and
$\,4000 \, {\rm GeV}$.
\begin{table}[H]
\begin{center}
\begin{tabular}{|c||c|c|c|c|c|c|}
\hline 
$\sqrt{s}$   & $400$ GeV & $500$ GeV & $800$ GeV 
           & $1000$ GeV & $1500$ GeV & $4000$ GeV \\ \hline
\hline
$\beta$    & $0.4841$  & $0.7141$  & $0.8992$  & $0.9368$ 
           & $0.9723$  & $0.9962$ \\ \hline
$\alpha_{s}$   
           & $0.09804$ & $0.09564$ & $0.09096$ & $0.08890$ 
           & $0.08539$ & $0.07794$ \\ \hline
$\sigma_{T,L}^{0}$ [{\rm pb}] 
           & $0.8707$ & $0.7728$ & $0.3531$  
           & $0.2312$ & $0.1047$ & $0.0149$ \\ \hline
$\sigma_{T,L}^{SGA}$ [{\rm pb}] 
           & $1.105$  & $0.8578$ & $0.3643$  
           & $0.2358$ & $0.1058$ & $0.0150$ \\ \hline
$\sigma_{T,L}^{1}$ [{\rm pb}] 
           & $1.113$  & $0.8719$ & $0.3734$  
           & $0.2418$ & $0.1084$ & $0.0153$ \\ \hline
$\sigma^{0}_{T,R}$ [{\rm Pb}]
           & $0.3827$ & $0.3562$ & $0.1710$ 
           & $0.1132$ & $0.0519$ & $0.0074$ \\ \hline
$\sigma^{SGA}_{T,R}$ [{\rm pb}]
           & $0.4865$ & $0.3964$ & $0.1768$ 
           & $0.1157$ & $0.0525$ & $0.0075$ \\ \hline
$\sigma^{1}_{T,R}$ [{\rm pb}]
           & $0.4898$ & $0.4031$ & $0.1813$ 
           & $0.1187$ & $0.0537$ & $0.0076$ \\ \hline
%$\kappa_{L}^{SGA} $
%           & $0.2691$ & $0.1283$ & $0.0575$ 
%           & $0.0457$ & $0.0347$ & $0.0259$ \\ \hline
%$\kappa_{R}^{SGA} $
%           & $0.2711$ & $0.1316$ & $0.0602$ 
%           & $0.0461$ & $0.0360$ & $0.0262$ \\ \hline
$\kappa_{L} $
           & $0.2778$ & $0.1283$ & $0.0575$ 
           & $0.0457$ & $0.0347$ & $0.0259$ \\ \hline
$\kappa_{R} $
           & $0.2799$ & $0.1316$ & $0.0602$ 
           & $0.0461$ & $0.0360$ & $0.0262$ \\ \hline
\end{tabular}
\caption[The values of $\beta$, $\alpha_s$, tree level and next to leading order
cross-sections and $\kappa_{L/R}$ for $e^{-}_{L/R} e^{+}$
 scattering.]{The values of $\beta$, $\alpha_s$, tree level
 cross-section, one-loop cross-section in soft-gluon
 approximation and full one-loop cross-section and
 $\kappa_{L/R}$ for $e^{-}_{L/R} e^{+}$ scattering.
 The superscript $0,1$ and SGA stand for tree level, full one-loop,
 soft-gluon approximation with $\omega_{\rm max}=(\sqrt{s}-2m)/5=~10$ GeV.}   
\label{tbl:tbl1}
\end{center}
\end{table}
\noindent
Table \ref{tbl:tbl1} contains the values of the maximum center of mass
speed $\beta$, the running $\alpha_s$,
and the tree level cross-section $\sigma^0_{T,L/R}$, 
the one-loop total cross-section  $\sigma^{SGA}_{T,L/R}$,
the full one-loop total cross-section $\sigma^1_{T,L/R}$
for $e^-_{L/R} e^+$ scattering.
$\kappa_{L/R}$ stands for the fractional $\cal{O}\mit(\alpha_{s})$
enhancement of the tree level cross-section for top quark pair 
production in $e^{-}_{L/R} e^{+}$ scattering. 
\[
 \kappa_{L/R} ~=~ 
 \frac{\sigma^1_{T,L/R}-\sigma^0_{T,L/R}}
      {\sigma^0_{T,L/R}}~.
\]
At $\sqrt{s} = 400$ GeV, the QCD corrections enhance the total
cross-section by $\sim 30 \%$ compared to the tree level results.
At higher energies, $800 \sim 4000$ GeV, the enhancements are at the $2
\sim 6 \%$ level.
  
%The energy dependence of the enhancement $\kappa_{L/R}$ is stronger
%than the dependence of running coupling constant $\alpha_s$.
%Therefore the energy dependence of enhancement is dominated by
%the coefficients of $\alpha_s$.
%\[
%\frac{(D_{klmn} + \hat{\alpha}_{s} C_{klmn})}{\hat{\alpha}_{s}}~,
%\qquad
%(klmn) ~\in~(0000,1000,2000)~.
%\]
 
Next we present the numerical values for the coefficients 
($D_{klmn} + \hat{\alpha}_{s} C_{klmn}$) in Eqn.(\ref{txsection}). 
Since the dominant effect of the $\cal{O}\mit(\alpha_s)$ corrections is
a multiplicative enhancement of the tree level result, we have
chosen to write
\[
D_{klmn}+\hat{\alpha}_{s} C_{klmn} 
~\equiv~ (1+\kappa_{L/R})D^0_{klmn} + S_{klmn}.
\]
The $(1+\kappa_{L/R})D^0_{klmn}$ terms are the multiplicative enhancement
of the tree level result whereas the $S_{klmn}$ give the $\alpha_{s}$
deviations to the spin correlations. 
The coefficients $D^0_{klmn}$ are given by the $D_{klmn}$ with limit,
$\alpha_s \to 0$.
The numerical value of these
coefficients are given in Table \ref{tbl:tbl2} and \ref{tbl:tbl3}. 
\begin{table}[H]
\begin{center}
\begin{tabular}{|c||c|c|c|c|c|c|}
\hline 
             & $400$  GeV & $500$  GeV & $800$ GeV  
             & $1000$ GeV & $1500$ GeV & $4000$GeV  \\ \hline
\hline
$(1 + \kappa_L)D^{0}_{0000}$ 
             & $ 1.511  $ & $ 1.724 $ & $ 1.722 $   
             & $ 1.700  $ & $ 1.671 $ & $ 1.644 $ \\ \hline
$(1 + \kappa_L)D^{0}_{2000}$ 
             & $ 0.2404 $ & $ 0.6724$ & $ 1.241 $   
             & $ 1.383  $ & $ 1.526 $ & $ 1.624 $ \\ \hline
$(1 + \kappa_L)D^{0}_{1000}$ 
             & $ 0.7809 $ & $ 1.466 $ & $ 2.126 $   
             & $ 2.269  $ & $ 2.406 $ & $ 2.494 $ \\ \hline
$(1 + \kappa_L)D^{0}_{0010}$ 
             & $ 0.3905 $ & $ 0.7330$ & $ 1.063 $   
             & $ 1.134  $ & $ 1.203 $ & $ 1.247 $ \\ \hline
$(1 + \kappa_L)D^{0}_{2010}$ 
             & $ 0.3905 $ & $ 0.7330$ & $ 1.063 $   
             & $ 1.134  $ & $ 1.203 $ & $ 1.247 $ \\ \hline
$(1 + \kappa_L)D^{0}_{1010}$ 
             & $ 1.751  $ & $ 2.396  $ & $ 2.963 $   
             & $ 3.083  $ & $ 3.197  $ & $ 3.268 $ \\ \hline
$(1 + \kappa_L)D^{0}_{0101}$ 
             & $ 1.452  $ & $ 1.502  $ & $ 1.100 $   
             & $ 0.9048 $ & $ 0.6186 $ & $ 0.2353$ \\ \hline
$(1 + \kappa_L)D^{0}_{1101}$ 
             & $ 0.3416 $ & $ 0.5131 $ & $ 0.4650$   
             & $ 0.3970 $ & $ 0.2807 $ & $ 0.1091$ \\ \hline
\hline
$S_{0000}$ & $ -0.002552  $ & $-0.005000 $ & $  0.0005645$ 
               & $  0.004836  $ & $  0.01172 $ & $  0.02039  $ \\ \hline
$S_{2000}$ & $  0.007655  $ & $ 0.01500  $ & $ -0.001682 $ 
               & $ -0.01450   $ & $ -0.03516 $ & $ -0.06116  $ \\ \hline
$S_{1000}$ & $  0.02154   $ & $ 0.03170  $ & $  0.01224  $ 
               & $ -0.0008102 $ & $ -0.02085 $ & $ -0.04516  $ \\ \hline
$S_{0010}$ & $  0.01094   $ & $ 0.01785  $ & $  0.01586  $ 
               & $ 0.01304    $ & $ 0.008357 $ & $  0.002274 $ \\ \hline
$S_{2010}$ & $  0.01059   $ & $ 0.01356  $ & $ -0.006158 $ 
               & $ -0.01791   $ & $ -0.03614 $ & $ -0.05987  $ \\ \hline
$S_{1010}$ & $  0.004433  $ & $ 0.005069 $ & $ -0.02030  $ 
               & $ -0.03590   $ & $ -0.06134 $ & $ -0.09742  $ \\ \hline
$S_{0101}$ & $ -0.006564  $ & $-0.02193  $ & $ -0.04413  $ 
               & $ -0.04852   $ & $ -0.04952 $ & $ -0.03639  $ \\ \hline
$S_{1101}$ & $  0.007920  $ & $ 0.004308 $ & $ -0.01270  $ 
               & $ -0.01734   $ & $ -0.02047 $ & $ -0.01642  $ \\ \hline
\end{tabular}
\caption{The values of $(1+\kappa_L)D^0_{klmn}$ and $S_{klmn}$ 
for $e^{-}_{L} e^{+}$ scattering.}
\label{tbl:tbl2}
\end{center}
\end{table}
\clearpage

\begin{table}[H]
\begin{center}
\begin{tabular}{|c||c|c|c|c|c|c|}
\hline 
$\sqrt{s}$ & $400$  GeV & $500$  GeV  
           & $800$  GeV & $1000$ GeV 
           & $1500$ GeV & $4000$ GeV \\ \hline \hline
$(1 + \kappa_R)D^{0}_{0000}$ &	$ 0.6613$ & $ 0.7901$ & $ 0.8313$ & $ 0.8307$ & $ 0.8266$ & $ 0.8204$ \\ \hline
$(1 + \kappa_R)D^{0}_{2000}$ &	$ 0.1174$ & $ 0.3318$ & $ 0.6180$ & $ 0.6899$ & $ 0.7622$ & $ 0.8112$ \\ \hline
$(1 + \kappa_R)D^{0}_{1000}$ &	$ 0.4408$ & $ 0.8357$ & $ 1.222 $ & $ 1.306 $ & $ 1.386 $ & $ 1.437 $ \\ \hline
$(1 + \kappa_R)D^{0}_{0010}$ &	$ 0.2204$ & $ 0.4178$ & $ 0.6109$ & $ 0.6529$ & $ 0.6931$ & $ 0.7187$ \\ \hline
$(1 + \kappa_R)D^{0}_{2010}$ &	$ 0.2204$ & $ 0.4178$ & $ 0.6109$ & $ 0.6529$ & $ 0.6931$ & $ 0.7187$ \\ \hline
$(1 + \kappa_R)D^{0}_{1010}$ &	$ 0.7787$ & $ 1.122 $ & $ 1.449 $ & $ 1.521 $ & $ 1.589 $ & $ 1.632 $ \\ \hline
$(1 + \kappa_R)D^{0}_{0101}$ &	$ 0.6216$ & $ 0.6547$ & $ 0.4875$ & $ 0.4024$ & $ 0.2759$ & $ 0.1052$ \\ \hline
$(1 + \kappa_R)D^{0}_{1101}$ &	$ 0.1929$ & $ 0.2925$ & $ 0.2673$ & $ 0.2285$ & $ 0.1617$ & $0.06289$ \\ \hline
\hline	
$S_{0000}$ &	$-0.001287$&$-0.002358$&$0.0006297$&$0.002742$&$0.006079$&$0.01024$ \\ \hline
$S_{2000}$ &	$ 0.003862$&$0.007074$&$-0.001884$&$-0.008222$&$-0.01824$&$-0.03072$ \\ \hline
$S_{1000}$ &	$0.01143 $&$ 0.01556 $&$ 0.003835$&$-0.003159$&$-0.01367$&$-0.02639$ \\ \hline
$S_{0010}$ &	$0.005814$&$ 0.008914$&$ 0.007500$&$ 0.006143$&$0.003972$&$0.001128$ \\ \hline
$S_{2010}$ &	$0.005612$&$0.006475 $&$-0.005121$&$-0.01164 $&$-0.02164$&$-0.03468$ \\ \hline
$S_{1010}$ &	$0.002248$&$0.002290 $&$-0.01078 $&$-0.01854 $&$-0.03107$&$-0.04878$ \\ \hline
$S_{0101}$ &	$-0.003764$&$-0.01122$&$-0.02028 $&$-0.02189 $&$-0.02208$&$-0.01640$ \\ \hline
$S_{1101}$ &	$0.004154 $&$0.001587$&$-0.007977$&$-0.01043 $&$-0.01198$&$-0.009478$ \\ \hline
\end{tabular}
\caption{The values of $(1+\kappa_R)D^0_{klmn}$ and $S_{klmn}$ 
for $e^{-}_{R} e^{+}$ scattering.}
\label{tbl:tbl3}
\end{center}
\end{table}
\noindent

\clearpage
\begin{table}[H]
\begin{center}
\begin{tabular}{|c||c|c|c|c|c|c|}
\hline 
$e^-_L e^+$           & $400$ GeV  & $500$ GeV  & $800$ GeV  
                      & $1000$ GeV & $1500$ GeV & $4000$ GeV \\ \hline
\hline
$R_{0000}$ & $ -0.001689  $  &  $-0.002900  $ & $ 0.0003278$ 
                  & $  0.002845  $  &  $  0.007014 $ & $ 0.01240  $ \\ \hline
$R_{2000}$ & $  0.03184   $  &  $ 0.02231   $ & $-0.001356 $ 
                  & $ -0.01048   $  &  $ -0.02304  $ & $-0.03767  $ \\ \hline
$R_{1000}$ & $  0.02758   $  &  $ 0.02163   $ & $ 0.005757 $ 
                  & $ -0.0003571 $  &  $ -0.008667 $ & $-0.01811  $ \\ \hline
$R_{0010}$ & $  0.02803   $  &  $ 0.02435   $ & $ 0.01492  $ 
                  & $  0.01150   $  &  $  0.006948 $ & $ 0.001823 $ \\ \hline
$R_{2010}$ & $  0.02711   $  &  $ 0.01849   $ & $-0.005793 $ 
                  & $ -0.01579   $  &  $ -0.03004  $ & $-0.04802  $ \\ \hline
$R_{1010}$ & $  0.0025310 $  &  $ 0.002115  $ & $-0.006849 $ 
                  & $ -0.01165   $  &  $ -0.01919  $ & $-0.02981  $ \\ \hline
$R_{0101}$ & $ -0.004520  $  &  $-0.01460   $ & $-0.04013  $ 
                  & $ -0.05362   $  &  $ -0.08005  $ & $-0.15465  $ \\ \hline
$R_{1101}$ & $  0.02318   $  &  $ 0.008395  $ & $-0.02731  $ 
                  & $ -0.04369   $  &  $ -0.07294  $ & $-0.15050  $ \\ 
\hline \hline 
$e^-_R e^+$           & $400$ GeV  & $500$ GeV  & $800$ GeV  
                      & $1000$ GeV & $1500$ GeV & $4000$ GeV \\ \hline
\hline
$R_{0000}$ & $ -0.001947$&$-0.002985$&$ 0.0007576$&$ 0.003301$&$ 0.007354$&$ 0.01248$ \\ \hline
$R_{2000}$ & $  0.03289 $&$ 0.02132 $&$-0.003048 $&$-0.01192 $&$-0.02393 $&$-0.03787$ \\ \hline
$R_{1000}$ & $  0.02593 $&$ 0.01862 $&$ 0.003139 $&$-0.002419$&$-0.009864$&$-0.01836$ \\ \hline
$R_{0010}$ & $  0.02638 $&$ 0.02133 $&$ 0.01228  $&$ 0.009409$&$ 0.005731$&$0.001570$ \\ \hline
$R_{2010}$ & $ 0.02546 $&$ 0.01550 $&$-0.008382 $ & $ -0.01782$ & $ -0.03122$&$ -0.04826$ \\ \hline
$R_{1010}$ & $0.002886 $&$ 0.002041$&$ -0.007438$&$ -0.01219 $&$ -0.01955$&$ -0.02989 $ \\ \hline
$R_{0101}$ & $-0.006056$&$ -0.01714$&$ -0.04161$&$ -0.05439$&$ -0.08003 $&$ -0.1559 $ \\ \hline
$R_{1101}$ & $ 0.02154$&$ 0.005427$&$ -0.02984$&$-0.04566$&$ -0.07406$&$ -0.1507$ \\ \hline
\end{tabular}
\caption{The ratios, $R_{klmn} = S_{klmn}/(1+\kappa)D^0_{klmn}$,
for $e^{-}_{L/R} e^{+}$ scattering.}
\label{tbl:tbl4}
\end{center}
\end{table}
\noindent
Table \ref{tbl:tbl4} shows the ratios, 
\[
R_{klmn} ~=~{S_{klmn}}/{(1+\kappa)D^0_{klmn}}. 
\]
The ratios are never larger than $10 \%$ and are typically of order a
few percent.
Hence the $\cal{O}\mit(\alpha_s)$ corrections make only small
changes to the spin orientation of the top quark. 

To illustrate the cross-sections in different spin bases, we present the top quark
production cross-section in the three different spin bases
discussed in the previous Section.
One is the usual helicity basis which corresponds to $\cos \xi = + 1$. 
The second is the beamline basis, in which the top quark spin is aligned
with the positron momentum in the top quark rest frame.
In this basis, $\xi$ is obtained by Eqn.(\ref{eqn:beam2-def}).
The third corresponds to the off-diagonal basis which has been defined
in Eqn.(\ref{offdiaxi}).  
Note that as $\beta \to 1$, the helicity and off-diagonal bases coincide.
Therefore, at an extremely high energy collider, there will be no
significant difference between these bases. 
But a remarkable differences among these bases exists at moderate
energies.
Now let us show the polarized differential cross-sections at
$\sqrt{s}=400$ and $500~{\rm GeV}$.
\clearpage
%----------------------------------------------------------------
\begin{figure}[H]
\begin{center}
        \leavevmode\psfig{file=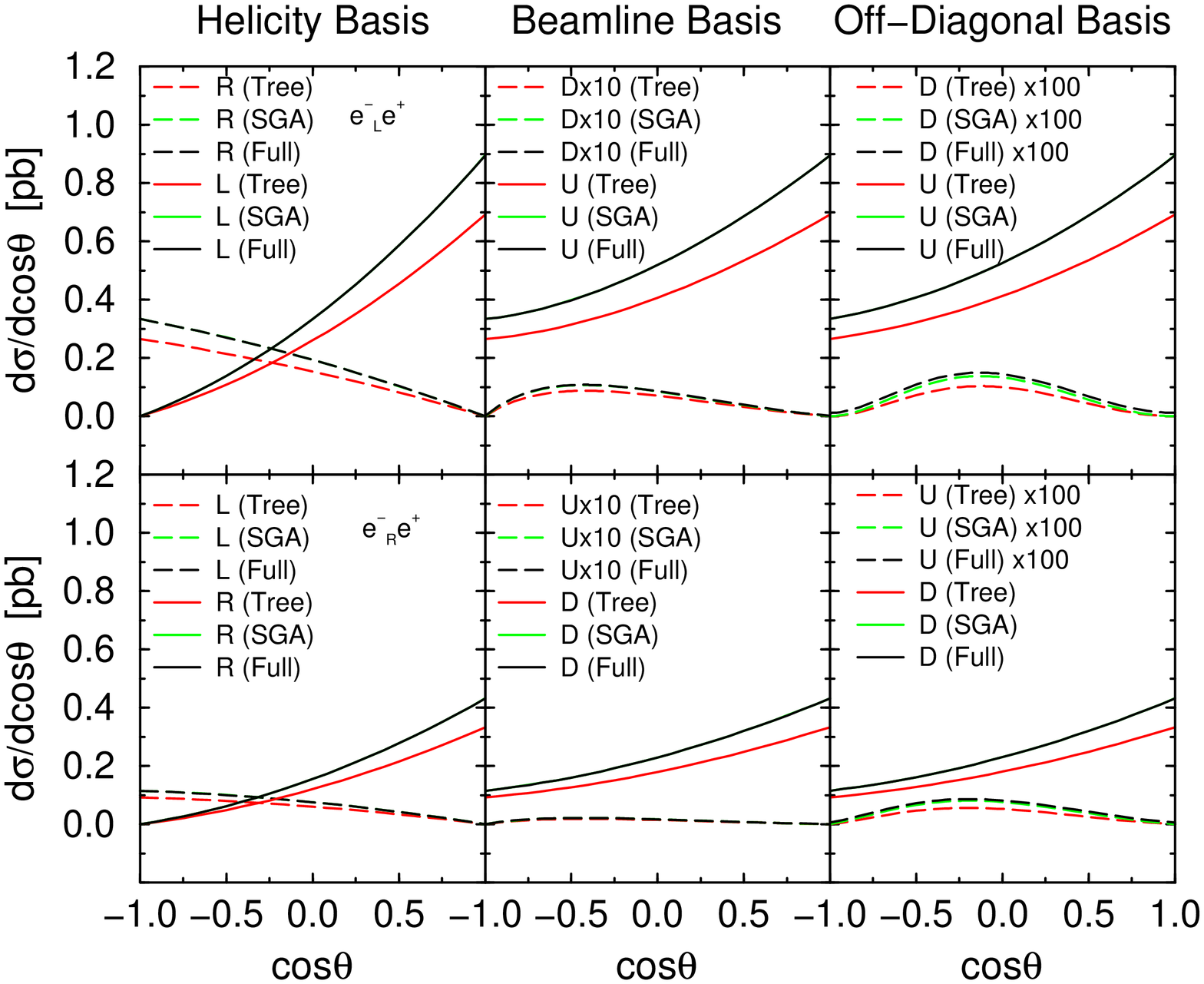,angle=-90,width=15cm}
\caption[The cross-sections in the helicity, beamline and off-diagonal bases
at $\sqrt{s} = 400 \, {\rm GeV}$.]{The cross-sections in the helicity,
 beamline and off-diagonal bases at $\sqrt{s} = 400 \, {\rm GeV}$.
Here we use a ``beamline basis'', in which the top quark axis is the
 positron direction in the top rest frame, for each $e^{-}_{L}e^{+}$ and 
$e^{-}_{R}e^{+}$ scattering.
For the soft gluon approximation (SGA) 
$\omega_{\rm max}=(\sqrt{s}-2m)/5=~10$ GeV.}
\label{fig:fig8-1}
\end{center} 
\end{figure}
\clearpage
%----------------------------------------------------------------
\begin{figure}[H]
\begin{center}
        \leavevmode\psfig{file=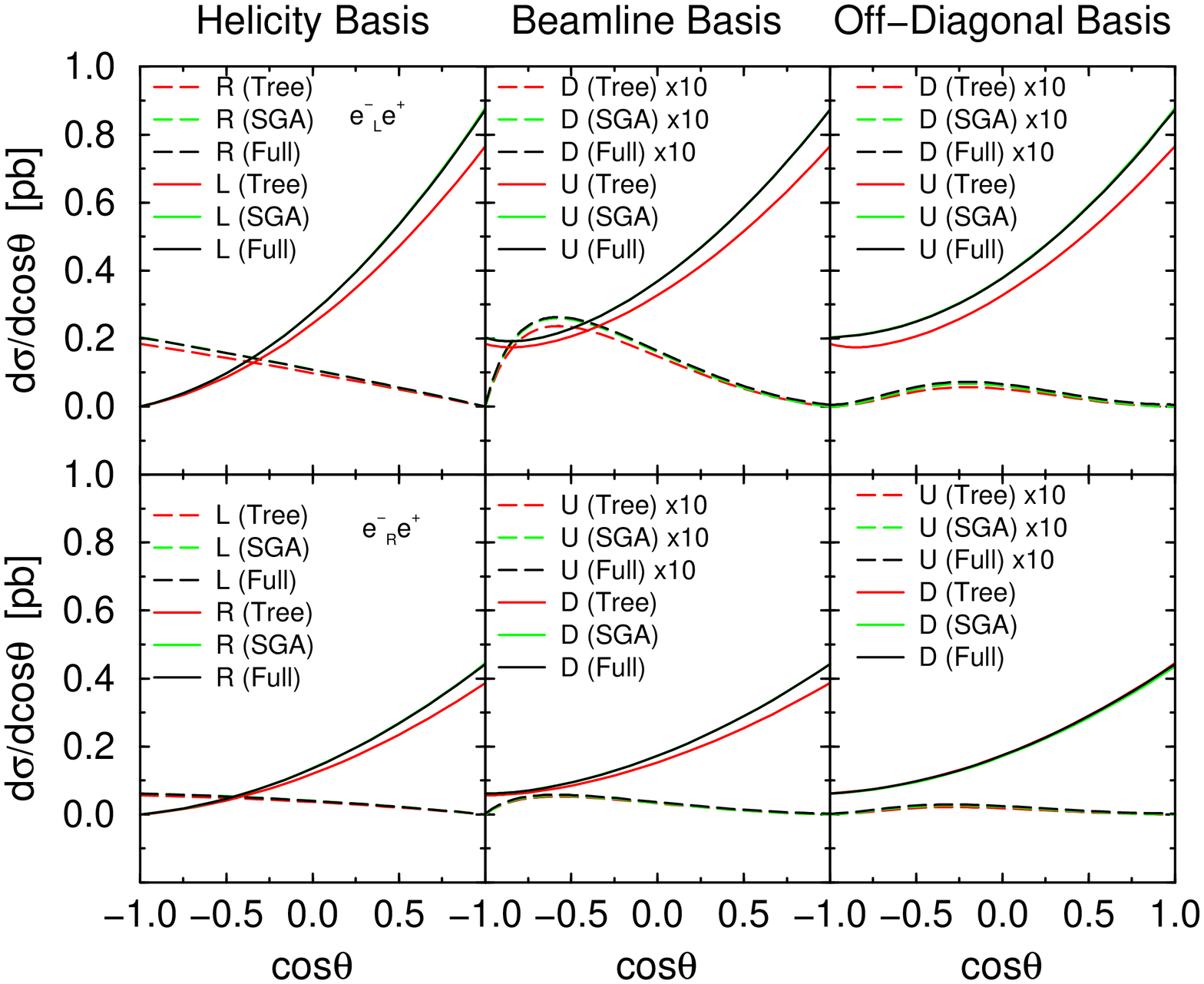,angle=-90,width=15cm}
\caption[The cross-sections in the helicity, beamline and off-diagonal
bases at $\sqrt{s} = 500 \, {\rm GeV}$.]{The cross-sections in the helicity, beamline and off-diagonal
bases at $\sqrt{s} = 500 \, {\rm GeV}$.
Here we use a ``beamline basis'', in which the top quark axis is the
positron direction in the top rest frame, for each $e^{-}_{L}e^{+}$ and 
$e^{-}_{R}e^{+}$ scattering.
For the soft gluon approximation (SGA) 
$\omega_{\rm max}=(\sqrt{s}-2m)/5=~30$ GeV.}
\label{fig:fig8-2}
\end{center} 
\end{figure}
\noindent
%%%%%%%%%%%%%%%%%%%%%%%%%%%%
In Fig.\ref{fig:fig8-1} and Fig.\ref{fig:fig8-2}, we give the results
for $\sqrt{s} = 400$ and $500$ GeV for both $e^-_L e^+$ and $e^-_R e^+$
scattering using the helicity, beamline and off-diagonal spin
bases. 
These figures show the tree level, the SGA and the full QCD results for all
three spin bases. 
Since the SGA results almost coincide with the full QCD results, 
the probability that hard gluon emission flips the spin of the top quark
is very small. 
Clearly, the qualitative features of
the cross-sections remain the same as those in the leading order
analysis.
That is the top quarks are produced with  very high
polarization in polarized $e^+e^-$ scattering.
In Table \ref{tbl:tbl5-1} and \ref{tbl:tbl5-2}, we give the fraction of
the top quarks in the sub-dominant spin configuration for $e^-_{L/R}
e^+$ scattering, 
\[
\sigma 
\left(
  e^{-}_{L} e^{+} \rightarrow t_{\uparrow} X(\bar{t} {\rm ~or~} \bar{t} g) 
\right) / \sigma^1_{T,L}
\]
for the three bases. 
Similar results also hold for $e^-_R e^+$ scattering.
%and their results are given by interchanging suffix $L,~R$ as well as
%$\uparrow,~\downarrow$. 

\begin{table}[H]
\begin{center}
\begin{tabular}{|c||c|c|c|}
\hline 
$e^-_L e^+$& Helicity & Beamline & Off-Diagonal \\ \hline
\hline
Tree       & $0.336$ & $0.0119 $ & $0.00124$ \\ \hline
SGA        & $0.332$ & $0.0113 $ & $0.00129$ \\ \hline
$\cal{O}\mit (\alpha_{s})$       
           & $0.332$ & $0.0115 $ & $0.00150$ \\ \hline
\hline 
$e^-_R e^+$& Helicity & Beamline & Off-Diagonal \\ \hline
\hline
Tree       & $0.290  $ & $0.00564 $ & $0.00167$ \\ \hline
SGA        & $0.285  $ & $0.00510 $ & $0.00191$ \\ \hline
$\cal{O}\mit (\alpha_{s})$       
           & $0.285  $ & $0.00530 $ & $0.00214$ \\ \hline
\end{tabular}
\caption[The fraction of the $e^-_{L/R} e^+$ cross-section in the
 sub-dominant spin at $\sqrt{s} = 400$ GeV for the helicity,
 beamline and off-diagonal bases.]{The fraction of the $e^-_{L/R} e^+$
 cross-section in the sub-dominant spin at $\sqrt{s} = 400$ GeV for the
 helicity, beamline and off-diagonal bases.
 For the soft gluon approximation (SGA) $\omega_{\rm max}=(\sqrt{s}-2m)/5=~10$ GeV.}
\label{tbl:tbl5-1}
\end{center}
\end{table}

%\clearpage

\begin{figure}[H]
\begin{center}
        \leavevmode\psfig{file=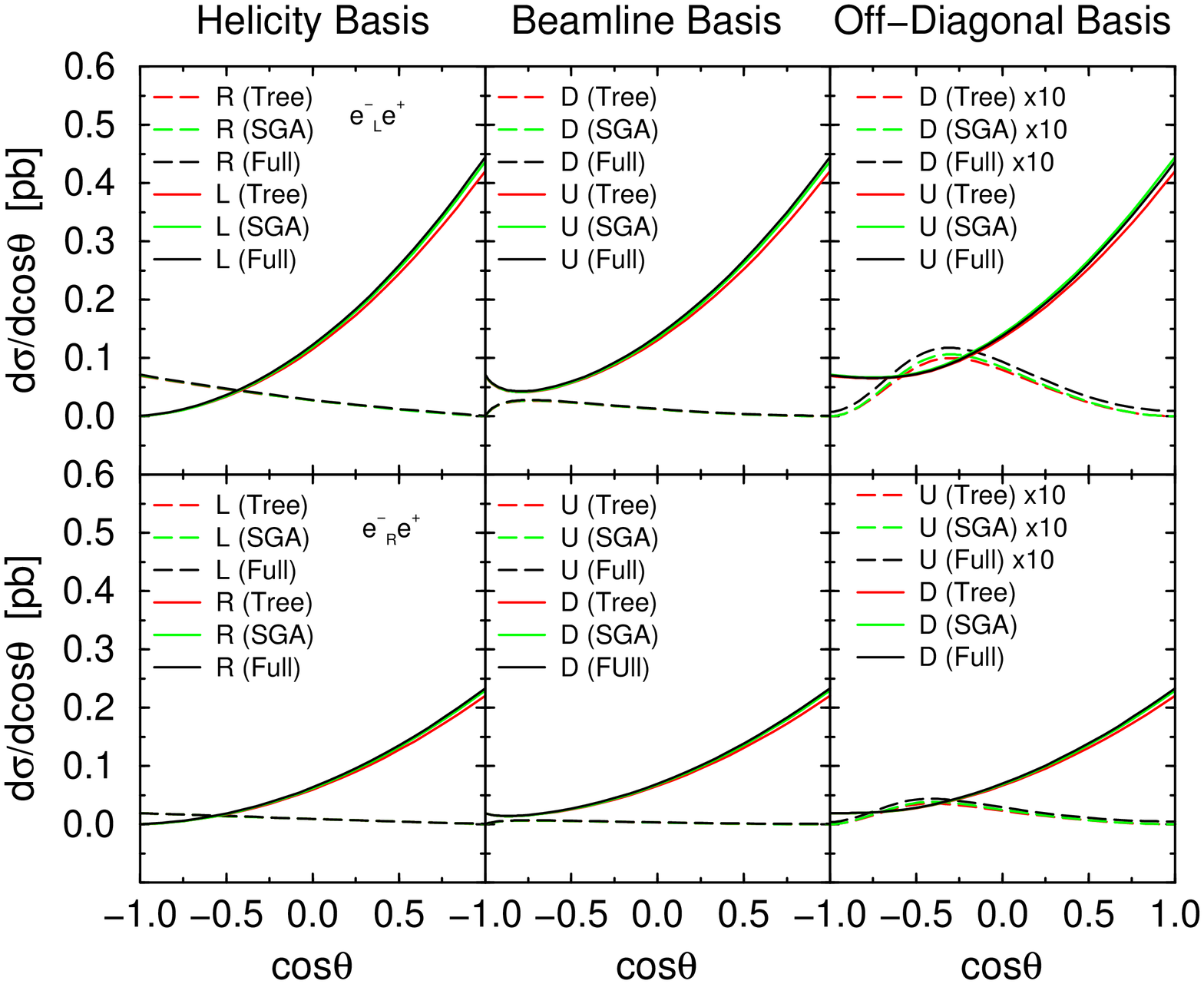,angle=-90,width=15cm}
\caption[The cross-sections in the helicity, beamline and off-diagonal bases
at $\sqrt{s} = 800  \, {\rm GeV}$.]{The cross-sections in the helicity,
 beamline and off-diagonal bases at $\sqrt{s} = 800  \, {\rm GeV}$.
Here we use a ``beamline basis'', in which the top quark axis is the
positron direction in the top rest frame, for each $e^{-}_{L}e^{+}$ and 
$e^{-}_{R}e^{+}$ scattering.
For the soft gluon approximation (SGA) 
$\omega_{\rm max}=(\sqrt{s}-2m)/5=~90$ GeV.}
\label{fig:fig9-1}
\end{center} 
\end{figure}

\begin{table}[H]
\begin{center}
\begin{tabular}{|c||c|c|c|}
\hline 
$e^-_L e^+$ & Helicity & Beamline & Off-Diagonal \\ \hline
\hline
Tree       & $0.249$ & $0.0323 $ & $0.00745 $ \\ \hline
SGA        & $0.243$ & $0.0321 $ & $0.00775 $ \\ \hline
$\cal{O}\mit (\alpha_{s})$       
           & $0.243$ & $0.0311 $ & $0.00907 $ \\ \hline
\hline 
$e^-_R e^+$ & Helicity & Beamline & Off-Diagonal \\ \hline
\hline
Tree       & $0.191$ & $0.0157$ & $0.00629$ \\ \hline
SGA        & $0.185$ & $0.0145$ & $0.00682$ \\ \hline
$\cal{O}\mit (\alpha_{s})$       
           & $0.185$ & $0.0156$ & $0.00816$ \\ \hline
\end{tabular}
\caption[The fraction of the $e^-_{L/R} e^+$ cross-section in the
sub-dominant spin at $\sqrt{s} = 500$ GeV for the helicity, beamline and
 off-diagonal bases.]{The fraction of the $e^-_{L/R} e^+$ cross-section in the
sub-dominant spin at $\sqrt{s} = 500$ GeV for the helicity, beamline and
 off-diagonal bases.
For the soft gluon approximation (SGA) 
$\omega_{\rm max}=(\sqrt{s}-2m)/5=~30$ GeV.}
\label{tbl:tbl5-2}
\end{center}
\end{table}
%
%\newpage
%%%%%%%%%%%%%%%%%%%%%%%%%%%%
\noindent
In Fig.\ref{fig:fig9-1} $\sim$ \ref{fig:fig10-2}, we have plotted the
similar results for $800$,$1000$,$1500$ and $4000$ GeV colliders.
The fraction of top quarks in the sub-dominant spin component for 
$e^-_L e^+$ is given in Tables \ref{tbl:tbl6-1} and \ref{tbl:tbl7-2}.
Our numerical studies demonstrate that the QCD corrections have a small
effect on the spin configuration of the produced top (or anti-top) quark
for any spin basis.
The off-diagonal and beamline bases are clearly more sensitive to
the radiative corrections than the helicity basis. 
However, in the off-diagonal basis,
the top (and/or anti-top) quarks are produced in an essentially unique
spin configuration even after including the lowest order QCD
corrections.

\clearpage
%----------------------------------------------------------------
%----------------------------------------------------------------
\begin{figure}[H]
\begin{center}
        \leavevmode\psfig{file=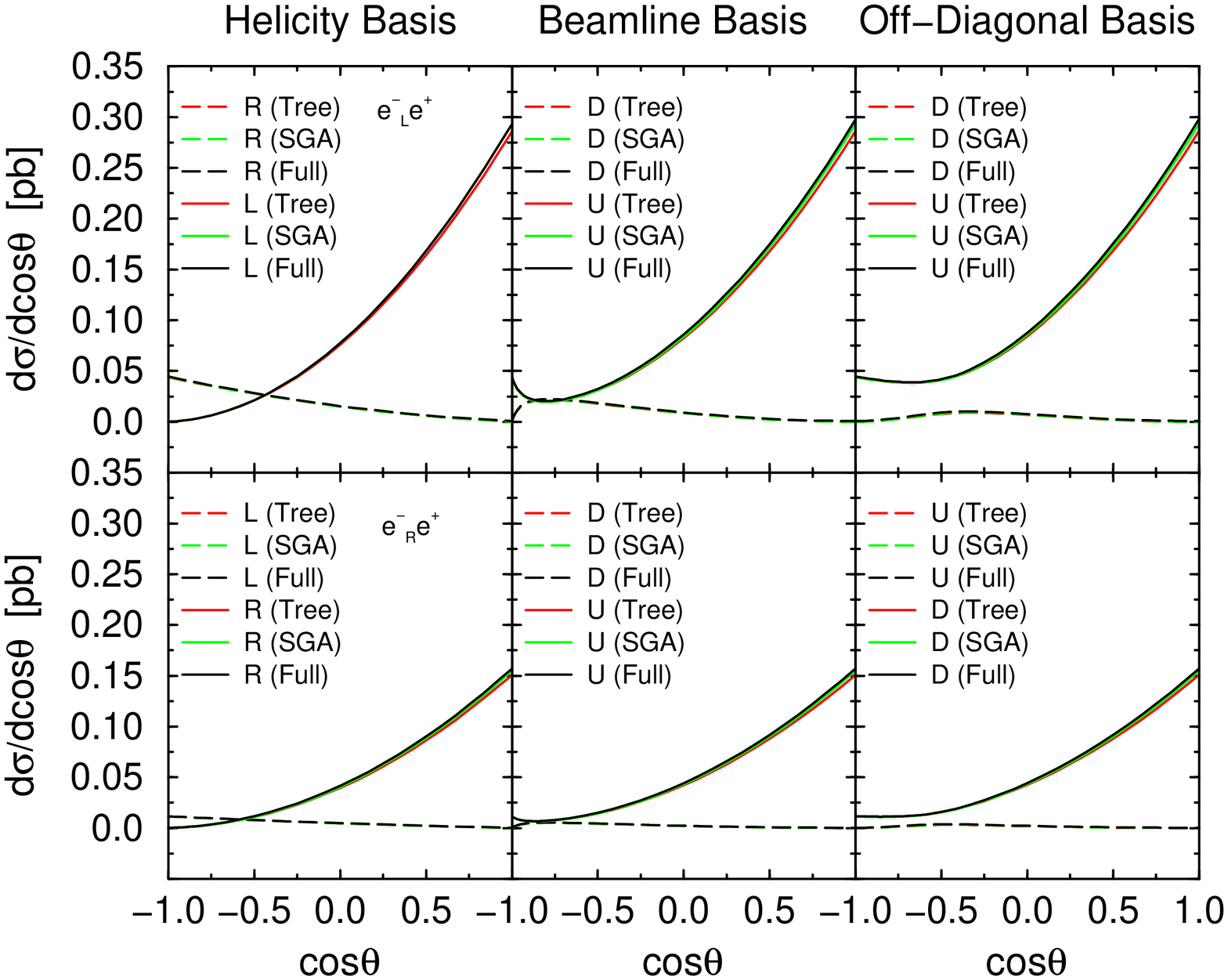,angle=-90,width=15cm}
\caption[The cross-sections in the helicity, beamline and off-diagonal bases
at $\sqrt{s} = 1000  \, {\rm GeV}$.]{The cross-sections in the helicity,
 beamline and off-diagonal bases at $\sqrt{s} = 1000 \, {\rm GeV}$.
Here we use a ``beamline basis'', in which the top quark axis is the
positron direction in the top rest frame, for each $e^{-}_{L}e^{+}$ and 
$e^{-}_{R}e^{+}$ scattering.
For the soft gluon approximation (SGA) 
$\omega_{\rm max}=(\sqrt{s}-2m)/5=~130$ GeV.}
\label{fig:fig9-2}
\end{center} 
\end{figure}
%----------------------------------------------------------------
\begin{figure}[H]
\begin{center}
        \leavevmode\psfig{file=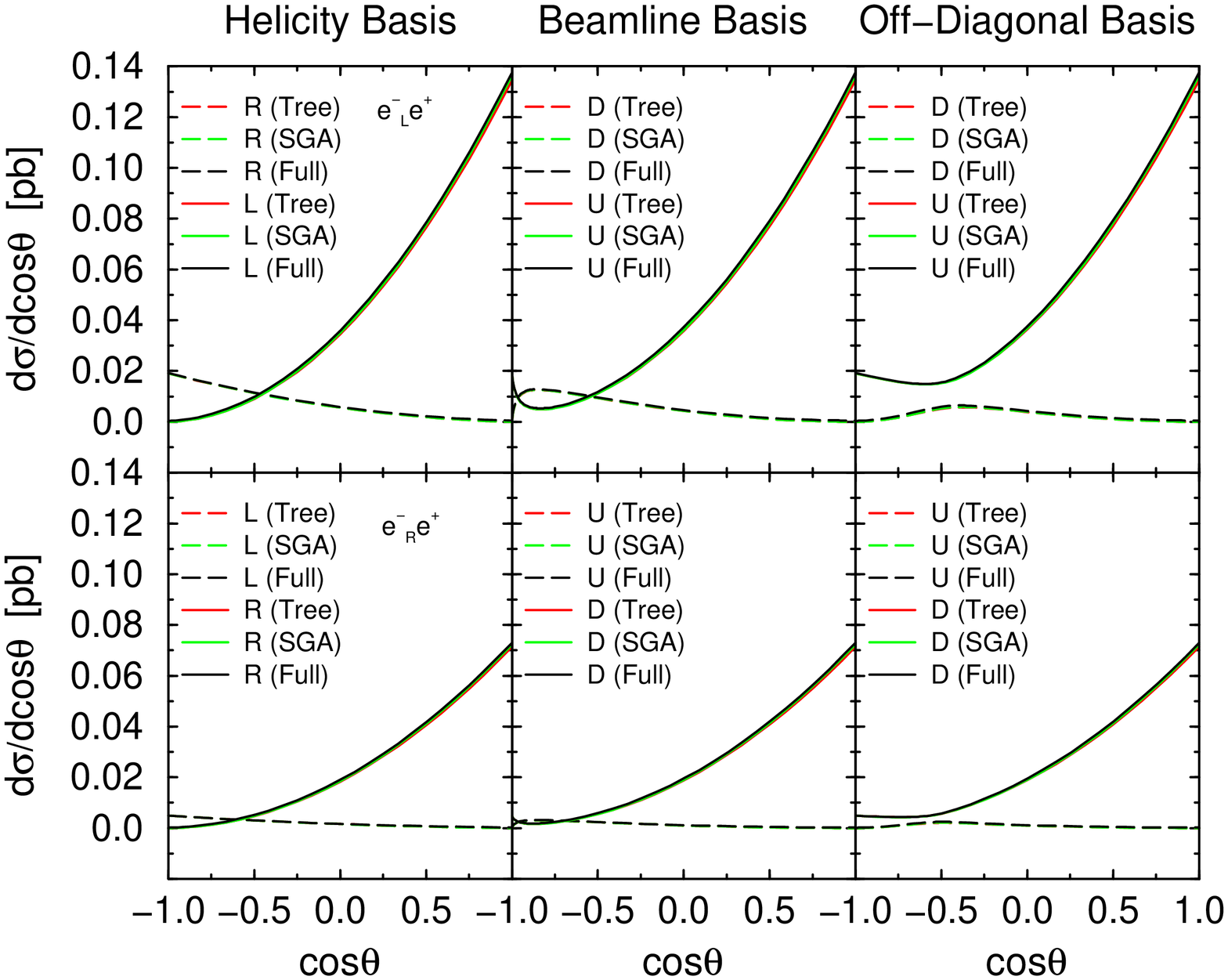,angle=-90,width=15cm}
\caption[The cross-sections in the helicity, beamline and off-diagonal bases
at $\sqrt{s} = 1500 \, {\rm GeV}$.]{The cross-sections in the helicity, beamline and off-diagonal bases
at $\sqrt{s} = 1500 \, {\rm GeV}$.
Here we use a ``beamline basis'', in which the top quark axis is the
positron direction in the top rest frame, for each $e^{-}_{L}e^{+}$ and 
$e^{-}_{R}e^{+}$ scattering.
For the soft gluon approximation (SGA) 
$\omega_{\rm max}=(\sqrt{s}-2m)/5=~230$ GeV.}
\label{fig:fig10-1}
\end{center} 
\end{figure}
%----------------------------------------------------------------
\begin{figure}[H]
\begin{center}
        \leavevmode\psfig{file=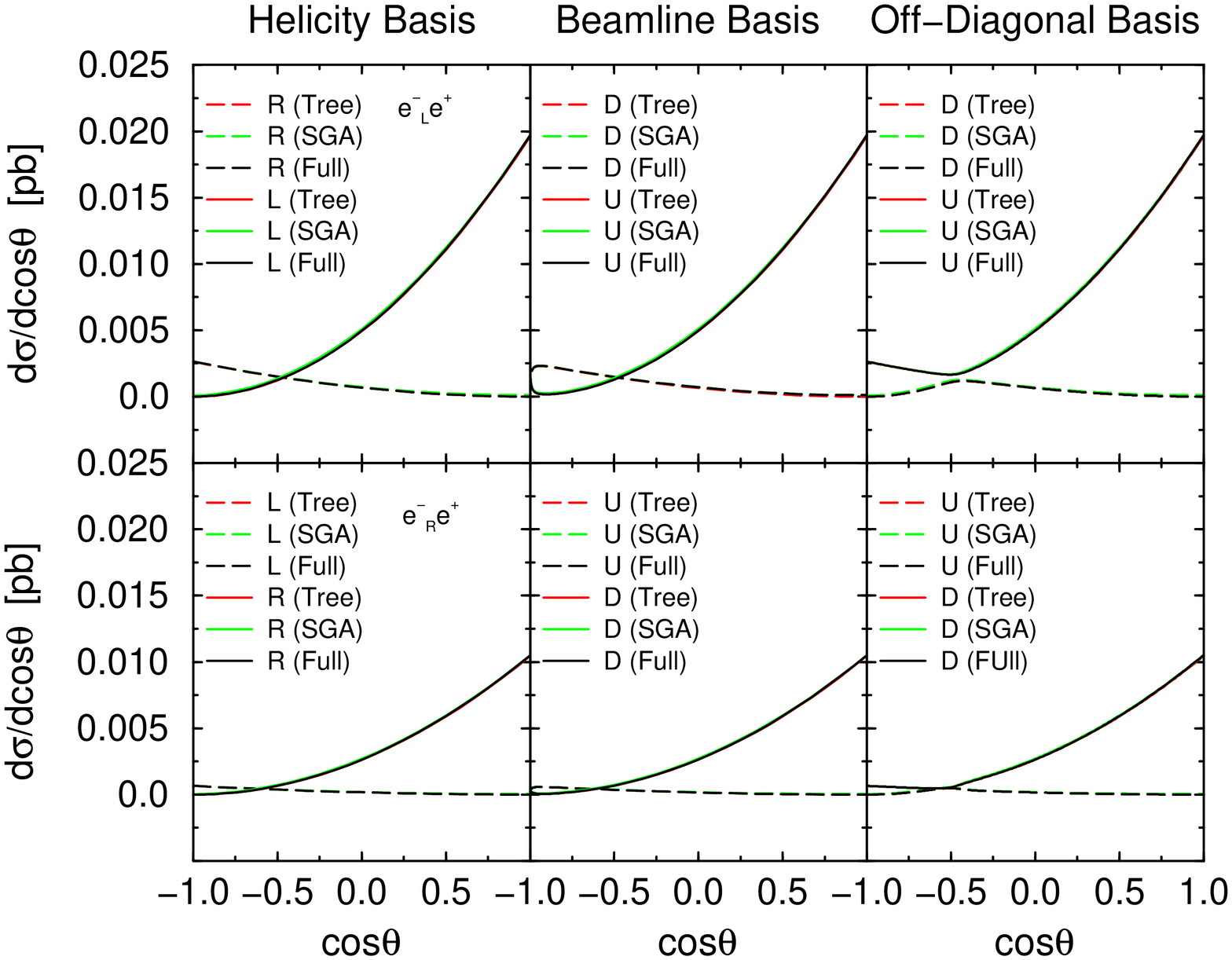,angle=-90,width=15cm}
\caption[The cross-sections in the helicity, beamline and off-diagonal bases
at $\sqrt{s} = 4000 \,  {\rm GeV}$.]{The cross-sections in the helicity,
 beamline and off-diagonal bases at $\sqrt{s} = 4000 \,  {\rm GeV}$.
Here we use a ``beamline basis'', in which the top quark axis is the
positron direction in the top rest frame, for each $e^{-}_{L}e^{+}$ and 
$e^{-}_{R}e^{+}$ scattering.
For the soft gluon approximation (SGA) 
$\omega_{\rm max}=(\sqrt{s}-2m)/5=~730$ GeV.}
\label{fig:fig10-2}
\end{center} 
\end{figure}
%%%%%%%%%%%%%%%%%%%%%%%%%%%%
\begin{table}[H]
\begin{center}
\begin{tabular}{|c||c|c|c|}
\hline 
$e^-_L e^+$& Helicity & Beamline & Off-Diagonal \\ \hline
\hline
Tree       & $0.168  $ & $0.0690 $ & $0.0265 $ \\ \hline
SGA        & $0.164  $ & $0.0679 $ & $0.0272 $ \\ \hline
$\cal{O}\mit (\alpha_{s})$       
           & $0.165  $ & $0.0708 $ & $0.0319 $ \\ \hline
\hline 
$e^-_R e^+$ & Helicity & Beamline & Off-Diagonal \\ \hline
\hline
Tree       & $0.107$ & $0.0342$ & $0.0186$ \\ \hline
SGA        & $0.103$ & $0.0331$ & $0.0194$ \\ \hline
$\cal{O}\mit (\alpha_{s})$       
           & $0.105$ & $0.0360$ & $0.0239$ \\ \hline
\end{tabular}
\caption[The fraction of the $e^-_{L/R} e^+$ cross-section in the
sub-dominant spin at $\sqrt{s} = 800$ GeV for the helicity, beamline and
 off-diagonal bases.]{The fraction of the $e^-_{L/R} e^+$ cross-section in the
sub-dominant spin at $\sqrt{s} = 800$ GeV for the helicity, beamline and
 off-diagonal bases.
For the soft gluon approximation (SGA) 
$\omega_{\rm max}=(\sqrt{s}-2m)/5=~90$ GeV.}
\label{tbl:tbl6-1}
\end{center}
\end{table}
%----------------------------------------------------------------
\begin{table}[H]
\begin{center}
\begin{tabular}{|c||c|c|c|}
\hline 
$e^-_L e^+$& Helicity & Beamline & Off-Diagonal \\ \hline
\hline
Tree       & $0.150$ & $0.0818$ & $0.0349$ \\ \hline
SGA        & $0.147$ & $0.0809$ & $0.0357$ \\ \hline
$\cal{O}\mit (\alpha_{s})$       
           & $0.148$ & $0.0844$ & $0.0417$ \\ \hline
\hline 
$e^-_R e^+$& Helicity & Beamline & Off-Diagonal \\ \hline
\hline
Tree       & $0.0896$ & $0.0407$ & $0.0238$ \\ \hline
SGA        & $0.0867$ & $0.0398$ & $0.0246$ \\ \hline
$\cal{O}\mit (\alpha_{s})$       
           & $0.0886$ & $0.0433$ & $0.0304$ \\ \hline
\end{tabular}
\caption[The fraction of the $e^-_{L/R} e^+$ cross-section in the
sub-dominant spin at $\sqrt{s} = 1000$ GeV for the helicity, beamline
 and off-diagonal bases.]{The fraction of the $e^-_{L/R} e^+$ cross-section in the
sub-dominant spin at $\sqrt{s} = 1000$ GeV for the helicity, beamline
 and off-diagonal bases.
For the soft gluon approximation (SGA) 
$\omega_{\rm max}=(\sqrt{s}-2m)/5=~130$ GeV.}
\label{tbl:tbl6-2}
\end{center}
\end{table}
\clearpage
\begin{table}[H]
\begin{center}
\begin{tabular}{|c||c|c|c|}
\hline 
$e^-_L e^+$ & Helicity & Beamline & Off-Diagonal \\ \hline
\hline
Tree       & $0.132  $ & $0.0978 $ & $0.0466 $ \\ \hline
SGA        & $0.130  $ & $0.0973 $ & $0.0472 $ \\ \hline
$\cal{O}\mit (\alpha_{s})$       
           & $0.133  $ & $0.101  $ & $0.0552 $ \\ \hline
\hline 
$e^-_R e^+$ & Helicity & Beamline & Off-Diagonal \\ \hline
\hline
Tree       & $0.0724$ & $0.0488$ & $0.0310$ \\ \hline
SGA        & $0.0709$ & $0.0483$ & $0.0316$ \\ \hline
$\cal{O}\mit (\alpha_{s})$       
           & $0.0739$ & $0.0527$ & $0.0394$ \\ \hline
\end{tabular}
\caption[The fraction of the $e^-_{L/R} e^+$ cross-section in the
sub-dominant spin at $\sqrt{s} = 1500$ GeV for the helicity, beamline
 and off-diagonal bases.]{The fraction of the $e^-_{L/R} e^+$ cross
 section in the sub-dominant spin at $\sqrt{s} = 1500$ GeV for the
 helicity, beamline and off-diagonal bases.
For the soft gluon approximation (SGA) 
$\omega_{\rm max}=(\sqrt{s}-2m)/5=~230$ GeV.}
\label{tbl:tbl7-1}
\end{center}
\end{table}
%-------------------------------------------------------------------
\begin{table}[H]
\begin{center}
\begin{tabular}{|c||c|c|c|}
\hline 
$e^-_L e^+$ & Helicity & Beamline & Off-Diagonal \\ \hline
\hline
Tree       & $0.120$ & $0.113$ & $0.0594$ \\ \hline
SGA        & $0.119$ & $0.113$ & $0.0596$ \\ \hline
$\cal{O}\mit (\alpha_{s})$       
           & $0.124$ & $0.118$ & $0.0698$ \\ \hline
\hline 
$e^-_R e^+$ & Helicity & Beamline & Off-Diagonal \\ \hline
\hline
Tree       & $0.0607$ & $0.0568$ & $0.0388$ \\ \hline
SGA        & $0.0604$ & $0.0566$ & $0.0391$ \\ \hline
$\cal{O}\mit (\alpha_{s})$       
           & $0.0655$ & $0.0622$ & $0.0492$ \\ \hline
\end{tabular}
\caption[The fraction of the $e^-_{L/R} e^+$ cross-section in the
sub-dominant spin at $\sqrt{s} = 4000$ GeV for the helicity, beamline
 and off-diagonal bases.]{The fraction of the $e^-_{L/R} e^+$ cross-section in the
sub-dominant spin at $\sqrt{s} = 4000$ GeV for the helicity, beamline
 and off-diagonal bases.
For the soft gluon approximation (SGA) 
$\omega_{\rm max}=(\sqrt{s}-2m)/5=~730$ GeV.}
\label{tbl:tbl7-2}
\end{center}
\end{table}
%-------------------------------------------------------------------
\clearpage
\section{$f_{LR}~=~f_{RL}=~0$ Limit}
In the previous chapter, we investigated
why the polarized cross-sections for the beamline and off-diagonal bases give
similar behaviors. 
At the leading order, the polarized cross-sections under the approximation
$f_{LR}=f_{RL}=0$ in the beamline and off-diagonal bases coincide, 
the top spin-down configuration for the $e^-_L e^+$ scattering process
is identically zero, alternatively, the top spin-up configuration for the $e^-_R 
e^+$ scattering process is identically zero.
The smallness of the parameters $f_{LR}\sim f_{RL}\sim 0$ is a key point 
to understand the relation between the beamline and the off-diagonal bases.

\begin{figure}[H]
\begin{center}
        \leavevmode\psfig{file=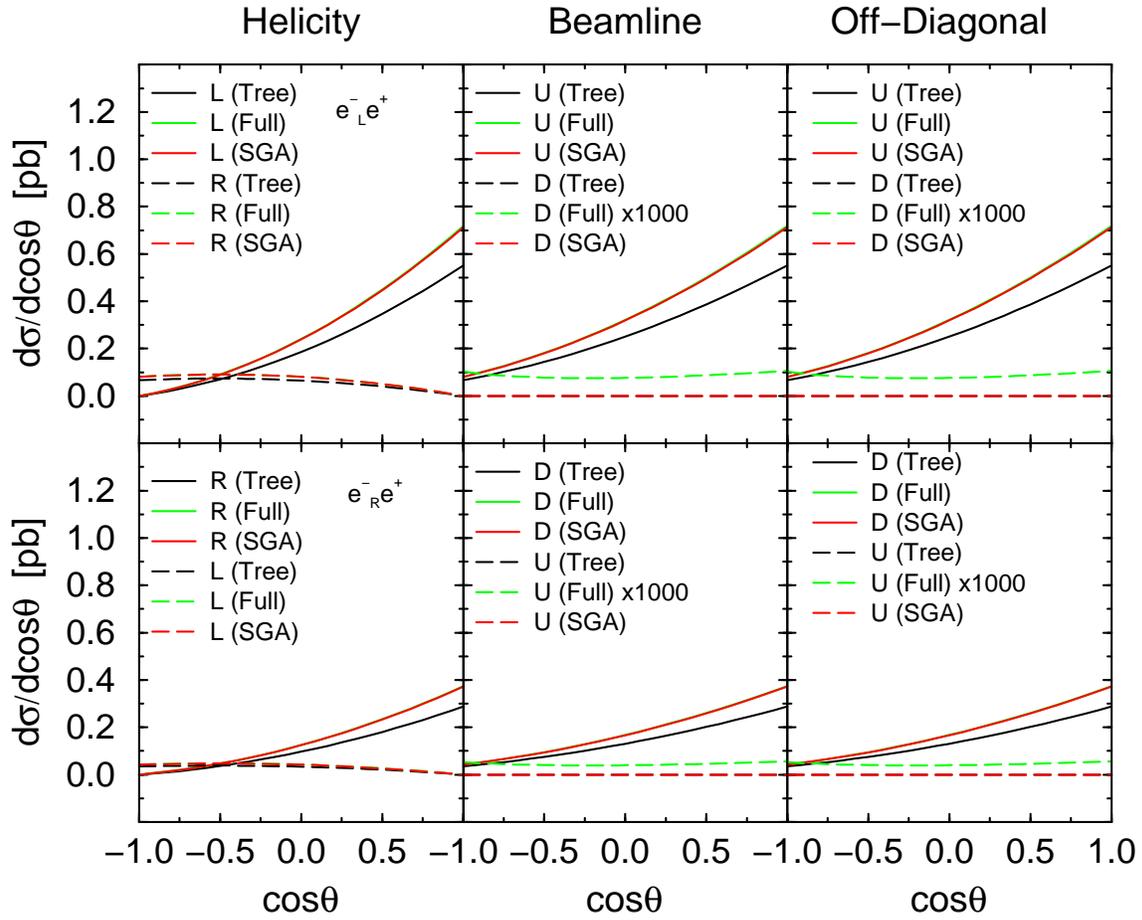,angle=-90,width=15cm}
\caption{The cross-sections for $f_{LR}=f_{RL}=0$ in the helicity, beamline and
 off-diagonal bases at $\sqrt{s} = 400 \, {\rm GeV}$.}
\label{fig:flr-01}
\end{center} 
\end{figure}

Since these analyses were useful to get further understanding on the off-diagonal
and the beamline bases,
we examine the polarized cross-sections at the one-loop level by setting
$f_{LR}=f_{RL}=0$. 
\begin{table}[H]
\begin{center}
\begin{tabular}{|c||c|c|c|c|c|}
\hline 
$\sqrt{s}$   & $400$ GeV & $500$ GeV & $800$ GeV 
             & $1000$ GeV& $1500$ GeV \\ \hline
\hline
Tree         & $1$  & $1$ & $1$ & $1$ & $1$  \\ \hline
${\cal O}(\alpha_s)$ &
$0.9998$ & $0.9988$ & $0.9969$ & $0.9962$ & $0.9952$  \\ \hline
\end{tabular}
\caption[The fraction of the $e^-_{L/R} e^+ \to t_{\uparrow/\downarrow} \bar{t}$ cross-section with $f_{LR}=f_{RL}=0$ in the dominant spin at $\sqrt{s} = 400,~500,~800,~1000$ and $1500$ GeV for the beamline and off-diagonal bases.]{The fraction of the $e^-_{L/R} e^+ \to t_{\uparrow/\downarrow} \bar{t}$ cross-section with $f_{LR}=f_{RL}=0$ in the dominant spin at $\sqrt{s} = 400,~500,~800,~1000$ and $1500$ GeV for the beamline and off-diagonal bases. Note that cross-section in the beamline basis coincides with that in the off-diagonal basis with $f_{LR}=0$.}   
\label{tbl:tbl-flr-zero}
\end{center}
\end{table}
\noindent
Table \ref{tbl:tbl-flr-zero} shows the fraction of the $e^-_{L/R}e^+$
cross-section with $f_{LR}=f_{RL}=0$ in the dominant spin
at the energy region $400 ~\sim 1500$ GeV. 
The QCD corrections to these processes slightly change the tree level
result. 
Although the top spin-up configuration for the $e^-_L 
e^+$ scattering process is $100 \%$ at the tree level,
its configuration at the QCD one-loop level is less than $100 \%$.
However this change is of order $0.002 \sim 0.05 \%$ at energy region $400 ~\sim
1500$ GeV and the QCD corrections are almost negligible.
We plot the differential cross-sections with $f_{LR}=f_{RL}=0$
in the helicity, beamline and off-diagonal bases in Fig.\ref{fig:flr-01}
$\sim$ Fig.\ref{fig:flr-05}.
The characteristic features of the polarized cross-sections in the beamline and
off-diagonal bases do not change even after including the lowest 
order QCD corrections.
\clearpage
%%%%%%%%%%%%%%%%%%%%%%%%%%%%
%%%%%%%%%%%%%%%%%%%%%%%%%%%%
\begin{figure}[H]
\begin{center}
        \leavevmode\psfig{file=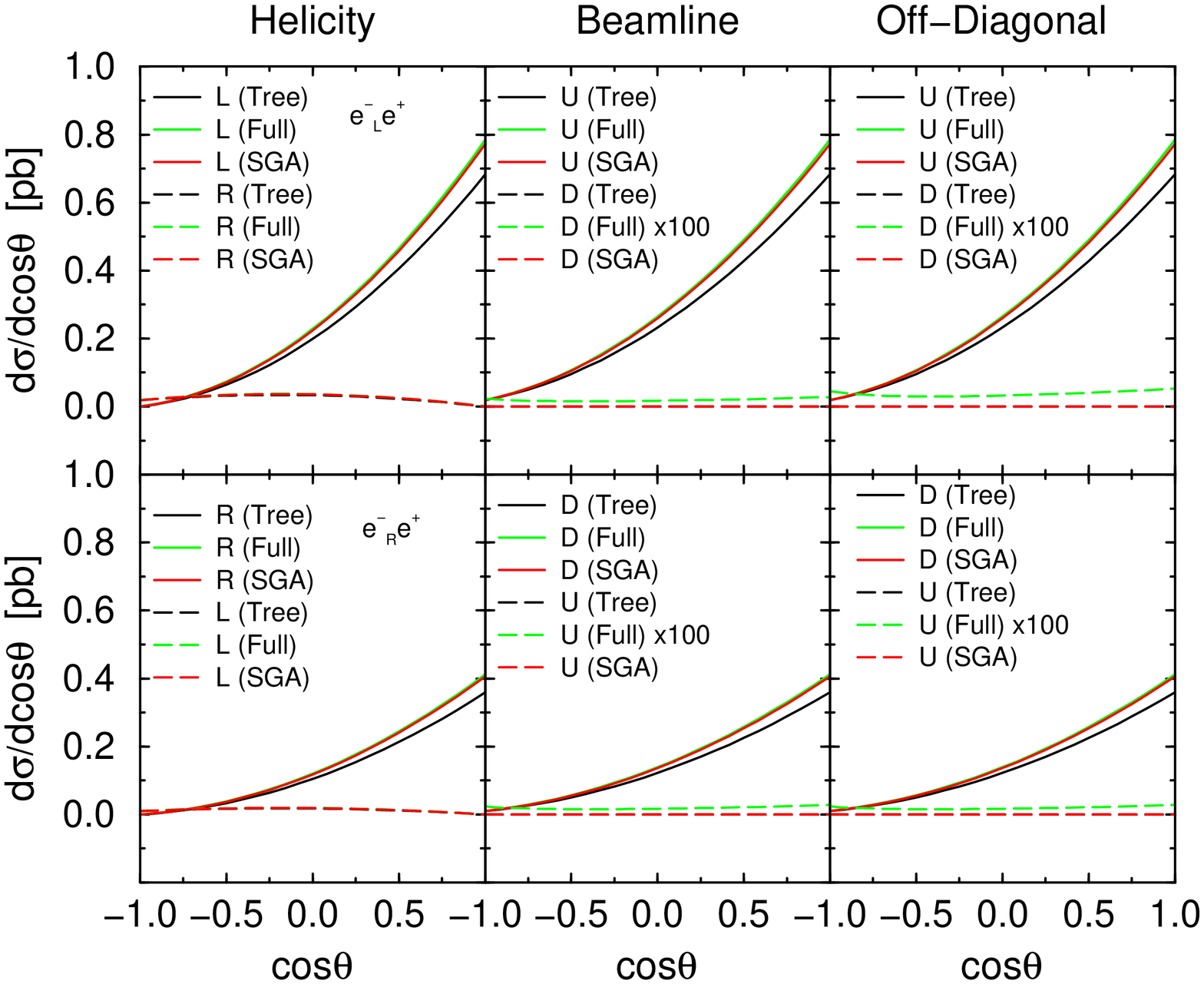,angle=-90,width=15cm}
\caption{The cross-sections for $f_{LR}=f_{RL}=0$ in the helicity, beamline and
 off-diagonal bases at $\sqrt{s} = \,500 \, {\rm GeV}$.}
\label{fig:flr-02}
\end{center} 
\end{figure}

%%%%%%%%%%%%%%%%%%%%%%%%%%%%
\begin{figure}[H]
\begin{center}
        \leavevmode\psfig{file=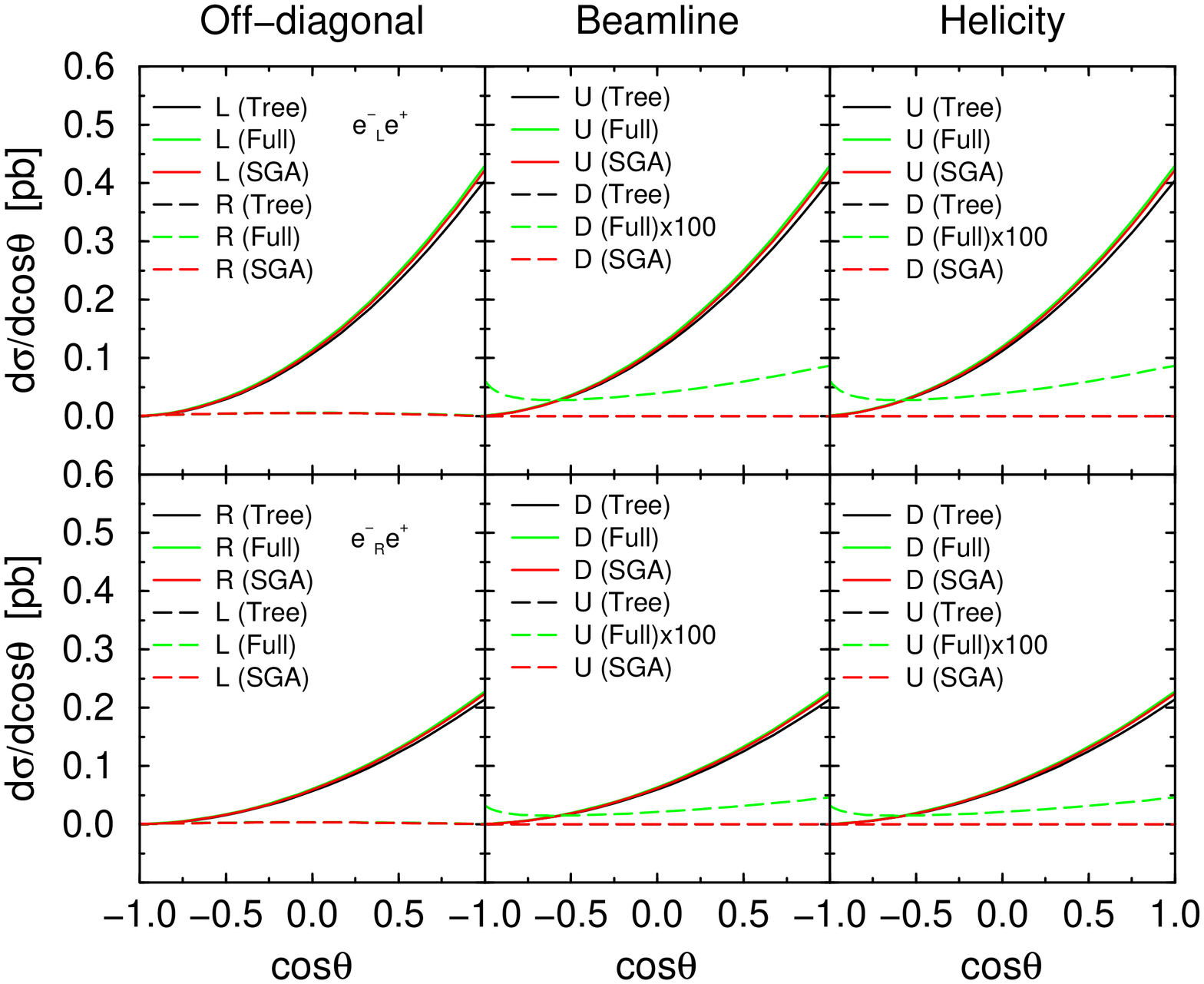,angle=-90,width=15cm}
\label{fig:flr-03}
\caption{The cross-sections for $f_{LR}=f_{RL}=0$ in the helicity, beamline
 and off-diagonal bases at $\sqrt{s} = \, 800 \,{\rm GeV}$.}

\end{center} 
\end{figure}
%%%%%%%%%%%%%%%%%%%%%%%%%%%%
\begin{figure}[H]
\begin{center}
        \leavevmode\psfig{file=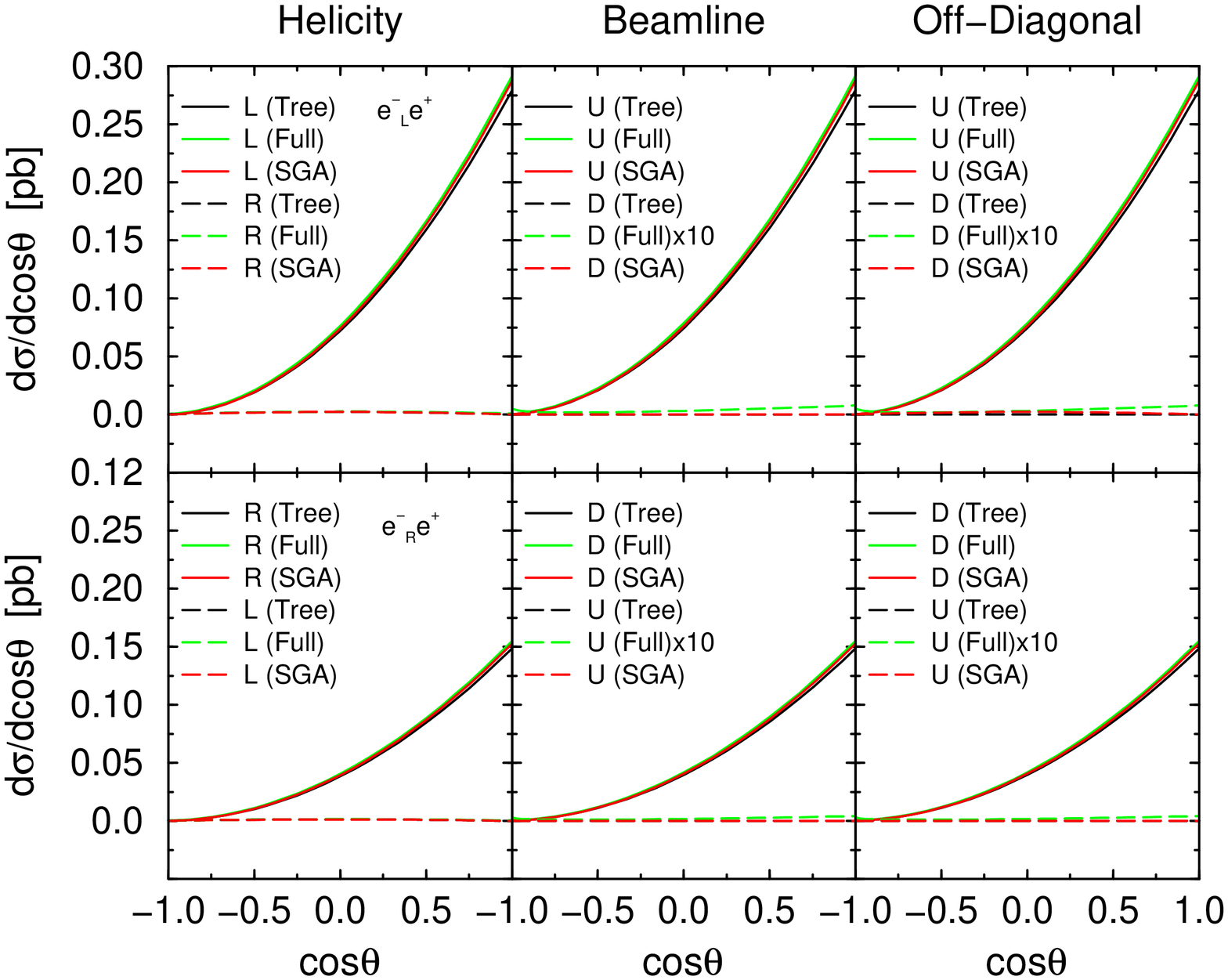,angle=-90,width=15cm}
\label{fig:flr-04}
\caption{The cross-sections for $f_{LR}=f_{RL}=0$ in the helicity, beamline
 and off-diagonal bases at $\sqrt{s} = \, 1000 \, {\rm GeV}$.}
\end{center} 
\end{figure}
%%%%%%%%%%%%%%%%%%%%%%%%%%%%
\begin{figure}[H]
\begin{center}
        \leavevmode\psfig{file=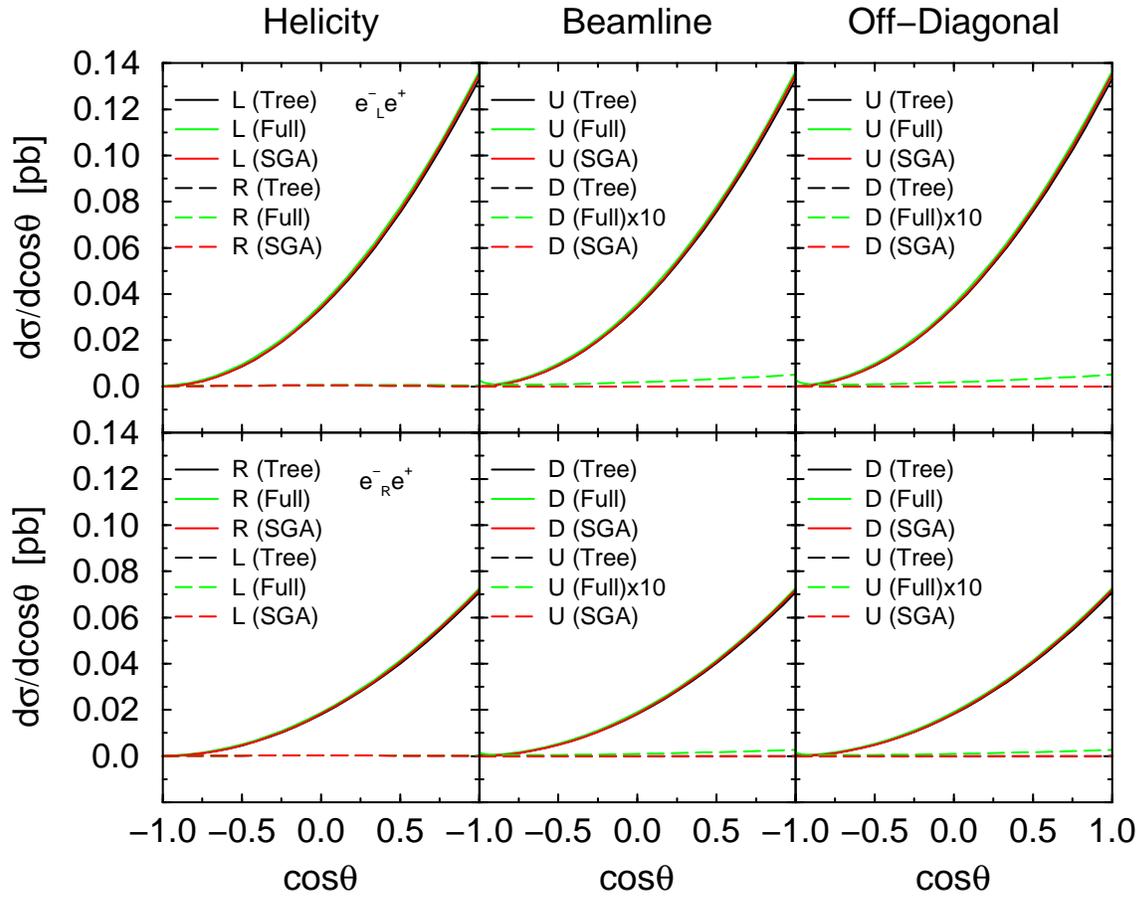,angle=-90,width=15cm}
\caption{The cross-sections for $f_{LR}=f_{RL}=0$ in the helicity,
 beamline and off-diagonal bases at $\sqrt{s} = 1500 \, {\rm GeV}$.}
\label{fig:flr-05}
\end{center} 
\end{figure}
%%%
\clearpage
\section{Validity of Soft Gluon Approximation}

In the previous section, we have arrived at the following results.
The spin-flip effects caused by the hard gluon are quit small, and the
QCD corrections produce only the multiplicative enhancement to the leading 
order results.
Moreover, this enhancement is almost attributed to the soft-gluon effects.
In this Section, we consider the possibility that the full QCD
corrections are reproduced by the soft-gluon approximation with
effective cut-off $\omega_{\rm max}$.
We repeat the same procedure as in the previous analyses in the
soft-gluon approximation.
At first, let us rewrite the spin dependent cross-sections in the generic spin 
basis as follows,
\bea
 \frac{d \sigma^{\rm SGA}}{d \cos \theta}
    ( e_L^-  e_R^+ \to t_{\uparrow} X (\bar{t} g) )
  = \frac{3 \pi \alpha^2}{4 s}
    \sum_{klmn} 
       \left( 
         D^0_{klmn} + B^1_{klmn} 
       \right)
              \cos^k \theta \sin^l \theta
             \cos^m \xi \sin^n \xi \ ,
\eea
with 
\begin{eqnarray}
  B^{1}_{0000} &=& \beta [ f_{LL}^2 + f_{LR}^2 + 2 a f_{LL} f_{LR} ]
                    S_{I} \nonumber \\
           & & \qquad\qquad\qquad - \beta [ 2 (f_{LL} + f_{LR})^2 - \beta^2
                (f_{LL} - f_{LR})^2 ] S_{II} \ ,\nonumber \\
  B^{1}_{2000} &=& \beta^3 (f_{LL}^2 + f_{LR}^2 )
                    S_{I} + \beta^3
              (f_{LL} - f_{LR})^2 S_{II} \ ,\nonumber \\
  B^{1}_{1000} &=& 2 \beta^2 (f_{LL}^2 - f_{LR}^2 )
                    S_{I} \ ,\nonumber \\
  B^{1}_{0010} &=& \beta^2 (f_{LL}^2 - f_{LR}^2 )
                    S_{I} \ ,\nonumber \\
  B^{1}_{2010} &=& \beta^2 (f_{LL}^2 - f_{LR}^2 )
                    S_{I} \ ,\nonumber \\
  B^{1}_{1010} &=& \beta [ (f_{LL} + f_{LR} )^2 + 
              \beta^2 (f_{LL} - f_{LR} )^2  ]
                    S_{I}\nonumber \\
           & & \qquad\qquad\qquad - 2 \beta [ (f_{LL} + f_{LR})^2 - \beta^2
                (f_{LL} - f_{LR})^2 ] S_{II} \ ,\nonumber \\
  B^{1}_{0101} &=& \frac{\beta}{\sqrt{a}} (f_{LL} + f_{LR} )^2 
                 [ a S_{I} - (1 + a)
                       S_{II} ] \ ,\nonumber \\
  B^{1}_{1101} &=& \frac{\beta^2}{\sqrt{a}} (f_{LL}^2 - f_{LR}^2 ) 
                 [ a S_{I} - (1 - a)
                       S_{II}]~. \nonumber 
\end{eqnarray}
Now we separate the multiplicative enhancement factor from the coefficients,
$D^0_{klmn} + B^1_{klmn}$.
\begin{eqnarray}
  D^0_{klmn} + B^1_{klmn}
  & \equiv &
  (1 + \kappa )D^{0}_{klmn} + S^{\rm SGA}_{klmn}~. 
\nonumber 
\end{eqnarray}
The deviations $S^{\rm SGA}_{klmn}$ from the multiplicative
enhancement and the ratio $R^{\rm SGA}_{klmn}$ are defined by 
\bea
  S^{\rm SGA}_{klmn} & = & B^{1}_{klmn} - 
  \kappa  D^{0}_{klmn} ~. \nonumber \\
  R^{\rm SGA}_{klmn} & = & 
  \frac{S^{\rm SGA}_{klmn}}{(1 + \kappa) D^{0}_{klmn}}~.\nonumber 
\eea
%We constrain the soft-gluon approximation in order to compare this
%deviation to the full ${\cal O}(\alpha_s)$ calculation. 
%The constraint is that the enhancement in the soft-gluon approximation is
%equivalent to the the full ${\cal O}(\alpha_s)$.
%Then we determine the energy cut $\omega_{\rm max}$ of gluon under this
%constraint,
At first we determine the effective energy cut-off $\omega_{\rm max}$ of 
gluon such that the total cross-section in the soft-gluon approximation 
reproduces the full ${\cal O}(\alpha_s)$ total cross-section.
\begin{eqnarray}
& ~ &
\int_{-1}^{1} d \cos \theta 
\sum_{s = \uparrow, \downarrow } 
\frac{ d \sigma_{T,L/R} }
     { d \cos \theta }
     (e^{-}_{L/R} e^{+} \to t_{s} X (\bar{t}~{\rm or}~\bar{t}g)) \nonumber \\
& = &
\int_{-1}^{1} d \cos \theta 
\sum_{s = \uparrow, \downarrow } 
\frac{ d \sigma_{SGA,L/R} }
     { d \cos \theta }
     (e^{-}_{L/R} e^{+} \to t_{s} \bar{t}).
\nonumber 
\end{eqnarray}
%where $\sigma_{SGA,L/R}$ is the total cross-section in the soft-gluon
%approximation.  
Next using this cut-off $\omega_{\rm max}$, we compare the above
deviation $R_{klmn}^{SGA}$ with $R_{klmn}$ in the full calculation.

We show the effective $\omega_{\rm max}$ and the numerical results of
$B^1_{klmn},~S^{\rm SGA}_{klmn}$ and $R^{\rm SGA}_{klmn}$ in
Table.\ref{tbl:VSG1} and \ref{tbl:VSG2}. 
Compared to Table \ref{tbl:tbl2}$\sim$\ref{tbl:tbl4} in the previous
Section, we find that all values are in good agreement with the full
calculation results at $\sqrt{s}=400$ and $500$ GeV.
As the total energy get larger, the difference of results between the
soft-gluon approximation and in the full ${\cal O}(\alpha_s)$
calculation become more remarkable.
%This difference from hard gluons, 
%show the contribution get large as energy get larger.
However the spin-flip effects, $S^{SGA}_{klmn}$, are still small.
These become smaller and negligible, as the total energy get larger.
Therefore we conclude that the soft-gluon approximation can be an effective
tool to represent the spin dependent cross-section at 
${\cal O}(\alpha_s)$.
We also present the polarized cross-sections in the helicity, beamline
and off-diagonal bases.
\clearpage
\begin{table}[H]
\begin{center}
\begin{tabular}{|c||c|c|c|c|c|c|}
\hline 
$e^-_L e^+$           & $400$ GeV  & $500$ GeV  & $800$ GeV  
                      & $1000$ GeV & $1500$ GeV & $4000$ GeV \\ \hline
\hline	
$\omega_{max}$ & $ 13.49 $ & $ 38.60 $ & $ 106.7 $ 
               & $ 149.7 $ & $ 254.8 $ & $ 775.9 $ \\ \hline
\hline
$B^{1}_{0000}$ &	$0.3259$ & $0.1900$ & $0.08785$ & $0.06979$ & $0.05347$ & $0.04103$ \\ \hline
$B^{1}_{2000}$ &	$0.06009$ & $0.09431$ & $0.08469$ & $0.07414$ & $0.05908$ & $0.04258$ \\ \hline
$B^{1}_{1000}$ &	$0.1911$ & $0.1996$ & $0.1407$ & $0.1183$ & $0.09133$ & $0.06506$ \\ \hline
$B^{1}_{0010}$ &	$0.09556$ & $0.09979$ & $0.07034$ & $0.05916$ & $0.04567$ & $0.03253$ \\ \hline
$B^{1}_{2010}$ &	$0.09556$ & $0.09979$ & $0.07034$ & $0.05916$ & $0.04567$ & $0.03253$ \\ \hline
$B^{1}_{1010}$ &	$0.3860$ & $0.2843$ & $0.1725$ & $0.1439$ & $0.1126$ & $0.08361$ \\ \hline
$B^{1}_{0101}$ &	$0.3099$ & $0.1540$ & $0.03373$ & $0.01373$ & $-0.001588$ & $-0.005863$ \\ \hline
$B^{1}_{1101}$ &	$0.08219$ & $0.06395$ & $0.01957$ & $0.009227$ & $0.0004544$ & $-0.002634$ \\ \hline
\hline	
$S^{\rm SGA}_{0000}$ &	$-0.002607$ & $-0.005957$ & $-0.005747$ & $-0.004551$ & $-0.002615$ & $-0.0005169$ \\ \hline
$S^{\rm SGA}_{2000}$ &	$0.007821$ & $0.01787$ & $0.01724$ & $0.01365$ & $0.007846$ & $0.001551$ \\ \hline
$S^{\rm SGA}_{1000}$ &	$0.02134$ & $0.03293$ & $0.02514$ & $0.01910$ & $0.01057$ & $0.002038$ \\ \hline
$S^{\rm SGA}_{0010}$ &	$0.01067$ & $0.01646$ & $0.01257$ & $0.009549$ & $0.005284$ & $0.001019$ \\ \hline
$S^{\rm SGA}_{2010}$ &	$0.01067$ & $0.01646$ & $0.01257$ & $0.009549$ & $0.005284$ & $0.001019$ \\ \hline
$S^{\rm SGA}_{1010}$ &	$0.005214$ & $0.01191$ & $0.01149$ & $0.009101$ & $0.005231$ & $0.001034$ \\ \hline
$S^{\rm SGA}_{0101}$ &	$-0.005870$ & $-0.01676$ & $-0.02604$ & $-0.02585$ & $-0.02235$ & $-0.01181$ \\ \hline
$S^{\rm SGA}_{1101}$ &	$0.007915$ & $0.005621$ & $-0.005706$ & $-0.008136$ & $-0.008968$ & $-0.005391$ \\ \hline
\hline		
$R^{SGA}_{0000}$ &	$-0.001725$ & $-0.003455$ & $-0.003337$ & $-0.002677$ & $-0.001566$ & $-0.0003144$ \\ \hline
$R^{SGA}_{2000}$ &	$0.03253$ & $0.02658$ & $0.01389$ & $0.009870$ & $0.005141$ & $0.0009552$ \\ \hline
$R^{SGA}_{1000}$ &	$0.02733$ & $0.02246$ & $0.01183$ & $0.008418$ & $0.004393$ & $0.0008172$ \\ \hline
$R^{SGA}_{0010}$ &	$0.02733$ & $0.02246$ & $0.01183$ & $0.008418$ &
 $0.004393$ & $0.0008172$ \\ \hline
$R^{SGA}_{2010}$ &	$0.02733$ & $0.02246$ & $0.01183$ & $0.008418$ & $0.004393$ & $0.0008172$ \\ \hline
$R^{SGA}_{1010}$ &	$0.002977$ & $0.004972$ & $0.003879$ & $0.002952$ & $0.001636$ & $0.0003164$ \\ \hline
$R^{SGA}_{0101}$ &	$-0.004042$ & $-0.01115$ & $-0.02368$ & $-0.02857$ & $-0.03614$ & $-0.05018$ \\ \hline
$R^{SGA}_{1101}$ &	$0.02317$ & $0.01096$ & $-0.01227$ & $-0.02049$ & $-0.03195$ & $-0.04941$ \\ \hline
\end{tabular}
\caption{The values of $B^1_{klmn}$, $S_{klmn}$ and $R^{\rm SGA}_{klmn}$
for $e^{-}_{L} e^{+}$ scattering}
\label{tbl:VSG1}
\end{center}
\end{table}
\begin{table}[H]
\begin{center}
\begin{tabular}{|c||c|c|c|c|c|c|}
\hline 
	              & $400$ GeV  & $500$ GeV  & $800$ GeV  
                      & $1000$ GeV & $1500$ GeV & $4000$ GeV \\ \hline
\hline
$\omega_{max}$ & $ 13.55 $ & $ 38.84 $ & $ 107.2 $ 
               & $ 150.2 $ & $ 255.2 $ & $ 776.1 $ \\ \hline	
\hline		
$B^{1}_{0000}$ &	$0.1433$ & $0.1907$ & $0.08886$ & $0.07067$ & $0.05405$ & $0.04117$ \\ \hline
$B^{1}_{2000}$ &	$0.02964$ & $0.09457$ & $0.08541$ & $0.07486$ & $0.05962$ & $0.04272$ \\ \hline
$B^{1}_{1000}$ &	$0.1078$ & $0.2002$ & $0.1419$ & $0.1195$ & $0.09217$ & $0.06527$ \\ \hline
$B^{1}_{0010}$ &	$0.05388$ & $0.1001$ & $0.07096$ & $0.05975$ & $0.04609$ & $0.03264$ \\ \hline
$B^{1}_{2010}$ &	$0.05388$ & $0.1001$ & $0.07096$ & $0.05975$ & $0.04609$ & $0.03264$ \\ \hline
$B^{1}_{1010}$ &	$0.1729$ & $0.2853$ & $0.1743$ & $0.1455$ & $0.1137$ & $0.08389$ \\ \hline
$B^{1}_{0101}$ &	$0.1325$ & $0.1546$ & $0.03437$ & $0.01420$ & $-0.001372$ & $-0.005843$ \\ \hline
$B^{1}_{1101}$ &	$0.04635$ & $0.06415$ & $0.01984$ & $0.009433$ & $0.0005521$ & $-0.002624$ \\ \hline
\hline
$S^{\rm SGA}_{0000}$ &	$-0.001319$ & $-0.0103680$ & $-0.009239$ & $-0.007183$ & $-0.002033$ & $-0.0003753$ \\ \hline
$S^{\rm SGA}_{2000}$ &	$0.003957$ & $0.01615$ & $0.01473$ & $0.01151$ & $0.008378$ & $0.001691$ \\ \hline
$S^{\rm SGA}_{1000}$ &	$0.01137$ & $0.02918$ & $0.02083$ & $0.01558$ & $0.01141$ & $0.002253$ \\ \hline
$S^{\rm SGA}_{0010}$ &	$0.005683$ & $0.01459$ & $0.01042$ & $0.007792$ & $0.005703$ & $0.001126$ \\ \hline
$S^{\rm SGA}_{2010}$ &	$0.005683$ & $0.01459$ & $0.01042$ & $0.007792$ & $0.005703$ & $0.001126$ \\ \hline
$S^{\rm SGA}_{1010}$ &	$0.002638$ & $0.005783$ & $0.005487$ & $0.004327$ & $0.006345$ & $0.001315$ \\ \hline
$S^{\rm SGA}_{0101}$ &	$-0.003445$ & $-0.02060$ & $-0.02827$ & $-0.02725$ & $-0.02214$ & $-0.011790$ \\ \hline
$S^{\rm SGA}_{1101}$ &	$0.004170$ & $0.004308$ & $-0.006649$ & $-0.008751$ & $-0.008870$ & $-0.005381$ \\ \hline
\hline		
$R^{SGA}_{0000}$ &	$-0.001994$ & $-0.005996$ & $-0.005351$ & $-0.004217$ & $-0.001217$ & $-0.0002282$ \\ \hline
$R^{SGA}_{2000}$ &	$0.03369$ & $0.02395$ & $0.01183$ & $0.008305$ & $0.005490$ & $0.001041$ \\ \hline
$R^{SGA}_{1000}$ &	$0.02578$ & $0.01984$ & $0.009774$ & $0.006855$ & $0.004741$ & $0.0009033$ \\ \hline
$R^{SGA}_{0010}$ &	$0.02578$ & $0.01984$ & $0.009774$ & $0.006855$ & $0.004741$ & $0.0009033$ \\ \hline
$R^{SGA}_{2010}$ &	$0.02578$ & $0.01984$ & $0.009774$ & $0.006855$ & $0.004741$ & $0.0009033$ \\ \hline
$R^{SGA}_{1010}$ &	$0.003387$ & $0.002406$ & $0.001847$ & $0.001401$ & $0.001985$ & $0.0004025$ \\ \hline
$R^{SGA}_{0101}$ &	$-0.005543$ & $-0.01367$ & $-0.02564$ & $-0.03005$ & $-0.03579$ & $-0.05010$ \\ \hline
$R^{SGA}_{1101}$ &	$0.02162$ & $0.008372$ & $-0.01426$ & $-0.02200$ & $-0.03161$ & $-0.04932$ \\ \hline
\end{tabular}
\caption{The values of $B^1_{klmn}$, $S_{klmn}$ and $R^{\rm SGA}_{klmn}$
for $e^{-}_{R} e^{+}$ scattering}
\label{tbl:VSG2}
\end{center}
\end{table}
%\clearpage
%%%%%%%%%%%%%%%%%%%%%%%%%%%%
\begin{figure}[H]
\begin{center}
\begin{tabular}{c}
        \leavevmode\psfig{file=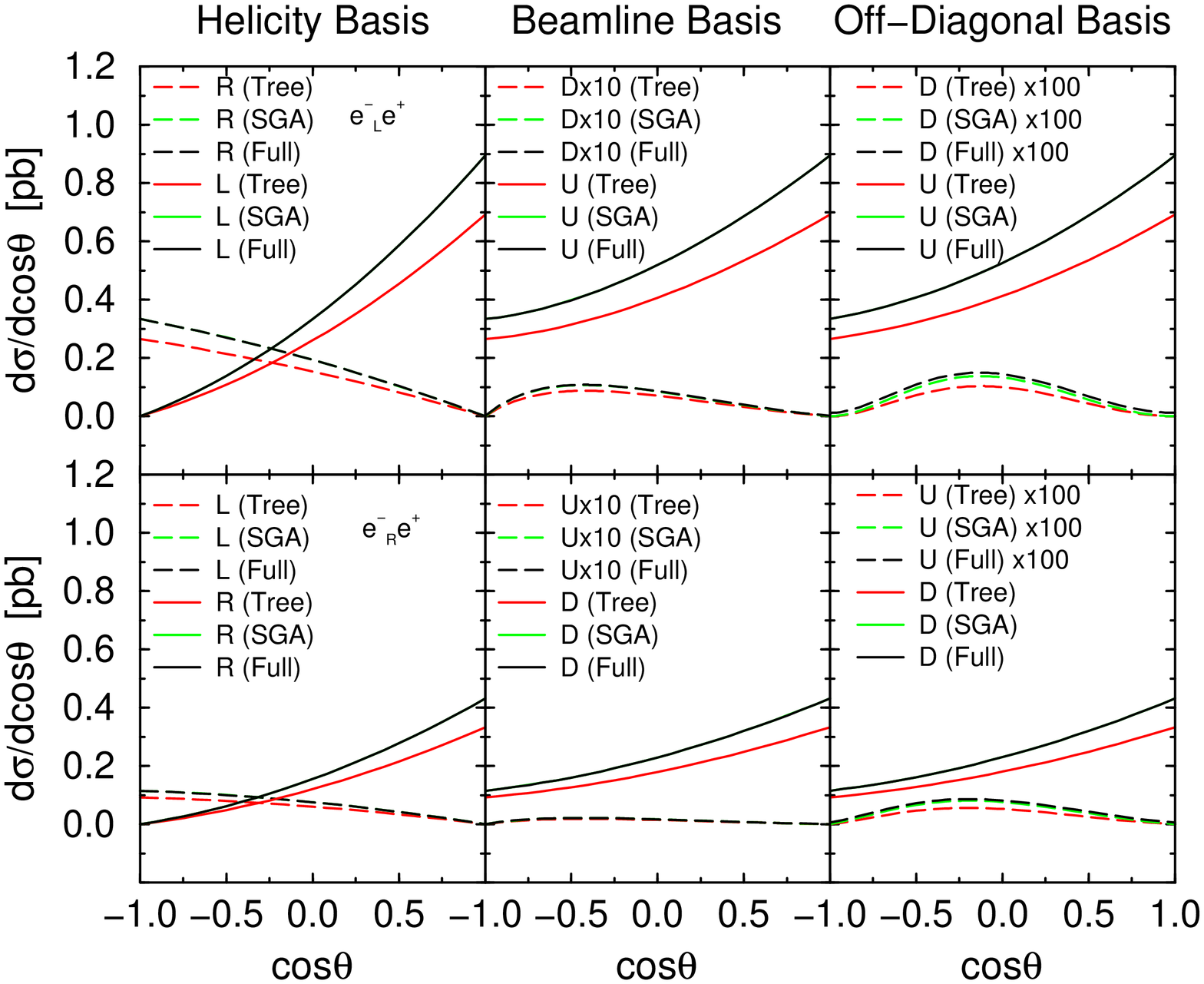,angle=-90,width=15.cm}% \\
%        \leavevmode\psfig{file=all-3-500.eps,angle=-90,width=15.cm}
\end{tabular}
\caption[The cross-sections in the helicity, beamline and off-diagonal bases
at $\sqrt{s} = 400~ \,{\rm GeV}$.]{The cross-sections in the helicity,
 beamline and off-diagonal bases at $\sqrt{s} = 400~ \, {\rm GeV}$.
Here we use a ``beamline basis'', in which the top quark axis is the
positron direction in the top rest frame, for each $e^{-}_{L}e^{+}$ and 
$e^{-}_{R}e^{+}$ scattering.
For the soft gluon approximation (SGA) 
$\omega_{\rm max}=13.49 $ GeV.}
\end{center} 
\end{figure}
%%%%%%%%%%%%%%%%%%%%%%%%%%%%%
\begin{figure}[H]
\begin{center}
        \leavevmode\psfig{file=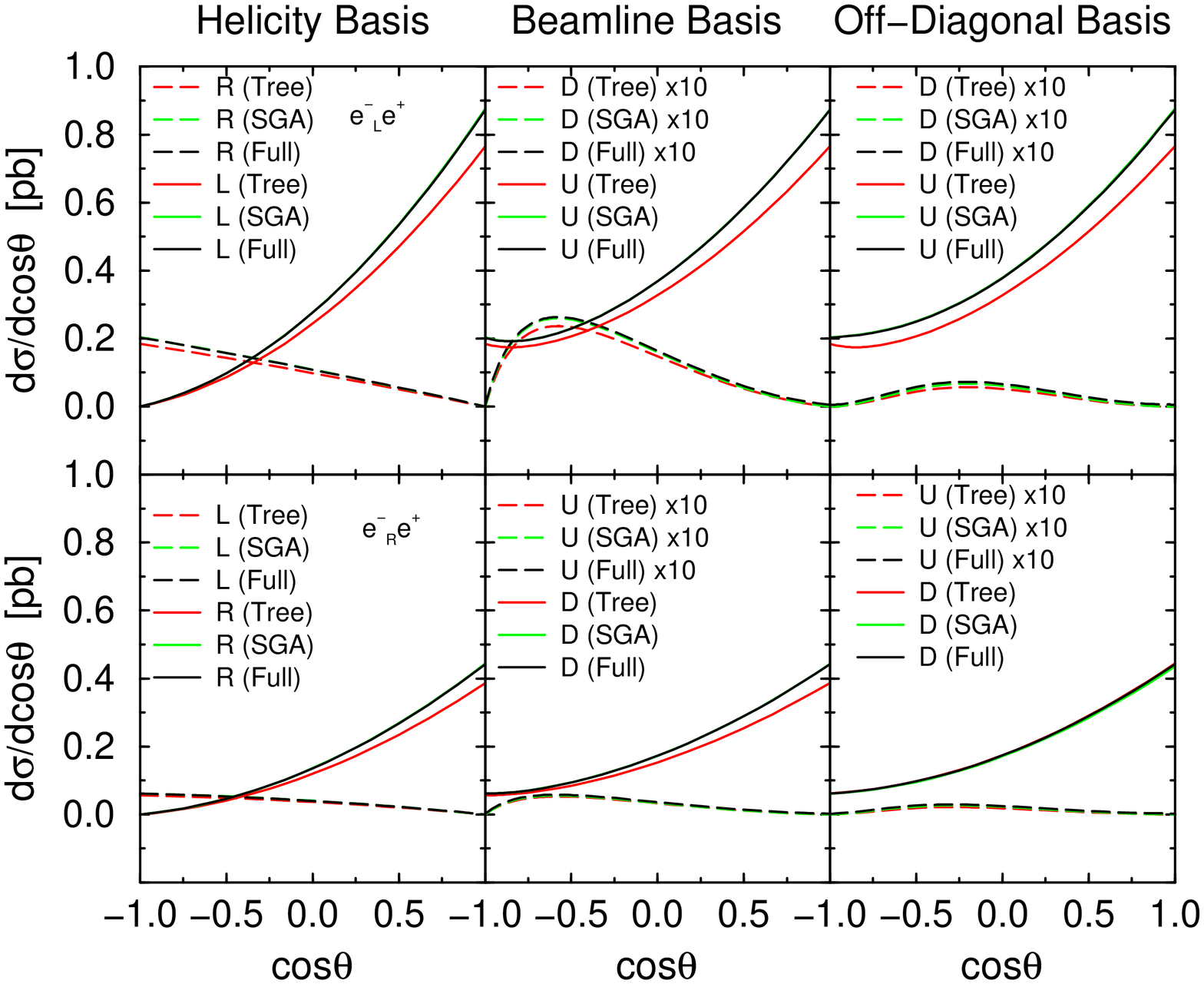,angle=-90,width=13cm}
\caption[The cross-sections in the helicity, beamline and off-diagonal bases
at $\sqrt{s} = 500 \,{\rm GeV}$.]{The cross-sections in the helicity,
 beamline and off-diagonal bases at $\sqrt{s} = 500 \,  {\rm GeV}$.
Here we use a ``beamline basis'', in which the top quark axis is the
positron direction in the top rest frame, for each $e^{-}_{L}e^{+}$ and 
$e^{-}_{R}e^{+}$ scattering.
For the soft gluon approximation (SGA) 
$\omega_{\rm max}=38.60 $ GeV.}
\end{center} 
\end{figure}
%%%%%%%%%%%%%%%%%%%%%%%%%%%%%
\begin{figure}[H]
\begin{center}
\begin{tabular}{c}
        \leavevmode\psfig{file=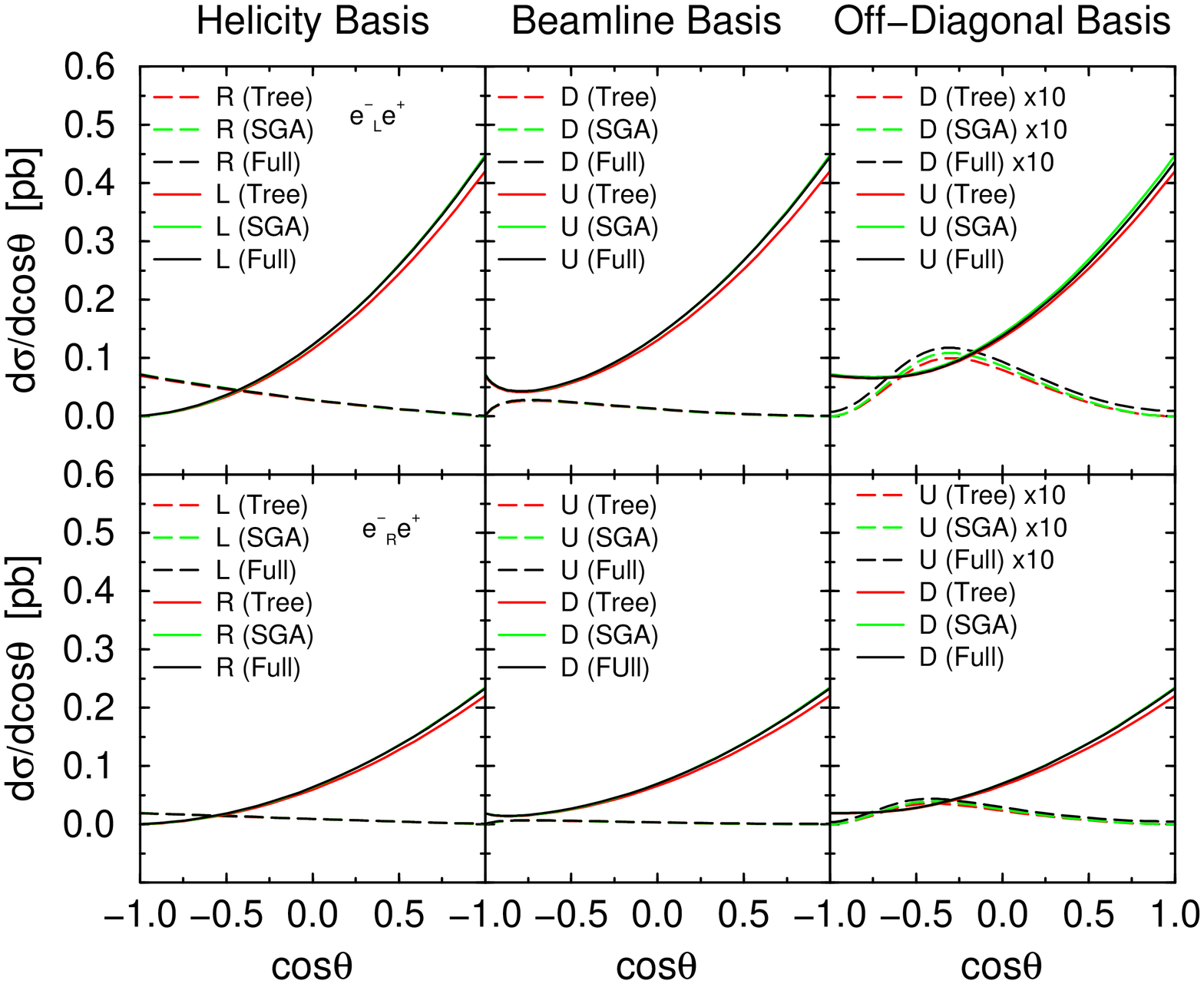,angle=-90,width=15cm} 
%        \leavevmode\psfig{file=all-3-1000.eps,angle=-90,width=15cm} \\
\end{tabular}
\caption[The cross-sections in the helicity, beamline and off-diagonal bases
at $\sqrt{s} = 800~ \,{\rm GeV}$.]{The cross-sections in the helicity,
 beamline and off-diagonal bases at $\sqrt{s} = 800~ \, {\rm GeV}$.
Here we use a ``beamline basis'', in which the top quark axis is the
positron direction in the top rest frame, for each $e^{-}_{L}e^{+}$ and 
$e^{-}_{R}e^{+}$ scattering.
For the soft gluon approximation (SGA) 
$\omega_{\rm max}=106.7$ GeV.}
\end{center} 
\end{figure}
%%%%%%%%%%%%%%%%%%%%%%%%%%%%%
\begin{figure}[H]
\begin{center}
        \leavevmode\psfig{file=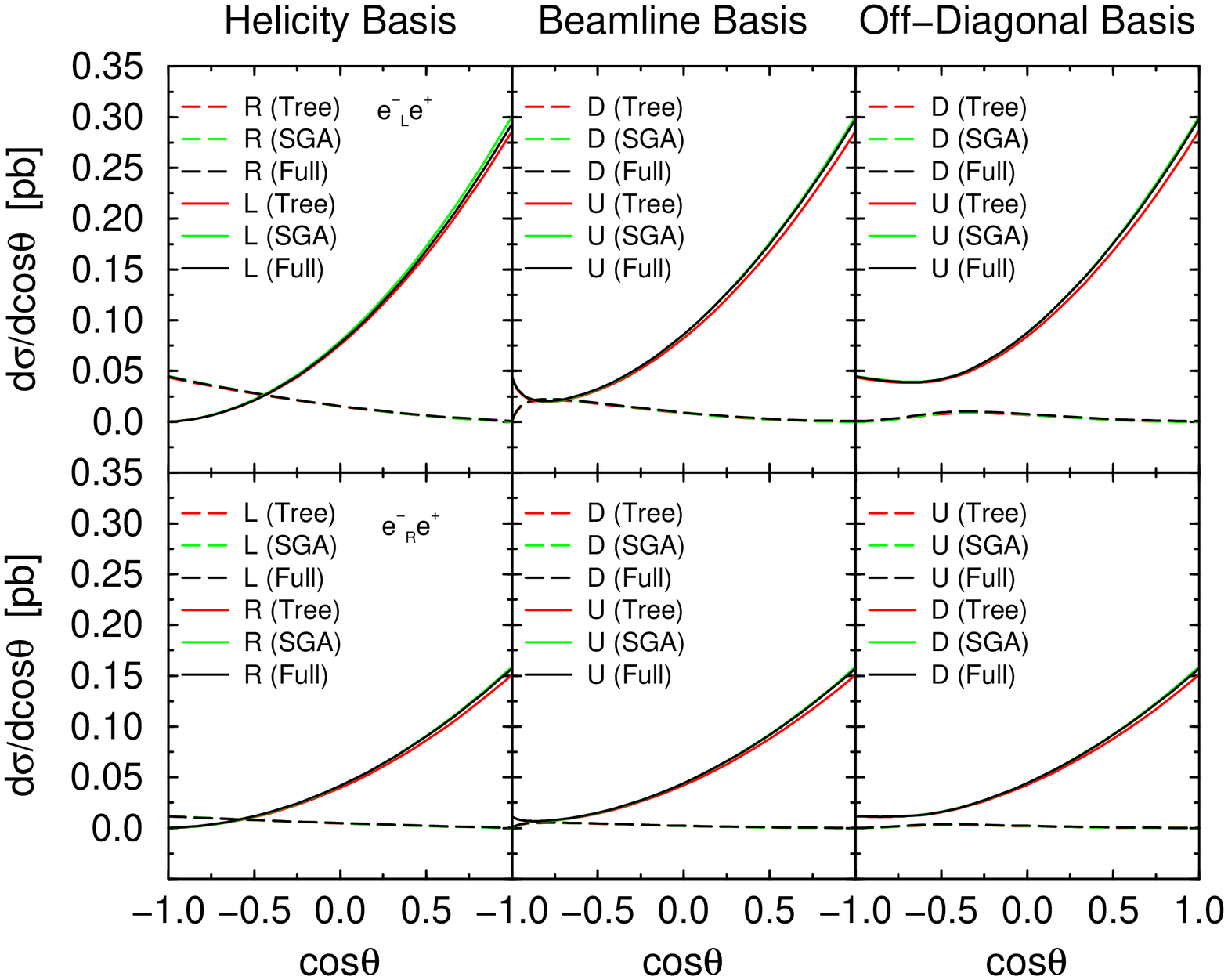,angle=-90,width=13cm}
\caption[The cross-sections in the helicity, beamline and off-diagonal bases
at $\sqrt{s} = 1000 \,{\rm GeV}$.]{The cross-sections in the helicity,
 beamline and off-diagonal bases at $\sqrt{s} = 1000 \,  {\rm GeV}$.
Here we use a ``beamline basis'', in which the top quark axis is the
positron direction in the top rest frame, for each $e^{-}_{L}e^{+}$ and 
$e^{-}_{R}e^{+}$ scattering.
For the soft gluon approximation (SGA) 
$\omega_{\rm max}=149.7 $ GeV.}
\end{center} 
\end{figure}
%%%%%%%%%%%%%%%%%%%%%%%%%%%%%
\begin{figure}[H]
\begin{center}
%\begin{tabular}{c}
        \leavevmode\psfig{file=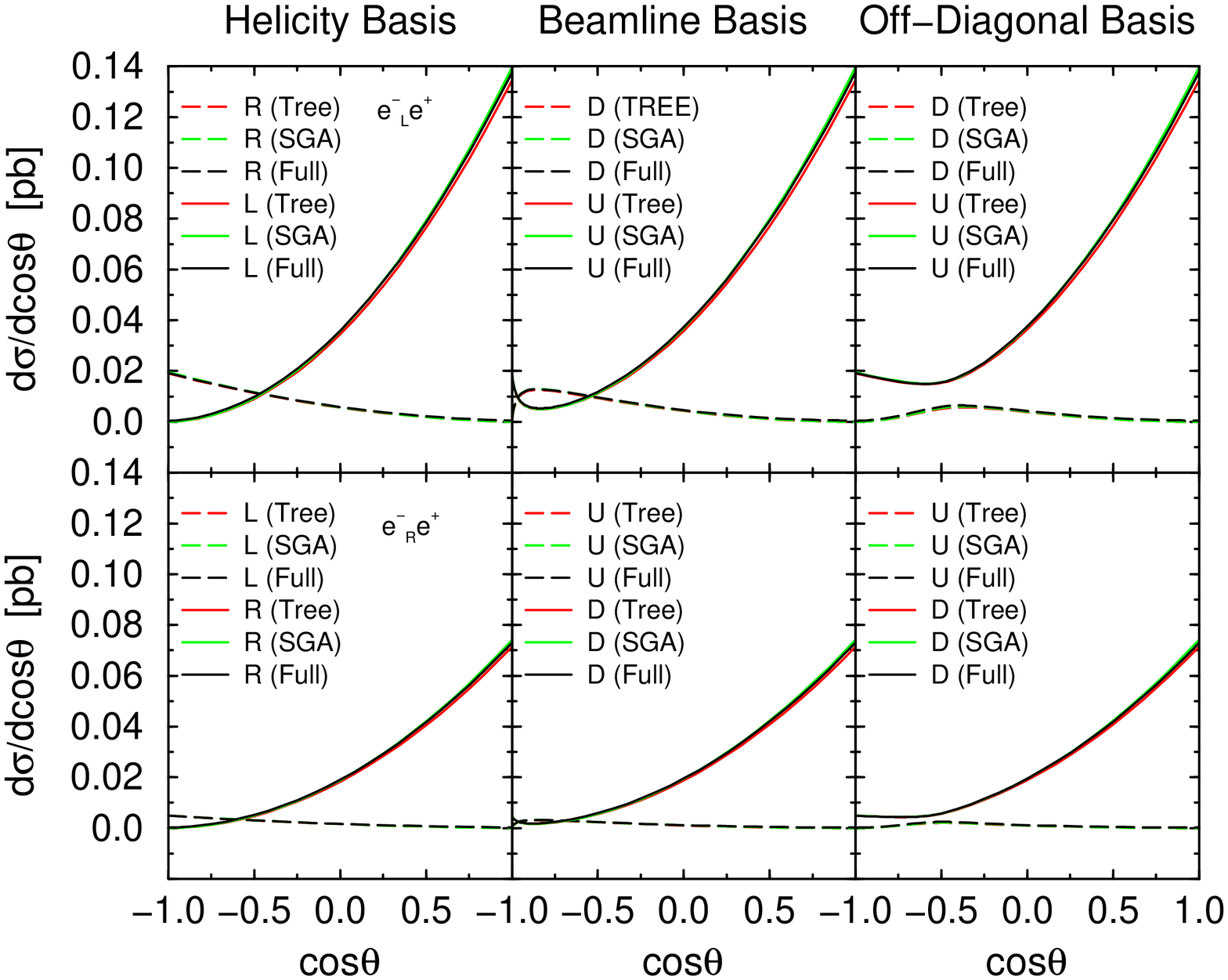,angle=-90,width=15cm} %\\
%        \leavevmode\psfig{file=all-3-4000.eps,angle=-90,width=15cm} \\
%\end{tabular}
\caption[The cross-sections in the helicity, beamline and off-diagonal bases
at $\sqrt{s} = 1500 \,{\rm GeV}$.]{The cross-sections in the helicity,
 beamline and off-diagonal bases at $\sqrt{s} = 1500 \,  {\rm GeV}$.
Here we use a ``beamline basis'', in which the top quark axis is the
positron direction in the top rest frame, for each $e^{-}_{L}e^{+}$ and 
$e^{-}_{R}e^{+}$ scattering.
For the soft gluon approximation (SGA) 
$\omega_{\rm max}=254.8$ GeV.}
\end{center} 
\end{figure}
%%%%%%%%%%%%%%%%%%%%%%%%%%%%%
\begin{figure}[H]
\begin{center}
        \leavevmode\psfig{file=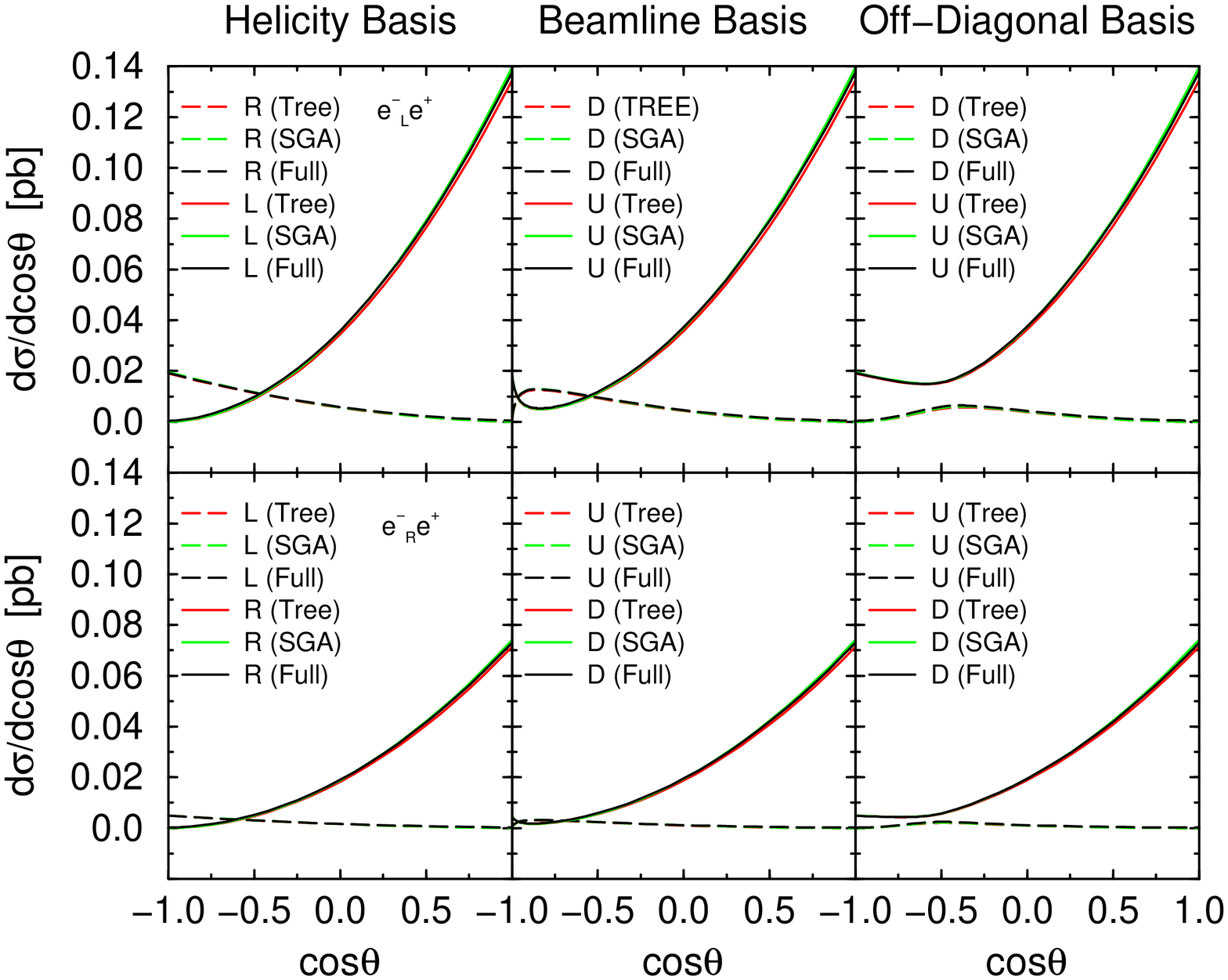,angle=-90,width=15cm}
\caption[The cross-sections in the helicity, beamline and off-diagonal bases
at $\sqrt{s} = 4000 \,{\rm GeV}$.]{The cross-sections in the helicity,
 beamline and off-diagonal bases at $\sqrt{s} = 4000 \,  {\rm GeV}$.
Here we use a ``beamline basis'', in which the top quark axis is the
positron direction in the top rest frame, for each $e^{-}_{L}e^{+}$ and 
$e^{-}_{R}e^{+}$ scattering.
For the soft gluon approximation (SGA) 
$\omega_{\rm max}=775.9 $ GeV.}
\end{center} 
\end{figure}
%%%%%%%%%%%%%%%%%%%%%%%%%%%%%
\newpage
\section{Fourth Generations Quarks}

The existence of three generations of quarks and leptons is well
established in the Standard Model.
Although there is no strong expectation of additional quarks and leptons, 
there is no reason that additional generations are ruled out.
Several models with new generations or arguments flavoring new
generations have been proposed~\cite{bprime}.

We have shown the efficiency and the usefulness of the study of spin
correlations in the previous Sections and Chapters.
Therefore, we consider the pair production of a fourth
generation quark ($b'$) with charge $-1/3$. 
and investigate the polarized cross-sections at polarized $e^+e^-$
linear colliders. 
Here we assume that the $b'$ mass is between $150$ GeV and $300$ GeV,
and take the CM energy at $500$ and $1000$ GeV.
In this analyses, we use three bases as shown in the previous Sections.
One is the usual helicity basis which corresponds to $\cos \xi = + 1$. 
The second is the beamline basis, in which the top quark spin is aligned
with the positron momentum in the top quark rest frame.
The third is the off-diagonal basis which makes like-spin configuration 
vanish in the $e^-_L e^+$ scattering process. 

To present the differential cross-section, we investigate the $b'$ quark
speed, $\beta$, dependence of couplings $f_{IJ}$'s. 
These couplings are given by,
\bea
0.97 \leq f_{LL} \leq  1.02 &,& 0.208 \leq f_{LR}\leq 0.218 ~,
\nonumber \\
0.433 \leq f_{RR}\leq 0.442 &,& - 0.260 \leq f_{RL} \leq - 0.215~.
\nonumber
\eea  
The couplings $f_{IJ}$'s have weak $\beta$ dependence.
Without $Z$ boson couplings, the couplings become $f_{LL}=f_{LR}=1/3$.
So the contribution from the $Z$ boson is constructive to the
$f_{LL}$ and $f_{RR}$, and destructive to the $f_{LR}$ and $f_{RL}$.

\begin{figure}[H]
\begin{center}
        \leavevmode\psfig{file=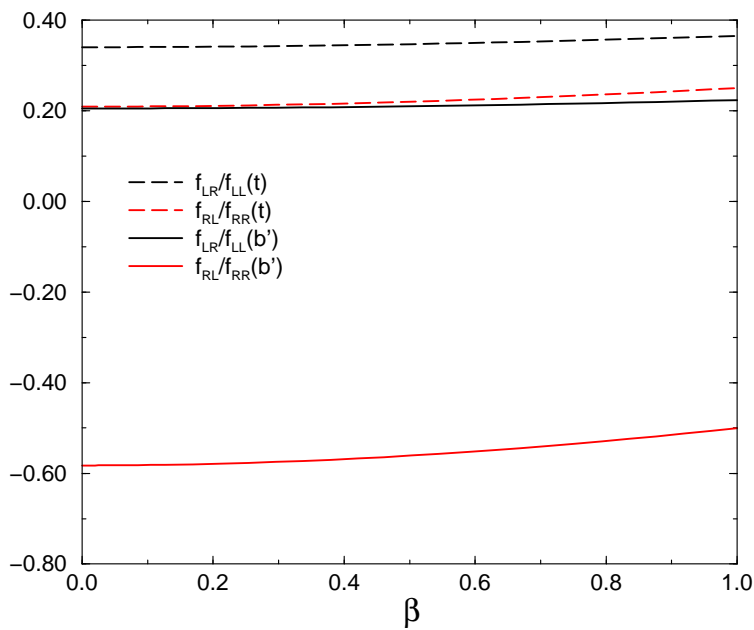,angle=-90,width=10cm}
\caption{$b'$ quark speed, $\beta$, dependence of ratios, $f_{LR}/f_{LL}$ and
 $f_{RL}/f_{RR}$.}
\label{fig:bpr-ratio}
\end{center} 
\end{figure}

Next, we plot the $\beta$ dependence of the ratios, $f_{LR}/f_{LL}$ and
$f_{RL}/f_{RR}$, in Fig.\ref{fig:bpr-ratio}. 
The ratio $f_{LR}/f_{LL}$ has a weak $\beta$ dependence, and 
it shows the similar behavior as the ratios of the top quark
couplings, $f_{LR}/f_{LL}$ and $f_{RL}/f_{RR}$. 
However, the ratio $f_{RL}/f_{RR}$ for $b'$ productions gives the large
contribution $\sim 0.5$ compared to the top quark case.
As discussed previously,
the parameters $f_{LL}$ and $f_{RR}$ control the behavior of the
dominant configurations in the beamline and off-diagonal bases, and 
the parameters $f_{LR}/f_{LL}$ and $f_{RL}/f_{RR}$ decide the behavior
of the subdominant configuration.
Therefore, in the $e^-_R e^+ \to b'X (\bar{b'}~{\rm or}~\bar{b'}g)$
process, the subdominant configuration gives larger contribution to the
total cross-section than $e^-_L e^+ \to b'X (\bar{b'}~{\rm
or}~\bar{b'}g)$ process. 

\clearpage
%
%%%%%%%%%%%%%%%%%%%%%%%%%%%%
\begin{figure}[H]
\begin{center}
        \leavevmode\psfig{file=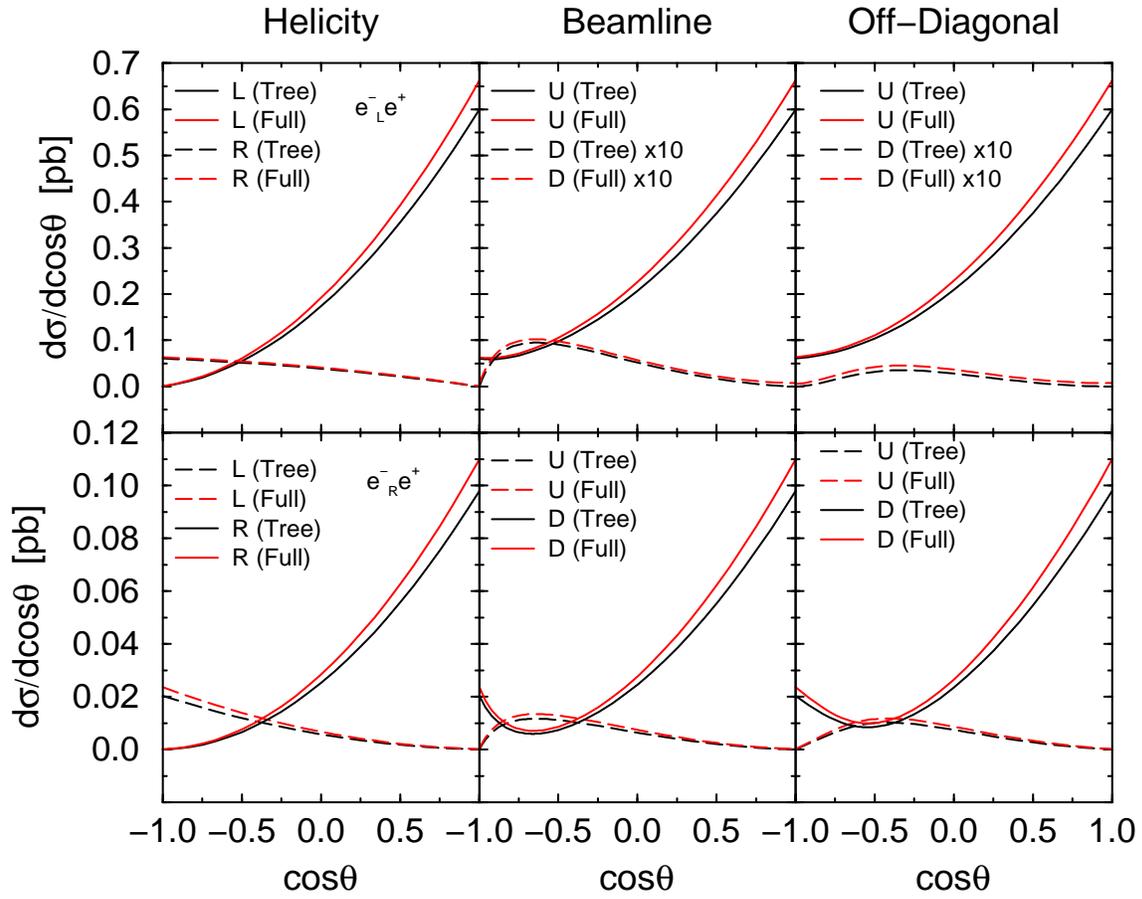,angle=-90,width=15cm}
\caption[The cross-sections for the process $e^+ e^- \rightarrow b' X (\bar{b}'~{\rm or}~ \bar{b}' g)$ 
in the helicity, beamline and off-diagonal bases at 
$\sqrt{s} = 500 \, {\rm GeV}$.]{The cross-sections for the process $e^+ e^- \rightarrow b' X (\bar{b}'~{\rm or}~ \bar{b}' g)$ 
in the helicity, beamline and off-diagonal bases at 
$\sqrt{s} = 500 \, {\rm GeV}$. 
The value of $b'$ mass is $150${\rm GeV}.}
\label{fig:bpr1}
\end{center} 
\end{figure}

%%%%%%%%%%%%%%%%%%%%%%%%%%%%
\noindent
The differential cross-sections in three bases are shown in
Fig.\ref{fig:bpr1} $\sim$ Fig.\ref{fig:bpr5},
and these figures contain the results of the tree level and the full QCD.
For the $e^-_L e^+ \to b'X (\bar{b}'~{\rm or}~\bar{b}'g)$ process, 
the dominant component is top quark spin-up configuration. 
In the beamline and off-diagonal bases, this configuration makes up more
than $96 \%$ of total cross-section in Fig.\ref{fig:bpr1} $\sim$
Fig.\ref{fig:bpr5}.
While, in the helicity basis, the dominant configuration occupies $77 \%$
of the total cross-section with $b'$ mass $m=200$ GeV at $\sqrt{s}=500$
GeV, and is less than $92 \%$ of the total cross-section in Fig.\ref{fig:bpr1} $\sim$ Fig.\ref{fig:bpr5}.
The off-diagonal and the beamline bases represent very
strong spin correlations. 
Thus, $b'$ quark is produced in an essentially unique spin configuration
even after including the lowest order QCD corrections.
This result is the same as the top quark case.

For the $e^-_R e^+ \to b'X (\bar{b}'~{\rm or}~\bar{b}'g)$ process, 
the dominant component is top quark spin-down configuration. 
In the helicity basis, the dominant configuration contributes to $80
\sim 87 \%$ of the total cross-section.
In the beamline and off-diagonal bases, this configuration 
contributes to the $83 \sim 88 \%$ of total cross-section.
This results show that the dominant configuration gives the same amount
of contribution to the total cross-section in these three basis.
\clearpage
%%%%%%%%%%%%%%%%%%%%%%%%%%%%
%%%%%%%%%%%%%%%%%%%%%%%%%%%%
\clearpage
\begin{figure}[H]
\begin{center}
        \leavevmode\psfig{file=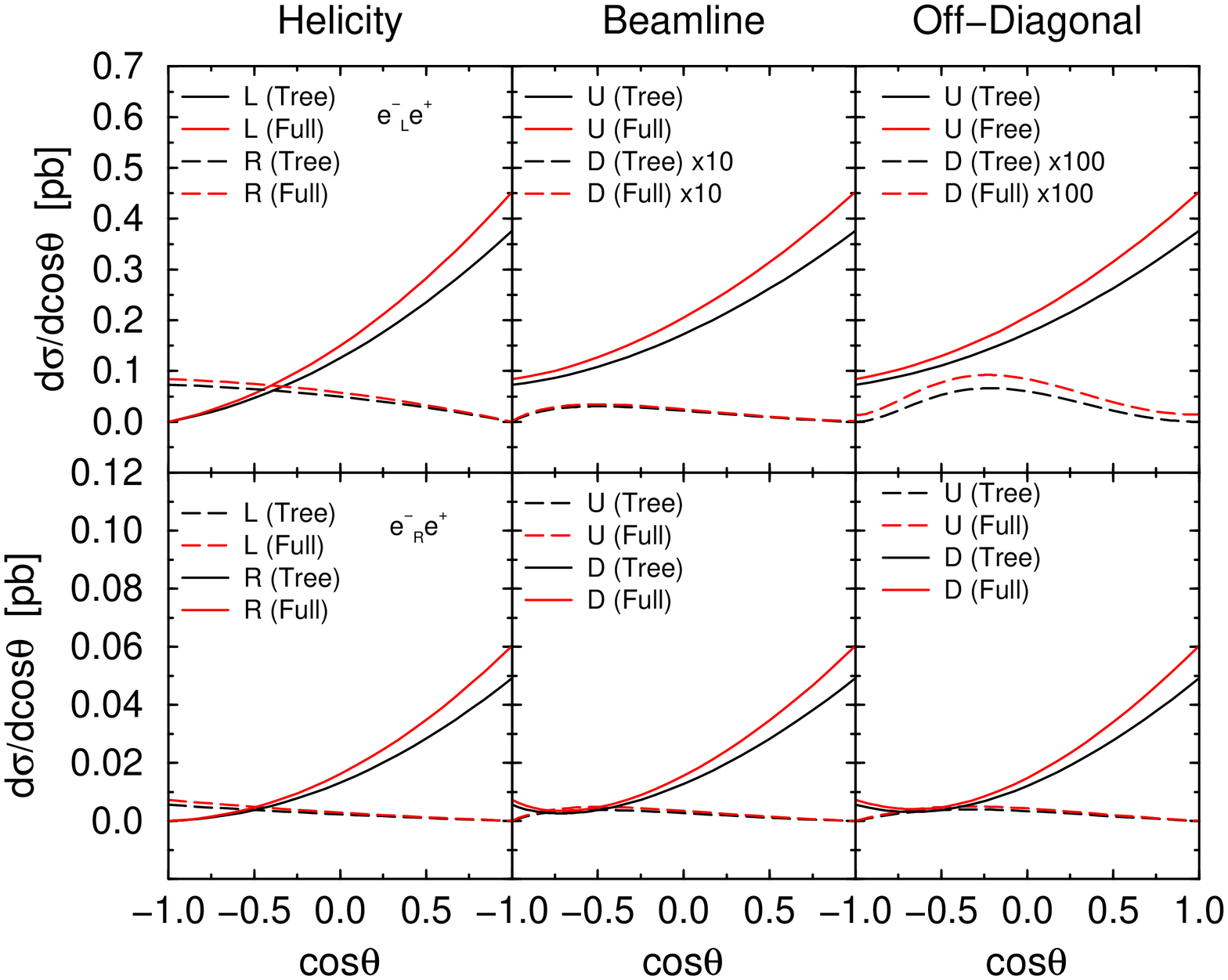,angle=-90,width=15cm}
\caption[The cross-sections for the process $e^+ e^- \rightarrow b' X (\bar{b}'~{\rm or}~ \bar{b}' g)$ 
in the helicity, beamline and off-diagonal bases at $\sqrt{s} = 500 \, {\rm GeV}$.]{The cross-sections for the process $e^+ e^- \rightarrow b'  X (\bar{b}'~{\rm or}~ \bar{b}' g)$ in the helicity, beamline and off-diagonal bases at $\sqrt{s} = 500 \, {\rm GeV}$. The value of $b'$ mass is $200$ {\rm GeV}.}
\label{fig:bpr2}
\end{center} 
\end{figure}
%%%%%%%%%%%%%%%%%%%%%%%%%%%%
\begin{figure}[H]
\begin{center}
        \leavevmode\psfig{file=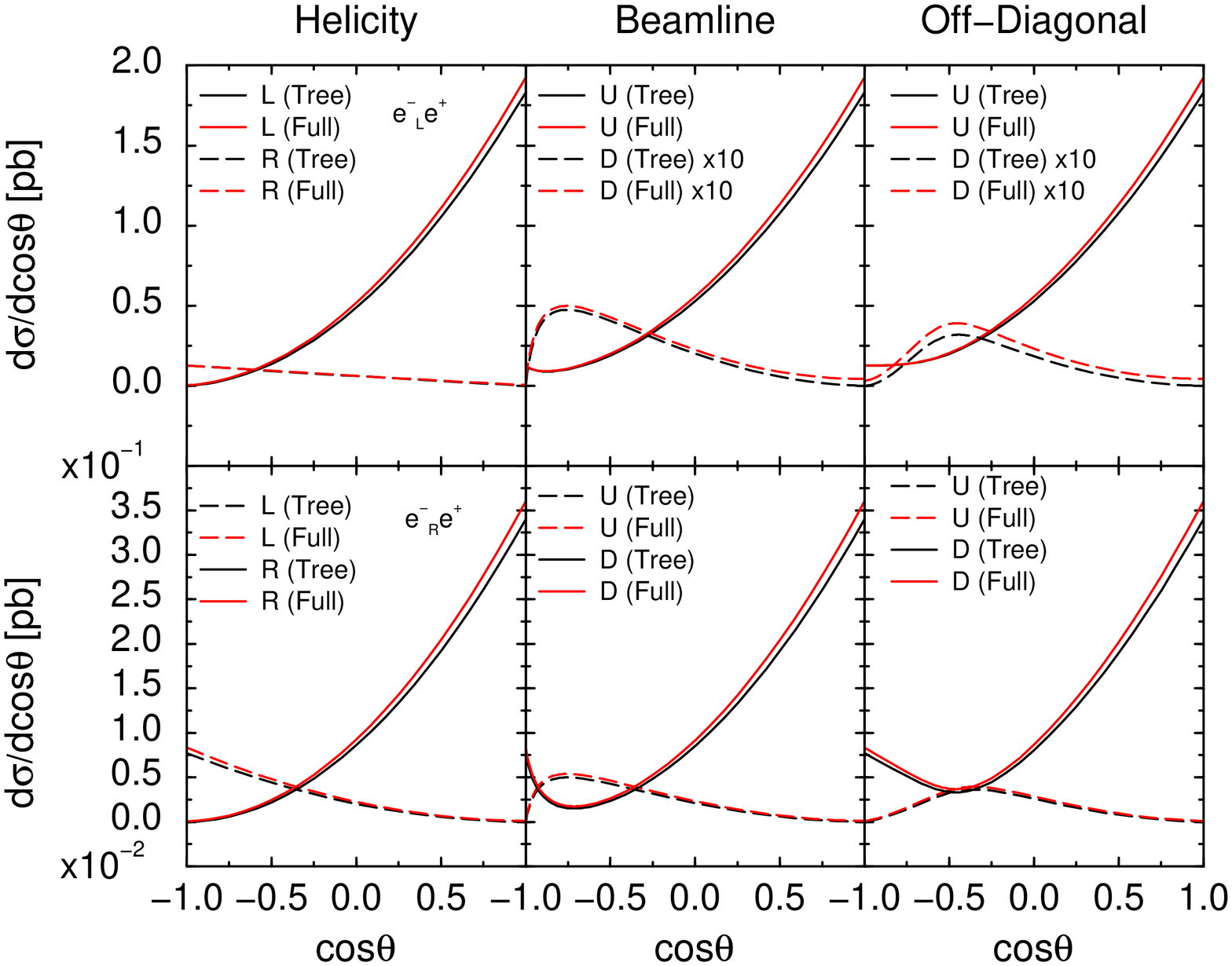,angle=-90,width=15cm}
\caption[The cross-sections for the process $e^+e^- \rightarrow b' X (\bar{b}'~{\rm or}~ \bar{b}' g)$ 
in the helicity, beamline and off-diagonal bases 
at $\sqrt{s} = 1000 {\rm GeV}$.]{The cross-sections for the process
 $e^+e^- \rightarrow b' X (\bar{b}'~{\rm or}~ \bar{b}' g)$ in the
 helicity, beamline and off-diagonal bases at $\sqrt{s} = 1000 {\rm GeV}$.
The value of $b'$ mass is $200$ {\rm GeV}.}
\label{fig:bpr3}
\end{center} 
\end{figure}
%%%%%%%%%%%%%%%%%%%%%%%%%%%%
\begin{figure}[H]
\begin{center}
        \leavevmode\psfig{file=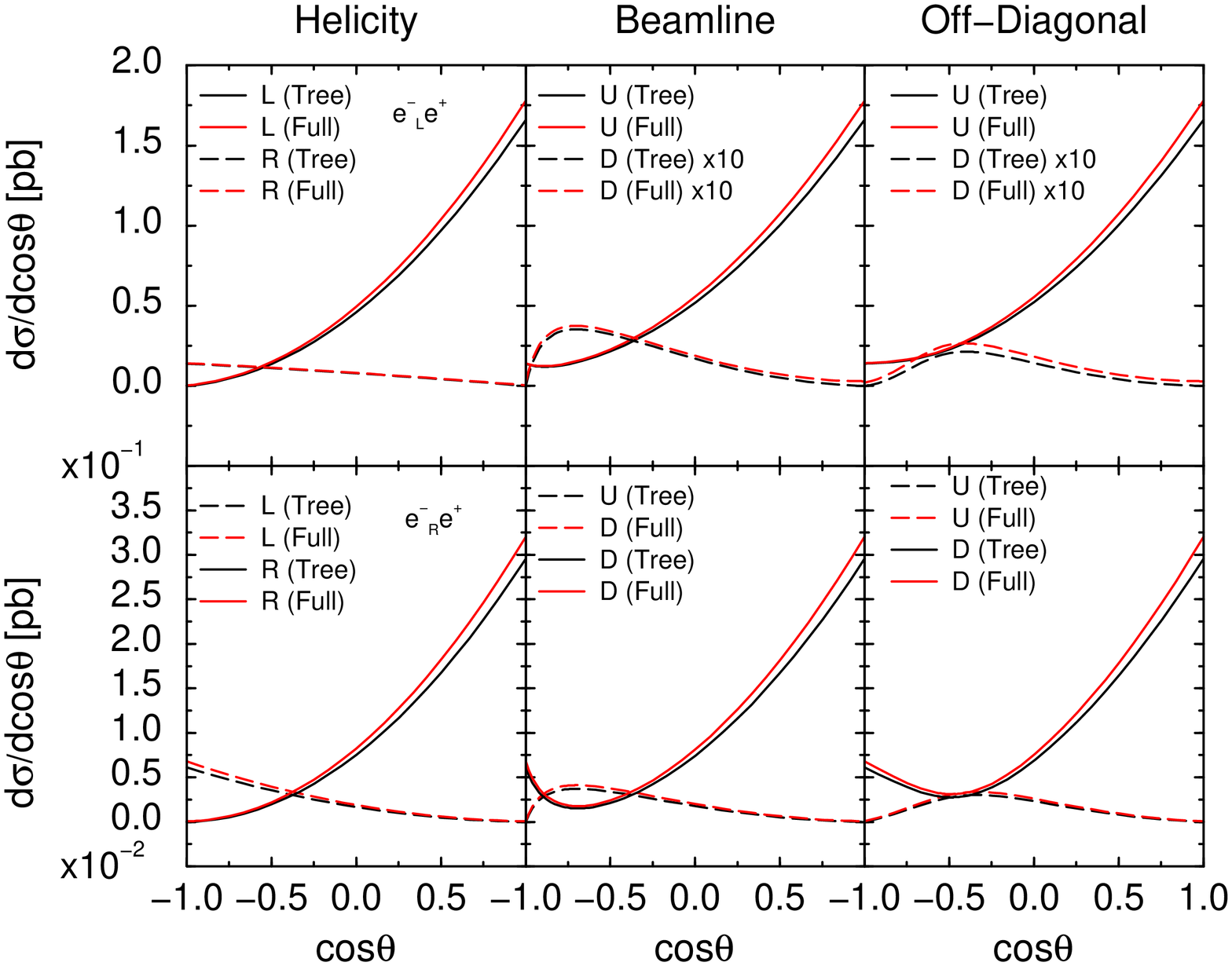,angle=-90,width=15cm}
\caption[The cross-sections for the process $e^+e^- \rightarrow b' X (\bar{b}'~{\rm or}~ \bar{b}' g)$ 
in the helicity, beamline and off-diagonal bases 
at $\sqrt{s} = 1000 {\rm GeV}$.]{The cross-sections for the process
 $e^+e^- \rightarrow b' X (\bar{b}'~{\rm or}~ \bar{b}' g)$ in the
 helicity, beamline and off-diagonal bases at $\sqrt{s} = 1000 {\rm GeV}$.
The value of $b'$ mass is $250$ {\rm GeV}.}
\label{fig:bpr4}
\end{center} 
\end{figure}
%%%%%%%%%%%%%%%%%%%%%%%%%%%%
\begin{figure}[H]
\begin{center}
        \leavevmode\psfig{file=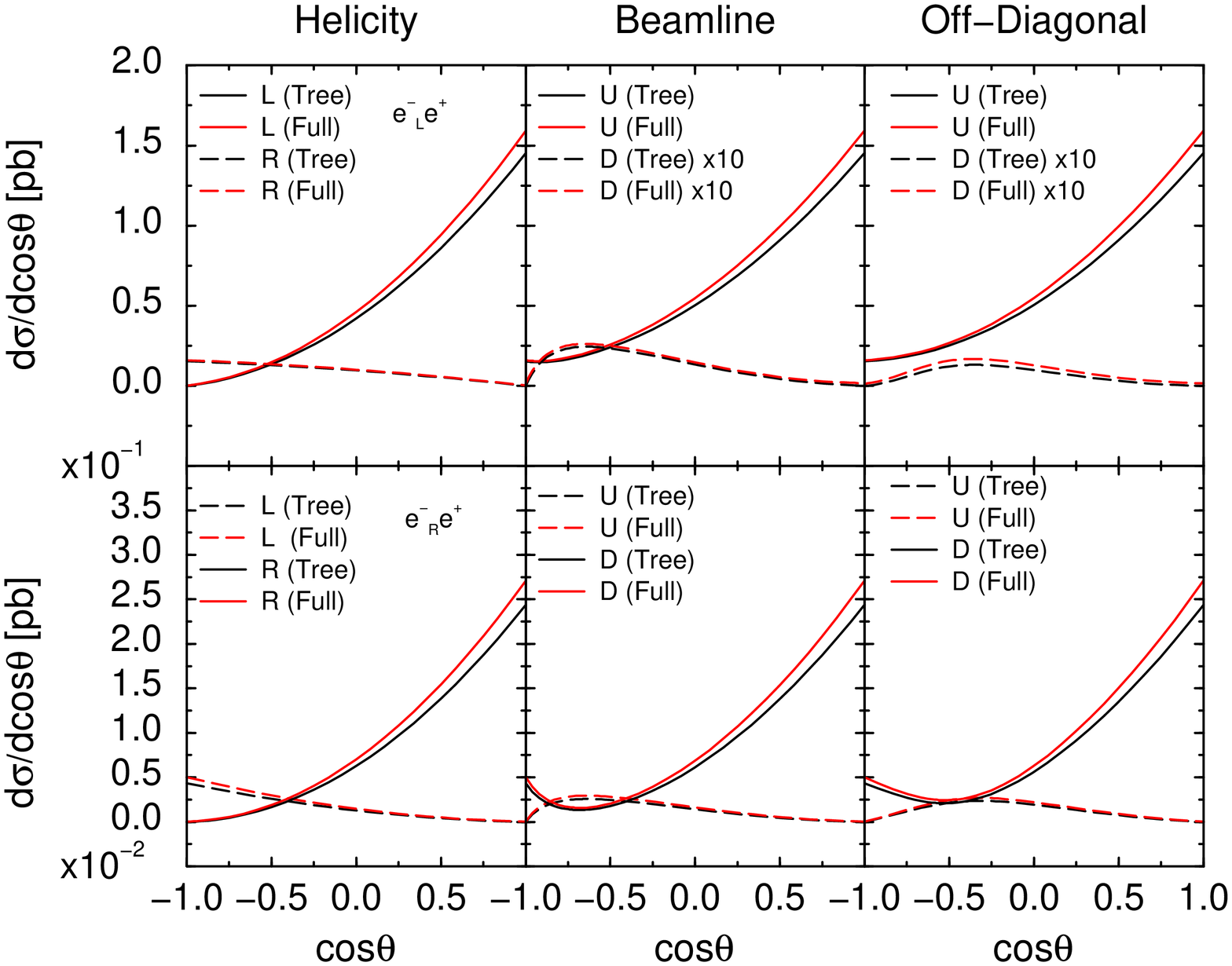,angle=-90,width=15cm}
\caption[The cross-sections for the process $e^+e^- \rightarrow b' X (\bar{b}'~{\rm or}~ \bar{b}' g)$ 
in the helicity, beamline and off-diagonal bases at $\sqrt{s} = 1000 {\rm GeV}$.]{The cross-sections for the process $e^+e^- \rightarrow b' X (\bar{b}'~{\rm or}~ \bar{b}' g)$ in the helicity, beamline and off-diagonal bases at $\sqrt{s} = 1000 {\rm GeV}$. The value of $b'$ mass is $300$ {\rm GeV}.}  
\label{fig:bpr5}
\end{center} 
\end{figure}
%
%----------------------- Summary ------------------------------------
%\input summ.tex
\chapter{Summary and Outlook}

We have studied in detail the spin-spin correlations for top quark
productions and spin-angular correlations for its decay at polarized
$e^+e^-$ linear colliders as well as an brief 
introduction to the spinor helicity method which is a powerful
calculational technique to obtain analytically compact expression for
the complicated process. 

For the top quark productions,
we presented the polarized cross-sections using a ``generic'' spin basis. 
We introduced three characteristic bases; ``helicity'',
``beamline'' and ``off-diagonal'' bases, 
and discussed the differences among these bases.
At threshold energy, the beamline basis is equivalent to the off-diagonal
basis.
In the high energy limit, the beamline and the helicity bases
coincide.
However, at moderate energies, there are remarkable differences among
these bases.
We first analyzed the polarized cross-sections in these bases at the center of 
mass energy far above the threshold at the leading order of perturbation 
theory.
We have shown that the most optimal spin basis is the off-diagonal basis.
In this basis, 
the like spin configurations, UU ($t_{\uparrow},~\bar{t}_{\uparrow}$)
and DD ($t_{\downarrow},~\bar{t}_{\downarrow}$), vanish, and
UD ($t_{\uparrow},~\bar{t}_{\downarrow}$) configuration 
dominates to total cross-section, for instance, occupies $99.88 \%$ at
the $400$ GeV $e^+ e^-$ collider.
This result show that the top (anti-top) quark spin is strongly
correlated with the positron (electron) spin and top quark pairs are
produced in a unique spin configuration in the off-diagonal basis.  
We have also discussed why the top (anti-top) quark spin is associated
with the positron (electron) spin and the relation between the beamline
and off-diagonal bases.
We have clarified that the smallness of parameters $f_{LR} \sim
f_{RL}\sim 0$ is a key point to answer these questions.

For the top quark decay, we investigated the correlation between the
direction of top quark spin and the direction of motion of the
decay products. 
The consequence was that the anti-lepton and $d$-type quark are maximally
correlated with the top quark spin.
The longitudinal and the transverse $W$ boson contribution to the top
quark decay was also explained.
The interference effect between the longitudinal and the transverse $W$ boson 
produces significant contributions to the angular distribution of the
decay products. 
Thus the topology of the top quark decay provides us important
information to obtain a constraint to the Standard Model.
We also suggested that the spin correlations can be measured by analyzing
the decay products of the top and ant-top quark.

We have extended the analyses of top quark productions at the leading
order to the next to leading order of QCD, and
studied the ${\cal O} (\alpha_s )$ QCD 
corrections to top quark productions in a generic spin basis. 
The QCD corrections introduce two effects not included in the tree level
approximation.  
One is the modification of the coupling of the top and anti-top quarks
to $\gamma$ and $Z$ bosons due to the virtual corrections. 
The other is the presence of the real gluon emission process. 
First, we consider the QCD corrections in the soft gluon
approximation to see the effects of the modified 
$\gamma / Z - t - \bar{t}$ vertex.
Using this approximation, we found that the tree level 
off-diagonal basis continues to make the like spin components vanish 
and that the effects of the induced anomalous magnetic moment are small.
Next we analyzed the full QCD corrections at the one-loop level. 
When we consider the three particle final state, 
the top and anti-top quarks are not necessarily
produced back to back. 
So we have calculated the inclusive top (anti-top) quark production.  
We have given an exact analytic form for the
differential cross-section with an arbitrary orientation of the top
quark spin. 
Our numerical studies show that
the ${\cal O} (\alpha_s )$ QCD corrections enhance the 
tree level results and modify only slightly the spin
orientation of the produced top quark.
In the kinematical region where the emitted gluon has small
energy, it is natural to expect that the real gluon emission effects
introduce only a multiplicative correction to the tree level result.
Therefore only \lq\lq hard\rq\rq\ gluon emission could possibly modify
the top quark spin orientation. 
What we have found, by explicit calculation, is that this effect
is numerically very small. 
The size of the QCD corrections to the total cross-section is summarized
in Table 4.2 of Chapter 4.
At $\sqrt{s} = 400 \, {\rm GeV}$, the enhancement from the tree level
result is $\sim 30 \%$ whereas at higher energies, $800 \sim 4000$ GeV,
it is at the $2 \sim 6 \%$ level. 
Near the threshold, the QCD corrections have a singular
behavior in $\beta$, the speed of the produced quark. 
This factor enhances the value of the correction at smaller energies.
The size of these corrections is reasonable for QCD.
On the other hand, the change of the orientation of top quark spin is
quite small. The deviation from the enhanced tree level result is less
than a few percent.
Actually, the fraction of the top quarks in the dominant (up) spin
configuration for $e_L^- e^+$ scattering in the off-diagonal basis is
more than $93 \%$ at all energies we have considered.
We can, therefore, conclude that the results of the tree level analysis
are not altered even after including QCD radiative corrections except
for a multiplicative enhancement. 
Since the the dominant enhancement can be described by the soft-gluon
approximation,
we have also analyzed the validity of the soft-gluon approximation and
shown the efficiency of this approximation.

Finally two additional discussions regarding the spin correlations have
been presented.
One is the analysis of the polarized cross-sections under the
approximation $f_{LR}/f_{LL}= f_{RL}/f_{RR} = 0$ at the QCD one-loop level.
The other is the pair production of the ``fourth'' generation quark $b'$
with the charge $-1/3$.
We have estimated the polarized cross-sections for the $b'$ quark
productions with $b'$ mass between $150$ GeV and $300$ GeV,
at the center of mass energy $500$ and $1000$ GeV.
For the $e^-_L e^+$ scattering process, the differential cross-sections
in three bases show the same behavior as the top quark productions
except for magnitude of total cross-section.
The dominant configuration is the spin-up $b'$ quark, which is strongly
correlated with the positron spin.
However, for the $e^-_R e^+$ scattering process, the dominant
configuration in three bases almost give the same contribution to the total
cross-section. 
This fact is caused by the ratio of the coupling, $|f_{RL}/f_{RR}|~\sim
0.5$.
This situation is the characteristic feature of $b'$ quark productions.

As has been discussed in this thesis and many articles,
there are strong correlations between the orientation of the
spin of produced top (anti-top) quark and the angular distribution
of its decay products.
And the radiative corrections to these correlation seem to be very small 
as we have explicitly shown for the production process.
Therefore, measuring the top quark spin orientation will give us 
important information on the top quark sector of the Standard Model
as well as possible physics beyond the Standard Model.

\newpage
%----------------------- Appendix ---------------------------------
%\input app.tex
%----------------------- Appendix ---------------------------------
\appendix
\chapter{ Notation and Conventions }
%
%%%%%%%%%%%%%%%%%%%%%%%%%%%%%%%%%%%%%%%%%%%%%%%%%%%%%%%%%%%
\section{ Natural Unit and Convention }
%%%%%%%%%%%%%%%%%%%%%%%%%%%%%%%%%%%%%%%%%%%%%%%%%%%%%%%%%%%
%
\begin{itemize}
\item
%%%%%%%%%%%%%%%%%%%%%%%%%%
Natural Unit:
%%%%%%%%%%%%%%%%%%%%%%%%%%  
\begin{eqnarray}
  \hbar ~ = ~ c~=~1,~~
( \hbar ~ = ~ \frac{h}{ 2 \pi } )~. \nonumber
\end{eqnarray}
\item
%%%%%%%%%%%%%%%%%%%%%%%%%%
Transform rules of Unit:
%%%%%%%%%%%%%%%%%%%%%%%%%%
\begin{eqnarray}
1~{\rm kg} & = & 5.61 \times 10^{26} ~{\rm GeV} :
~\left(
\frac{{\rm GeV}}{{\rm c}^2}
\right)
,\nonumber \\
1~m & = & ~ 5.07 \times 10^{15} ~{\rm GeV}^{-1} :
~\left(
\frac{ \hbar c }{{\rm GeV}}
\right),
\nonumber \\
1~{\rm sec} & = & 1.52 \times 10^{24} ~{\rm GeV}^{-1} :
\left(
\frac{ \hbar }{{\rm GeV}}
\right),
\nonumber \\
1~{\rm TeV} & = & 10 ^{3}  ~{\rm GeV} 
    ~ = ~ 10^{6}   ~{\rm MeV} 
    ~ = ~ 10 ^{9}  ~{\rm KeV}
    ~ = ~ 10 ^{12} ~{\rm eV}, \nonumber \\ 
1~{\rm fermi} & = & 1~{\rm fm}
      ~ = ~ 10 ^{-13} ~{\rm cm}
      ~ = ~ 10 ^{-15} ~{\rm m}
      ~ = ~ 5.07 ~{\rm GeV}^{-1}, \nonumber \\ 
\left( 1~{\rm fermi} \right) ^{2} 
      & = & 10 ~   {\rm mb} 
      ~=~ 10 ^{4} ~\mu {\rm b}
      ~=~ 10 ^{7}  ~{\rm n b}
      ~=~ 10 ^{10} ~{\rm p b}
      ~=~ 10 ^{13} ~{\rm f b}, \nonumber \\
1~b & = & 10 ^{2}~{\rm fm} ^{2}
~ = ~ 10^{-24}  ~{\rm cm}^{2} 
~ = ~ 10 ^{-28} ~{\rm m}^{2}, \nonumber \\
\left(  1~{\rm GeV}  \right)^{ -2 } 
& = &  0.389~{\rm mb }  
~ = ~  3.89~\times~10^{8} ~{\rm pb}~. \nonumber 
\end{eqnarray}
\end{itemize}
%
%%%%%%%%%%%%%%%%%%%%%%%%%%%%%%%%%%%%%%%%%%%%%%%%%%%%%%%%%%%
\section{Metric Tensor}
%%%%%%%%%%%%%%%%%%%%%%%%%%%%%%%%%%%%%%%%%%%%%%%%%%%%%%%%%%%
%
Our Metric Tensor $ g_{ \mu \nu } $ in  Minkowski Space 
$ \{ x^\mu~:(~\mu~=~0,1,2,3~) \}$ ,
\begin{eqnarray}
g_{ \mu \nu }~=~g^{ \mu \nu }~=~\left(
\begin{array}{cccc}
+1 & 0  & 0  & 0  \\
0  & -1 & 0  & 0  \\
0  & 0  & -1 & 0  \\
0  & 0  & 0  & -1 
\end{array}
\right)~. 
\nonumber 
\end{eqnarray}
%
%%%%%%%%%%%%%%%%%%%%%%%%%%%%%%%%%%%%%%%%%%%%%%%%%%%%%%%%%%%
\section{Coordinates and Momenta}
%%%%%%%%%%%%%%%%%%%%%%%%%%%%%%%%%%%%%%%%%%%%%%%%%%%%%%%%%%%
%
The space-time coordinates $( t, x, y, z) \equiv ( t, \mbox{\boldmath{$x$}})$ are
denoted by the covariant four-vector:
\begin{eqnarray}
x^{\mu} & \equiv & ( x^{0}, x^{1}, x^{2}, x^{3}) \nonumber \\
& \equiv & (t,x,y,z)~. \nonumber
\end{eqnarray}
The covariant four-vector $x^{\mu}$ is obtained by the sign of the space 
components:
\begin{eqnarray}
x_{\mu} & \equiv & ( x_{0}, x_{1}, x_{2}, x_{3}) \nonumber \\
& \equiv & (t,-x,-y,-z)    \nonumber \\
& = & g_{\mu \nu} x^{\nu} ~.\nonumber
\end{eqnarray}
The inner product is given by
\begin{eqnarray}
x \cdot x & \equiv & x^{\mu} x_{\mu} \nonumber \\
  & = &      t^2 - \mbox{{\boldmath $x$}}^2  ~. \nonumber 
\end{eqnarray}
Momentum vectors are similarly defined by
\begin{eqnarray}
p^{\mu} & \equiv & ( p^{0}, p^{1}, p^{2}, p^{3}) \nonumber \\
& \equiv & (E,p_{x},p_{y},p_{z}) \nonumber \\
& = & g^{\mu \nu} p_{\nu}       ~.\nonumber 
\end{eqnarray}
The inner product is given by
\begin{eqnarray}
p_{1} \cdot p_{2} 
& \equiv & p_{1}^{\mu} p_{2 \mu} \nonumber \\
& = & E_{1}E_{2} - \mbox{{\boldmath $p$}}_{1} 
		       \cdot 
		   \mbox{{\boldmath $p$}}_{2} ~. \nonumber 
\end{eqnarray}
%
%%%%%%%%%%%%%%%%%%%%%%%%%%%%%%%%%%%%%%%%%%%%%%%%%%%%%%%%%%%
\section{ Pauli Matrices}
%%%%%%%%%%%%%%%%%%%%%%%%%%%%%%%%%%%%%%%%%%%%%%%%%%%%%%%%%%%
%
\begin{itemize}
\item
%%%%%%%%%%%%%%%%%%%%%%%%%%
Pauli $ \sigma $ Matrices:
%%%%%%%%%%%%%%%%%%%%%%%%%%
\[   \sigma_{1}~=~\left( 
\begin{array}{cc}
0  & 1  \\
1  & 0 
\end{array}
	  \right)~~,~~
\sigma_{2}~=~\left( 
\begin{array}{cc}
0  & -i  \\
i  & 0 
\end{array}
     \right)~~,~~
\sigma_{3}~=~\left( 
\begin{array}{cc}
1  & 0  \\
0  & -1 
\end{array}
     \right)~.
\]
\item
%%%%%%%%%%%%%%%%%%%%%%%
The Relations of Pauli Matrices:
%%%%%%%%%%%%%%%%%%%%%%%
\begin{eqnarray}
\{ \sigma_{i} , \sigma_{j} \} & = & 2 \delta_{ij} ,
\nonumber \\
\left [ \sigma_{i} ,  \sigma_{j} \right] & = & 2 i \varepsilon _{ijk} \sigma_{k}~.
\nonumber
\end{eqnarray}
\item
%%%%%%%%%%%%%%%%%%%%%%%
The Properties of Pauli Matrices:
%%%%%%%%%%%%%%%%%%%%%%%
\[ 
\sigma_{i}^{\dagger} ~=~ \sigma_{i} ,~
{\rm det}(\sigma_{i}) ~=~ -1 ,~
\Tr[ \sigma_{i} ] ~=~ 0~.
\] 
\item
%%%%%%%%%%%%%%%%%%%%%%%%%%%
Useful Relation:
%%%%%%%%%%%%%%%%%%%%%%%%%%%
\begin{eqnarray}
( \mbox{\boldmath{ $\sigma$ }} \cdot \mbox{\boldmath{ $a$ }} )
( \mbox{\boldmath{ $\sigma$ }} \cdot \mbox{\boldmath{ $b$ }} )
~=~
( \mbox{\boldmath{ $a$ }} \cdot \mbox{\boldmath{ $b$ }} ) +
i \mbox{\boldmath{ $\sigma$ }} \cdot (\mbox{\boldmath{ $a$ }} 
\times \mbox{\boldmath{ $b$ }})~.
\nonumber 
\end{eqnarray}
\end{itemize}
%%%%%%%%%%%%%%%%%%%%%%%%%%%%%%%%%%%%%%%%%%%%%%%%%%%%%%%%%%%
\section{$ \gamma $ Matrices}
%%%%%%%%%%%%%%%%%%%%%%%%%%%%%%%%%%%%%%%%%%%%%%%%%%%%%%%%%%%
%
\begin{itemize}
\item
%%%%%%%%%%%%%%%%%%%%%%%%%%
Dirac $\gamma$ Matrices,
%%%%%%%%%%%%%%%%%%%%%%%%%%
~$ \gamma^{ \mu }~=~( \gamma^{0} , \mbox{\boldmath{ $\gamma$ }} )$:
\begin{eqnarray}
\{ \gamma^{\mu} , \gamma^{\nu} \} 
& = &     \gamma^{\mu} \gamma^{\nu} ~+~  \gamma^{\nu} \gamma^{\mu}
~=~      2~g^{\mu \nu} , \nonumber \\
\left[ \gamma^{\mu} , \gamma^{\nu} \right] 
& = &     \gamma^{\mu} \gamma^{\nu} ~-~  \gamma^{\nu} \gamma^{\mu}
~=~    - 2 i~ \sigma^{\mu \nu}   , \nonumber \\
\gamma_{\mu } & = &  g_{\mu \nu} \gamma^{\nu}
      ~ = ~
( \gamma^{0},- \mbox{\boldmath{ $\gamma$ }}) ~. \nonumber 
\end{eqnarray}
\item
%%%%%%%%%%%%%%%%%%%%%%%%%%
$\gamma_{5}$ Matrix:
%%%%%%%%%%%%%%%%%%%%%%%%%%
\begin{eqnarray}
\lefteqn{
\gamma _{5}
}~  & \equiv  & \gamma^{5} 
    ~ = ~  i \gamma^{0} \gamma^{1} \gamma^{2} \gamma^{3} 
    ~ = ~ - \frac{i}{ 4 ! } \varepsilon _{ \mu \nu \lambda \sigma }
\gamma ^{ \mu } \gamma ^{ \nu } \gamma ^{ \lambda } \gamma ^{ \sigma }~,
\nonumber \\
\varepsilon ^{0 1 2 3}~ & = & - \varepsilon _{0 1 2 3}~=~ + 1 ~. \nonumber 
\end{eqnarray}
\item
%%%%%%%%%%%%%%%%%%%%%%%%%%%%%%%%%%%%%%%%%%%%%%%%%%%%
The Relation of $ \gamma _{5} $~Matrix:  
%%%%%%%%%%%%%%%%%%%%%%%%%%%%%%%%%%%%%%%%%%%%%%%%%%%%
\begin{eqnarray}
\gamma _{5}^{2} & = & 1 , \nonumber \\
\{  \gamma_{5} , \gamma^{\mu} \} & = & 0 ~.\nonumber
\end{eqnarray}
\item
%%%%%%%%%%%%%%%%%%%%%%%%%%%%%%%%%%%%%%%%%%%%%%%%%%%%
Familiar Presentation of $ \gamma $:
%%%%%%%%%%%%%%%%%%%%%%%%%%%%%%%%%%%%%%%%%%%%%%%%%%%%
%
\[ 
\gamma^{0}~=~\left(  
\begin{array}{cc}
  1  &  0  \\
  0  & -1 
\end{array}
\right)~,~
\mbox{{\boldmath $\gamma$}}~=~
\left( 
\begin{array}{cc}
0                          & \mbox{ \boldmath$\sigma$}  \\
-\mbox{ \boldmath$\sigma$}  &    0 
\end{array}
\right)~~,~~
\gamma_{5}~=~ 
\left(
\begin{array}{cc}
  0  &  1  \\
  1  &  0 
\end{array}
\right)~,
\]
where {\boldmath $\sigma$} is given by 
\[
\mbox{{\boldmath $\sigma$}}~=~
(\sigma_{1},\sigma_{2},\sigma_{3}). 
\]

\end{itemize}
%
%%%%%%%%%%%%%%%%%%%%%%%%%%%%%%%%%%%%%%%%%%%%%%%%%%%%%%%%%%%
\section{ Dirac Spinors }
%%%%%%%%%%%%%%%%%%%%%%%%%%%%%%%%%%%%%%%%%%%%%%%%%%%%%%%%%%%
%
Dirac spinors for fermion with momentum $p$, $u_{\lambda}(p)$ and
for anti-fermion with momentum $p$ $v_{\lambda}(p)$ satisfy following 
equations.
($p$: momentum , $\lambda$: spin components , $m$: fermion mass.)
\begin{itemize}
\item 
%%%%%%%%%%%%%%%%%%%%%%%%%%
Equations of Motion: 
%%%%%%%%%%%%%%%%%%%%%%%%%%
%
\begin{eqnarray}
( p\hspace{-6pt}/~-~m ) u_{\lambda}(p)
& = &
0 ,\nonumber \\
( p\hspace{-6pt}/~+~m ) v_{\lambda}(p)
& = &
0 ~. 
\nonumber
\end{eqnarray}
\item
%%%%%%%%%%%%%%%%%%%%%%%%%%
Normalization: 
%%%%%%%%%%%%%%%%%%%%%%%%%%
%
\begin{eqnarray}
\bar{u}_{ \lambda }(p) u_{ \lambda' }(p)
& = &
2m ~ \delta_{ \lambda \lambda' } ,
\nonumber \\
\bar{v}_{ \lambda }(p) v_{ \lambda' }(p)
& = &
-2m ~ \delta_{ \lambda \lambda' }~. 
\nonumber 
\end{eqnarray}
\item
%%%%%%%%%%%%%%%%%%%%%%%%%%
Projection Operators (In the case of Spin Sum): 
%%%%%%%%%%%%%%%%%%%%%%%%%%
%
\begin{eqnarray}
\sum_{ \lambda } u_{ \lambda }(p) \bar{u}_{ \lambda }(p)
& = & p \hspace{-6pt}/~+~m  , \nonumber \\
\sum_{ \lambda } v_{ \lambda }(p) \bar{v}_{ \lambda }(p)
& = & p \hspace{-6pt}/~-~m  ~. \nonumber 
\end{eqnarray}
\end{itemize}
%
%%%%%%%%%%%%%%%%%%%%%%%%%%%%%%%%%%%%%%%%%%%%%%%%%%%%%%%%%%%
\section{ Spin Vector}
%%%%%%%%%%%%%%%%%%%%%%%%%%%%%%%%%%%%%%%%%%%%%%%%%%%%%%%%%%%
%
spin vector $s^{\mu}$ for fermions is taken to be,
\begin{eqnarray}
s^{\mu} ~\equiv~ \left(
    \frac{
	   \vec{p} \cdot \vec{s}
	 }{m},~
\vec{s} + \frac{ 
	   \vec{p} ( \vec{p} \cdot \vec{s} )
	}{ m( p^{0} + m ) }
	 \right)~. 
\nonumber 
\end{eqnarray}
where 
\begin{eqnarray}
\left( \vec{s} \right)^{2} ~=~1,~
~s^{2} ~=~ s^{ \mu } s_{ \mu } ~=~ -1,~
s^{\mu}~p_{\mu} ~=~ s \cdot p~=~0 ~. \nonumber 
\end{eqnarray}
\begin{itemize}
\item
%%%%%%%%%%%%%%%%%%%%%%%%%%%%%%%%%%%%%%%%%%%%%%%%%%%%
The Relations between Spin Vector and Spinor:
%%%%%%%%%%%%%%%%%%%%%%%%%%%%%%%%%%%%%%%%%%%%%%%%%%%%
%
\begin{eqnarray}
\bar{u}( p , s ) \gamma^{\mu} \gamma_{5} u( p , s )
~=~ 2 m ~ s^{ \mu }~. 
\nonumber
\end{eqnarray}
\item
%%%%%%%%%%%%%%%%%%%%%%%%%%
Projection Operators:
%%%%%%%%%%%%%%%%%%%%%%%%%%
%
\begin{eqnarray}
u( p , s ) \bar{u}( p , s )
& = &
( p \hspace{-6pt}/ ~+~ m )
\frac{ 1 + \gamma_{5} s \hspace{-6pt}/ }{2} 
,\nonumber \\
v( p , s ) \bar{v}( p , s )
& = &
( p \hspace{-6pt}/ ~-~ m )
\frac{ 1 + \gamma_{5} s \hspace{-6pt}/}{2} 
~.\nonumber
\end{eqnarray}
\end{itemize}
\newpage
%
%%%%%%%%%%%%%%%%%%%%%%%%%%%%%%%%%%%%%%%%%%%%%%%%%%%%%%%%%%%
%
\chapter{ Useful Formulae }
%
%%%%%%%%%%%%%%%%%%%%%%%%%%%%%%%%%%%%%%%%%%%%%%%%%%%%%%%%%%%
\section{Dirac $\gamma$ Matrices}
%%%%%%%%%%%%%%%%%%%%%%%%%%%%%%%%%%%%%%%%%%%%%%%%%%%%%%%%%%%
%
\begin{itemize}
\item
$\gamma^{\dagger}$ and $\overline{\gamma}$:
\begin{eqnarray}
\gamma^{\mu  \dagger} & = & \gamma ^{0} \gamma ^{ \mu } \gamma ^{0} 
~, \nonumber  \\
\gamma ^{ 0 \dagger } & = &   \gamma ^{0} ~,~
\gamma ^{ k \dagger } ~ = ~ - \gamma ^{k} ~~
({\rm  where }~~k~=~1,~2,~3 )~, \nonumber  \\
\gamma ^{ 5 \dagger } & = &   \gamma ^{5} ~,~
\sigma _{ \mu \nu }^{ \dagger } ~ = ~ \sigma _{ \mu \nu } ~, \nonumber  \\
\overline{ \Gamma }   &\equiv& \gamma ^{0} \Gamma ^{ \dagger } \gamma ^{0}
~, \nonumber  \\
\overline{ \gamma }^{ \mu } & = & \gamma ^{0} \gamma ^{ \mu \dagger } \gamma ^{0}
~ = ~ \gamma ^{ \mu } ~,~\overline{ \gamma ^{ \mu } \gamma ^{5} }
~=~ \gamma ^{ \mu } \gamma^{ 5 } ~, \nonumber  \\
\overline{ \sigma } ^{ \mu \nu }
& = &  \gamma ^{0} \sigma ^{ \mu \nu  \dagger } \gamma ^{0}
~ = ~  \sigma ^{ \mu \nu } ~, \nonumber  \\
\overline{ i \gamma _{5} }
& = & \gamma ^{0} \left(
		 i \gamma _{5}
	  \right)  \gamma ^{0}
~=~ i \gamma _{5} ~, \nonumber  \\
\overline{ \slashs{a} \slashs{b} \slashs{c} \slashs{d} \cdots  \slashs{p}}
& = & \slashs{p} \cdots \slashs{c} \slashs{b} \slashs{a}~.\nonumber
\end{eqnarray} 
\item 
%%%%%%%%%%%%%%%%%%%%%%%%%%
Matrix Element:
%%%%%%%%%%%%%%%%%%%%%%%%%%
\begin{eqnarray}
\left[ \bar{ u } (p',s') ~ \Gamma ~u(p,s) \right] ^{ \dagger }
~=~ \bar{ u } (p,s) ~ \overline{ \Gamma } ~u(p',s') ~, \nonumber
\end{eqnarray}
\item
%%%%%%%%%%%%%%%%%%%%%%%%%%%%%%%
Trace Theorems:
%%%%%%%%%%%%%%%%%%%%%%%%%%%%%%%
\begin{eqnarray}
\Tr \left[ \gamma^{ {\mu}_{1} } \gamma^{ {\mu}_{2} } 
   \cdots \gamma^{ {\mu}_{n} }
\right]
& = &  0~~ ( \mbox{ for $ n = $ odd }) ~, \nonumber \\
\Tr~1 & = & 4 ~, \nonumber \\
\Tr \left[ \gamma ^{ \mu } \gamma ^{ \nu } \right] 
& = & 4 g ^{ \mu \nu } ~, \nonumber \\
\Tr \left[ \gamma ^{ \mu } \gamma ^{ \nu }  
   \gamma ^{ \lambda } \gamma ^{ \sigma }
\right]
& = & 4 \left ( g^{ \mu \nu } g^ { \lambda \sigma } ~-~
	g^{ \mu \lambda } g^{ \nu \sigma } ~+~
	g^{ \mu \sigma } g^ { \nu \lambda } \right ) ~, \nonumber \\
\Tr \left[  \gamma^{ \lambda } \gamma^{ \mu } \gamma^{ \nu }
    \gamma^{ \rho } \gamma^{ \sigma } \gamma^{ \tau } \right]
& = & ~g^{ \lambda \mu } 
\Tr\left[  
      \gamma^{ \nu } \gamma^{ \rho } \gamma^{ \sigma } \gamma^{ \tau }
  \right] -
g^{ \lambda \nu }
\Tr \left[ 
      \gamma^{ \mu } \gamma^{ \rho } \gamma^{ \sigma } \gamma^{ \tau }
   \right] + 
g^{ \lambda \rho } 
\Tr \left[ 
      \gamma^{ \mu } \gamma^{ \nu } \gamma^{ \sigma } \gamma^{ \tau }
   \right] \nonumber \\
& & -  g^{ \lambda \sigma }
\Tr \left[ 
      \gamma^{ \mu } \gamma^{ \nu } \gamma^{ \rho } \gamma^{ \tau }
   \right] +
g^{ \lambda \tau }
\Tr \left[ 
      \gamma^{ \mu } \gamma^{ \nu } \gamma^{ \rho } \gamma^{ \sigma }
  \right] ~, \nonumber \\
\Tr \left[ 
      \gamma^{{\mu}_{1}} \gamma^{{\mu}_{2}} \cdots \gamma^{{\mu}_{n}}
  \right]
& = & 
g^{ {\mu}_{1} {\mu}_{2} }
\Tr \left[
      \gamma^{{\mu}_{3}} \gamma^{{\mu}_{4}} \cdots \gamma^{{\mu}_{n}}
  \right] ~-~
g^{{\mu}_{1} {\mu}_{3}}
\Tr \left[
      \gamma^{{\mu}_{2}} \gamma^{{\mu}_{4}} \cdots \gamma^{{\mu}_{n}}
  \right] \nonumber \\
& & + \cdots  ~ + ~
g^{ {\mu}_{1} {\mu}_{n} }
\Tr \left[
      \gamma^{{\mu}_{1}} \gamma^{{\mu}_{2}} \cdots \gamma^{{\mu}_{n-1}}
  \right] ~~( \mbox{ for $ n = $ even }) ~, \nonumber \\
\Tr \left[ 
\gamma _{5} \gamma^{{\mu}_{1}} \gamma^{{\mu}_{2}} \cdots \gamma^{{\mu}_{n}}
\right]  
& = & 0~~ ( \mbox{ for $ n \leq $ 3 } ) \nonumber \\
\Tr \left[
\gamma _{5} \gamma ^{ \mu } \gamma ^{ \nu }
\gamma ^{ \lambda } \gamma ^{ \sigma }
\right]
& = & - 4i \varepsilon ^{ \mu \nu \lambda \sigma } ~, \nonumber \\
\Tr \left[ 
\gamma _{5} \gamma ^{ \lambda } \gamma ^{ \mu } 
\gamma ^{ \nu } \gamma ^{ \rho } \gamma ^{ \sigma }
\gamma ^{ \tau }
\right]
& = &
~- 4i \bigl(
 \varepsilon ^{ \lambda \mu \nu \rho }g^{ \sigma \tau } ~-~
 \varepsilon ^{ \lambda \mu \nu \sigma }g^{ \rho \tau } ~+~
 \varepsilon ^{ \lambda \mu \nu \tau }g^{ \rho \sigma } \nonumber \\
& & \hspace*{1cm} ~-~
 \varepsilon ^{ \nu \tau \sigma \rho }g^{ \lambda \mu } ~ + ~
 \varepsilon ^{ \mu \tau \sigma \rho }g^{ \lambda \nu } ~ - ~
 \varepsilon ^{ \lambda \tau \sigma  \rho }g^{ \mu \nu }
\bigr) ~. \nonumber
\end{eqnarray}
\item
%%%%%%%%%%%%%%%%%%%%%%%%%%%%%%%
$ \gamma $ Identities: 
%%%%%%%%%%%%%%%%%%%%%%%%%%%%%%%
\begin{eqnarray}
\gamma_{\mu} \gamma^{\mu} & = & 4 ~, \nonumber \\
\gamma_{\mu} \gamma^{\nu} \gamma^{\mu} 
& = & -2 \gamma^{\nu} ~, \nonumber \\
\gamma_{\mu} \gamma^{\lambda} \gamma^{\nu} \gamma^{\mu} 
& = & 4 g_{\lambda \nu}  ~, \nonumber \\
\gamma_{\mu} \gamma^{\lambda} \gamma^{\nu} \gamma^{\sigma} \gamma^{\mu}
& = & -2 \gamma^{\sigma} \gamma^{\nu} \gamma^{\lambda} ~, \nonumber \\
\gamma_{\mu} \gamma^{\lambda} \gamma^{\nu} 
\gamma^{\sigma} \gamma^{\rho} \gamma^{\mu}
& = & 
2 ( \gamma^{\rho} \gamma^{\lambda} \gamma^{\nu} \gamma^{\sigma}
+   \gamma^{\sigma} \gamma^{\nu} \gamma^{\lambda} \gamma^{\rho}
) ~, \nonumber 
\end{eqnarray}
\item
%%%%%%%%%%%%%%%%%%%%%%%%%%%%%%%
Version:
%%%%%%%%%%%%%%%%%%%%%%%%%%%%%%%
\begin{eqnarray}
\gamma _{\mu}  \slashs{a}  \gamma^{\mu} 
& = & - 2 \slashs{a} ~, \nonumber \\ 
\gamma_{\mu} \slashs{a} \slashs{b} \gamma^{\mu}
& = &  4 a \cdot b  ~, \nonumber \\
\gamma_{\mu} \slashs{a} \slashs{b} \slashs{c} \gamma^{\mu}
& = & - 2 \slashs{c} \slashs{b} \slashs{a} ~, \nonumber \\ 
\gamma_{\mu} \slashs{a} \slashs{b} \slashs{c} \slashs{d} \gamma^{\mu}
& = &   2 \left(
\slashs{d} \slashs{a} \slashs{b} \slashs{c} + 
\slashs{c} \slashs{b} \slashs{a} \slashs{d}
   \right ) \nonumber
\end{eqnarray}
\item
%%%%%%%%%%%%%%%%%%%%%%%%%%%%%%%
Other Relations:
%%%%%%%%%%%%%%%%%%%%%%%%%%%%%%%
\begin{eqnarray}
g^{\mu \nu} \varepsilon^{\lambda \sigma \tau \rho}-
g^{\mu \lambda} \varepsilon^{\nu \sigma \tau \rho}+
g^{\mu \sigma} \varepsilon^{\nu \lambda \tau \rho}-
g^{\mu \tau} \varepsilon^{\nu \lambda \sigma \rho}+
g^{\mu \rho} \varepsilon^{\nu \lambda \sigma \tau}~=~0 ~, \nonumber \\
( g^{\mu \nu} \varepsilon^{\lambda \sigma \tau \rho}~ = ~
g^{\mu \lambda} \varepsilon^{\nu \sigma \tau \rho}+
g^{\mu \sigma} \varepsilon^{\lambda \nu \tau \rho}+
g^{\mu \tau} \varepsilon^{\lambda \sigma  \nu \rho}+
g^{\mu \rho} \varepsilon^{\lambda \sigma \tau \nu} )~. \nonumber
\end{eqnarray}
\end{itemize}
%
%
%%%%%%%%%%%%%%%%%%%%%%%%%%%%%%%%%%%%%%%%%%%%%%%%%%%%%%%%%%%
\section{Conventions for Dimensional Regularization}
%%%%%%%%%%%%%%%%%%%%%%%%%%%%%%%%%%%%%%%%%%%%%%%%%%%%%%%%%%%
%
\begin{itemize}
\item
%%%%%%%%%%%%%%%%%%%%%%%%%%%%%%%%%%%%%%%%%
$ \gamma $ Identities~(~D-dimension~):
%%%%%%%%%%%%%%%%%%%%%%%%%%%%%%%%%%%%%%%%%
\begin{eqnarray}
\gamma_{\mu} \gamma^{\mu} & = & D ~, \nonumber \\
\gamma_{\mu} \gamma^{\nu} \gamma^{\mu}
& = & (2 - D) \gamma^{\nu} ~, \nonumber \\
\gamma_{\mu} \gamma^{\lambda} \gamma^{\nu} \gamma^{\mu} & = & 
4 g _{\lambda \nu} + (D - 4) \gamma^{\lambda} \gamma^{\nu} ~, \nonumber \\
\gamma_{\mu} \gamma^{\lambda} \gamma^{\nu} \gamma^{\sigma} \gamma^{\mu}
& = & -2 \gamma^{\sigma} \gamma^{\nu} \gamma^{\lambda}
- (D - 4) \gamma^{\lambda} \gamma^{\nu} \gamma^{\sigma}~, \nonumber \\
\gamma_{\mu} \gamma^{\lambda} \gamma^{\nu} 
\gamma^{\sigma} \gamma^{\rho} \gamma^{\mu} & = & 
2 (\gamma^{\rho} \gamma^{\lambda} \gamma^{\nu} \gamma^{\sigma} + 
\gamma^{\sigma} \gamma^{\nu} \gamma^{\lambda} \gamma^{\rho} ) +
(D - 4) \gamma^{\lambda} \gamma^{\nu} \gamma^{\sigma} \gamma^{\rho} ~, \nonumber 
\end{eqnarray}
\item
%%%%%%%%%%%%%%%%%%%%%%%%%%%%%%%
Version:
%%%%%%%%%%%%%%%%%%%%%%%%%%%%%%%
\begin{eqnarray}
\gamma_{\mu}  \slashs{a} \gamma ^{ \mu } 
& = &  (2-D) \slashs{a} ~, \nonumber \\ 
\gamma_{\mu} \slashs{a} \slashs{b} \gamma^{\mu}
& = & 4 a \cdot b  ~+~
(D - 4) \slashs{a}\slashs{b} ~, \nonumber \\
\gamma_{\mu} \slashs{a} \slashs{b} \slashs{c} \gamma^{\mu}
& = &  - 2  \slashs{c} \slashs{b} \slashs{a} 
- (D-4) \slashs{a} \slashs{b} \slashs{c} ~, \nonumber \\ 
\gamma_{\mu} \slashs{a} \slashs{b} \slashs{c} \slashs{d} \gamma^{\mu}
& = &  2 \left(  
	 \slashs{d} \slashs{a} \slashs{b} \slashs{c} +
	 \slashs{c} \slashs{b} \slashs{a} \slashs{d}
\right ) + 
(D-4) \slashs{a} \slashs{b} \slashs{c} \slashs{d} ~.\nonumber
\end{eqnarray}
\item
%%%%%%%%%%%%%%%%%%%%%%%%%%%%%%%
Expansion in $ \varepsilon $:
%%%%%%%%%%%%%%%%%%%%%%%%%%%%%%%
%
\begin{eqnarray}
a ^{ \varepsilon } & = & e ^{ \varepsilon \ln a }
~=~ 1 + \varepsilon \ln a + \frac{\varepsilon^{2}}{2!}(\ln a)^{2}
+ \frac{\varepsilon^{3}}{3!} (\ln a)^{3} 
+ {\cal O}(\varepsilon^4) ~, \nonumber \\
\Gamma( \varepsilon ) & = &  \frac{1}{\varepsilon} - \gamma_{E}
+ \frac{\varepsilon}{2} ( \gamma_{E}^{2} + \frac{\pi^{2}}{6}
		) + {\cal O}(\varepsilon^2) ~, \nonumber \\
\Gamma(\varepsilon - 1) & = & - \frac{1}{\varepsilon} - (\gamma_{E} - 1)
- \frac{\varepsilon}{2} (\gamma_{E}^{2}- 2 \gamma_{E} + \frac{\pi^{2}}{6}
		 ) + {\cal O}(\varepsilon^2) ~, \nonumber \\
\Gamma(\varepsilon - n) & = & \frac{(-1)^{n}}{n!}
\left[  \frac{1}{\varepsilon} + (\sum_{k=1}^{n} \frac{1}{k}) - \gamma_{E} 
\right] + {\cal O}(\varepsilon) ~. \nonumber
\end{eqnarray}
\item
%%%%%%%%%%%%%%%%%%%%%%%%%%%%%%%
$ \Gamma $ - Function:
%%%%%%%%%%%%%%%%%%%%%%%%%%%%%%%
\begin{eqnarray}
z \Gamma(z)  & = & \Gamma(z + 1) ~, \nonumber \\
\Gamma(z) \Gamma(1 - z) & = & \frac{\pi}{\sin \pi z} ~, \nonumber \\
\Gamma \left (z + \frac{1}{2} \right) 
\Gamma \left (-z + \frac{1}{2} \right) & = & 
\frac{\pi}{\cos \pi z} ~, \nonumber \\
\Gamma(2z) & = & \frac{2^{2 z - 1}}{\sqrt{\pi}}
\Gamma \left( z \right) \Gamma \left(z + \frac{1}{2}\right) ~, \nonumber \\
\Gamma(1) ~ = ~ \Gamma(2) ~=~1, \Gamma(3) & = & 2,~\Gamma( n+1 )~=~n!,\nonumber \\
\Gamma \left( \frac{1}{2} \right) ~ = ~ \sqrt{ \pi }~,
\Gamma \left( \frac{3}{2} \right) & = & \frac{ \sqrt{ \pi } }{2}~,
\Gamma \left( \frac{5}{2} \right) ~ = ~ \frac{ 3 \sqrt{ \pi } }{4}~,
\nonumber \\
\Gamma \left(n + \frac{1}{2} \right) 
& = & \frac{(2n - 1)!!}{2^{ n }} \sqrt{\pi}
~ = ~ \frac{(2n)!}{4^{n} n!} \sqrt{ \pi },  \nonumber \\
\Gamma \left( - n~+~\frac{1}{2} \right) 
& = & \frac{(-)^{n} 2^{n}}{(2n - 1)!!} \sqrt{\pi}
~ = ~ \frac{(^4)^{n} n!}{(2n)!} \sqrt{\pi} ~. \nonumber 	
\end{eqnarray}
\item
%
%%%%%%%%%%%%%%%%%%%%%%%%%%%%%
$\psi$ - Function:
%%%%%%%%%%%%%%%%%%%%%%%%%%%%%
%
\begin{eqnarray}
\psi(1) & = & - \gamma~,~\psi(2) ~ = ~ 1 ~-~ \gamma~,
\psi(n) ~ = ~ \sum_{i=1}^{n-1} \frac{1}{i} ~-~ \gamma ~, \nonumber \\
\psi \left( \frac{1}{2} \right)& = &
- \ln 4 ~-~ \gamma~,\psi \left( \pm n ~ + ~ \frac{1}{2} \right)
~ = ~ 2 \sum_{ i=1 }^{n} \frac{1}{2i-1} - \ln 4 ~-~ \gamma ~. \nonumber
\end{eqnarray}
\item
%
%%%%%%%%%%%%%%%%%%%%%%%%%%%%%%
$\beta$ - function:
%%%%%%%%%%%%%%%%%%%%%%%%%%%%%%
%
\begin{eqnarray}
B(x,y) & \equiv & \int_{0}^{1} \xi^{x-1} (1-\xi)^{y-1}
\hspace{1cm} (x,y~>~0)
~, \nonumber \\
B(x,y) & = & B(y,x) ~, \nonumber \\
B(x,y) & = & 2 \int_{0}^{\pi /2} d \theta \cos^{2x-1}\theta ~
\sin^{2y-1}\theta ~, \nonumber \\
B(x,x) & = & 2^{1-2x}B(x,\frac{1}{2}) ~, \nonumber \\
B(x,y) & = & \frac{\Gamma(x) \Gamma(y)}{\Gamma(x+y)} ~. \nonumber 
\end{eqnarray}
\item
%%%%%%%%%%%%%%%%%%%%%%%%%%%%%%%
Feynman Parametrization:
%%%%%%%%%%%%%%%%%%%%%%%%%%%%%%%
\begin{eqnarray}
\frac{1}{ a_{1} a_{2} \cdots a_{n} }& = &
( n-1 )!\int d \alpha _{1} d \alpha _{2} \cdots d \alpha _{n}
\frac{ \delta ( 1 - \alpha _{1} - \alpha _{2} - \cdots -  \alpha _{n}
      )
}{ \left( \alpha _{1} a_{1} + \alpha _{2} a_{2} + \cdots +  
       \alpha _{n} a_{n}
\right)^{n}      } ~, \nonumber \\
\frac{1}{ A B } & = & \int^{1}_{0} d x 
\frac{1}{ \left[ x A + ( 1- x )B \right]^{2} } ~, \nonumber \\
\frac{1}{A^{\alpha } B^{\beta }}
& = & \frac{\Gamma \left( \alpha + \beta \right) 
    }{ \Gamma \left( \alpha \right) 
       \Gamma \left( \beta  \right)  }
\int^{1}_{0} d x \frac{ x^{\alpha - 1} (1- x)^{\beta -1}
	      }{\left[ x A + ( 1- x )B \right]^{\alpha + \beta }
	       } ~, \nonumber \\
\frac{1}{ A B C } & = & \int ^{1}_{0} dx  \int ^{1-x}_{0} dy 
\frac{2}{ \left[ x A + y B + ( 1 - x - y )C \right]^{3} } ~. \nonumber
\end{eqnarray}
\item
%%%%%%%%%%%%%%%%%%%%%%%%%%%%%%%
D-Dimension Integral:
%%%%%%%%%%%%%%%%%%%%%%%%%%%%%%%
\begin{eqnarray}
\int
\frac{ d^{D} k }{( 2 \pi )^{D}}
\frac{1}{ \left[ k^{2} - L \right]^{n} 	}
& = & \frac{i}{ (4 \pi )^{ \frac{D}{2} } }
(-)^{n} \frac{ 
   \Gamma \left( n- \frac{D}{2} \right)
     } { \Gamma \left( n  \right) } L^{ \frac{D}{2} - n } ~, \nonumber \\
\int \frac{ d^{D} k }{ ( 2 \pi )^{D} } 
\frac{ k ^{2} }{ \left[ k^{2} - L \right]^{n} }
& = &
\frac{i}{ (4 \pi)^{\frac{D}{2}}} (-)^{ n + 1 }
\frac{\Gamma \left( n - 1 -  \frac{D}{2} \right)}
{\Gamma \left( n  \right)}\frac{D}{2} L^{ \frac{D}{2} + 1  - n } ~, \nonumber \\
\int \frac{d^{D} k}{( 2 \pi )^{D}} 
\frac{ k^{ \mu }k^{ \nu }}{\left[ k^{2} - L \right]^{n}}
& = &
\frac{i}{ (4 \pi )^{\frac{D}{2} }} (-)^{ n + 1 }
\frac{ \Gamma \left( n - 1 - \frac{D}{2} \right)} { \Gamma \left( n  \right) }
\frac{ g^{ \mu \nu } }{2} L^{ \frac{D}{2} + 1 - n } ~, \nonumber \\ 
\int \frac{ d^{D} k }{ ( 2 \pi )^{D} } 
\frac{ k^{ \mu }k^{ \nu }k^{ \lambda }k^{ \rho }}
{ \left[ k^{2} - L \right]^{n}		}
& = & \frac{i}{ (4 \pi ) ^{ \frac{D}{2} } }
(-)^{ n } \frac{ \Gamma \left( n - 2 - \frac{D}{2} \right)} 
       { \Gamma \left( n  \right) } 
\nonumber \\
&& \times \frac{1}{4}
\left(
g^{ \mu \nu } g^{ \lambda \rho } 
+ g^{ \mu \lambda } g^{ \nu \rho }
+ g^{ \mu \rho } g^{ \nu \lambda } 
\right)
L^{ \frac{D}{2} + 2 - n } ~, \nonumber \\
\int
\frac{ d^{D} k }{ ( 2 \pi )^{D} } k^{ \mu } k^{ \nu } f\left( k^{2} \right)
& = & \frac{g^{ \mu \nu } }{D} 
\int  \frac{ d^{D} k }{ ( 2 \pi )^{D} } k^{ 2 } f\left( k^{2} \right) 
~. \nonumber 
\end{eqnarray}
%%%%%%%%%%%%%%%%%%%%%%%%%%%%%%%%%%%%%%%%%%%%%%%%%%%%%%%%%
%
\end{itemize}
%
%%%%%%%%%%%%%%%%%%%%%%%%%%%%%%%%%%%%%%%%%%%%%%%%%%%%%%%%%%%
\section{SU($N$) Generators }
%%%%%%%%%%%%%%%%%%%%%%%%%%%%%%%%%%%%%%%%%%%%%%%%%%%%%%%%%%%
%
The SU($N$) Generators $T^{a}(a~=~ 1, 2 ,\ldots, N^{2} - 1 )$
are hermitian and  traceless matrices, which generate the closed SU($N$)
algebra.
\begin{eqnarray}
\left[ T^{a} ,T^{b} \right] ~ = ~ i f^{ a b c } T^{c} ~. \nonumber
\end{eqnarray}
The fundamental representation is N-dimensional where
$T^{a}$ satisfy an additional relation
\begin{eqnarray}
\left \{ T^{a} , T^{b} \right \} & = & 
\frac{1}{N}\delta_{ab} + d^{abc} T^{c} ~, \nonumber 
\end{eqnarray}
which is consistent with the normalization
\begin{eqnarray}
\Tr \left( T^{a} T^{b} \right) ~ = ~  \frac{1}{2} \delta_{ab}~. \nonumber
\end{eqnarray}
Here $d^{abc}$ is totally symmetric in $a,b$ and $c$.
\newline
It's  given by
\begin{eqnarray}
d^{abc} ~ = ~ 2 \Tr \left[ \{ T^{a} , T^{b} \} T^{c} \right]~.
\nonumber
\end{eqnarray}
According to above equations we get following relation.
\begin{eqnarray}
T^{a} T^{b} ~ = ~  \frac{1}{ 2 N } \delta_{ a b }
	 + \frac{1}{2} d^{ a b c } T^{c}
	 + \frac{1}{2} i f^{ a b c } T^{c} ~. \nonumber
\end{eqnarray}
The SU($N$) structure constants satisfy following relation.
\begin{eqnarray}
f^{ a d e } f^{ b e f } f^{ c f d } ~ = ~
\frac{N}{2} f^{ a b c } ~. \nonumber
\end{eqnarray}
and Jacobi Identities
\begin{eqnarray}
f^{ a b e } f^{ c d e }  +  
f^{ c b e } f^{ d a e }  + 
f^{ d b e } f^{ a c e }& = & 0  ~, \nonumber \\
f^{ a b e } d^{ c d e }  +  
f^{ c b e } d^{ d a e }  + 
f^{ d b e } d^{ a c e }& = & 0  ~. \nonumber 
\end{eqnarray}
For The SU($3$) Generators and structure constants 
are explicitly written down as follows.
\begin{eqnarray}
\lambda_{1} & = & 
\left(
\begin{array}{ccc}
0 & 1 & 0 \\
1 & 0 & 0 \\
0 & 0 & 0 
\end{array} 
\right) ~ , 
\lambda_{2} ~ = ~  
\left(
\begin{array}{ccc}
0 & -i & 0 \\ 
i &  0 & 0 \\
0 &  0 & 0 
\end{array} 
\right) ~, \nonumber \\
\lambda_{3} & = & 
\left(
\begin{array}{ccc}
1 &  0 & 0 \\
0 & -1 & 0 \\
0 &  0 & 0 
\end{array} 
\right) ~ , 
\lambda_{4} ~ = ~  
\left(
\begin{array}{ccc}
0 & 0 & 1 \\
0 & 0 & 0 \\
1 & 0 & 0 
\end{array}
\right)  ~, \nonumber \\
\lambda_{5} & = & 
\left(
\begin{array}{ccc}
0 & 0 & -i \\
0 & 0 & 0 \\
i & 0 & 0 
\end{array} 
\right) ~ ,
\lambda_{6} ~ = ~  
\left(
\begin{array}{ccc}
0 & 0 & 0 \\
0 & 0 & 1 \\
0 & 1 & 0 
\end{array} 
\right) ~, \nonumber \\
\lambda_{7} & = & 
\left(
\begin{array}{ccc}
0 & 0 & 0 \\
0 & 0 & -i \\
0 & i & 0 
\end{array} 
\right) ~ ,
\lambda_{8} ~ = ~  
\frac{1}{ \sqrt{3} }
\left(
\begin{array}{ccc}
1 & 0 & 0 \\
0 & 1 & 0 \\
0 & 0 & -2 
\end{array} 
\right) ~, \nonumber \\
{\rm where}
& ~ & 
T^{a} ~ = ~ \frac{ \lambda^{a} }{2}.
\nonumber
\end{eqnarray}
\newline
Non-zero components of structure constants are
\begin{eqnarray}
f^{ 1 2 3 } & = & 1  ~, \nonumber \\
f^{ 1 4 7 } & = & - f^{ 1 5 6 } ~ = ~ f^{ 2 4 6 } 
~ = ~ f^{ 2 5 7 } ~=~ f^{ 3 4 5 } 
~ = ~ -f^{ 3 6 7 } ~ = ~ \frac{1}{2} ~, \nonumber \\
f^{ 4 5 8 } & = &  f^{ 6 7 8 } ~ = ~ \frac{\sqrt{3}}{2} ~. \nonumber
\end{eqnarray}
Now we can get some formulae, of course these formulae are
independent of the explicit representation of SU($N$) Generators
and structure constants.
\begin{eqnarray}
    (T^{c} T^{c})_{ij} 
&=& C_{2}(R) \delta _{ij} ~, \nonumber \\ 
    T^{c} T^{a} T^{c} 
&=& \left[ - \frac{1}{2} C_{2}(G) + C_{2}(R) \right ] T^{a} ~, \nonumber \\
    \Tr \left[T^{a} T^{b} 	\right] 
&=& \frac{1}{2} \delta^{ a b } ~, \nonumber \\
    \Tr \left[ T^{a} T^{b} T^{c} \right] 
&=& \frac{1}{4} ( d^{ a b c } + i f^{ a b c } ) ~,\nonumber \\
    \Tr \left[ T^{a} T^{b} T^{c} T^{d} \right] 
&=& \frac{1}{4 N} \delta^{a b} \delta^{c d} + 
    \frac{1}{8} (d^{ a b e } + i f^{a b e})
                (d^{c d e} + i f^{c d e}) ~, \nonumber \\
    f^{ a b c } T^{b} T^{c} 
&=& \frac{i}{2} f^{ a b c } f^{ b c d } T^{d} \nonumber \\
&=& \frac{i}{2} C_{2}(G) T^{a} ~. \nonumber
\end{eqnarray}
\begin{eqnarray}
    C_{2}(R) 
&=& \frac{ N^{2} -1 }{ 2 N }  \nonumber \\  
    f^{ a c d } f^{ b c d } 
&=& C_{2}(G) \delta^{ a b }
~=~ N  \delta^{ a b }  ~. \nonumber 
\end{eqnarray}
%
%%
%%%%%%%%%%%%%%%%%%%%%%%%%%%%%%%%%%%%%%%%%%%%%%%%%%%%%%%%%%%
\chapter{ Feynman Rule }
%%%%%%%%%%%%%%%%%%%%%%%%%%%%%%%%%%%%%%%%%%%%%%%%%%%%%%%%%%%
%
%
%%%%%%%%%%%%%%%%%%%%%%%%%%%%%%%%%%%%%%%%%%%%%%%%%%%%%%%%%%%
\section{ QED Lagrangian } 
%%%%%%%%%%%%%%%%%%%%%%%%%%%%%%%%%%%%%%%%%%%%%%%%%%%%%%%%%%%
%
\begin{eqnarray}
    {\cal L}_{\mbox{QED}}
~=~ {\cal L }_{class} ~+~ {\cal L}_{gauge} ~. \nonumber
\end{eqnarray}
%%%%%%%%%%%%%%%%%%%%%%
\begin{eqnarray}
    {\cal L}_{class} 
&=& \bar{ \psi } \left( i \slashs{D} ~-~ m \right) \psi
~-~ \frac{1}{4} F_{ \mu \nu } F^{ \mu \nu }
~, \nonumber \\
%%%%%%%%%%%%%%%%%%%%%%
{\cal L }_{gauge}
&=& - \frac{1}{ 2 \alpha } 
      \left(  \partial ^{ \mu } A_{ \mu }  \right)^{2}
~. \nonumber 
%%%%%%%%%%%%%%%%%%%%%%
\end{eqnarray}
\begin{itemize}
\item {Covariant Derivative and Field Strength Tensor:}
\begin{eqnarray}
           D_{\mu} 
&\equiv&   \partial_{ \mu }~ - ~ i e A_{ \mu }
        ~, \nonumber \\
          F_{ \mu \nu } 
&\equiv&  \partial_{ \mu } A_{ \nu } - \partial_{ \nu } A_{ \mu }
~, \nonumber 
\end{eqnarray}
\end{itemize}
%%%%%%%%%%%%%%%%%%%%%%%%%%%%%%%%%%%%%%%%%%%%%%%%%%%%%%%%%%
We split up the above Lagrangian into the free part ${\cal L}_{0}$
and the interaction part ${\cal L}_{I}$.
\begin{eqnarray}
         {\cal L}_{\mbox{QED}}
&\equiv& {\cal  L}_{0} ~+~ {\cal L}_{I}
~, \nonumber \\
%%%%%%%%%%%%%%%%%%%%%%%%
    {\cal L}_{ 0 } 
&=& \bar{\psi}
    ( i \partial \hspace{ -6pt }/ ~-~m )
    \psi
- \frac{1}{4}( 
\partial _{ \mu }A_{ \nu } -
\partial _{ \nu }A_{ \mu } )(
\partial ^{ \mu }A^{ \nu } -
\partial ^{ \nu }A^{ \mu } )  \nonumber \\
&-& \frac{1}{2 \alpha }
( \partial ^{ \mu }A_{ \mu })^{2}
~, \nonumber \\
%%%%%%%%%%%%%%%%%%%%%%%%%%%%%%%%%%
{\cal L}_{I} 
& = & e~ \bar{ \psi } \gamma ^{ \mu } \psi A_{ \mu }.
\nonumber
\end{eqnarray}  
\clearpage
%%%%%%%%%%%%%%%%%%%%%%%%%%%%%%%%%%%%%%%%%%%%%%%%%%%%%%%%%%%
\section{ QED Feynman Rules } 
%%%%%%%%%%%%%%%%%%%%%%%%%%%%%%%%%%%%%%%%%%%%%%%%%%%%%%%%%%%
%
\begin{itemize}
\item
Fermion Propagator:
\begin{figure}[H]
\begin{center}
\leavevmode\psfig{file=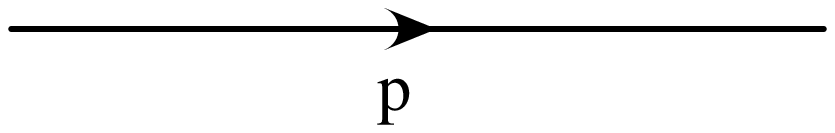,width=5cm}
\end{center}
\end{figure}
\vspace{-1.5cm}
\begin{eqnarray}
\frac{ i }{ \slashs{p}- m + i \varepsilon } \nonumber
\end{eqnarray}
\item
Photon Propagator:
\begin{figure}[H]
\begin{center}
\leavevmode\psfig{file=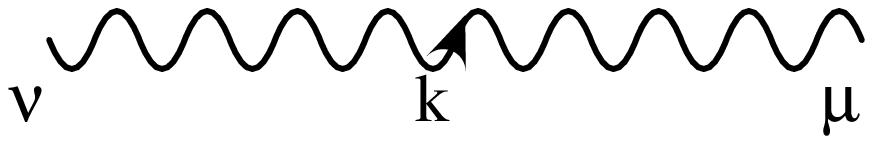,width=5cm}
\end{center}
\end{figure}
\vspace{-1.5cm}
\begin{eqnarray}
\frac{ - i }{ k^{2} + i \varepsilon }
\left[ g _{ \mu \nu } - ( 1 - \alpha )
\frac{ k_{ \mu } k_{ \nu } }{ k^{2} + i \varepsilon }
\right] \nonumber
\end{eqnarray}
\item
Photon - Fermion Vertex:
\begin{figure}[H]
\begin{center}
\leavevmode\psfig{file=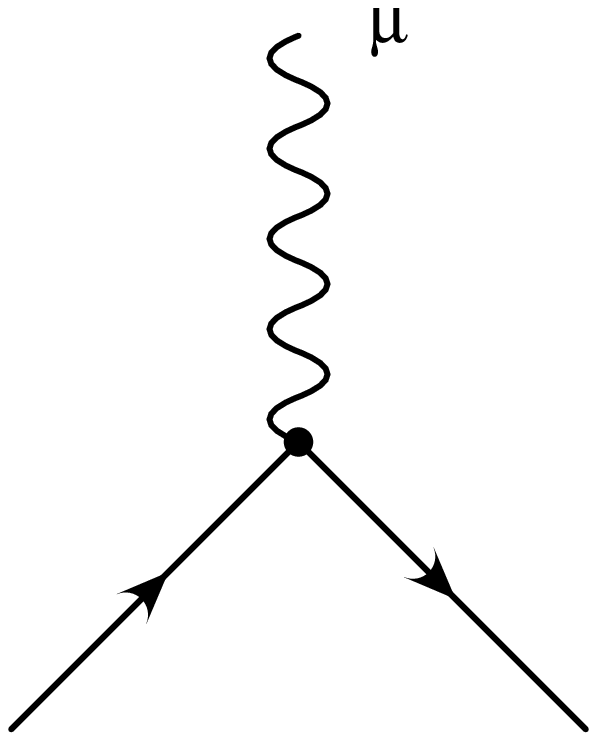,width=4cm}
\end{center}
\end{figure}
\vspace{-1.5cm}
\begin{eqnarray}
i e \gamma _{ \mu }\nonumber
\end{eqnarray}
\end{itemize}
\clearpage
%%%%%%%%%%%%%%%%%%%%%%%%%%%%%%%%%%%%%%%%%%%%%%%%%%%%%%%%%%%
\section{ QCD Lagrangian } 
%%%%%%%%%%%%%%%%%%%%%%%%%%%%%%%%%%%%%%%%%%%%%%%%%%%%%%%%%%%
%
\begin{eqnarray}
    {\cal L}_{\mbox{QCD}}
~=~ {\cal L }_{class}~+~{\cal L}_{gauge}~+~{\cal L}_{ghost}
~, \nonumber 
\end{eqnarray}
%%%%%%%%%%%%%%%%%%%%%%
\begin{eqnarray}
     {\cal L}_{class}
&=&  \bar{ \psi } \left( iD \hspace{ -7pt }/ ~-~ m \right) \psi
~-~ \frac{1}{4} F_{ \mu \nu }^{a} F^{ \mu \nu ,a}
~, \nonumber \\
%%%%%%%%%%%%%%%%%%%%%%
     {\cal L}_{gauge}
&=&  - \frac{1}{ 2 \alpha } \left(  G^{ \mu } A_{ \mu }^{a}  \right)^{2}
~, \nonumber \\
%%%%%%%%%%%%%%%%%%%%%%
      {\cal L}_{ghost}
& = & \left( \partial^{ \mu } \chi^{* a} \right)
      D_{ \mu }^{a b} { \chi }^{b}  
~.\nonumber
\end{eqnarray}
%%%%%%%%%%%%%%%%%%%%%%%%%%%%%%%%%%%%%%%%%%%%%%%%%%%%%%%%%%
\begin{itemize}
\item
$\chi$, $\chi^{*}$ represent ghost and anti-ghost field.
\item
$G^{ \mu }$ is given following,
\begin{enumerate}
\item
Coulomb Gauge:                     $ \cdots ~G^{ \mu } = 
				(0,\nabla)$
\item
Covariant Gauge:                   $ \cdots ~G^{ \mu } = 
				\partial^{ \mu }$
\begin{enumerate}
\item
Feynman Gauge:  			  $ \cdots \alpha = 1 $
\item
Landau Gauge:   			  $ \cdots \alpha = 0 $
\end{enumerate}
\item
Axial Gauge:                       $ \cdots ~G^{ \mu } = 
				n^{ \mu } $  ( Light-cone Vector )
\item
Temporal Gauge:                    $ \cdots ~G^{ \mu } = 
				(1, \mbox{ \boldmath{ $ 0 $ }})$
\end{enumerate}
\item
Covariant Derivative and Field Strength Tensor:
\begin{eqnarray}
D_{ \mu } & \equiv & 
\partial_{ \mu }~ - ~ ig_{s} A_{ \mu }^{a} T^{a}
~, \nonumber \\
F_{ \mu \nu } & \equiv & 
\partial_{ \mu } A_{ \nu } - \partial_{ \nu } A_{ \mu }
- i g_{s} \left[ A_{ \mu },A_{ \nu } \right]
\nonumber \\
&=&
\left(
\partial_{ \mu } A_{ \nu }^{a} - \partial_{ \nu } A_{ \mu }^{a}
+ g_{s} f^{a b c} A_{ \mu }^{b} A_{ \nu }^{c} 
\right) T^{a}
~. \nonumber 
\end{eqnarray}
\end{itemize}
%%%%%%%%%%%%%%%%%%%%%%%%%%%%%%%%%%%%%%%%%%%%%%%%%%%%%%%%%%
We split up the above Lagrangian into the free part $ {\cal L}_{0}$
and the interaction part $ {\cal L }_{I}$ .
\begin{eqnarray}
  {\cal L}_{\mbox{QCD}}
&\equiv&  
{\cal L}_{0} ~+~ {\cal L}_{I} ~, \nonumber \\
%%%%%%%%%%%%%%%%%%%%%%%%
    {\cal L}_{0} 
&=& \bar{ \psi } (i \partial \hspace{ -6pt }/ ~-~m )\psi
- \frac{1}{4} ( 
\partial _{ \mu }A^{a}_{ \nu }-
\partial _{ \nu }A^{a}_{ \mu })(
\partial ^{ \mu }A^{a \nu }-
\partial ^{ \nu }A^{a \mu })
\nonumber \\
&-& 
\frac{1}{2 \alpha }
( \partial ^{ \mu }A^{ a }_{ \mu })^{2}~+~
( \partial _{ \mu } { \chi }^{*a} )
( \partial ^{ \mu } { \chi }^{a} )
~, \nonumber \\
%%%%%%%%%%%%%%%%%%%%%%%%%%%%%%%%%%
     {\cal L}_{I} 
&=&  g_{s} \bar{ \psi } \gamma ^{ \mu } T^{a}\psi  A^{a}_{ \mu }
~-~
\frac{ g_{s} }{2} f^{ a b c }
( \partial _{ \mu }A_{ \nu }^{a} -
  \partial _{ \nu }A_{ \mu }^{a} )A^{ \mu b }A^{ \nu c }  
\nonumber \\
&-&
\frac{ g_{s}^{ 2 } }{4}
f^{ a b c } f^{ a d e }
A_{ \mu }^{b} A_{ \nu }^{c} A^{ \mu d } A^{ \nu  e } ~-~
g_{s} f^{ a b c }
( \partial ^{ \mu } \chi ^{* a} )
  \chi ^{b} A^{ \mu c } 
~.\nonumber
\end{eqnarray}  
%
%%%%%%%%%%%%%%%%%%%%%%%%%%%%%%%%%%%%%%%%%%%%%%%%%%%%%%%%%%%
\section{ Notation and Conventions} 
%%%%%%%%%%%%%%%%%%%%%%%%%%%%%%%%%%%%%%%%%%%%%%%%%%%%%%%%%%%
%
\begin{itemize}
\item Covariant Derivative:
%%%%%%%%%%%%%%%%%%%%%%%%%%%%%
\begin{eqnarray}
D_{ \mu ,ij }[A] 
& \equiv &
\partial_{ \mu } \delta _{ i j }~-~i g_{s} A_{ \mu }^{a} 
\left[ 	T^{a}
\right]_{ij}
~, \nonumber \\
D_{ \mu }^{ ab }[A] 
& \equiv &
\partial_{ \mu } \delta ^{ ab }~-~ g_{s} f^{ a b c } A_{ \mu }^{c} 
~. \nonumber   
\end{eqnarray}
%%%%%%%%%%%%%%%%%%%%%%%%%%%%%%
\item
Field Strength:
\begin{eqnarray}
\left[ D_{ \mu }, D_{ \nu } \right]_{ ij }
& \equiv & 
-ig_{s} F_{ \mu \nu }^{a} [ T^{a} ]_{ ij }
~, \nonumber \\ 
\left[ D_{ \mu }, D_{ \nu } \right]^{ ab }
& \equiv & 
-ig_{s} F_{ \mu \nu }^{c} [ T^{c} ]^{ ab }
\nonumber \\
& = &
-g_{s}f^{ a b c } F_{ \mu \nu }^{c}
~, \nonumber \\
F_{ \mu \nu }^{a}
& = &
\partial_{ \mu }A_{ \nu }^{a}~-~\partial_{ \nu }A_{ \mu }^{a}
~+~g_{s}f^{ a b c } A_{ \mu }^{b} A_{ \nu }^{c}
~. \nonumber  
\end{eqnarray}
%%%%%%%%%%%%%%%%%%%%%%%%%%%%%%%%%%%%%%%%%
\item
Dual Tensor:
\begin{eqnarray}
\tilde{ F }^{ \mu \nu }
& \equiv &
\frac{1}{2} \varepsilon^{ \mu \nu \rho \sigma } F_{ \rho \sigma }
~, \nonumber \\
\tilde{ F }^{ \mu \nu a }
& = &
\frac{1}{2} \varepsilon^{ \mu \nu \rho \sigma }
\left(
\partial_{ \rho }A_{ \sigma }^{a}
~-~\partial_{ \sigma }A_{ \rho }^{a}
~+~g_{s}f^{ a b c } A_{ \rho }^{b} A_{ \sigma }^{c}
\right)
~. \nonumber
\end{eqnarray}
\item
Bianchi Identity:
\begin{eqnarray}
\left[
D_{ \mu },[ D_{ \nu }, D_{ \rho } ]
\right] ~+~
\left[
D_{ \nu },[ D_{ \rho }, D_{ \mu } ]
\right] ~+~
\left[
D_{ \rho },[ D_{ \mu }, D_{ \nu } ]
\right]
~ = ~ 0
~. \nonumber 
\end{eqnarray}
Thus we get follow identity.
\begin{eqnarray}
\varepsilon^{ \mu \nu \rho \sigma }
\left[
D_{ \mu },[ D_{ \nu }, D_{ \rho } ]
\right] 
~ = ~ 0
~. \nonumber  
\end{eqnarray}
\end{itemize}
\clearpage
%%%%%%%%%%%%%%%%%%%%%%%%%%%%%%%%%%%%%%%%%%%%%%%%%%%%%%%%%%%
\section{ QCD Feynman Rules } 
%%%%%%%%%%%%%%%%%%%%%%%%%%%%%%%%%%%%%%%%%%%%%%%%%%%%%%%%%%%
%
\begin{itemize}
\item
Fermion Propagator:
\begin{figure}[H]
\begin{center}
\leavevmode\psfig{file=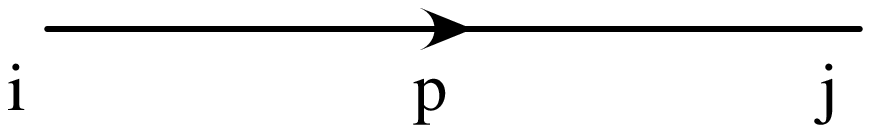,width=5cm}
\end{center}
\end{figure}
\vspace{-1.5cm}
\begin{eqnarray}
\frac{ i \delta^{ij} }{ \slashs{p} - m + i \varepsilon } \nonumber
\end{eqnarray}
\item
Gluon Propagator:
\begin{figure}[H]
\begin{center}
\leavevmode\psfig{file=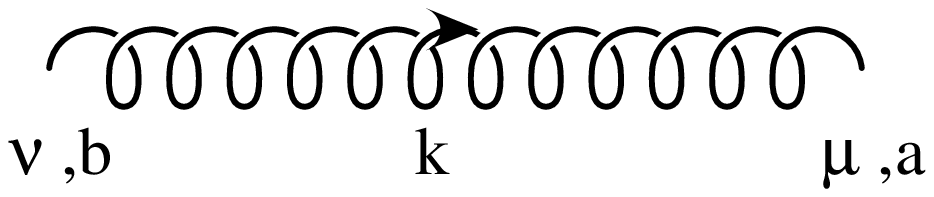,width=5cm}
\end{center}
\end{figure}
\vspace{-1.5cm}
\begin{eqnarray}
\frac{ - i \delta_{ a b } }{ k^{2} + i \varepsilon }
\left[ g _{ \mu \nu } - ( 1 - \alpha ) 
\frac{ k_{ \mu } k_{ \nu } }{ k^{2} + i \varepsilon }
\right] \nonumber
\end{eqnarray}
\item
Ghost Propagator:
\begin{figure}[H]
\begin{center}
\leavevmode\psfig{file=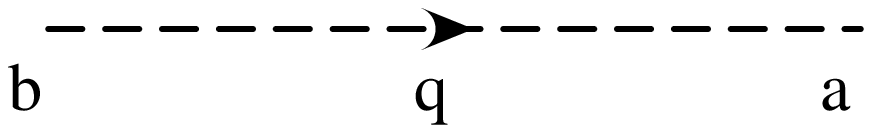,width=5cm}
\end{center}
\end{figure}
\vspace{-1.5cm}
\begin{eqnarray}
\frac{ - i \delta_{ a b } }{ q^{2} + i \varepsilon } \nonumber
\end{eqnarray}
\clearpage
\item
Gluon - Fermion Vertex:
\begin{figure}[H]
\begin{center}
\leavevmode\psfig{file=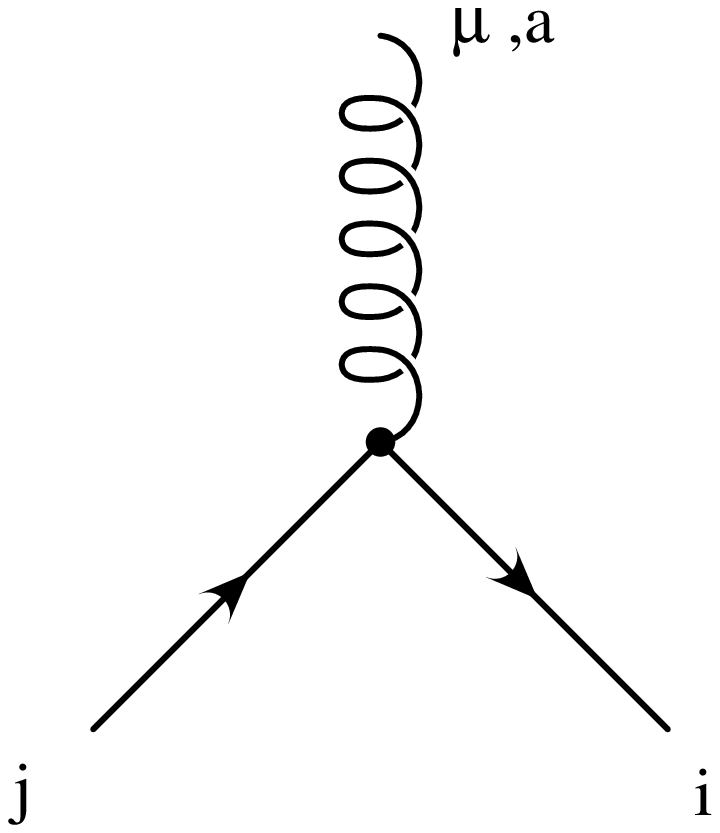,width=4cm}
\end{center}
\end{figure}
\vspace{-1.5cm}
\begin{eqnarray}
i g_{s} \gamma _{ \mu } T^{a}_{i j }
\nonumber
\end{eqnarray}
\end{itemize}
\begin{itemize}
\item
Gluon 3-point  Vertex:
\begin{figure}[H]
\begin{center}
\leavevmode\psfig{file=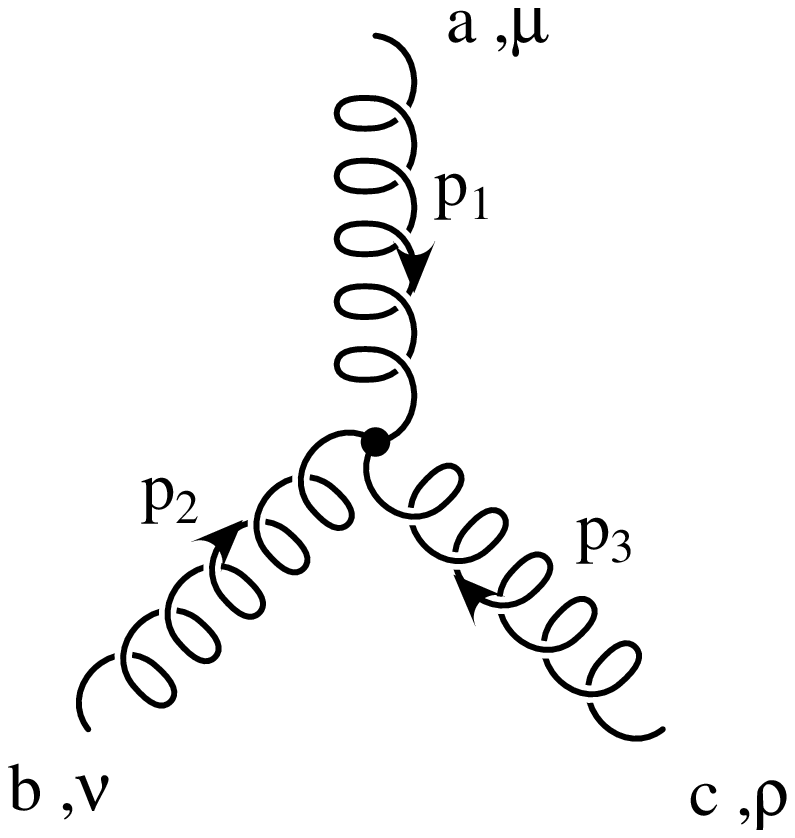,width=4cm}
\end{center}
\end{figure}
\vspace{-1.5cm}
\begin{eqnarray}
g_{s} f^{a b c}
\left[ g_{ \mu \nu }( p_{1} - p_{2} )_{ \rho } +
g_{ \nu \rho }( p_{2} - p_{3} )_{ \mu } +
g_{ \rho \mu }( p_{3} - p_{1} )_{ \nu } 
\right]       \nonumber
\end{eqnarray}
\clearpage
\item
Gluon - Ghost Vertex:
\begin{figure}[H]
\begin{center}
\leavevmode\psfig{file=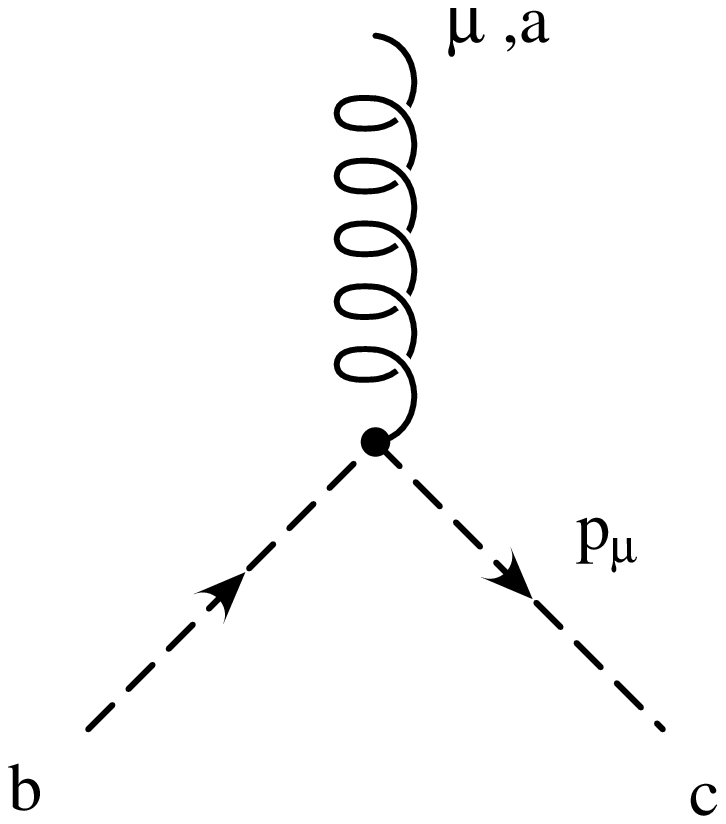,width=4cm}
\end{center}
\end{figure}
\vspace{-1.5cm}
\begin{eqnarray}
- g_{s} f^{a b c} p_{ \mu } \nonumber
\end{eqnarray}
\item
Gluon 4-point Vertex:
\begin{figure}[H]
\begin{center}
\leavevmode\psfig{file=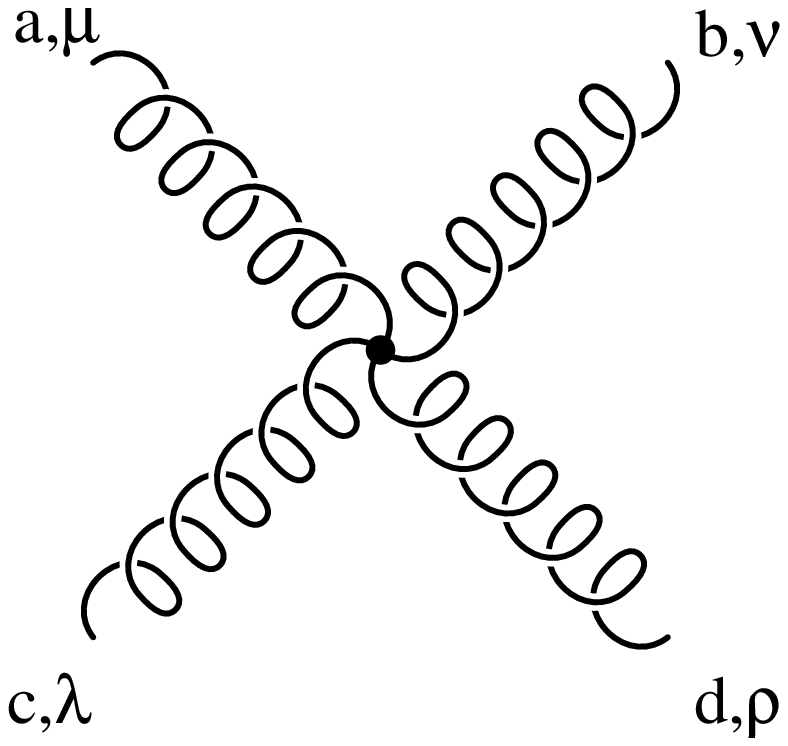,width=4cm}
\end{center}
\end{figure}
\vspace{-1.5cm}
\begin{eqnarray}
& - & i g_{s}^{2}f^{ a b e }f^{ c d e }
\left(  g_{ \mu \lambda }g_{ \nu \rho } -
g_{ \mu \rho }g_{ \nu \lambda }  \right)  \nonumber \\
& - & i g_{s}^{2}f^{ a c e }f^{ b d e }
\left(  g_{ \mu \nu }g_{ \lambda \rho } -
g_{ \mu \rho }g_{ \nu \lambda }  \right)  \nonumber \\
& - & i g_{s}^{2}f^{ a d e }f^{ c b e } 
\left(  g_{ \mu \lambda }g_{ \nu \rho } -
g_{ \mu \nu }g_{ \rho \lambda }  \right)  \nonumber 
\end{eqnarray}
\end{itemize}
\clearpage
\section{Electro-Weak Feynman Rule}
%%%%%%%%%%%%%%%%%%%%%%%%%%%%%%%%%%%%%%%%%%%%%%%%%%%%%%%%
%
\begin{itemize}
\item
$Z^{0}$ Boson Propagator:
\begin{figure}[H]
\begin{center}
\leavevmode\psfig{file=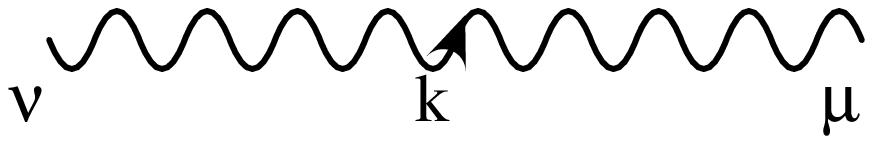,width=5cm}
\end{center}
\end{figure}
\vspace{-1.5cm}
\begin{eqnarray}
\frac{ - i }{ k^{2} - M_{Z}^2 + i \varepsilon }
\left[ g _{ \mu \nu } - ( 1 - \alpha_Z ) 
\frac{ k_{ \mu } k_{ \nu } }{ k^{2} - \alpha_Z M_{Z} + i \varepsilon }
\right] \nonumber
\end{eqnarray}
\item
$W$ Boson Propagator:
\begin{figure}[H]
\begin{center}
\leavevmode\psfig{file=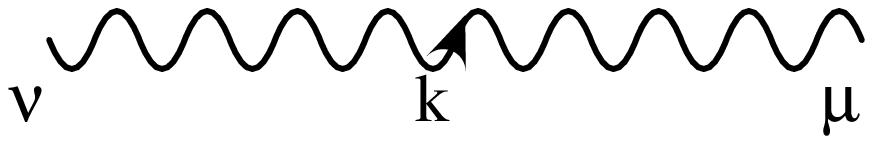,width=5cm}
\end{center}
\end{figure}
\vspace{-1.5cm}
\begin{eqnarray}
\frac{ - i }{ k^{2} - M_{W}^2 + i \varepsilon }
\left[ g _{ \mu \nu } - ( 1 - \alpha_W ) 
\frac{ k_{ \mu } k_{ \nu } }{ k^{2} - \alpha_W M_{W} + i \varepsilon }
\right] \nonumber
\end{eqnarray}
\clearpage
\item
Photon - Fermion Vertex:
\begin{figure}[H]
\begin{center}
\leavevmode\psfig{file=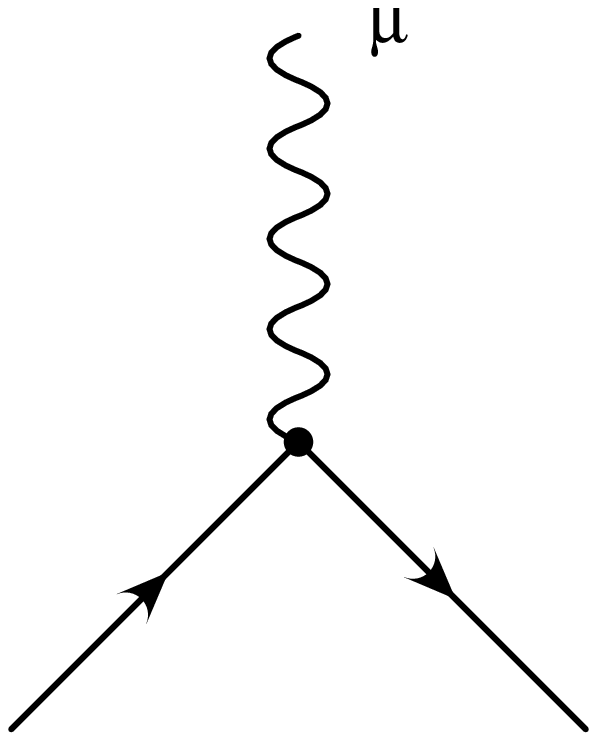,width=3.5cm}
\end{center}
\end{figure}
\vspace{-1.5cm}
\begin{eqnarray}
i e Q_{ \gamma } (f) \gamma ^{ \mu }
\nonumber
\end{eqnarray}
\item
$ Z^{0} $ Boson - Fermion Vertex:
\begin{figure}[H]
\begin{center}
\leavevmode\psfig{file=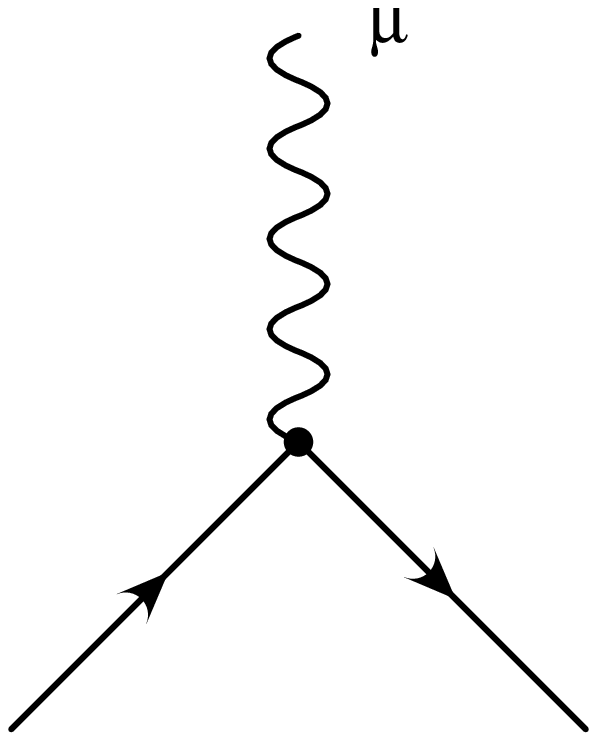,width=3.5cm}
\end{center}
\end{figure}
\vspace{-1.5cm}
\begin{eqnarray}
i e \left [  
Q_{Z}^{R}(f) \gamma^{\mu}_{R} + Q_{Z}^{L}(f) \gamma^{\mu}_{L} 
\right ]
\nonumber
\end{eqnarray}
\item
Coefficients $ Q_{ \gamma }(f), Q_{Z}^{R,L}(f) $:

When ``$f$'' stands for lepton $( e^{-},\mu ^{-}$ and $ \tau ^{-})$,
then coefficients $Q_{ \gamma }(f),~Q_{Z}^{R}(f)$ and $Q_{Z}^{L}(f)$ are 
\begin{eqnarray}
Q_{ \gamma }(f) & = & - 1~,~
Q_{Z}^{R}(f) ~ = ~ \frac{\sin \theta_{w} }{\cos \theta_{w}},~ 
Q_{Z}^{L}(f) ~ = ~ 
\frac{ 2 \sin ^{2} \theta_{w} - 1}{2 \sin \theta_{w} \cos \theta_{w}}~. 
\nonumber  
\end{eqnarray}
While ``$f$'' stands for Quark,
we get 
\begin{eqnarray}
Q_{ \gamma }(\mbox{U}) & = & \frac{2}{3},~
Q_{Z}^{R}(\mbox{U}) ~ = ~ 
\frac{ - 2 \sin \theta_{w} }{3 \cos \theta_{w}},~
Q_{Z}^{L}(\mbox{U}) ~ = ~ 
\frac{ 3 - 4 \sin^{2} \theta_{w} }{ 6 \sin \theta_{w} \cos \theta_{w}} ,~
\nonumber  \\
Q_{ \gamma }(\mbox{D}) & = & - \frac{1}{3},~
Q_{Z}^{R}(\mbox{D}) ~ = ~ 
\frac{ \sin \theta_{w} }{3 \cos \theta_{w}},~ 
Q_{Z}^{L}(\mbox{D}) ~ = ~ 
\frac{ - 3 + 2 \sin^{2} \theta_{w} }{ 6 \sin \theta_{w} \cos \theta_{w}} 
~, \nonumber 
\end{eqnarray}
where the suffices U, D mean up-type ($u,c,t$) and down-type ($d,s,b$) quarks.
\item
$W$ Boson - Fermion Vertex:
\begin{figure}[H]
\begin{center}
\leavevmode\psfig{file=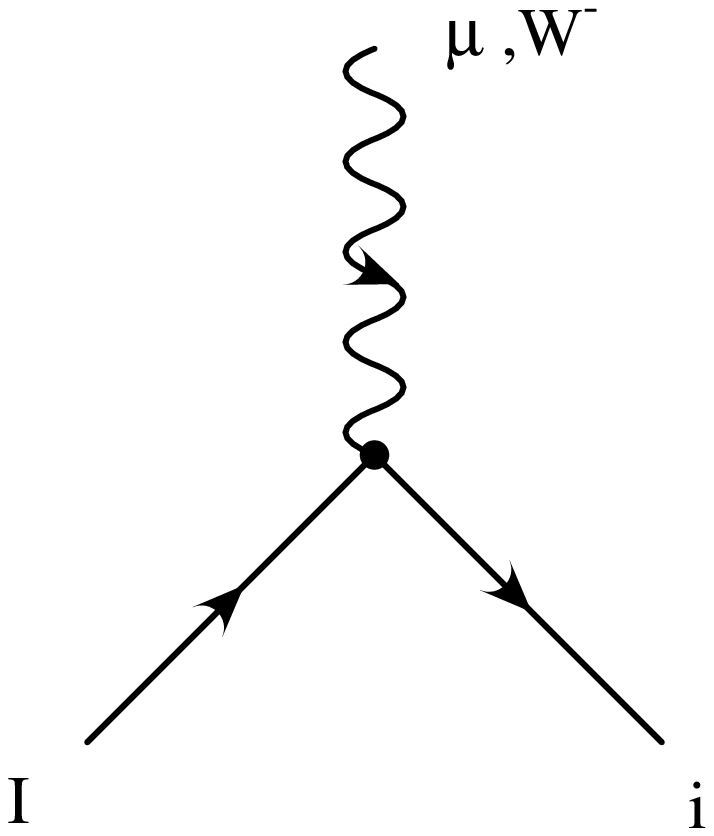,width=3.5cm}
\end{center}
\end{figure}
\vspace{-1.5cm}
\begin{eqnarray}
- i \frac{e}{\sqrt{2} \sin \theta_{W}}
C_{iI} \gamma^{\mu}_{L}.
\nonumber
\end{eqnarray}
Coefficients $C_{iI}$:

When ``$i,~I$'' stand for leptons $(e^{-},~\mu ^{-},~\tau^{-}$ and
($\nu_{e},~\nu_{\mu},~\nu_{\tau}$), 
then coefficients $C_{iI} = 1$.
While ``$i,~I$'' stand for Quark,
we get $C_{iI} ~=~ V_{iI}$, where $V_{iI}$ is Cabbibo-Kobayashi-Maskawa
mixing parameter.
\end{itemize}
\clearpage
%
%%%%%%%%%%%%%%%%%%%%%%%%%%%%%%%%%%%%%%%%%%%%%%%%%%%%%%%%
\chapter{Spinor Helicity Method}
%%%%%%%%%%%%%%%%%%%%%%%%%%%%%%%%%%%%%%%%%%%%%%%%%%%%%%%%
%
\section{Notation and Convention}
\begin{itemize}
\item
Pseudo-scalar, Axial-vector:
%%%%%%%%%%%%%%%%%%%%%%%%%%%%%%%%%
\bea
\gamma_{R} &\equiv& \frac{1}{2}( 1 + \gamma _{5} )~,~
\gamma_{L} ~\equiv~ \frac{1}{2}( 1 - \gamma _{5} )~, \nonumber \\
\gamma^{\mu}_{R} &\equiv&
    \gamma^{ \mu }\frac{ 1+ \gamma _{5} }{2} ~,~
\gamma^{\mu}_{L} ~\equiv~
    \gamma^{\mu}\frac{ 1 - \gamma _{5} }{2} ~. \nonumber 
\eea
\item  Massless Helicity States:
\bea
u(p,+) &=& \gamma_{R} u(p) ~\equiv~ \ket{p +}~, \nonumber \\
u(p,-) &=& \gamma_{L} u(p) ~\equiv~ \ket{p -}~. \nonumber 
\eea
\item
Dirac Equation:
\bea
\hat{p} \ket{p \pm} ~=~ \bra{p \pm} \hat{p} ~=~ 0, 
(\hat{p} \equiv \gamma^{\mu} p_{\mu}). \nonumber
\eea
\item
Chirality Conditions:
\bea
\begin{array}{lcr}
(1 \pm \gamma_{5}) \ket{p \mp} & = & 0, \nonumber \\
\bra{p \pm} (1 \pm \gamma_{5}) & = & 0. \nonumber
\end{array}
\eea
\item
Normalization and Completeness:
\bea
\bra{p \pm} \gamma^{\mu} \ket{p \pm} &=& 2 p^{\mu}, 
\nonumber \\
\ket{p +}\bra{p +} + \ket{p -}\bra{p -} &=& \hat{p}. 
\nonumber
\eea
\item
Massless Spinor Relations:
\bea
\ket{p \pm} \bra{p \pm} &=& \frac{1 \pm \gamma_{5}}{2} \hat{p}, 
\nonumber \\
\mbk{p + |q +} &=& \mbk{p -|q -} = 0, 
\nonumber \\
\mbk{p - |q +} &=& - \mbk{q -|p +}, 
\nonumber \\
\mbk{p - |p +} &=&  \mbk{p +|p -} = 0. \nonumber
\eea
\item
Fiertz rearrangement theorem:
\bea
\bra{p \pm} \gamma^{ \mu } \ket{q \pm} 
\gamma_{\mu} \frac{1 \pm \gamma_5}{2} & = &  
2 \ket{p \mp} \bra{q \mp}~, \nonumber \\
\bra{p \pm} \gamma ^{ \mu } \ket{q \pm} 
\gamma_{\mu} \frac{1 \mp \gamma_5}{2} & = &  
2 \ket{q \pm} \bra{p \pm}~, \nonumber \\
\mbk{A + |\gamma^{\mu}|B +}
\mbk{C - |\gamma_{\mu}|D -} &=&
2 \mbk{A + |D -} \mbk{C - |B +}~, \nonumber \\
\mbk{ p -|q + } \gamma_R &=& 
\ket{q +}\bra{p -} - \ket{p +}\bra{q -} ~, \nonumber \\
\mbk{A -| B +} \mbk{C -| D +} &=& 
\mbk{A -| D +} \mbk{C -| B +} + 
\mbk{A -| C +} \mbk{B -| D +}~.
\nonumber  
\eea

\item Properties:
\bea
\mbk{A \mp|B \pm} &=& - \mbk{B \mp| A \pm} ~, \nonumber \\
\mbk{A \pm| \gamma_{\mu}|B \pm} &=& 
\mbk{B \mp|\gamma_{\mu}|A \mp} ~, \nonumber \\
\mbk{p \pm | \gamma_{\mu_1} \cdots \gamma_{\mu_n}| q \mp} &=& 
- \mbk{1 \pm | \gamma_{\mu_1} \cdots \gamma_{\mu_n}| p \mp}~
(n~{\rm even}), \nonumber \\
\mbk{p \pm | \gamma_{\mu_1} \cdots \gamma_{\mu_n}| q \pm} &=& 
\mbk{p \pm | \gamma_{\mu_1} \cdots \gamma_{\mu_n}| p \pm}~
(n~{\rm odd}). \nonumber
\eea
\item Polarization Vectors:
\bea
\varepsilon^{\mu}_{\mp}(k) &=& \pm
      \frac{1}{\sqrt{2}} 
      \frac{ \bra{k \pm} \gamma^{\mu} \ket{p \pm}}
           { \mbk{p \mp| k \pm}} ~, \nonumber \\
\varepsilon^{\mu}_{\mp}(k) \gamma_{\mu} &=& 
\pm \frac{\sqrt{2}}{ \mbk{p \mp| k \pm}}
\left( \ket{k \mp} \bra{p \mp} + \ket{p \pm} \bra{k \pm} \right)~.
\nonumber 
\eea

\item Properties of Polarization Vector:
\bea
\varepsilon_{\pm}(k) \cdot k &=& 0 ~, 
\nonumber \\
\varepsilon_{\mp}^{\mu}(k) &=& [\varepsilon_{\pm}^{\mu}(k) ]^*~, 
\nonumber \\
\varepsilon_{\lambda}(k) \cdot \varepsilon_{\lambda'}^{*}(k)  
&=& - \delta_{\lambda \lambda'} ~(\lambda,\lambda' \in (\pm,0))~, \nonumber \\
\varepsilon_{\lambda}(k) \cdot k  &=& 0, 
\nonumber \\
\sum_{\lambda=\pm,0} 
\varepsilon_{\lambda}^{\mu}(k) \varepsilon_{\lambda}^{\nu *}(k) &=& 
- g^{\mu \nu} + \frac{k^{\mu} k^{\nu}}{m^2}~.
\nonumber 
\eea

\item Massive Spinor States:
The massive fermions with momentum $p$ and anti-fermion with momentum $\bar{p}$
\bea
u(p,\uparrow)   &=& \ket{p_1 +} + \ket{p_2 - }
             \frac{ \mbk{p_2~ p_1} }{m} , \nonumber \\    
u(p,\downarrow) &=& \ket{p_1 -} + \ket{p_2 + }
             \frac{ \pbk{p_2~ p_1} }{m}, \nonumber \\
\bar{u}(p,\uparrow)   &=& \bra{p_1 +} + 
             \frac{ \pbk{p_1~ p_2} }{m} \bra{p_2 -},\nonumber \\ 
\bar{u}(p,\downarrow) &=& \bra{p_1 - } + 
             \frac{ \mbk{p_1~ p_2} }{m} \bra{p_2 +},\nonumber \\
v(\bar{p},\uparrow)   &=& \ket{ \bar{p}_1 - } - \ket{ \bar{p}_2 +}
             \frac{ \pbk{ \bar{p}_2~ \bar{p}_1 } }{m}, \nonumber \\
v(\bar{p},\downarrow) &=& \ket{ \bar{p}_1 + } - \ket{ \bar{p}_2 -}
             \frac{ \mbk{ \bar{p}_2~ \bar{p}_1 } }{m}, \nonumber \\
\bar{v}(\bar{p},\uparrow)   &=& \bra{ \bar{p}_1 - } - 
             \frac{ \mbk{ \bar{p}_1~ \bar{p}_2 } }{m} 
             \bra{ \bar{p}_2 +}, \nonumber \\ 
\bar{v}(\bar{p},\downarrow) &=& \bra{ \bar{p}_1 + } - 
             \frac{ \pbk{ \bar{p}_1~ \bar{p}_2 } }{m} 
             \bra{\bar{p}_2 - }, \nonumber 
\eea 
with light like momenta $p_{1}$, $p_{2}$, $\bar{p}_{1}$ and
$\bar{p}_{2}$,
\bea
p_{1} &\equiv& \frac{p + m s_{p}}{2},~~
p_{2} ~\equiv~ \frac{p - m s_{p}}{2},~~
\nonumber  \\
\bar{p}_{1} &\equiv& \frac{\bar{p} + m s_{\bar{p}}}{2},~~
\bar{p}_{2} ~\equiv~ \frac{\bar{p} - m s_{\bar{p}}}{2},
\nonumber 
\eea 
where $s_{p}$ is spin vector of the massive fermion, and 
$s_{\bar{p}}$ is spin vector of the massive anti-fermion.

\item Polarization Vectors for Massive Boson:

Three polarization vectors $\varepsilon^{\mu}_{\lambda}(k)$
are given in terms of two light-like vectors,
\bea
k_{1}^{\mu} &=& \frac{k^{\mu} - m s^{\mu}}{2},  \nonumber \\
k_{2}^{\mu} &=& \frac{k^{\mu} + m s^{\mu}}{2}. \nonumber 
\eea
The polarization vectors for the incoming gauge boson:
\bea
\varepsilon_{\pm}^{\mu} &=& 
\frac{\bra{k_1 \pm} \gamma^{\mu} \ket{k_2 \pm} }{\sqrt{2} m}, 
\nonumber \\
\varepsilon_{0}^{\mu} &=& 
\frac{
        \bra{k_1 +} \gamma^{\mu} \ket{k_1 +} -
        \bra{k_2 +} \gamma^{\mu} \ket{k_2 +} }{2 m} ~=~
\frac{k_1^{\mu} - k_2^{\mu}}{m}. 
\nonumber 
\eea
\end{itemize}

\newpage

%%%%%%%%%%%%%%%%%%%%%%%%%%%%%%%%%%%%%%%%%%%%%%%%%%%%%%%%

\chapter{ Phase space integrals over $y$ and $z$ }

\noindent
The phase space integrals necessary to derive the cross section
are summarized in this Appendix. Although many of them have already appeared
in the literatures~\cite{tung}, we will list all of them below
for the convenience of the reader. After the integration over the angular
variables, we are left with the following four types of integrals:
\bean
J_i &=& \int dy dz f_i (y , z)\ , \qquad
N_i = \int \frac{dy dz}{\sqrt{(1 - y)^2 - a}} f_i (y , z)\ ,\\
L_i &=& \int \frac{dy dz}{(1 - y)^2 - a} f_i (y , z) \ , \quad
K_i = \int \frac{dy dz}{\{ (1 - y)^2 - a \}^{3/2}} f_i (y , z) \ .
\eean
The infrared divergences are regularized by the small gluon mass
$\lambda$ and $\beta = \sqrt{1 - a} = \sqrt{1 - 4 m^2 / s}$.
For the type $L_i$ integrals, a shorthand notation
$\omega = \sqrt{ (1 - \sqrt{ a})/(1 + \sqrt{ a}) }$ is used.
The $K_i$ integrals have a spurious singularity at the upper bound of
the $y$ integral, $y_+ = 1 - \sqrt{a}$. Since this singularity turns out
to be canceled out in the cross section, we regularize each integrals
by deforming the integration region
as $y_+ \to 1 - \sqrt{a} - \epsilon$~\cite{tung}. 
$\Li$ is the Spence function.

\vspace{1cm}
\noindent
Class {\boldmath{$J$}} Integrals:
\bean
\J{1} & = & 
\uint \ = \ 
\frac{1}{2} \beta \left(1+\frac{1}{2} a\right)-\frac{1}{2} a\left(
1-\frac{1}{4} a\right)\ln\left({1+ \beta \over 1- \beta }\right)\\
\J{2} & = &
\uint {y\over z}\ =\ \uint {z\over y}\\
& = &
-\frac{1}{4} \beta \left(5-\frac{1}{2} a\right)+\frac{1}{2}\left(
1+\frac{1}{8} a^2\right)\ln\left({1+ \beta \over 1- \beta }\right)\\
\J{3} & = &
\uint {1\over y}\ =\ \uint {1\over z} = 
- \beta +\left(1-\frac{1}{2} a\right)\ln\left({1+ \beta \over 1- 
\beta }\right) \\
\J{4} & = & \uint \frac{y}{z^{2}} = \uint \frac{z}{y^{2}} =
\frac{2}{ a} \beta  - \ln \left(\frac{1+ \beta }{1- \beta } \right) \\
\J{5} & = &
\uint {1\over y^2}\ =\ \uint {1\over z^2} \\
& = &
-{2 \beta \over a}\left(
\ln \frac{\lambda^2}{s} + 2 \ln a-4\ln \beta -4\ln 2+2\right)+
2\left(1-{3\over a}\right)\ln\left({1+ \beta \over 1-\beta }\right)\\
\J{6} & = &
\uint {1\over y z} \\
& = &
\left(- \ln \frac{\lambda^2}{s} -\ln a+4\ln \beta +2\ln 2\right)
\ln\left({1+ \beta \over 1- \beta }\right)  \\
& &
+2\left[\Li\left({1+ \beta \over 2}\right)
-\Li\left({1- \beta \over 2}\right)\right]
+3\left[\Li\left(-{2 \beta \over 1- \beta }\right)
-\Li\left({2 \beta \over 1+ \beta }\right)\right]
\eean

\vspace{1cm}
\noindent
Class {\boldmath{$N$}} Integrals:
\bean
\N{1} & = & \pint  = 
1-\sqrt{ a}-\frac{1}{2} a\ln\left({2-\sqrt{ a}\over\sqrt{ a}}\right)\\
\N{2} & = & \pint {z\over y} =
-\frac{1}{2}\ln a+\ln\left(2-\sqrt{ a}\right)+
{2\over 2-\sqrt{ a}}-2 \\
\N{3} & = &
\pint {y\over z}\\
& = &
2 \beta \ln\left({1- \beta \over 1+ \beta }\right)
-\ln\left({1+ \beta \over 2}\right)\ln\left({1- \beta \over 2}\right)
+\left(2-\frac{a}{2}\right)
\ln\left({2-\sqrt{ a}\over\sqrt{ a}}\right)\\
& & +\, \frac{1}{4} \ln^2 \frac{a}{4} -\sqrt{ a} + 1
+ \Li\left({1+ \beta \over 2}\right)+\Li\left({1- \beta \over 2}\right)-
2\,\Li\left({\sqrt{ a}\over 2}\right)\\ 
\N{4} & = & \pint {y^2\over z^2}  = 
{2\over a} \left(1-\sqrt{ a}\right)^2 \\
\N{5}& = & \pint y \\
& = &
{1\over 16} \left[ - a^2\ln a+2\, a^2\ln(2-\sqrt{ a})+
4(2-\sqrt{ a})^2-4\left(2- a^{3\over2}\right)\right]\\
\N{6} & = & \pint z \\
& = &
{1\over 32} \left[ 12-(2+ a)^2-{2+\sqrt{ a}\over 2-\sqrt{ a}}\, a^2
+2(8- a) a\,\ln{\sqrt{ a}\over 2-\sqrt{ a}} \right] \\
\N{7} & = & \pint {y^2\over z} \\
& = &
\left(1+\frac{1}{2} a\right) \left[ \Li\left({1+ \beta \over 2}\right)+
\Li\left({1- \beta \over 2}\right)-2\,\Li\left(\frac{1}{2}\sqrt{ a}\right)+
\frac{1}{4}\ln^2\left(\frac{1}{4} a\right) \right. \\
& &
\left. -\ln\left({1+ \beta \over 2}\right)
 \ln\left({1- \beta \over 2}\right) \right]+
3v \ln\left({1- \beta \over 1+ \beta }\right)+ \frac{1}{8}(18+ a) \\
& &
-\frac{1}{8}(20- a)\sqrt{ a} + \left(3- a+\frac{1}{16}
a^2\right) \ln\left({2-\sqrt{ a}\over\sqrt{ a}}\right) \\
\N{8} & = & \pint {1\over y}  =
2 \ln\left({2-\sqrt{ a}\over\sqrt{ a}}\right) \\
\N{9} & = &
\pint {1\over z} \\
& = &
\Li\left({1+ \beta \over 2}\right)+\Li\left({1- \beta \over 2}\right)+
2\,\Li\left(-{\sqrt{ a}\over 2-\sqrt{ a}}\right)+\frac{1}{4} \ln^2 a  \\
& &
+\ln^2\left({2-\sqrt{ a}\over 2}\right)-\ln(1+ \beta )\ln(1- \beta ) \\
\N{10} & = & \pint {y\over z^2}  = 
{4\over a} \left(1-\sqrt{ a}\right) \\
\N{11} & = & \pint {1\over y^2} \\
& = &
{2\over a} \left[ -\ln \frac{\lambda^2}{s} + \ln a +
2 \ln(1-\sqrt{ a})- 4\ln(2-\sqrt{ a})+ 2 \ln 2- 2 \right] \\
\N{12} & = & \pint {1\over z^2} \\
& = &
{2\over a} \left[ -\ln \frac{\lambda^2}{s} - \ln a +
2 \ln(1-\sqrt{ a})-{1+ \beta^2 \over \beta}
\ln\left({1+ \beta \over 1- \beta }\right)+ 2 \ln 2 \right] \\
\N{13} & = & \pint {1\over y z} \\
& = &
{1\over \beta}\ln\left({1- \beta \over 1+ \beta}\right)
\left[ \ln \frac{\lambda^2}{s} +
\frac{1}{2}\ln a+4\ln(2-\sqrt{ a})-4\ln (2 \beta ) -
2\ln\left({1- \beta \over 1+ \beta}\right) \right] \\
& &
+ \, {1\over \beta}\ln^2\left({(1-\beta )^2\over\sqrt{ a}
 (2-\sqrt{ a})} \right) +
{2\over \beta } \ln\left({\sqrt{ a}(2-\sqrt{ a})\over 2}\right)
\ln\left({2\sqrt{ a}(1-\sqrt{ a})\over (1-\sqrt{ a}- \beta )^2}\right) \\
& &
+ \, {2\over \beta} \left[ 
\Li\left({\sqrt{ a}(2-\sqrt{ a})\over (1+ \beta )^2}
\right)-\Li\left[\left({1- \beta \over 1+ \beta }\right)^2\right]+
\Li\left({(1- \beta )^2\over\sqrt{ a}(2-\sqrt{ a})}\right) \right] \\
& &
+ \, {1\over \beta } \left[ \Li\left({1+ \beta \over 2}\right)
+ \Li\left(-{2 \beta \over 1- \beta }\right) - (\beta \to -
    \beta ) - {\pi^2\over 3} \right]
\eean

\vspace{1cm}
\noindent
Class {\boldmath{$L$}} Integrals:
\bean
\L{1} & = & \tint =
2\,{1 - a \over 4 - a}\ln\left({1+ \beta \over 1- \beta }\right) \\
\L{2} & = & \tint  {z\over y} \\
& = &
\left(-1+{12\over 4- a}-{24\over(4- a)^2}\right)
\ln\left({1+ \beta \over 1- \beta }\right)
-\frac{2 \beta }{4- a} \\
\L{3} & = &
\tint {y\over z} \\
& = &
\frac{1}{2}\ln\left(\frac{1+ \beta }{1- \beta }\right)
\Biggl[\,\frac{1}{2}\ln a+\ln(2+\sqrt{a})-\ln(1+\sqrt{a})-2\ln2\,\Biggr]
\nonumber \\ &&
+ \, {1-\sqrt{a} \over\sqrt{a} }\Biggl[\,\Li(\omega )
+ \Li\left({2+\sqrt{a} \over 2-\sqrt{a}}\;\omega \right)
      - ( \omega \to - \omega )\,\Biggr] \\
      & &
      + \, \Biggl[ \, \Li\left({1+\omega \over 2}\right)
      + \Li\left((2+\sqrt{a} ){1+\omega \over 4}\right)
      + \Li\left({2\sqrt{a} \over (2+\sqrt{a} )(1+\omega )}\right)\\
      & &  \qquad\qquad\qquad\qquad - \, ( \omega \to - \omega ) \, \Biggr] \\
      \L{4} & = &  \tint {y^2\over z^2}  =
      \frac{2}{a}\left[\,\ln\left({1+ \beta \over 1-\beta }\right)
          -2 \beta \,\right] \\
      \L{5} & = & \tint y  = 
      -\left(1+\frac{1}{2} a-{6\over 4- a}\right)
      \ln\left({1+ \beta \over 1-\beta }\right)- \beta \\
      \L{6} & = & \tint z  =
      -{3 a\over 4- a}\left(1-{2\over 4- a}\right)
      \ln\left({1+ \beta \over 1-\beta }\right)+
      {2 \beta \over 4- a} \\
      \L{7} & = & \tint {\:1\over z} 
       = \frac{1}{\sqrt{a}} \, \Biggl[\,\Li(\omega )
      +\Li\left({2+\sqrt{a} \over 2-\sqrt{a} }\;\omega \right)
      - (\omega \to - \omega ) \,\Biggr] \\
      \L{8} & = & \tint {y\over z^2} =
      \frac2 a\ln\left({1+ \beta \over 1-\beta }\right)
\eean

\vspace{1cm}
\noindent
Class {\boldmath{$K$}} Integrals:
\bean
     \K{1} & = & \kint \\
     &=& \frac{2 (1-\sqrt{ a})}{\sqrt{ a}(2-\sqrt{ a})^2}
     \ln \frac{2 \sqrt{ a}(1-\sqrt{ a})}{\epsilon (1+\sqrt{ a})}-
     \frac{4 a}{(4- a)^2}
     \ln \frac{(1+\sqrt{ a })(2-\sqrt{ a })^2}{2 \sqrt{ a } a} \\
     \K{2} & = & \kint y \\
     &=& \frac{2 (1-\sqrt{ a})^2}{\sqrt{ a}(2-\sqrt{ a})^2}
     \ln \frac{2 \sqrt{ a}(1-\sqrt{ a})}{\epsilon (1+\sqrt{ a})}+
     \frac{  a ^2}{(4-  a)^2}
     \ln \frac{(1+\sqrt{ a})(2-\sqrt{ a})^2}{2\sqrt{ a} a)}\\
     & & +\, \ln\frac{2 \sqrt{ a }}{(1+\sqrt{ a })} \\
     \K{3} & = & \kint z \\
     &=& \frac{2 (1-\sqrt{ a})^2}{(2-\sqrt{ a})^3}
     \ln \frac{2 \sqrt{ a}(1-\sqrt{ a})}{\epsilon (1+\sqrt{ a})}+
     \frac{ a( a^2+20 a-32)}{2(4- a)^3}
     \ln \frac{(1+\sqrt{ a})(2-\sqrt{ a})^2}{2\sqrt{ a} a} \\
     && - \, \frac{1}{2} \ln\frac{2 \sqrt{ a}}{(1+\sqrt{ a})}+
     \frac{2 a(1-\sqrt{ a})}{(4- a)(2-\sqrt{ a})^2}  \\
     \K{4} & = & \kint yz \\
     &=& \frac{2 (1-\sqrt{ a})^3}{(2-\sqrt{ a})^3}
     \ln \frac{2 \sqrt{ a}(1-\sqrt{ a})}{\epsilon (1+\sqrt{ a})}+
     \frac{ a^2(12-7 a)}{2(4- a)^3}
     \ln \frac{(1+\sqrt{ a})(2-\sqrt{ a})^2}{2\sqrt{ a} a} \\
     && -\,  \frac{1}{2} \ln\frac{2 \sqrt{ a}}{(1+\sqrt{ a})}+
     \frac{2}{(4- a)} 
     \Biggl[  a \frac{(1-\sqrt{ a})^2}{(2-\sqrt{ a})^2}+
     \sqrt{ a}-1 \Biggr] \\
     \K{5} & = & \kint \frac{y}{z} \\
     &=&  \frac{2 (1-\sqrt{ a})}{ a(2-\sqrt{ a})}
     \ln \frac{2 \sqrt{ a}(1-\sqrt{ a})}{\epsilon (1+\sqrt{ a})}+
     \frac{(4-3 a)}{ a(4- a)}
     \ln \frac{(1+\sqrt{ a})(2-\sqrt{ a})^2}{2\sqrt{ a} a} \\
     && -\, \frac{1}{ a} \ln\frac{2 \sqrt{ a}}{(1+\sqrt{ a})}+
     \frac{4(1-\sqrt{ a})}{ a(2-\sqrt{ a})}-
     \frac{2\beta }{ a} \ln \left( \frac{1+\beta }{1-\beta } \right) \\
     \K{6} & = & \kint \frac{1}{z}\\
     &=& \frac{2}{ a(2-\sqrt{ a})}
     \ln \frac{2 \sqrt{ a}(1-\sqrt{ a})}{\epsilon (1+\sqrt{ a})}+
     \frac{(4+ a)}{ a(4- a)}
     \ln \frac{(1+\sqrt{ a})(2-\sqrt{ a})^2}{2 \sqrt{ a} a} \\
     && - \, \frac{1}{ a} \ln\frac{2 \sqrt{ a}}{(1+\sqrt{ a})}+
     \frac{4}{ a(2-\sqrt{ a})}-
     \frac{2}{ a \beta } \ln \left( \frac{1+ \beta }{1- \beta } \right) \\
     \K{7} & = & \kint \frac{y}{z^2} =
     \frac{2}{ a \sqrt{ a}} \ln\frac{2 \sqrt{ a}(1-\sqrt{ a})}
            {\epsilon (1+\sqrt{ a})} \\
     \K{8} & = & \kint \frac{y^2}{z^2} \\
     &=&  \frac{2(1 - \sqrt{ a})}{ a\sqrt{ a}}
     \ln \left(\frac{2 \sqrt{ a}(1-\sqrt{ a})}
                    {\epsilon (1+\sqrt{ a})}\right) +
     \frac{4}{  a } \ln \left(\frac{2 \sqrt{ a}}{1+\sqrt{ a}}\right)
\eean
%
%%%%%%%%%%%%%%%%%%%%%%%%%%%%%%%%%%%%%%%%%%%%%%%

%\vspace{2cm}  
\newpage
\chapter{Total cross section}

\noindent
Let us examine the unpolarized total
cross section including the top quark pairs using our formulae.
It is given by
\[  \sigma_T (e^- e^+ \to t + \bar{t} + X)
      = \frac{1}{4} \left[ \sigma_T (e^-_L e^+_R \to t + \bar{t} +X)
               + \sigma_T (e^-_R e^+_L \to t + \bar{t} + X) \right] \ .\]
Note that only $k=0, 2$ and $l,m,n = 0$ terms in Eq.(\ref{txsection})
contribute to the total cross section.
Integrating over the angle $\theta$, we get
\bean
  \lefteqn{\sigma_T (e^-_L e^+_R \to t + \bar{t} + X)}\\
    &=& \frac{\pi \alpha^2}{s} \beta \Biggl[
      \left( f_{LL} + f_{LR} \right)^2 \left( 3 - \beta^2 \right)\\
    & & \qquad \times \ 
         \left( 1 + \hat{\alpha}_s \left\{ V_I 
          - \frac{6}{3 - \beta^2} V_{II} + \frac{2}{\beta} J_{IR}^1
          -  \frac{8}{\beta} J_3 + \frac{8}{\beta (3 - \beta^2 )} J_2
            \right\} \right)\\
      & & + \  2 \left( f_{LL} - f_{LR} \right)^2 \beta^2\\
      & & \qquad \times \ 
           \left( 1 + \hat{\alpha}_s 
        \left\{ V_I + 2 V_{II} + \frac{2}{\beta} J_{IR}^1 -
             \frac{8}{\beta} J_3 
          + \frac{2 (3 - \beta^2 )}{\beta^3} J_2 + \frac{2 a}
                    {\beta^3} J_1 \right\} \right) \Biggr].
\eean
The cross section for the process $e^-_R e^+_L$ is obtained by
interchanging the coupling constant $L \leftrightarrow R$ in the above
expression.
Parametrizing the total cross section as
\bean
     R_t (s) &\equiv& \left( \frac{1}{\sigma_{\rm pt}} \right)
             \sigma_T (e^- e^+ \to t + \bar{t} + X)\\
            &=& \ R_t^{(0)} (s) + \frac{\alpha_s (s)}{\pi}
           C_2 (R) R_t^{(1)} (s) + \cdots \ ,
\eean
where $\sigma_{\rm pt} = 4 \pi \alpha^2 / 3s$, we get the following
numerical results at the CM energies $\sqrt{s} =
400\,,\,500\,,\,800\,,\,1000\,,\,1500$ GeV.
\begin{center}
\begin{tabular}{|c||c|c|}
\hline
  \quad $\sqrt{s}$ ({\rm GeV}) \quad & \quad $R_t^{(0)} (s)$ \quad &
          \quad $C_2 (R) R_t^{(1)} (s)$ \quad\\ \hline
\hline
 $ 400$ & $1.0083$   &  $8.9963$ \\ \hline 
 $ 500$ & $1.4190$   &  $6.0267$ \\ \hline
 $ 800$ & $1.6864$   &  $3.3999$ \\ \hline     
 $1000$ & $1.7317$   &  $2.8422$ \\ \hline   
 $1500$ & $1.7714$   &  $2.2911$ \\ 
   \hline
\end{tabular} 
\end{center}
\noindent
These are consistent with the results in Ref.~\cite{hs}.
\chapter{Numerical Tables}
\noindent
We have shown the numerical tables for $e^-_{L/R} e^+ \to t X
(\bar{t}~{\rm or}~\bar{t} g)$ process in the previous Chapters and Sections.
In this Appendix, we present the numerical tables, we used to analyze
the spin correlations because of its usefulness.

Now we show the numerical value of the coefficients $C_{klmn}$,
$D_{klmn}$ using following inputs.
\begin{eqnarray}
m & = & 175 \hspace{1.5cm}(\mbox{GeV})
\nonumber \\
M_{z} & = & 91.187 \hspace{1.cm}(\mbox{GeV})
\nonumber \\
\alpha & = & 1/128 
\nonumber \\
\alpha_{s} (M_{z}^{2}) & = & 0.118 
\nonumber \\
\sin^{2} \theta_{W} & = & 0.2315 \nonumber 
\end{eqnarray}
The multiplicative enhancement, $\kappa_{L}/\kappa_{L}$, the
top quark speed $\beta$, and QCD strong coupling constant $\alpha_s$,
$\hat{\alpha}_s \equiv C_2(R)/(4 \pi) \alpha_s$.
\begin{center}
\begin{tabular}{|c||c|c|c|c|c|c|}
\hline 
$\sqrt{s}$ & $400$ GeV & $500$ GeV & $800$ GeV & $1000$ GeV 
           & $1500$ GeV& $4000$ GeV  \\ \hline \hline
$\kappa(L)$   & $0.2778 $ & $0.1283 $ & $0.05747$
              & $0.04574$ & $0.03474$ & $0.02593$ \\ \hline
$\kappa(R)$   & $0.2799 $ & $0.1316 $ & $0.06023$         
              & $0.04790$ & $0.03599$ & $0.02618$ \\ \hline  
$\beta$    & $0.4841$  & $0.7141$  & $0.8992$  & $0.9368$ 
           & $0.9723$  & $0.9962$  \\ \hline
$\alpha_{s}$   
           & $0.09804$ & $0.09564$ & $0.09096$ & $0.08890$ 
           & $0.08539$ & $0.07794$ \\ \hline
$\hat{\alpha}_{s}$       
           & $0.01040$ & $0.01015$ & $0.009652$ & $0.009433$ 
           & $0.009060$& $0.008270$ \\ \hline
\end{tabular}
\end{center}
\newpage

\section{ $e^{-}_{L} e^{+} \to t_{\uparrow} + X$ }

Now we define coefficients $D^{0}_{klmn}, D^{1}_{klmn},S_{klmn}, 
S^{1}_{klmn}$ and $S^{2}_{klmn}$.  
\bea
 \frac{d \sigma}{d \cos \theta}
                  ( e_L^-  e_R^+ \to t_{\uparrow} X )
  = \frac{3 \pi \alpha^2}{4 s}
        \sum_{klmn} \left( D_{klmn} + \hat{\alpha}_s C_{klmn} \right)
              \cos^k \theta \sin^l \theta
             \cos^m \xi \sin^n \xi \ ,
\nonumber
\eea
with
\bea
D_{klmn} \equiv D_{klmn}^{0} + D_{klmn}^{1}.
\nonumber
\eea

\begin{eqnarray}
  D^{0}_{0000} &=& \beta [ f_{LL}^2 + f_{LR}^2 + 2 a f_{LL} f_{LR} ] ,\nonumber \\ 
  D^{0}_{2000} &=& \beta^3 (f_{LL}^2 + f_{LR}^2 )   ,\nonumber \\ 
  D^{0}_{1000} &=& 2 \beta^2 (f_{LL}^2 - f_{LR}^2 ) ,\nonumber \\ 
  D^{0}_{0010} &=& \beta^2 (f_{LL}^2 - f_{LR}^2 )   ,\nonumber \\ 
  D^{0}_{2010} &=& \beta^2 (f_{LL}^2 - f_{LR}^2 )   ,\nonumber \\ 
  D^{0}_{1010} &=& \beta [ (f_{LL} + f_{LR} )^2 + 
   \beta^2 (f_{LL} - f_{LR} )^2  ]                 , \nonumber \\ 
  D^{0}_{0101} &=& \frac{\beta}{\sqrt{a}} a (f_{LL} + f_{LR} )^2  ,\nonumber \\  
  D^{0}_{1101} &=& \frac{\beta^2}{\sqrt{a}} a (f_{LL}^2 - f_{LR}^2 ) .\nonumber 
\end{eqnarray}

\begin{eqnarray}
 D^{1}_{0000} &=& \beta [ f_{LL}^2 + f_{LR}^2 + 2 a f_{LL} f_{LR} ]
                    ( \hat{\alpha}_s V_{I} ) \nonumber \\
           & & \qquad\qquad\qquad - \beta [ 2 (f_{LL} + f_{LR})^2 - \beta^2
                (f_{LL} - f_{LR})^2 ] \hat{\alpha}_s V_{II} \ , \nonumber \\
  D^{1}_{2000} &=& \beta^3 (f_{LL}^2 + f_{LR}^2 )
                    ( \hat{\alpha}_s V_{I} ) + \beta^3
              (f_{LL} - f_{LR})^2 \hat{\alpha}_s V_{II} \ , \nonumber \\
  D^{1}_{1000} &=& 2 \beta^2 (f_{LL}^2 - f_{LR}^2 )
                    ( \hat{\alpha}_s V_{I} ) \ , \nonumber \\
  D^{1}_{0010} &=& \beta^2 (f_{LL}^2 - f_{LR}^2 )
                    ( \hat{\alpha}_s V_{I} ) \ , \nonumber \\
  D^{1}_{2010} &=& \beta^2 (f_{LL}^2 - f_{LR}^2 )
                    ( \hat{\alpha}_s V_{I} ) \ , \nonumber \\
  D^{1}_{1010} &=& \beta [ (f_{LL} + f_{LR} )^2 + 
              \beta^2 (f_{LL} - f_{LR} )^2  ]
                    ( \hat{\alpha}_s V_{I} ) \nonumber \\
           & & \qquad\qquad\qquad - 2 \beta [ (f_{LL} + f_{LR})^2 - \beta^2
                (f_{LL} - f_{LR})^2 ] \hat{\alpha}_s V_{II} \ , \nonumber \\
  D^{1}_{0101} &=& \frac{\beta}{\sqrt{a}} (f_{LL} + f_{LR} )^2 
                 [ a ( \hat{\alpha}_s V_{I} ) - (1 + a)
                       \hat{\alpha}_s V_{II} ] \ , \nonumber \\
  D^{1}_{1101} &=& \frac{\beta^2}{\sqrt{a}} (f_{LL}^2 - f_{LR}^2 ) 
                 [ a ( \hat{\alpha}_s V_{I} ) - (1 - a)
                       \hat{\alpha}_s V_{II}]\ . \nonumber 
\end{eqnarray}

\clearpage
\bea
 \frac{d \sigma^{SGA}}{d \cos \theta}
                  ( e_L^-  e_R^+ \to t_{\uparrow} \bar{t})
  &=& \frac{3 \pi \alpha^2}{4 s}
        \sum_{klmn} B_{klmn}
              \cos^k \theta \sin^l \theta
             \cos^m \xi \sin^n \xi \ , \nonumber \\
  &=& \frac{3 \pi \alpha^2}{4 s}
        \sum_{klmn} \left( D^{0}_{klmn} + B^1_{klmn} \right)
              \cos^k \theta \sin^l \theta
             \cos^m \xi \sin^n \xi \ , \nonumber
\eea
with
\begin{eqnarray}
  B^{1}_{0000} &=& \beta [ f_{LL}^2 + f_{LR}^2 + 2 a f_{LL} f_{LR} ]
                    S_{I} \nonumber \\
           & & \qquad\qquad\qquad - \beta [ 2 (f_{LL} + f_{LR})^2 - \beta^2
                (f_{LL} - f_{LR})^2 ] S_{II} \ ,\nonumber \\
  B^{1}_{2000} &=& \beta^3 (f_{LL}^2 + f_{LR}^2 )
                    S_{I} + \beta^3
              (f_{LL} - f_{LR})^2 S_{II} \ ,\nonumber \\
  B^{1}_{1000} &=& 2 \beta^2 (f_{LL}^2 - f_{LR}^2 )
                    S_{I} \ ,\nonumber \\
  B^{1}_{0010} &=& \beta^2 (f_{LL}^2 - f_{LR}^2 )
                    S_{I} \ ,\nonumber \\
  B^{1}_{2010} &=& \beta^2 (f_{LL}^2 - f_{LR}^2 )
                    S_{I} \ ,\nonumber \\
  B^{1}_{1010} &=& \beta [ (f_{LL} + f_{LR} )^2 + 
              \beta^2 (f_{LL} - f_{LR} )^2  ]
                    S_{I}\nonumber \\
           & & \qquad\qquad\qquad - 2 \beta [ (f_{LL} + f_{LR})^2 - \beta^2
                (f_{LL} - f_{LR})^2 ] S_{II} \ ,\nonumber \\
  B^{1}_{0101} &=& \frac{\beta}{\sqrt{a}} (f_{LL} + f_{LR} )^2 
                 [ a S_{I} - (1 + a)
                       S_{II} ] \ ,\nonumber \\
  B^{1}_{1101} &=& \frac{\beta^2}{\sqrt{a}} (f_{LL}^2 - f_{LR}^2 ) 
                 [ a S_{I} - (1 - a)
                       S_{II}] \nonumber 
\end{eqnarray}

\begin{eqnarray}
  D_{klmn} + \hat{\alpha}_{s} C_{klmn}
  & \equiv &
  D^{0}_{klmn} + S_{klmn} \nonumber \\
  & \equiv &
  (1 + \kappa )D^{0}_{klmn} + S^{1}_{klmn} 
  \mbox{\hspace*{1cm} $\cdots$ Full}
  \nonumber \\
  & \equiv &
  (1 + \kappa )D^{0}_{klmn} + S^{2}_{klmn} 
  \mbox{\hspace*{1cm} $\cdots$ S.G.A}
  \nonumber 
\end{eqnarray}
Therefore, $S_{klmn}, S^{1}_{klmn}$ and $S^{2}_{klmn}$ are given by 
\begin{eqnarray}
  S_{klmn} & = & D^{1}_{klmn} + \hat{\alpha}_{s} C_{klmn} \nonumber \\
  S^{1}_{klmn} & = & D^{1}_{klmn} + \hat{\alpha}_{s} C_{klmn} 
                - \kappa D^{0}_{klmn} \nonumber \\
  S^{2}_{klmn} & = & B^{1}_{klmn} - \kappa D^{0}_{klmn} ~.\nonumber 
\end{eqnarray}

\begin{eqnarray}
R^{Full}_{klmn} & = & \frac{S^{1}_{klmn}}{(1 + \kappa) D^{0}_{klmn}}
\nonumber \\
R^{SGA}_{klmn} & = & \frac{S^{2}_{klmn}}{(1 + \kappa) D^{0}_{klmn}}
\nonumber 
\end{eqnarray}

\clearpage
\begin{itemize}
\item{$\cal{O}\mit( \alpha_{s} )$ Analysis}
\begin{center}
\begin{tabular}{|c||c|c|c|c|c|c|}
\hline 
$\sqrt{s}$ & $400$  GeV  & $500$  GeV  
           & $800$  GeV  & $1000$ GeV 
           & $1500$ GeV  & $4000$ GeV \\ \hline \hline
$C_{0000}$ & $ -11.66  $ & $ -30.13 $ & $ -58.68 $ 
           & $ -71.01  $ & $ -94.84 $ & $ -167.3 $ \\ \hline 
$C_{2000}$ & $  -1.872 $ & $ -12.07 $ & $ -44.72 $ 
           & $ -61.57  $ & $ -92.84 $ & $ -175.3 $ \\ \hline 
$C_{1000}$ & $  -6.010 $ & $ -25.83 $ & $ -74.58 $ 
           & $ -98.212 $ & $ -142.3 $ & $ -263.3 $ \\ \hline 
$C_{0010}$ & $  -2.988 $ & $ -12.72 $ & $ -36.28 $ 
           & $ -47.68  $ & $ -69.09 $ & $ -128.7 $ \\ \hline 
$C_{2010}$ & $  -3.023 $ & $ -13.14 $ & $ -38.56 $ 
           & $ -50.97  $ & $ -74.00 $ & $ -136.2 $ \\ \hline 
$C_{1010}$ & $ -13.60  $ & $ -42.69 $ & $ -105.4 $ 
           & $-135.4   $ & $ -191.9 $ & $ -349.5 $ \\ \hline 
$C_{0101}$ & $ -11.28  $ & $ -26.85 $ & $ -39.76 $ 
           & $ -40.73  $ & $ -38.70 $ & $ -27.28 $ \\ \hline 
$C_{1101}$ & $  -2.637 $ & $ -9.126 $ & $ -16.75 $ 
           & $ -17.79  $ & $ -17.47 $ & $ -12.60 $ \\ \hline
\hline 
$D_{0000}$ & $ 1.630  $ & $ 2.025  $ & $ 2.289 $ 
           & $ 2.374  $ & $ 2.542  $ & $ 3.048 $ \\ \hline 
$D_{2000}$ & $ 0.2675 $ & $ 0.8100 $ & $ 1.671 $ 
           & $ 1.949  $ & $ 2.332  $ & $ 3.012 $ \\ \hline 
$D_{1000}$ & $ 0.8650 $ & $ 1.760  $ & $ 2.858 $ 
           & $ 3.194  $ & $ 3.674  $ & $ 4.626 $ \\ \hline 
$D_{0010}$ & $ 0.4325 $ & $ 0.8799 $ & $ 1.429 $ 
           & $ 1.597  $ & $ 1.837  $ & $ 2.313 $ \\ \hline 
$D_{2010}$ & $ 0.4325 $ & $ 0.8799 $ & $ 1.429 $ 
           & $ 1.597  $ & $ 1.837  $ & $ 2.313 $ \\ \hline 
$D_{1010}$ & $ 1.890  $ & $ 2.835  $ & $ 3.960 $ 
           & $ 4.324  $ & $ 4.874  $ & $ 6.060 $ \\ \hline 
$D_{0101}$ & $ 1.563  $ & $ 1.753  $ & $ 1.439 $ 
           & $ 1.241  $ & $ 0.9196 $ & $ 0.4246$ \\ \hline 
$D_{1101}$ & $ 0.3770 $ & $ 0.6100 $ & $ 0.6139$ 
           & $ 0.5475 $ & $ 0.4185 $ & $ 0.1969$ \\ \hline 
\end{tabular}
\end{center}

%\clearpage

\begin{center}
\begin{tabular}{|c||c|c|c|c|c|c|}
\hline 
$\sqrt{s}$ & $400$  GeV & $500$ GeV & $800$ GeV & $1000$ GeV 
           & $1500$ GeV & $4000$ GeV \\ \hline \hline
$D^{0}_{0000}$ & $ 1.183  $ & $ 1.528  $ & $ 1.629 $ 
               & $ 1.625  $ & $ 1.614  $ & $ 1.603 $ \\ \hline
$D^{0}_{2000}$ & $ 0.1882 $ & $ 0.5960 $ & $ 1.174 $ 
               & $ 1.323  $ & $ 1.475  $ & $ 1.583 $ \\ \hline
$D^{0}_{1000}$ & $ 0.6111 $ & $ 1.299  $ & $ 2.010 $ 
               & $ 2.169  $ & $ 2.325  $ & $ 2.431 $ \\ \hline
$D^{0}_{0010}$ & $ 0.3056 $ & $ 0.6497 $ & $ 1.005 $ 
               & $ 1.085  $ & $ 1.163  $ & $ 1.215 $ \\ \hline
$D^{0}_{2010}$ & $ 0.3056 $ & $ 0.6497 $ & $ 1.005 $ 
               & $ 1.085  $ & $ 1.163  $ & $ 1.215 $ \\ \hline
$D^{0}_{1010}$ & $ 1.371  $ & $ 2.124  $ & $ 2.802 $ 
               & $ 2.948  $ & $ 3.089  $ & $ 3.185 $ \\ \hline
$D^{0}_{0101}$ & $ 1.137  $ & $ 1.331  $ & $ 1.040 $ 
               & $ 0.8652 $ & $ 0.5978 $ & $ 0.2294$ \\ \hline
$D^{0}_{1101}$ & $ 0.2674 $ & $ 0.4548 $ & $ 0.4397$ 
               & $ 0.3797 $ & $ 0.2712 $ & $ 0.1063$ \\ \hline
\hline 
$S_{0000}$ & $ 0.3260  $ & $ 0.1910  $ & $ 0.09417 $ 
           & $ 0.07918 $ & $ 0.06780 $ & $ 0.06193 $ \\ \hline
$S_{2000}$ & $ 0.05993 $ & $ 0.09144 $ & $ 0.06577 $ 
           & $ 0.04599 $ & $ 0.01608 $ & $-0.02013 $ \\ \hline
$S_{1000}$ & $ 0.1913  $ & $ 0.1984  $ & $ 0.1278  $ 
           & $ 0.09841 $ & $ 0.05992 $ & $ 0.01786 $ \\ \hline
$S_{0010}$ & $ 0.09583 $ & $ 0.1012  $ & $ 0.07363 $ 
           & $ 0.06265 $ & $ 0.04874 $ & $ 0.03378 $ \\ \hline
$S_{2010}$ & $ 0.09548 $ & $ 0.09688 $ & $ 0.05161 $ 
           & $ 0.03170 $ & $ 0.004244$ & $-0.02836 $ \\ \hline
$S_{1010}$ & $ 0.3852  $ & $ 0.2775  $ & $ 0.1408  $ 
           & $ 0.09893 $ & $ 0.04598 $ & $-0.01484 $ \\ \hline
$S_{0101}$ & $ 0.3092  $ & $ 0.1488  $ & $ 0.01564 $ 
           & $-0.008947$ & $-0.02875 $ & $-0.03045 $ \\ \hline
$S_{1101}$ & $ 0.08220 $ & $ 0.06263 $ & $ 0.01257 $ 
           & $0.00001984$& $-0.01105 $ & $-0.01366 $ \\ \hline
\end{tabular}
\end{center}
\clearpage

\begin{center}
\begin{tabular}{|c||c|c|c|c|c|c|}
\hline 
             & $400$  GeV & $500$  GeV & $800$ GeV  
             & $1000$ GeV & $1500$ GeV & $4000$GeV  \\ \hline
\hline
$(1 + \kappa)D^{0}_{0000}$ 
             & $ 1.511  $ & $ 1.724 $ & $ 1.722 $   
             & $ 1.700  $ & $ 1.671 $ & $ 1.644 $ \\ \hline
$(1 + \kappa)D^{0}_{2000}$ 
             & $ 0.2404 $ & $ 0.6724$ & $ 1.241 $   
             & $ 1.383  $ & $ 1.526 $ & $ 1.624 $ \\ \hline
$(1 + \kappa)D^{0}_{1000}$ 
             & $ 0.7809 $ & $ 1.466 $ & $ 2.126 $   
             & $ 2.269  $ & $ 2.406 $ & $ 2.494 $ \\ \hline
$(1 + \kappa)D^{0}_{0010}$ 
             & $ 0.3905 $ & $ 0.7330$ & $ 1.063 $   
             & $ 1.134  $ & $ 1.203 $ & $ 1.247 $ \\ \hline
$(1 + \kappa)D^{0}_{2010}$ 
             & $ 0.3905 $ & $ 0.7330$ & $ 1.063 $   
             & $ 1.134  $ & $ 1.203 $ & $ 1.247 $ \\ \hline
$(1 + \kappa)D^{0}_{1010}$ 
             & $ 1.751  $ & $ 2.396  $ & $ 2.963 $   
             & $ 3.083  $ & $ 3.197  $ & $ 3.268 $ \\ \hline
$(1 + \kappa)D^{0}_{0101}$ 
             & $ 1.452  $ & $ 1.502  $ & $ 1.100 $   
             & $ 0.9048 $ & $ 0.6186 $ & $ 0.2353$ \\ \hline
$(1 + \kappa)D^{0}_{1101}$ 
             & $ 0.3416 $ & $ 0.5131 $ & $ 0.4650$   
             & $ 0.3970 $ & $ 0.2807 $ & $ 0.1091$ \\ \hline
\hline
$S^{1}_{0000}$ & $ -0.002552  $ & $-0.005000 $ & $  0.0005645$ 
               & $  0.004836  $ & $  0.01172 $ & $  0.02039  $ \\ \hline
$S^{1}_{2000}$ & $  0.007655  $ & $ 0.01500  $ & $ -0.001682 $ 
               & $ -0.01450   $ & $ -0.03516 $ & $ -0.06116  $ \\ \hline
$S^{1}_{1000}$ & $  0.02154   $ & $ 0.03170  $ & $  0.01224  $ 
               & $ -0.0008102 $ & $ -0.02085 $ & $ -0.04516  $ \\ \hline
$S^{1}_{0010}$ & $  0.01094   $ & $ 0.01785  $ & $  0.01586  $ 
               & $ 0.01304    $ & $ 0.008357 $ & $  0.002274 $ \\ \hline
$S^{1}_{2010}$ & $  0.01059   $ & $ 0.01356  $ & $ -0.006158 $ 
               & $ -0.01791   $ & $ -0.03614 $ & $ -0.05987  $ \\ \hline
$S^{1}_{1010}$ & $  0.004433  $ & $ 0.005069 $ & $ -0.02030  $ 
               & $ -0.03590   $ & $ -0.06134 $ & $ -0.09742  $ \\ \hline
$S^{1}_{0101}$ & $ -0.006564  $ & $-0.02193  $ & $ -0.04413  $ 
               & $ -0.04852   $ & $ -0.04952 $ & $ -0.03639  $ \\ \hline
$S^{1}_{1101}$ & $  0.007920  $ & $ 0.004308 $ & $ -0.01270  $ 
               & $ -0.01734   $ & $ -0.02047 $ & $ -0.01642  $ \\ \hline
\end{tabular}
\end{center}
%\clearpage
%
\begin{center}
\begin{tabular}{|c||c|c|c|c|c|c|}
\hline 
	              & $400$ GeV  & $500$ GeV  & $800$ GeV  
                      & $1000$ GeV & $1500$ GeV & $4000$ GeV \\ \hline
\hline
$R^{Full}_{0000}$ & $ -0.001689  $  &  $-0.002900  $ & $ 0.0003278$ 
                  & $  0.002845  $  &  $  0.007014 $ & $ 0.01240  $ \\ \hline
$R^{Full}_{2000}$ & $  0.03184   $  &  $ 0.02231   $ & $-0.001356 $ 
                  & $ -0.01048   $  &  $ -0.02304  $ & $-0.03767  $ \\ \hline
$R^{Full}_{1000}$ & $  0.02758   $  &  $ 0.02163   $ & $ 0.005757 $ 
                  & $ -0.0003571 $  &  $ -0.008667 $ & $-0.01811  $ \\ \hline
$R^{Full}_{0010}$ & $  0.02803   $  &  $ 0.02435   $ & $ 0.01492  $ 
                  & $  0.01150   $  &  $  0.006948 $ & $ 0.001823 $ \\ \hline
$R^{Full}_{2010}$ & $  0.02711   $  &  $ 0.01849   $ & $-0.005793 $ 
                  & $ -0.01579   $  &  $ -0.03004  $ & $-0.04802  $ \\ \hline
$R^{Full}_{1010}$ & $  0.0025310 $  &  $ 0.002115  $ & $-0.006849 $ 
                  & $ -0.01165   $  &  $ -0.01919  $ & $-0.02981  $ \\ \hline
$R^{Full}_{0101}$ & $ -0.004520  $  &  $-0.01460   $ & $-0.04013  $ 
                  & $ -0.05362   $  &  $ -0.08005  $ & $-0.15465  $ \\ \hline
$R^{Full}_{1101}$ & $  0.02318   $  &  $ 0.008395  $ & $-0.02731  $ 
                  & $ -0.04369   $  &  $ -0.07294  $ & $-0.15050  $ \\ \hline
\end{tabular}
\end{center}

\clearpage
\item{ Soft Gluon App. }

We define the $\omega_{max}$ to satisfy the following relation.

\begin{eqnarray}
& ~ &
\int_{-1}^{1} d \cos \theta 
\sum_{s = \uparrow, \downarrow } 
\frac{ d \sigma_{T,L/R} }
     { d \cos \theta }
     (e^{-}_{L/R} e^{+} \rightarrow t_{s}  X 
      (\bar{t} ~{\rm or}~ \bar{t} g) ) \nonumber \\ 
& = &
\int_{-1}^{1} d \cos \theta 
\sum_{s = \uparrow, \downarrow } 
\frac{ d \sigma_{SGA,L/R} }
     { d \cos \theta }
     (e^{-}_{L/R}  e^{+} \rightarrow t_{s}  \bar{t})
\nonumber 
\end{eqnarray}

\begin{center}
\begin{tabular}{|c||c|c|c|c|c|c|}
\hline 
	              & $400$ GeV  & $500$ GeV  & $800$ GeV  
                      & $1000$ GeV & $1500$ GeV & $4000$ GeV \\ \hline
\hline	
$\omega_{max}$ & $ 13.49 $ & $ 38.60 $ & $ 106.7 $ 
               & $ 149.7 $ & $ 254.8 $ & $ 775.9 $ \\ \hline
$B_{0000}$ &	$1.508$ & $1.718$ & $1.716$ & $1.695$ & $1.668$ & $1.644$ \\ \hline
$B_{2000}$ &	$0.2482$ & $0.6903$ & $1.258$ & $1.397$ & $1.534$ & $1.625$ \\ \hline
$B_{1000}$ &	$0.8023$ & $1.499$ & $2.151$ & $2.288$ & $2.416$ & $2.496$ \\ \hline
$B_{0010}$ &	$0.4011$ & $0.7495$ & $1.075$ & $1.144$ & $1.208$ & $1.248$ \\ \hline
$B_{2010}$ &	$0.4011$ & $0.7495$ & $1.075$ & $1.144$ & $1.208$ & $1.248$ \\ \hline
$B_{1010}$ &	$1.757$ & $2.408$ & $2.975$ & $3.092$ & $3.202$ & $3.269$ \\ \hline
$B_{0101}$ &	$1.446$ & $1.486$ & $1.0749$ & $0.8790$ & $0.5962$ & $0.2235$ \\ \hline
$B_{1101}$ &	$0.3496$ & $0.5187$ & $0.4593$ & $0.3889$ & $0.2717$ & $0.1037$ \\ \hline
\hline 		
$B^{1}_{0000}$ &	$0.3259$ & $0.1900$ & $0.08785$ & $0.06979$ & $0.05347$ & $0.04103$ \\ \hline
$B^{1}_{2000}$ &	$0.06009$ & $0.09431$ & $0.08469$ & $0.07414$ & $0.05908$ & $0.04258$ \\ \hline
$B^{1}_{1000}$ &	$0.1911$ & $0.1996$ & $0.1407$ & $0.1183$ & $0.09133$ & $0.06506$ \\ \hline
$B^{1}_{0010}$ &	$0.09556$ & $0.09979$ & $0.07034$ & $0.05916$ & $0.04567$ & $0.03253$ \\ \hline
$B^{1}_{2010}$ &	$0.09556$ & $0.09979$ & $0.07034$ & $0.05916$ & $0.04567$ & $0.03253$ \\ \hline
$B^{1}_{1010}$ &	$0.3860$ & $0.2843$ & $0.1725$ & $0.1439$ & $0.1126$ & $0.08361$ \\ \hline
$B^{1}_{0101}$ &	$0.3099$ & $0.1540$ & $0.03373$ & $0.01373$ & $-0.001588$ & $-0.005863$ \\ \hline
$B^{1}_{1101}$ &	$0.08219$ & $0.06395$ & $0.01957$ & $0.009227$ & $0.0004544$ & $-0.002634$ \\ \hline
\end{tabular}
\end{center}
\clearpage
\begin{center}
\begin{tabular}{|c||c|c|c|c|c|c|}
\hline 
	              & $400$ GeV  & $500$ GeV  & $800$ GeV  
                      & $1000$ GeV & $1500$ GeV & $4000$ GeV \\ \hline
\hline	
$S^{2}_{0000}$ &	$-0.002607$ & $-0.005957$ & $-0.005747$ & $-0.004551$ & $-0.002615$ & $-0.0005169$ \\ \hline
$S^{2}_{2000}$ &	$0.007821$ & $0.01787$ & $0.01724$ & $0.01365$ & $0.007846$ & $0.001551$ \\ \hline
$S^{2}_{1000}$ &	$0.02134$ & $0.03293$ & $0.02514$ & $0.01910$ & $0.01057$ & $0.002038$ \\ \hline
$S^{2}_{0010}$ &	$0.01067$ & $0.01646$ & $0.01257$ & $0.009549$ & $0.005284$ & $0.001019$ \\ \hline
$S^{2}_{2010}$ &	$0.01067$ & $0.01646$ & $0.01257$ & $0.009549$ & $0.005284$ & $0.001019$ \\ \hline
$S^{2}_{1010}$ &	$0.005214$ & $0.01191$ & $0.01149$ & $0.009101$ & $0.005231$ & $0.001034$ \\ \hline
$S^{2}_{0101}$ &	$-0.005870$ & $-0.01676$ & $-0.02604$ & $-0.02585$ & $-0.02235$ & $-0.01181$ \\ \hline
$S^{2}_{1101}$ &	$0.007915$ & $0.005621$ & $-0.005706$ & $-0.008136$ & $-0.008968$ & $-0.005391$ \\ \hline
\hline		
$R^{SGA}_{0000}$ &	$-0.001725$ & $-0.003455$ & $-0.003337$ & $-0.002677$ & $-0.001566$ & $-0.0003144$ \\ \hline
$R^{SGA}_{2000}$ &	$0.03253$ & $0.02658$ & $0.01389$ & $0.009870$ & $0.005141$ & $0.0009552$ \\ \hline
$R^{SGA}_{1000}$ &	$0.02733$ & $0.02246$ & $0.01183$ & $0.008418$ & $0.004393$ & $0.0008172$ \\ \hline
$R^{SGA}_{0010}$ &	$0.02733$ & $0.02246$ & $0.01183$ & $0.008418$ &
 $0.004393$ & $0.0008172$ \\ \hline
$R^{SGA}_{2010}$ &	$0.02733$ & $0.02246$ & $0.01183$ & $0.008418$ & $0.004393$ & $0.0008172$ \\ \hline
$R^{SGA}_{1010}$ &	$0.002977$ & $0.004972$ & $0.003879$ & $0.002952$ & $0.001636$ & $0.0003164$ \\ \hline
$R^{SGA}_{0101}$ &	$-0.004042$ & $-0.01115$ & $-0.02368$ & $-0.02857$ & $-0.03614$ & $-0.05018$ \\ \hline
$R^{SGA}_{1101}$ &	$0.02317$ & $0.01096$ & $-0.01227$ & $-0.02049$ & $-0.03195$ & $-0.04941$ \\ \hline
\end{tabular}
\end{center}

%\clearpage
\item{SGA I}

\begin{center}
\begin{tabular}{|c||c|c|c|c|c|c|}
\hline 
$\sqrt{s}$     &  $400$ GeV & $500$ GeV  & $800$ GeV 
               & $1000$ GeV & $1500$ GeV & $4000$ GeV \\ \hline
\hline
$\omega_{max}$ &  $10$ GeV  &  $30$ GeV  & $90$ GeV  
               &  $130$ GeV &  $230$ GeV & $730$ GeV \\ \hline  
$B_{0000}$     & $ 1.498  $ & $ 1.690  $ & $ 1.675  $ 
               & $ 1.653  $ & $ 1.629  $ & $ 1.610  $ \\ \hline 
$B_{2000}$     & $ 0.2466 $ & $ 0.6794 $ & $ 1.228  $ 
               & $ 1.363  $ & $ 1.498  $ & $ 1.592  $ \\ \hline 
$B_{1000}$     & $ 0.7970 $ & $ 1.475  $ & $ 2.099  $ 
               & $ 2.232  $ & $ 2.360  $ & $ 2.444  $ \\ \hline 
$B_{0010}$     & $ 0.3984 $ & $ 0.7376 $ & $ 1.050  $ 
               & $ 1.116  $ & $ 1.180  $ & $ 1.222  $ \\ \hline 
$B_{2010}$     & $ 0.3984 $ & $ 0.7376 $ & $ 1.050  $ 
               & $ 1.116  $ & $ 1.180  $ & $ 1.222  $ \\ \hline 
$B_{1010}$     & $ 1.745  $ & $ 2.370  $ & $ 2.903  $ 
               & $ 3.017  $ & $ 3.127  $ & $ 3.201  $ \\ \hline 
$B_{0101}$     & $ 1.436  $ & $ 1.461  $ & $ 1.047  $ 
               & $ 0.8566 $ & $ 0.5817 $ & $ 0.2187 $ \\ \hline 
$B_{1101}$     & $ 0.3472 $ & $ 0.5104 $ & $ 0.4480 $ 
               & $ 0.3790 $ & $ 0.2651 $ & $ 0.1015 $ \\ \hline 
\hline
$B^{1}_{0000}$     & $ 0.3157  $ & $  0.1621   $ & $ 0.04600  $ 
                   & $ 0.02769 $ & $  0.01430  $ & $ 0.007045 $ \\ \hline 
$B^{1}_{2000}$     & $ 0.05846 $ & $  0.08341  $ & $ 0.05453  $ 
                   & $ 0.03989 $ & $  0.02330  $ & $ 0.009018 $ \\ \hline 
$B^{1}_{1000}$     & $ 0.1858  $ & $  0.1758   $ & $ 0.08901  $ 
                   & $ 0.06213 $ & $  0.03493  $ & $ 0.01351  $ \\ \hline 
$B^{1}_{0010}$     & $ 0.09292 $ & $  0.08791  $ & $ 0.04451  $ 
                   & $ 0.03107 $ & $  0.01746  $ & $ 0.006754 $ \\ \hline 
$B^{1}_{2010}$     & $ 0.09292 $ & $  0.08791  $ & $ 0.04451  $ 
                   & $ 0.03107 $ & $  0.01746  $ & $ 0.006754 $ \\ \hline 
$B^{1}_{1010}$     & $ 0.3742  $ & $  0.2455   $ & $ 0.1005   $ 
                   & $ 0.06758 $ & $  0.03760  $ & $ 0.01606  $ \\ \hline 
$B^{1}_{0101}$     & $ 0.3000  $ & $  0.1297   $ & $ 0.007001 $ 
                   & $-0.008684$ & $ -0.01609  $ & $-0.01073  $ \\ \hline 
$B^{1}_{1101}$     & $ 0.07988 $ & $  0.05563  $ & $ 0.008266 $ 
                   & $-0.0006056$& $ -0.006126 $ & $-0.004889 $ \\ \hline 
\end{tabular}
\end{center}
\end{itemize}

\clearpage

\section{$e^-_{R}~e^+ \to t_{\downarrow} + X $}
\begin{itemize}
\item{$\cal{O}\mit( \alpha_{s} )$ Analysis}

\begin{center}
\begin{tabular}{|c||c|c|c|c|c|c|}
\hline 
$\sqrt{s}$ & $400$  GeV & $500$  GeV  
           & $800$  GeV & $1000$ GeV 
           & $1500$ GeV & $4000$ GeV \\ \hline \hline
$C_{0000}$ & 	$ -5.088 $ & $ -13.73 $ & $ -28.17 $ & $ -34.56 $ & $ -46.82 $ & $ -83.45 $ \\ \hline
$C_{2000}$ & 	$ -0.9132$ & $ -5.943 $ & $ -22.22 $ & $ -30.66 $ & $ -46.31 $ & $ -87.57 $ \\ \hline
$C_{1000}$ & 	$ -3.387 $ & $ -14.68 $ & $ -42.75 $ & $ -56.41 $ & $ -81.91 $ & $ -151.7 $ \\ \hline
$C_{0010}$ & 	$ -1.684 $ & $ -7.227 $ & $ -20.80 $ & $ -27.39 $ & $ -39.76 $ & $ -74.14 $ \\ \hline
$C_{2010}$ & 	$ -1.704 $ & $ -7.467 $ & $ -22.11 $ & $ -29.27 $ & $ -42.59 $ & $ -78.47 $ \\ \hline
$C_{1010}$ & 	$ -6.033 $ & $ -19.91 $ & $ -51.38 $ & $ -66.60 $ & $ -95.22 $ & $ -174.4 $ \\ \hline
$C_{0101}$ & 	$ -4.816 $ & $ -11.64 $ & $ -17.53 $ & $ -18.03 $ & $ -17.21 $ & $ -12.20 $ \\ \hline
$C_{1101}$ & 	$ -1.486 $ & $ -5.187 $ & $ -9.600 $ & $ -10.22 $ & $ -10.05 $ & $ -7.263 $ \\ \hline
\hline	
$D_{0000}$ &	$ 0.7129 $ & $ 0.9271 $ & $ 1.104 $ & $ 1.159 $ & $ 1.257 $ & $ 1.521 $ \\ \hline
$D_{2000}$ &	$ 0.1308 $ & $ 0.3992 $ & $ 0.8305$ & $ 0.9708$ & $ 1.164 $ & $ 1.505 $ \\ \hline
$D_{1000}$ &	$ 0.4875 $ & $ 1.000  $ & $ 1.638 $ & $ 1.835 $ & $ 2.115 $ & $ 2.666 $ \\ \hline
$D_{0010}$ &	$ 0.2438 $ & $ 0.5001 $ & $ 0.8192$ & $ 0.9174$ & $ 1.057 $ & $ 1.333 $ \\ \hline
$D_{2010}$ &	$ 0.2438 $ & $ 0.5001 $ & $ 0.8192$ & $ 0.9174$ & $ 1.057 $ & $ 1.333 $ \\ \hline
$D_{1010}$ &	$ 0.8437 $ & $ 1.3263 $ & $ 1.9343$ & $ 2.130 $ & $ 2.420 $ & $ 3.025 $ \\ \hline
$D_{0101}$ &	$ 0.6679 $ & $ 0.7616 $ & $ 0.6364$ & $ 0.5506$ & $ 0.4097$ & $ 0.1897$ \\ \hline
$D_{1101}$ &	$ 0.2125 $ & $ 0.3467 $ & $ 0.3520$ & $ 0.3145$ & $ 0.2408$ & $ 0.1135$ \\ \hline
\end{tabular}
\end{center}
\begin{center}

%\clearpage

\begin{tabular}{|c||c|c|c|c|c|c|}
\hline 
$\sqrt{s}$ & $400$  GeV & $500$  GeV  
           & $800$  GeV & $1000$ GeV 
           & $1500$ GeV & $4000$ GeV \\ \hline \hline
$D^{0}_{0000}$ &	$ 0.5167 $ & $ 0.6982 $ & $ 0.7840$ & $ 0.7927$ & $ 0.7978$ & $ 0.7995$ \\ \hline
$D^{0}_{2000}$ &	$0.09176 $ & $ 0.2932 $ & $ 0.5829$ & $ 0.6583$ & $ 0.7357$ & $ 0.7905$ \\ \hline
$D^{0}_{1000}$ &	$ 0.3444 $ & $ 0.7385 $ & $ 1.152 $ & $ 1.246 $ & $ 1.338 $ & $ 1.401 $ \\ \hline
$D^{0}_{0010}$ &	$ 0.1722 $ & $ 0.3692 $ & $ 0.5762$ & $ 0.6231$ & $ 0.6690$ & $ 0.7004$ \\ \hline
$D^{0}_{2010}$ &	$ 0.1722 $ & $ 0.3692 $ & $ 0.5762$ & $ 0.6231$ & $ 0.6690$ & $ 0.7004$ \\ \hline
$D^{0}_{1010}$ &	$ 0.6085 $ & $ 0.9915 $ & $ 1.367 $ & $ 1.451 $ & $ 1.534 $ & $ 1.590 $ \\ \hline
$D^{0}_{0101}$ &	$ 0.4856 $ & $ 0.5786 $ & $ 0.4598$ & $ 0.3840$ & $ 0.2663$ & $ 0.1025$ \\ \hline
$D^{0}_{1101}$ &	$ 0.1507 $ & $ 0.2585 $ & $ 0.2521$ & $ 0.2181$ & $ 0.1561$ & $0.06128$ \\ \hline
\hline	
$S_{0000}$ &	$ 0.1433$ & $ 0.08952$ & $0.04786$ & $0.04071$ & $0.03479$ &$0.03117  $ \\ \hline
$S_{2000}$ &	$ 0.02954$& $ 0.04566$ & $0.03323$ & $0.02331$ & $0.008236$&$-0.01002 $\\ \hline
$S_{1000}$ &	$ 0.1078$ & $ 0.1127 $ & $0.07325$ & $0.05653$ & $0.03448$ &$0.01029  $\\ \hline
$S_{0010}$ &	$0.05401$ & $ 0.05750$ & $0.04221$ & $0.03599$ & $0.02805$ &$ 0.01947 $\\ \hline
$S_{2010}$ &	$0.05381$ & $ 0.05506$ & $0.02959$ & $0.01821$ & $0.002443$&$-0.01634 $\\ \hline
$S_{1010}$ &	$0.1725 $ & $ 0.1328 $ & $0.07156$ & $0.05096$ & $0.02413 $&$-0.007143$\\ \hline
$S_{0101}$ &	$0.1322 $ & $ 0.06490$ & $0.007412$&$-0.003494$& $-0.01250$&$-0.01371 $ \\ \hline
$S_{1101}$ &	$0.04633$ & $ 0.03560$ & $0.007208$&$0.000011398$&$-0.006358$&$-0.007873$ \\ \hline
\end{tabular}
\end{center}	
\clearpage
\begin{center}
\begin{tabular}{|c||c|c|c|c|c|c|}
\hline 
$\sqrt{s}$ & $400$  GeV & $500$  GeV  
           & $800$  GeV & $1000$ GeV 
           & $1500$ GeV & $4000$ GeV \\ \hline \hline
$(1 + \kappa)D^{0}_{0000}$ &	$ 0.6613$ & $ 0.7901$ & $ 0.8313$ & $ 0.8307$ & $ 0.8266$ & $ 0.8204$ \\ \hline
$(1 + \kappa)D^{0}_{2000}$ &	$ 0.1174$ & $ 0.3318$ & $ 0.6180$ & $ 0.6899$ & $ 0.7622$ & $ 0.8112$ \\ \hline
$(1 + \kappa)D^{0}_{1000}$ &	$ 0.4408$ & $ 0.8357$ & $ 1.222 $ & $ 1.306 $ & $ 1.386 $ & $ 1.437 $ \\ \hline
$(1 + \kappa)D^{0}_{0010}$ &	$ 0.2204$ & $ 0.4178$ & $ 0.6109$ & $ 0.6529$ & $ 0.6931$ & $ 0.7187$ \\ \hline
$(1 + \kappa)D^{0}_{2010}$ &	$ 0.2204$ & $ 0.4178$ & $ 0.6109$ & $ 0.6529$ & $ 0.6931$ & $ 0.7187$ \\ \hline
$(1 + \kappa)D^{0}_{1010}$ &	$ 0.7787$ & $ 1.122 $ & $ 1.449 $ & $ 1.521 $ & $ 1.589 $ & $ 1.632 $ \\ \hline
$(1 + \kappa)D^{0}_{0101}$ &	$ 0.6216$ & $ 0.6547$ & $ 0.4875$ & $ 0.4024$ & $ 0.2759$ & $ 0.1052$ \\ \hline
$(1 + \kappa)D^{0}_{1101}$ &	$ 0.1929$ & $ 0.2925$ & $ 0.2673$ & $ 0.2285$ & $ 0.1617$ & $0.06289$ \\ \hline
\hline	
$S^{1}_{0000}$ &	$-0.001287$&$-0.002358$&$0.0006297$&$0.002742$&$0.006079$&$0.01024$ \\ \hline
$S^{1}_{2000}$ &	$ 0.003862$&$0.007074$&$-0.001884$&$-0.008222$&$-0.01824$&$-0.03072$ \\ \hline
$S^{1}_{1000}$ &	$0.01143 $&$ 0.01556 $&$ 0.003835$&$-0.003159$&$-0.01367$&$-0.02639$ \\ \hline
$S^{1}_{0010}$ &	$0.005814$&$ 0.008914$&$ 0.007500$&$ 0.006143$&$0.003972$&$0.001128$ \\ \hline
$S^{1}_{2010}$ &	$0.005612$&$0.006475 $&$-0.005121$&$-0.01164 $&$-0.02164$&$-0.03468$ \\ \hline
$S^{1}_{1010}$ &	$0.002248$&$0.002290 $&$-0.01078 $&$-0.01854 $&$-0.03107$&$-0.04878$ \\ \hline
$S^{1}_{0101}$ &	$-0.003764$&$-0.01122$&$-0.02028 $&$-0.02189 $&$-0.02208$&$-0.01640$ \\ \hline
$S^{1}_{1101}$ &	$0.004154 $&$0.001587$&$-0.007977$&$-0.01043 $&$-0.01198$&$-0.009478$ \\ \hline
	\end{tabular}
\end{center}
%\clearpage
\begin{center}
\begin{tabular}{|c||c|c|c|c|c|c|}
\hline 
$\sqrt{s}$ & $400$  GeV & $500$  GeV  
           & $800$  GeV & $1000$ GeV 
           & $1500$ GeV & $4000$ GeV \\ \hline \hline
$R^{Full}_{0000}$ & 	$ -0.001947$&$-0.002985$&$ 0.0007576$&$ 0.003301$&$ 0.007354$&$ 0.01248$ \\ \hline
$R^{Full}_{2000}$ & 	$  0.03289 $&$ 0.02132 $&$-0.003048 $&$-0.01192 $&$-0.02393 $&$-0.03787$ \\ \hline
$R^{Full}_{1000}$ & 	$  0.02593 $&$ 0.01862 $&$ 0.003139 $&$-0.002419$&$-0.009864$&$-0.01836$ \\ \hline
$R^{Full}_{0010}$ & 	$  0.02638 $&$ 0.02133 $&$ 0.01228  $&$ 0.009409$&$ 0.005731$&$0.001570$ \\ \hline
$R^{Full}_{2010}$ &	$ 0.02546 $&$ 0.01550 $&$-0.008382 $ & $ -0.01782$ & $ -0.03122$&$ -0.04826$ \\ \hline
$R^{Full}_{1010}$ &	$0.002886 $&$ 0.002041$&$ -0.007438$&$ -0.01219 $&$ -0.01955$&$ -0.02989 $ \\ \hline
$R^{Full}_{0101}$ &	$-0.006056$&$ -0.01714$&$ -0.04161$&$ -0.05439$&$ -0.08003 $&$ -0.1559 $ \\ \hline
$R^{Full}_{1101}$ &	$ 0.02154$&$ 0.005427$&$ -0.02984$&$-0.04566$&$ -0.07406$&$ -0.1507$ \\ \hline
\end{tabular}
\end{center}
\clearpage
\item{ Soft Gluon App. }

We define the $\omega_{max}$ to satisfy the following relation.

\begin{eqnarray}
& ~ &
\int_{-1}^{1} d \cos \theta 
\sum_{s = \uparrow, \downarrow } 
\frac{ d \sigma_{T,R} }
     { d \cos \theta }
     (e^{-}_{L/R} e^{+} \rightarrow 
      t_{s} X (\bar{t}~{\rm or}~\bar{t}g)) \nonumber \\
& = &
\int_{-1}^{1} d \cos \theta 
\sum_{s = \uparrow, \downarrow } 
\frac{ d \sigma_{SGA,R} }
     { d \cos \theta }
     (e^{-}_{L/R} e^{+} \rightarrow t_{s} \bar{t})
\nonumber 
\end{eqnarray}
\begin{center}
\begin{tabular}{|c||c|c|c|c|c|c|}
\hline 
	              & $400$ GeV  & $500$ GeV  & $800$ GeV  
                      & $1000$ GeV & $1500$ GeV & $4000$ GeV \\ \hline
\hline
$\omega_{max}$ & $ 13.55 $ & $ 38.84 $ & $ 107.2 $ 
               & $ 150.2 $ & $ 255.2 $ & $ 776.1 $ \\ \hline	
$B_{0000}$ &	$0.6600$ & $1.719$ & $1.717$ & $1.696$ & $1.668$ & $1.644$ \\ \hline
$B_{2000}$ &	$0.1214$ & $0.6906$ & $1.259$ & $1.397$ & $1.535$ & $1.625$ \\ \hline
$B_{1000}$ &	$0.4522$ & $1.500$ & $2.152$ & $2.289$ & $2.417$ & $2.496$ \\ \hline
$B_{0010}$ &	$0.2261$ & $0.7498$ & $1.076$ & $1.144$ & $1.209$ & $1.248$ \\ \hline
$B_{2010}$ &	$0.2261$ & $0.7498$ & $1.076$ & $1.144$ & $1.209$ & $1.248$ \\ \hline
$B_{1010}$ &	$0.7814$ & $2.409$ & $2.976$ & $3.094$ & $3.203$ & $3.269$ \\ \hline
$B_{0101}$ &	$0.6181$ & $1.486$ & $1.074$ & $0.879$ & $0.5964$ & $0.2235$ \\ \hline
$B_{1101}$ &	$0.1970$ & $0.5189$ & $0.4596$ & $0.3891$ & $0.2718$ & $0.1037$ \\ \hline
\hline		
$B^{1}_{0000}$ &	$0.1433$ & $0.1907$ & $0.08886$ & $0.07067$ & $0.05405$ & $0.04117$ \\ \hline
$B^{1}_{2000}$ &	$0.02964$ & $0.09457$ & $0.08541$ & $0.07486$ & $0.05962$ & $0.04272$ \\ \hline
$B^{1}_{1000}$ &	$0.1078$ & $0.2002$ & $0.1419$ & $0.1195$ & $0.09217$ & $0.06527$ \\ \hline
$B^{1}_{0010}$ &	$0.05388$ & $0.1001$ & $0.07096$ & $0.05975$ & $0.04609$ & $0.03264$ \\ \hline
$B^{1}_{2010}$ &	$0.05388$ & $0.1001$ & $0.07096$ & $0.05975$ & $0.04609$ & $0.03264$ \\ \hline
$B^{1}_{1010}$ &	$0.1729$ & $0.2853$ & $0.1743$ & $0.1455$ & $0.1137$ & $0.08389$ \\ \hline
$B^{1}_{0101}$ &	$0.1325$ & $0.1546$ & $0.03437$ & $0.01420$ & $-0.001372$ & $-0.005843$ \\ \hline
$B^{1}_{1101}$ &	$0.04635$ & $0.06415$ & $0.01984$ & $0.009433$ & $0.0005521$ & $-0.002624$ \\ \hline
\end{tabular}
\end{center}
\clearpage
\begin{center}
\begin{tabular}{|c||c|c|c|c|c|c|}
\hline 
	              & $400$ GeV  & $500$ GeV  & $800$ GeV  
                      & $1000$ GeV & $1500$ GeV & $4000$ GeV \\ \hline
\hline
$S^{2}_{0000}$ &	$-0.001319$ & $-0.0103680$ & $-0.009239$ & $-0.007183$ & $-0.002033$ & $-0.0003753$ \\ \hline
$S^{2}_{2000}$ &	$0.003957$ & $0.01615$ & $0.01473$ & $0.01151$ & $0.008378$ & $0.001691$ \\ \hline
$S^{2}_{1000}$ &	$0.01137$ & $0.02918$ & $0.02083$ & $0.01558$ & $0.01141$ & $0.002253$ \\ \hline
$S^{2}_{0010}$ &	$0.005683$ & $0.01459$ & $0.01042$ & $0.007792$ & $0.005703$ & $0.001126$ \\ \hline
$S^{2}_{2010}$ &	$0.005683$ & $0.01459$ & $0.01042$ & $0.007792$ & $0.005703$ & $0.001126$ \\ \hline
$S^{2}_{1010}$ &	$0.002638$ & $0.005783$ & $0.005487$ & $0.004327$ & $0.006345$ & $0.001315$ \\ \hline
$S^{2}_{0101}$ &	$-0.003445$ & $-0.02060$ & $-0.02827$ & $-0.02725$ & $-0.02214$ & $-0.011790$ \\ \hline
$S^{2}_{1101}$ &	$0.004170$ & $0.004308$ & $-0.006649$ & $-0.008751$ & $-0.008870$ & $-0.005381$ \\ \hline
\hline		
$R^{SGA}_{0000}$ &	$-0.001994$ & $-0.005996$ & $-0.005351$ & $-0.004217$ & $-0.001217$ & $-0.0002282$ \\ \hline
$R^{SGA}_{2000}$ &	$0.03369$ & $0.02395$ & $0.01183$ & $0.008305$ & $0.005490$ & $0.001041$ \\ \hline
$R^{SGA}_{1000}$ &	$0.02578$ & $0.01984$ & $0.009774$ & $0.006855$ & $0.004741$ & $0.0009033$ \\ \hline
$R^{SGA}_{0010}$ &	$0.02578$ & $0.01984$ & $0.009774$ & $0.006855$ & $0.004741$ & $0.0009033$ \\ \hline
$R^{SGA}_{2010}$ &	$0.02578$ & $0.01984$ & $0.009774$ & $0.006855$ & $0.004741$ & $0.0009033$ \\ \hline
$R^{SGA}_{1010}$ &	$0.003387$ & $0.002406$ & $0.001847$ & $0.001401$ & $0.001985$ & $0.0004025$ \\ \hline
$R^{SGA}_{0101}$ &	$-0.005543$ & $-0.01367$ & $-0.02564$ & $-0.03005$ & $-0.03579$ & $-0.05010$ \\ \hline
$R^{SGA}_{1101}$ &	$0.02162$ & $0.008372$ & $-0.01426$ & $-0.02200$ & $-0.03161$ & $-0.04932$ \\ \hline
\end{tabular}
\end{center}
%\clearpage

\item{SGA I}

\begin{center}
\begin{tabular}{|c||c|c|c|c|c|c|}
\hline 
$\sqrt{s}$     &  $400$ GeV & $500$ GeV  & $800$ GeV 
               & $1000$ GeV & $1500$ GeV & $4000$ GeV\\ \hline
\hline
$\omega_{max}$ &  $10$ GeV  &  $30$ GeV  & $90$ GeV  
               &  $130$ GeV &  $230$ GeV & $730$\\ \hline  
$B_{0000}$     & $ 0.6554 $ & $ 0.7741 $ & $ 0.8080 $ 
               & $ 0.8077 $ & $ 0.8057 $ & $ 0.8031 $ \\ \hline 
$B_{2000}$     & $ 0.1206 $ & $ 0.3350 $ & $ 0.6106 $ 
               & $ 0.6787 $ & $ 0.7476 $ & $ 0.7950 $ \\ \hline 
$B_{1000}$     & $ 0.4492 $ & $ 0.8384 $ & $ 1.2035 $ 
               & $ 1.282  $ & $ 1.358  $ & $ 1.4086 $\\ \hline 
$B_{0010}$     & $ 0.2246 $ & $ 0.4192 $ & $ 0.6017 $ 
               & $ 0.6409 $ & $ 0.6791 $ & $ 0.7043 $\\ \hline 
$B_{2010}$     & $ 0.2246 $ & $ 0.4192 $ & $ 0.6017 $ 
               & $ 0.6409 $ & $ 0.6791 $ & $ 0.7043 $\\ \hline 
$B_{1010}$     & $ 0.7760 $ & $ 1.109  $ & $ 1.419 $ 
               & $ 1.486  $ & $ 1.553  $ & $ 1.598 $\\ \hline 
$B_{0101}$     & $ 0.6138 $ & $ 0.6349 $ & $ 0.4629 $ 
               & $ 0.3802 $ & $ 0.2592 $ & $ 0.09768$\\ \hline 
$B_{1101}$     & $ 0.1957 $ & $ 0.2901 $ & $ 0.2568 $ 
               & $ 0.2177 $ & $ 0.1526 $ & $ 0.05847$\\ \hline 
\hline
$B^{1}_{0000}$     & $ 0.1388   $ & $ 0.07592  $ & $ 0.02394  $ 
                   & $ 0.01493  $ & $ 0.007892 $ & $ 0.003678 $\\ \hline 
$B^{1}_{2000}$     & $ 0.02883  $ & $ 0.04174  $ & $ 0.02772  $ 
                   & $ 0.02035  $ & $ 0.01190  $ & $ 0.004559 $\\ \hline 
$B^{1}_{1000}$     & $ 0.1047   $ & $ 0.09992  $ & $ 0.05103  $ 
                   & $ 0.03569  $ & $ 0.02010  $ & $ 0.007784 $\\ \hline 
$B^{1}_{0010}$     & $ 0.05237  $ & $ 0.04996  $ & $ 0.02552  $ 
                   & $ 0.01784  $ & $ 0.01005  $ & $ 0.003892 $\\ \hline 
$B^{1}_{2010}$     & $ 0.05237  $ & $ 0.04996  $ & $ 0.02552  $ 
                   & $ 0.01784  $ & $ 0.01005  $ & $ 0.003892 $\\ \hline 
$B^{1}_{1010}$     & $ 0.1676   $ & $ 0.1177   $ & $ 0.05166  $ 
                   & $ 0.03528  $ & $ 0.01979  $ & $ 0.008237 $\\ \hline 
$B^{1}_{0101}$     & $ 0.1282   $ & $ 0.05634  $ & $ 0.003095 $ 
                   & $-0.003854 $ & $-0.007169 $ & $-0.004792 $ \\ \hline 
$B^{1}_{1101}$     & $ 0.04502  $ & $ 0.03162  $ & $ 0.004739 $ 
                   & $-0.0003479$ & $-0.003526 $ & $-0.002817 $ \\ \hline 
\end{tabular}
\end{center}
%\end{enumerate}

\clearpage

\section{Off-Diagonal Basis}

\[
\sigma 
\left(
     e^{-}_{L} e^{+} \rightarrow t_{\uparrow} + X 
\right) / \sigma_{T,L}
\]

\begin{center}
\begin{tabular}{|c||c|c|c|c|c|}
\hline 
$\sqrt{s}$  & $400$ GeV & $500$ GeV & $800$ GeV 
            & $1000$ GeV & $1500$ GeV \\ \hline
\hline
Tree        & $0.99876$ & $0.99255$ & $0.97350$ & $0.96514$ & $0.95345$ \\ \hline
$\cal{O}\mit (\alpha_{s})$       
            & $0.99850$ & $0.99093$ & $0.96815$  & $0.95833$ & $0.94480$ \\ \hline
SGA    
            & $0.99871$ & $0.99225$ & $0.97276$ & $0.96434$ & $0.95275$  \\ \hline
\end{tabular}
\end{center}

\[
\sigma 
\left(
     e^{-}_{L} e^{+} \rightarrow t_{\downarrow} + X 
\right) / \sigma_{T,L}
\]

\begin{center}
\begin{tabular}{|c||c|c|c|c|c|}
\hline 
$\sqrt{s}$  & $400$ GeV & $500$ GeV & $800$ GeV 
            & $1000$ GeV & $1500$ GeV \\ \hline
\hline
Tree   & $0.00124$ & $0.00745$  & $0.02650$  & $0.03486$ & $0.04655$  \\ \hline
$\cal{O}\mit (\alpha_{s})$       
       & $0.00150$ & $0.00907$  & $0.03185$  & $0.04167$ & $0.05520$  \\ \hline
SGA    
       & $0.00129$ & $0.00775$  & $0.02724$  & $0.03566$ & $0.04725$  \\ \hline
\end{tabular}
\end{center}

\[
\sigma 
\left(
     e^{-}_{R} e^{+} \rightarrow t_{\downarrow} + X 
\right) / \sigma_{T,R}
\]

\begin{center}
\begin{tabular}{|c||c|c|c|c|c|}
\hline 
$\sqrt{s}$  & $400$ GeV & $500$ GeV & $800$ GeV 
            & $1000$ GeV & $1500$ GeV \\ \hline
\hline
Tree  & $0.99833$ & $0.99371$ & $0.98141$ & $0.97619$ & $0.96897$ \\ \hline
$\cal{O}\mit (\alpha_{s})$       
      & $0.99786$ & $0.99184$ & $0.97615$ & $0.96958$ & $0.96057$ \\ \hline
SGA 
      & $0.99809$ & $0.99318$ & $0.98065$ & $0.97545$ & $0.96837$ \\ \hline
\end{tabular}
\end{center}

\[
\sigma 
\left(
     e^{-}_{R} e^{+} \rightarrow t_{\uparrow} + X 
\right) / \sigma_{T,R}
\]

\begin{center}
\begin{tabular}{|c||c|c|c|c|c|}
\hline 
$\sqrt{s}$  & $400$ GeV & $500$ GeV & $800$ GeV 
            & $1000$ GeV & $1500$ GeV \\ \hline
\hline
Tree   & $0.00167$ & $0.00629$ & $0.01859$ & $0.02381$ & $0.03103$  \\ \hline
$\cal{O}\mit (\alpha_{s})$       
       & $0.00214$ & $0.00816$ & $0.02385$ & $0.03042$ & $0.03943$  \\ \hline
SGA    
       & $0.00191$ & $0.00682$ & $0.01935$ & $0.02455$ & $0.03163$  \\ \hline
\end{tabular}
\end{center}

%----------------------------------------------------------
\section{Helicity Basis}
\[
\sigma \left(
     e^{-}_{L} e^{+} \rightarrow t_{L} + X 
       \right) / \sigma_{T,L}
\]

\begin{center}
\begin{tabular}{|c||c|c|c|c|c|}
\hline 
$\sqrt{s}$  & $400$ GeV & $500$ GeV & $800$ GeV 
            & $1000$ GeV & $1500$ GeV \\ \hline
\hline
Tree   & $0.66359$ & $0.75084$ & $0.83177$ & $0.84997$ & $0.86797$ \\ \hline
$\cal{O}\mit (\alpha_{s})$       
       & $0.66814$ & $0.75658$ & $0.83500$ & $0.85160$ & $0.86712$ \\ \hline
SGA    
       & $0.66809$ & $0.75657$ & $0.83579$ & $0.85299$ & $0.86962$ \\ \hline
\end{tabular}
\end{center}

\[
 \sigma \left(
     e^{-}_{L} e^{+} \rightarrow t_{R} + X 
       \right) / \sigma_{T,L}
\]

\begin{center}
\begin{tabular}{|c||c|c|c|c|c|}
\hline 
$\sqrt{s}$  & $400$ GeV & $500$ GeV & $800$ GeV 
            & $1000$ GeV & $1500$ GeV \\ \hline
\hline
Tree  & $0.33641$ & $0.24916$ & $0.16823$ & $0.15003$ & $0.13203$ \\ \hline
$\cal{O}\mit (\alpha_{s})$       
      & $0.33187$ & $0.24342$ & $0.16500$ & $0.14840$ & $0.13288$ \\ \hline
SGA    
      & $0.33191$ & $0.24343$ & $0.16421$ & $0.14701$ & $0.13038$  \\ \hline
\end{tabular}
\end{center}

\[
\sigma \left(
     e^{-}_{R} e^{+} \rightarrow t_{R} + X 
       \right) / \sigma_{T,R}
\]

\begin{center}
\begin{tabular}{|c||c|c|c|c|c|}
\hline 
$\sqrt{s}$  & $400$ GeV & $500$ GeV & $800$ GeV 
            & $1000$ GeV & $1500$ GeV \\ \hline
\hline
Tree   & $0.70979$ & $0.80926$ & $0.89267$ & $0.91039$ & $0.92758$ \\ \hline
$\cal{O}\mit (\alpha_{s})$       
       & $0.71528$ & $0.81541$ & $0.89546$ & $0.91145$ & $0.92608$ \\ \hline
SGA    
       & $0.71524$ & $0.81550$ & $0.89660$ & $0.91327$ & $0.92913$  \\ \hline
\end{tabular}
\end{center}

\[
\sigma \left(
     e^{-}_{R} e^{+} \rightarrow t_{L} + X 
       \right) / \sigma_{T,R}
\]

\begin{center}
\begin{tabular}{|c||c|c|c|c|c|}
\hline 
$\sqrt{s}$  & $400$ GeV & $500$ GeV & $800$ GeV 
            & $1000$ GeV & $1500$ GeV \\ \hline
\hline
Tree  & $0.29021$ & $0.19074$ & $0.10733$ & $0.08961$ & $0.072419$ \\ \hline
$\cal{O}\mit (\alpha_{s})$       
      & $0.28472$ & $0.18459$ & $0.10454$ & $0.08855$ & $0.073918$ \\ \hline
SGA   
      & $0.28476$ & $0.18450$ & $0.10340$ & $0.08673$ & $0.07087$  \\ \hline
\end{tabular}
\end{center}
\end{itemize}
%----------------------------------------------------
\section{Beamline Basis}

\[
\sigma 
\left(
     e^{-}_{L} e^{+} \rightarrow t_{\uparrow} + X 
\right) / \sigma_{T,L}
\]

\begin{center}
\begin{tabular}{|c||c|c|c|c|c|}
\hline 
$\sqrt{s}$  & $400$ GeV & $500$ GeV & $800$ GeV 
            & $1000$ GeV & $1500$ GeV \\ \hline
\hline
Tree   & $0.98811$ & $0.96769$ & $0.93097$ & $0.91822$ & $0.90218$ \\ \hline
$\cal{O}\mit (\alpha_{s})$       
       & $0.98851$ & $0.96784$ & $0.92922$ & $0.91565$ & $0.89853$ \\ \hline
SGA  
       & $0.98869$ & $0.96889$ & $0.93210$ & $0.91913$ & $0.90272$  \\ \hline
\end{tabular}
\end{center}

\[
\sigma 
\left(
     e^{-}_{L} e^{+} \rightarrow t_{\downarrow} + X 
\right) / \sigma_{T,L}
\]

\begin{center}
\begin{tabular}{|c||c|c|c|c|c|}
\hline 
$\sqrt{s}$  & $400$ GeV & $500$ GeV & $800$ GeV 
            & $1000$ GeV & $1500$ GeV \\ \hline
\hline
Tree   & $0.01189$ & $0.03231$ & $0.06903$ & $0.08178$ & $0.09782$ \\ \hline
$\cal{O}\mit (\alpha_{s})$       
       & $0.01149$ & $0.03216$ & $0.07078$ & $0.08435$ & $0.10147$ \\ \hline
SGA    
       & $0.01131$ & $0.03111$ & $0.06790$ & $0.08087$ & $0.09728$  \\ \hline
\end{tabular}
\end{center}

\[
\sigma 
\left(
     e^{-}_{R} e^{+} \rightarrow t_{\downarrow} + X 
\right) / \sigma_{T,R}
\]

\begin{center}
\begin{tabular}{|c||c|c|c|c|c|}
\hline 
$\sqrt{s}$  & $400$ GeV & $500$ GeV & $800$ GeV 
            & $1000$ GeV & $1500$ GeV \\ \hline
\hline
Tree   & $0.99436$ & $0.98432$ & $0.96581$ & $0.95934$ & $0.95119$ \\ \hline
$\cal{O}\mit (\alpha_{s})$       
       & $0.99470$ & $0.98437$ & $0.96400$ & $0.95666$ & $0.94727$ \\ \hline
SGA 
       & $0.99490$ & $0.98545$ & $0.96693$ & $0.96025$ & $0.95174$  \\ \hline
\end{tabular}
\end{center}

\[
\sigma 
\left(
     e^{-}_{R} e^{+} \rightarrow t_{\downarrow} + X 
\right) / \sigma_{T,R}
\]

\begin{center}
\begin{tabular}{|c||c|c|c|c|c|}
\hline 
$\sqrt{s}$  & $400$ GeV & $500$ GeV & $800$ GeV 
            & $1000$ GeV & $1500$ GeV \\ \hline
\hline
Tree  & $0.00564$ & $0.01568$ & $0.03419$ & $0.04066$ & $0.04881$ \\ \hline
$\cal{O}\mit (\alpha_{s})$       
      & $0.00530$ & $0.01563$ & $0.03600$ & $0.04334$ & $0.05273$ \\ \hline
SGA 
      & $0.00510$ & $0.01455$ & $0.03307$ & $0.03975$ & $0.04826$  \\ \hline
\end{tabular}
\end{center}
%----------------------- Text ---------------------------------
%\input ref.tex


\baselineskip 20pt

\begin{thebibliography}{99}

\bibitem{QCD}
     H. Fritzsh, M. Gell-Mann and H. Leutwyler, {\sl Phys. Lett.}
     {\bf B47} (1973) 365.
\bibitem{EW}
     S.L. Glashow,{\sl Nucl. Phys.} {\bf 22} (1961)579;
     S. Weinberg, {\sl Phys. Rev. Lett} {\bf 19} (1967) 1264;
     A. Salam, in: N. Svartholm (Ed.), {\it Elementary particle Theory}
     Almqvistand Wiksells, Stockholm, (1968).
\bibitem{HIGGS}
     P.W.Higgs, {\sl Phys. Rev. Lett.} {\bf 12} (1964) 132;
                {\sl Phys. Rev.} {\bf 145} (1966) 1156;
     F. Englert and R. Brout, {\sl Phys. Rev. Lett.} {\bf 13} (1964)
                 321;
     G.S. Guralnik, C.R. Hagen and T.W. Kibble, {\sl Phys. Rev. Lett.}
                {\bf 13} (1964) 585.
\bibitem{CDF}
     F. Abe et al., CDF Collaboration, {\sl Phys. Rev. Lett.}
     {\bf 74} (1995) 2626.
\bibitem{D0}
     S. Abachi et al., D0 Collaboration, {\sl Phys. Rev. Lett.}
     {\bf 74} (1995) 2632.
\bibitem{PDG}
     C. Caso et al, {\sl The European Physical Journal} 
     {\bf C3} (1998) 1.
\bibitem{SEE} 
        See, for example,\\ 
        W. Bernreuther, O. Nachtmann, P. Overmann and T. Schr\" oder, 
        {\sl Nucl. Phys.} {\bf B388} (1992) 53-80;\\ 
        erratum {\sl Nucl. Phys.} {\bf B406} (1993) 516;\\
        S. Y. Choi and K. Hagiwara, 
        {\sl Phys. Lett.} {\bf B359} (1995) 369-374;\\
        B. Grz\c adkowski and Z. Hioki, 
        {\sl Nucl. Phys.} {\bf B484} (1997) 17-32, 
        and references therein.
\bibitem{k}
     J.~H.~K\"uhn, {\sl Nucl. Phys.} {\bf B237} (1984) 77.
\bibitem{bdkkz}
     I. Bigi, Y. Dokshizer, V. Khoze, J. K\"uhn and P. Zerwas,
     {\sl Phys. Lett.} {\bf B181} (1986) 157.
\bibitem{Decays}
     M. Jezabek, J. H. K\"uhn and T. Teubner, {\sl Z. Phys.} {\bf C56}
     (1992) 653.\\
     M. Jezabek and J. H. K\"uhn, {\sl Phys. Lett.} {\bf B329}
     (1994) 317. 
\bibitem{bop}
     V.~Barger, J.~Ohnemus and R.J.N~Phillips, {\sl Int. J. Mod.
     Phys.} {\bf A4} (1989) 617.
\bibitem{kly}
     G.~L.~Kane, J.~Pumplin and W.~Repko, {\sl Phys. Rev. Lett.}
     {\bf 41} (1978) 1689.\\
     G.~L.~Kane, G.~A.~Ladinsky and C.~-P.~Yuan, {\sl Phys. Rev.}
     {\bf D45} (1992) 124.
\bibitem{eecollider}
     M.~Anselmino, P.~Kroll and B.~Pire, {\sl Phys. Lett.} {\bf B167}
     (1986) 113.\\
     D.~Atwood and A.~Soni, {\sl Phys. Rev.} {\bf D45} (1992) 2405.\\
     C.~- P.~Yuan, {\sl Phys. Rev.} {\bf D45} (1992) 318.\\
     G.~A.~Ladinsky, {\sl Phys. Rev.} {\bf D46} (1992) 3789.\\
     W.~Bernreuther, O.~Nachtmann, P.~Overmann and T.~Schr\"oder,
     {\sl Nucl. Phys.} {\bf B388} (1992) 53; erratum {\bf B406} (1993)
     516.\\
     M.~E.~Peskin, in {\sl Physics and Experiments at Linear
     Colliders}, R.~Orava, P.~Eeorla and M.~Nordberg, eds. (World
     Scientific, 1992).\\
     T. Arens and L.~M.~Seghal, {\sl Nucl. Phys.} {\bf B393} (1993) 46.\\
     F.~Cuypers and S.~D.~Rindani, {\sl Phys. Lett.} {\bf B343} (1995) 
     333.\\
     P.~Poulose and S.~D.~Rindani, {\sl Phys. Lett.} {\bf B349} (1995)
     379.
\bibitem{hcollider}
     C. R. Schmidt and M. E. Peskin, {\sl Phys. Rev. Lett.} {\bf 69}
     (1992) 410.\\
     D. Atwood, A. Aeppli and A. Soni, {\sl Phys. Rev. Lett.} {\bf 69}
     (1992) 2754.\\
     Y. Hara, {\sl Prog. Theo. Phys.} {\bf 86} (1991) 779.\\
     W. Bernreuther and A. Brandenburg, {\sl Phys. Rev.} {\bf D49}
     (1994) 4481.  
\bibitem{mp}
     G.~Mahlon and S.~Parke, {\sl Phys. Rev.} {\bf D53} (1996)
     4886, 
     {\sl Phys. Lett.} {\bf B411} (1997) 173.  
\bibitem{ps}
     S.~Parke and Y.~Shadmi, {\sl Phys. Lett.} {\bf B387} (1996) 
     199.
\bibitem{gnt}
     G.~Grunberg, Y.~J.~Ng and S.~-H.~H.~Tye, {\sl Phys. Rev.}
     {\bf D21} (1980) 62.\\
     J.~Jers\'ak, E.~Laermann and P.~Zerwas, {\sl Phys. Rev.}
     {\bf D25} (1982) 1218.
\bibitem{gkl}
     S.~Groote, J.~G.~K\"orner and J.~A.~Leyva, hep-ph/9801255. 
\bibitem{s}
     C.~R.~Schmidt, {\sl Phys. Rev.} {\bf D54} (1996) 3250.
\bibitem{tung}
     S.~D.~Rindani and M.~M.~Tung, {\sl Phys. Lett.} {\bf B424} (1998) 125.\\
     S.~Groote, J.~G.~K\"orner and M.~M.~Tung, {\sl Z. Phys.}
     {\bf C70} (1996) 281.\\
     M.~M.~Tung, J. Bernabeu and J.~Penarrocha, {\sl Nucl. Phys.}
     {\bf B470} (1996) 41.\\
     M.~M.~Tung, {\sl Phys. Rev.} {\bf D52} (1995) 1353.\\
     J.~G.~K\"orner, A.~Pilaftsis and M.~M.~Tung, {\sl Z. Phys.}
     {\bf C63} (1994) 575.
\bibitem{TRS}
     W. Bernreuther, J.P. Ma and T. Schr\"oder, {\sl Phys. Lett.}
     {\bf B297} (1992) 318-326,
     S. Groorte and J.G. K\"orner, {\sl Z. Phys.} {\bf C72} (1996) 199.
\bibitem{THRES}
     R. Harlander, M. Jezabek, J.H. Kuhn and T. Teubner, 
     {\sl Phys. Lett.} {\bf B346} (1995) 137-142,
     R. Harlander, M. Jezabek, J.H. K\"uhn and M. Peter, {\sl Z. Phys.}
     {\bf C73} (1997) 477-494. 
\bibitem{BJ}
     J.D. Bjorken and M.C. Chen, {\sl Phys. Rev.}
     {\bf 154} (1966) 1335.
\bibitem{CC}
     P. de Causmaecker, R. Gastmans, W. Troost, and T.T. Wu,
     {\sl Phys. Lett.} {\bf 105B} (1981) 215;
     {\sl Nucl. Phys.} {\bf B206} (1982) 53; 
     F. A. Berends, R. Kleiss, P. De Causmaecker, R. Gastmans,
     W. Troost, and T. T. Wu, {\sl Nucl. Phys.} {\bf B206} (1982) 61;
     {\it ibid.} {\bf B239} (1984) 382; {\it ibid.} {\bf B239} 395;
     {\it ibid.} {\bf B264} (1986) 243; {\it ibid.} {\bf B266} 265.
\bibitem{XZC}
     Z. Xu, D.-H. Zhang and L. Chang, {\sl Nucl. Phys.} {\bf B291}
     (1987) 392.
\bibitem{SUSY}
     S. J. Parke and T. R. Taylor, {\sl Phys. Lett.} {\bf 157B} (1985)
     81;{\sl Nucl. Phys.} {\bf B269} (1986) 410.
\bibitem{PARKE1}
     M.L. Mangano and S.J. Parke, {\sl Phys. Rep.} {\bf 200}
     (1991) 301.
\bibitem{MASSIVE}
     F.A. Berends, P.H. Daverveldt and R. Kleiss,
     {\sl Nucl. Phys.} {\bf B253} (1985) 441;
     R. Kleiss and W.J. Stirling, {\sl Nucl. Phys.}
     {\bf B262} (1985) 235; C. Mana and M. Martinez, 
     {\sl Nucl. Phys.} {\bf B287} (1987) 601. 
\bibitem{HABER}
     Howard E. Haber, {\sl preprint} SCIPP-93-49, NSF-ITP-94-30 (1994).
\bibitem{BJ&Drell}
     J. Bjorken and S. Drell, {\sl Relativistic Quantum Mechanics},
     (McGraw-Hill Book Company, New York, 1964).
\bibitem{KOD-PARKE}
     J. Kodaira and S. Parke, in private communication.
\bibitem{PARKE2}
     G. Mahlon and S. Parke, {\sl Phys. Rev.} {\bf D58} (1998) 054015.
     (1991) 301.
\bibitem{TP}
     J.H. K\"uhn, A. Reiter and P.M. Zerwas, {\sl Nucl. Phys} {\bf B272}
     (1986) 560,
     M. Anselmino, P. Kroll adn B. Pire, {\sl Phys. Lett} {\bf B167} (1986)
     113.
\bibitem{pk1}
     S.~Parke,
     {\sl preprint} hep-ph/9802279, FERMILAB-Conf-98/056-T (1998).
\bibitem{BIGI} 
        I. Bigi, Y. Dokshitzer, V. Khoze, J. K\"uhn 
        and P. Zerwas, {\sl Phys. Lett.} {\bf B181} (1986) 157.
%
\bibitem{JEZA} 
        M. Je\. zabek and J. H. K\" uhn,
        {\sl Phys. Lett.} {\bf B329} (1994) 317-324, 
        and references therein.
%
\bibitem{KUHN}
        J. H. K\"uhn,
        {\sl Nucl. Phys.} {\bf B237} (1984) 77-85.
%
\bibitem{GREG} 
        G. Mahlon and S. Parke, 
        {\sl Phys. Rev.} {\bf D53} (1996) 4886-4896; 
        {\sl Phys. Lett.} {\bf B411} (1997) 173-179.
\bibitem{pk2}
     S.~Parke, 
     {\sl preprint} hep-ph/9807573, FERMILAB-Conf-98/212-T (1998).
\bibitem{HKN} 
         M. Hori, Y. Kiyo and T. Nasuno, {\sl Phys. Rev.} 
         {\bf D58} (1998) 014005.
\bibitem{KNP1}
        J. Kodaira, T. Nasuno and S. Parke, {\sl Physical Review} 
        {\bf D 59} (1999) 014023.
\bibitem{KNP2}
        M. Hori, Y. Kiyo, J. Kodaira, T. Nasuno and S. Parke
        {\sl preprint} hep-ph/9801370, HUPD-9801, FERMILAB-Conf-98/022-T 
        (1998).
\bibitem{bprime} 
        G. Bhattacharyya and R.N. Mohapatra,
        {\sl Phys. Rev.} {\bf D54} (1996) 4204;
        J.F. Gunion, D.W. McKay and H. Pois, 
        {\sl Phys. Rev.} {\bf D53} (1996) 1616;
        M. Carena, H.E. Habar and E. Nardi, 
        {\sl Phys. Lett.} {\bf B355} (1995) 199.
\bibitem{hs}
     R.~Harlander and M.~Steinhauser, {\sl Eur. Phys. J.} {\bf C2}
	(1998) 151.
\end{thebibliography}
\end{document}